%% file: CDR-eSPS.tex
\pdfoutput=1 
\documentclass[11pt,a4paper, twoside]{scrartcl}

\usepackage{CLICdp}

\usepackage{CLICdp_definitions}
\usepackage{lipsum} 
\usepackage{mathptmx}
\usepackage{afterpage}
\usepackage{graphicx}
\usepackage{tabularx}
    \newcolumntype{L}{>{\raggedright\arraybackslash}X}
  
\numberwithin{table}{section}
\numberwithin{figure}{section}

\fancypagestyle{mainmatter}{}
\renewcommand{\sectionmark}[1]{\markright{#1}}
\renewcommand{\subsectionmark}[1]{}
\begin{document}

\title{A primary electron beam facility at CERN --- eSPS \protect\\ Conceptual design report}
\pagenumbering{roman}
\setcounter{page}{1}

\thispagestyle{empty}
\setlength{\unitlength}{1mm}
\begin{picture}(0.001,0.001)
\put(-8,8){CERN Yellow Reports: Monographs}
\put(110,8){CERN-2020-008}

\put(-8,-40){\Huge \bfseries A primary electron beam facility} 
\put(-8,-50){\Huge \bfseries at CERN --- eSPS}
\put(-8,-60){\Large \bfseries Conceptual design report}

\put(-5,-170){\includegraphics[width=16cm,trim={0.0cm 0.0cm 0.0cm 0.5cm},clip]
{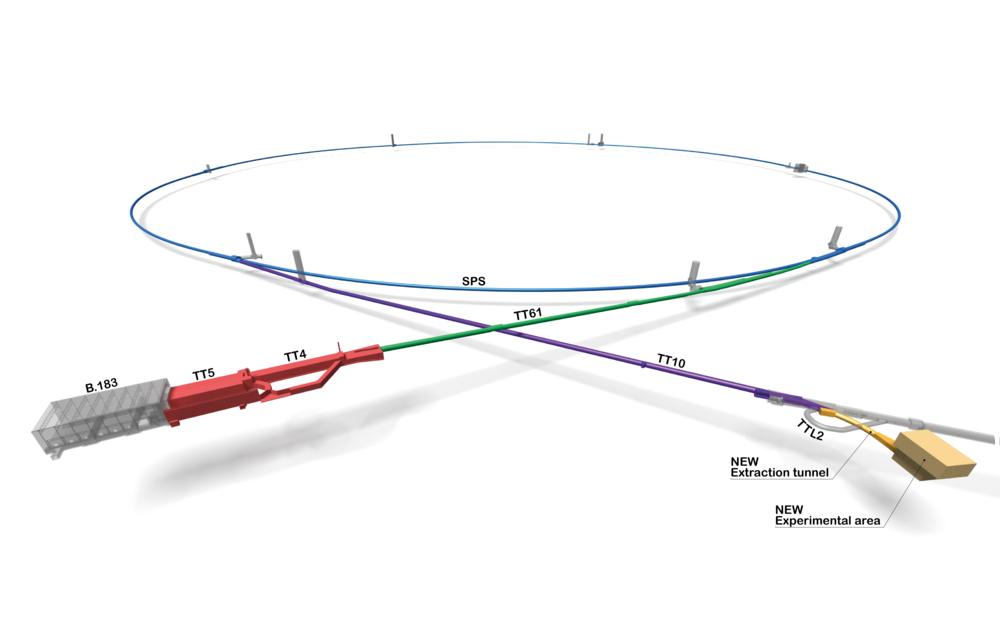}}

\put(3,-190){\Large Corresponding editors:}
\put(8,-198){\Large Torsten \AA kesson, Lund University}
\put(8,-206){\Large Steinar Stapnes, CERN}

\put(65,-250){\includegraphics{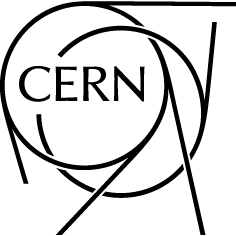}}

\end{picture}
\newpage

\thispagestyle{empty}
\mbox{}
\vfill

\begin{flushleft}
CERN Yellow Reports: Monographs\\
Published by CERN, CH-1211 Geneva 23, Switzerland\\[3mm]

\begin{tabular}{@{}l@{~}l}
 ISBN & 978-92-9083-584-4 (paperback) \\
 ISBN & 978-92-9083-585-1 (PDF) \\
 ISSN & 2519-8068 (Print)\\ 
 ISSN & 2519-8076 (Online)\\ 
 DOI & \url{https://doi.org/10.23731/CYRM-2020-008}\\
\end{tabular}\\[5mm]

Copyright \copyright{} CERN, 2020\\[1mm]
\raisebox{-1mm}{\includegraphics[height=12pt]{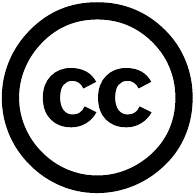}}
Creative Commons Attribution 4.0\\[5mm]

This volume should be cited as:\\[1mm]
A primary electron beam facility at CERN --- eSPS: Conceptual design report,\\ Torsten \AA kesson, Steinar Stapnes (eds.)\\ 
CERN Yellow Reports: Monographs, CERN-2020-008 (CERN, Geneva, 2020)\\ 
\url{https://doi.org/10.23731/CYRM-2020-008}.\\[3mm]

Editorial team:
\href{mailto:markus.aicheler@cern.ch}{Markus.Aicheler@cern.ch},
\href{mailto:torsten.akesson@cern.ch}{Torsten.Akesson@cern.ch},
\href{mailto:yann.dutheil@cern.ch}{Yann.Dutheil@cern.ch},
\href{mailto:luke.dyks@cern.ch}{Luke.Dyks@cern.ch},
\href{mailto:jonathan.gall@cern.ch}{Jonathan.Gall@cern.ch},
\href{mailto:Steinar.Stapnes@cern.ch}{Steinar.Stapnes@cern.ch}.\\[1mm]
Accepted in December 2020 by the \href{http://library.cern/about_us/editorial_board}{CERN Reports Editorial Board} 
(contact \href{mailto:Carlos.Lourenco@cern.ch}{Carlos.Lourenco@cern.ch}).\\[1mm]
Published by the CERN Scientific Information Service (contact \href{mailto:Jens.Vigen@cern.ch}{Jens.Vigen@cern.ch}).\\[1mm]
Indexed in the \href{https://cds.cern.ch/collection/CERN\%20Yellow\%20Reports?ln=en}{CERN Document Server} and in \href{https://inspirehep.net/}{INSPIRE}.\\[1mm]
Published Open Access to permit its wide dissemination, as knowledge transfer is an integral part of the mission of CERN.
\end{flushleft}





\addauthor{M.~Aicheler}{\institute{6}\hcomma\institute{1}}
\addauthor{T.~\AA kesson}{\institute{2}}
\addauthor{F.~Antoniou}{\institute{1}}
\addauthor{A.~Arnalich}{\institute{1}}
\addauthor{P.~A.~Arrutia~Sota}{\institute{1}\hcomma\institute{3}}
\addauthor{P.~Bettencourt~Moniz~Cabral}{\institute{1}}
\addauthor{D.~Bozzini}{\institute{1}}
\addauthor{M.~Brugger}{\institute{1}}
\addauthor{O.~Brunner}{\institute{1}}
\addauthor{P.~N.~Burrows}{\institute{4}}
\addauthor{R.~Calaga}{\institute{1}}
\addauthor{M.~J.~Capstick}{\institute{1}}
\addauthor{R.~Corsini}{\institute{1}}
\addauthor{S.~Doebert}{\institute{1}}
\addauthor{L.~A.~Dougherty}{\institute{1}}
\addauthor{Y.~Dutheil}{\institute{1}}
\addauthor{L.~A.~Dyks}{\institute{4}\hcomma\institute{1}}
\addauthor{O.~Etisken}{\institute{1}\hcomma\institute{5}}
\addauthor{L.~Evans}{\institute{1}}
\addauthor{A.~Farricker}{\institute{1}}
\addauthor{R.~Fernandez~Ortega}{\institute{1}}
\addauthor{M.~A.~Fraser}{\institute{1}}
\addauthor{J.~Gall}{\institute{1}}
\addauthor{S.~J.~Gessner}{\institute{1}}
\addauthor{B.~Goddard}{\institute{1}}
\addauthor{J-L.~Grenard}{\institute{1}}
\addauthor{A.~Grudiev}{\institute{1}}
\addauthor{E.~Gschwendtner}{\institute{1}}
\addauthor{J.~Gulley}{\institute{1}}
\addauthor{L.~Jensen}{\institute{1}}
\addauthor{R.~Jones}{\institute{1}}
\addauthor{M.~Lamont}{\institute{1}}
\addauthor{A.~Latina}{\institute{1}}
\addauthor{T.~Lef\`evre}{\institute{1}}
\addauthor{R.~Lopes}{\institute{1}}
\addauthor{H.~Mainaud~Durand}{\institute{1}}
\addauthor{S.~Marsh}{\institute{1}}
\addauthor{G.~Mcmonagle}{\institute{1}}
\addauthor{E.~Montesinos}{\institute{1}}
\addauthor{R.~Morton}{\institute{1}}
\addauthor{P.~Muggli}{\institute{1}}
\addauthor{A.~Navascues~Cornago}{\institute{1}}
\addauthor{M.~Nonis}{\institute{1}}
\addauthor{J.~A.~Osborne}{\institute{1}}
\addauthor{Y.~Papaphilippou}{\institute{1}}
\addauthor{A.~M.~Rossi}{\institute{1}}
\addauthor{C.~Rossi}{\institute{1}}
\addauthor{I.~Ruehl}{\institute{1}}
\addauthor{S.~Schadegg}{\institute{1}}
\addauthor{E.~Shaposhnikova}{\institute{1}}
\addauthor{D.~Schulte}{\institute{1}}
\addauthor{S.~Stapnes}{\institute{1}}
\addauthor{M.~Widorski}{\institute{1}}
\addauthor{O.~E.~Williams}{\institute{1}}
\addauthor{W.~Wuensch}{\institute{1}}


\addinstitute{1}{CERN, Switzerland}
\addinstitute{2}{Lund University, Sweden}
\addinstitute{3}{John Adams Institute, Royal Holloway, University of London, UK}
\addinstitute{4}{John Adams Institute, University of Oxford, UK}
\addinstitute{5}{Ankara University, Turkey}
\addinstitute{6}{University of Helsinki, Finland}


\abstract{The design of a primary electron beam facility at CERN is described. The study has been carried out within the framework of the wider Physics Beyond Colliders study. It re-enables the Super Proton Synchrotron (SPS) as an electron accelerator, and leverages the development invested in Compact Linear Collider (CLIC) technology for its injector and as an accelerator research and development infrastructure.
The facility would be relevant for several of the key priorities in the 2020 update of the European Strategy for Particle Physics, such as an electron-positron Higgs factory, accelerator R\&D, dark sector physics, and neutrino physics. In addition, it could serve experiments in nuclear physics. 
The electron beam delivered by this facility would provide access to light dark matter production significantly beyond the targets predicted by a thermal dark matter origin, and for natures of dark matter particles that are not accessible by direct detection experiments. It would also enable electro-nuclear measurements crucial for precise modelling the energy dependence of neutrino-nucleus interactions, which is needed to precisely measure neutrino oscillations as a function of energy.
The implementation of the facility is the natural next step in the development of X-band high-gradient acceleration technology, a key technology for compact and cost-effective electron/positron linacs.  It would also become the only facility with multi-GeV drive bunches and truly independent electron witness bunches for plasma wakefield acceleration.  A second phase capable to deliver positron witness bunches would make it a complete facility for plasma wakefield collider studies. 
The facility would be used for the development and studies of a large number of components and phenomena for a future electron-positron Higgs and electroweak factory as the first stage of a next circular collider at CERN, and its cavities in the SPS would be the same type as foreseen for such a future collider. The operation of the SPS with electrons would train a new generation of CERN staff on circular electron accelerators.
The facility could start operation in about five years, and would operate in parallel and without interference with Run 4 of the LHC.}



\graphicspath{ {./logos/} }



\nocolourlinks





\titlepage

\tableofcontents

\cleardoublepage
\input{include/01-Introduction/introduction.tex}

\cleardoublepage
\input{include/02-Physics_Goals/PhysicsGoals.tex}

\cleardoublepage
\input{include/03-LINAC/OverviewLINAC.tex}
\input{include/03-LINAC/BeamDynamics.tex}
\input{include/03-LINAC/GunInjector.tex}
\input{include/03-LINAC/X-BandLINAC.tex}
\input{include/03-LINAC/LINACIntegration.tex}

\input{include/03-LINAC/BeamInstrumentation.tex}

\input{include/03-LINAC/LINACPositrons.tex}
\input{include/03-LINAC/LINACPlasma.tex}

\cleardoublepage
\input{include/04-SPS/OverviewSPS.tex}
\input{include/04-SPS/Linac-SPS.tex}
\input{include/04-SPS/BI_TL.tex}

\input{include/04-SPS/SPS_RF.tex}
\input{include/04-SPS/SPS_Acceleration.tex}

\input{include/04-SPS/BI_SPS.tex}

\input{include/04-SPS/Extraction.tex}
\input{include/04-SPS/BI_Extr.tex}

\cleardoublepage
\input{include/05-ICE/OverviewICE.tex}
\input{include/05-ICE/GeneralConsiderations.tex}
\input{include/05-ICE/InjectionTT4_TT5.tex}
\input{include/05-ICE/TransferTT61.tex}
\input{include/05-ICE/AccelerationSPS.tex}

\input{include/05-ICE/ExtractionTT10_TT2.tex}
\input{include/05-ICE/ExperimentalArea.tex}
\cleardoublepage
\input{include/06-Acc_RnD/OverviewAcc_RnD.tex}
\input{include/06-Acc_RnD/Linac_RnD.tex}
\input{include/06-Acc_RnD/Ring_RnD.tex}
\input{include/06-Acc_RnD/Plasma_RnD.tex}
\input{include/06-Acc_RnD/CLEARER.tex}
\input{include/06-Acc_RnD/Positron_RnD.tex}

\cleardoublepage
\input{include/07-Implementation/Implementation.tex}

\cleardoublepage
\input{include/08-Conclusion/Conclusion.tex}

\clearpage

\printbibliography[title=References]

\end{document}

%% file: include/01-Introduction/introduction.tex
\section{Introduction}
\pagenumbering{arabic}
\setcounter{page}{1}
\label{sec:Introduction}
Recent interest in light dark matter searches using GeV electrons has stimulated a new study of how such a beam could be provided at CERN~\cite{Akesson2019}. This study has been carried out within the framework of the wider Physics Beyond Colliders study. The basic requirement is a very long spill of low intensity electrons to a missing-energy/missing-momentum experiment~\cite{whitepaper, Torstenakesson2018} as described in Section~\ref{sec:LDM-at-eSPS}.

\begin{figure}[!hbt]
\begin{center}
   \includegraphics[width=16cm,trim={0.0cm 0.0cm 0.0cm 0.5cm},clip]{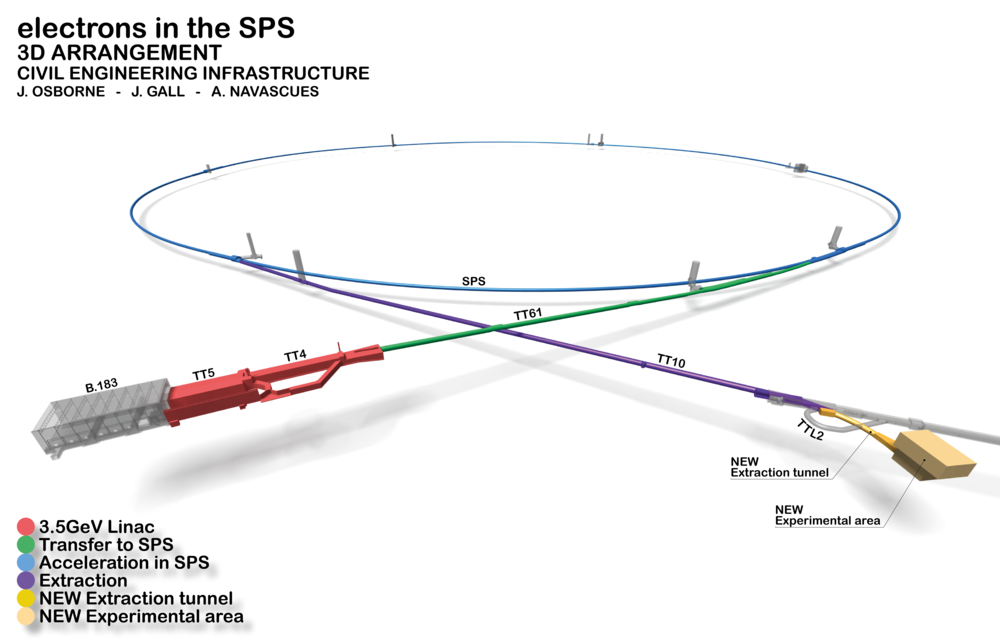}
   \caption{Schematic of the primary electron beam facility.} 
\label{fig:eSPS_layout}
\end{center}
\end{figure}

The present proposal as illustrated in Fig.~\ref{fig:eSPS_layout} is to use the Super Proton Synchrotron (SPS) simultaneously as an accelerator and as a very long pulse stretcher to provide such a beam. The SPS has, in the past, accelerated electrons and positrons from 3.5\,GeV to 22\,GeV when it was used as the injector to the Large Electron-Positron collider (LEP)~\cite{LepInjectorStudy:1983aa}, although most of the equipment required has now been dismantled. The electron injector would be replaced by a 3.5\,GeV compact high-gradient linac based on Compact Linear Collider (CLIC)~\cite{CLIC} technology injecting pulses of 200\,ns duration into the SPS, filling the ring at 100\,Hz on a 700\,msec duration plateau.
The beam would then be accelerated to 16\,GeV, using a 800\,MHz SC RF system, similar to what is needed for FCC-ee. SPS-LSS6 is the preferred location for the RF system in order to exploit the existing infrastructure from the crab cavity installation. 
The electrons would then be extracted at 16\,GeV using slow resonant extraction. 
The extracted beam is transported along an existing beamline to an experimental area on the Meyrin site. 

Use of the bypass for the SPS RF system requires a 10\,min changeover period. If a more rapid change-over from protons to electrons is interesting or  necessary, for example during a SPS supercycle, an extended pulsed bypass beamline can be considered in this area. 

The spacing of the SPS RF buckets is 1.25~ns given its 800~MHz SC RF system. Therefore, the filling of the SPS has to be a multiple of this frequency.
This multiple is determined by the matching of the S-band injector linac to the SPS RF frequency, and is thus limited to a multiple of 5 ns between bunches. The default number of extracted electrons per filled bucket in the SPS is 1--10, depending on the experimental capabilities. Around 2/3 of the ring is filled, and the filling and acceleration time is negligible compared to the extraction time.  Since the beam can be distributed over a relatively large area up to 30\,$\times$\,0.4\,cm$^2$ and the SPS RF bucket is significantly wider than the timing resolution of the detector there is scope of being able to deal with several extracted electrons per filled bucket. The experiment can be provided with 10$^{16}$ electrons in around 10$^7$ seconds of beamtime assuming an average of 6 electrons per delivered bunch spaced within 5~ns. 

Changes to the S-band linac would allow the filling of all 1.25 ns spaced filled buckets in the SPS giving a factor 4 higher rate. The RF system and the beamline for delivery to the experiment is compatible with a beam up to 18 GeV. 

A fast extraction is also possible, when the whole beam is extracted from the machine in one revolution (23\,$\mu$s) to feed a possible beam dump type experiment. This could be repeated every 2 seconds if such an operation was to be given priority. 



During the eSPS operation the beam from the linac is used for less than 5\% of the time for injection into the SPS and is available for other uses the rest of the time. Two experimental areas will be available for a broad range of accelerator R\&D using the injector and/or the full linac beam. 
The possibilities range from studies very relevant for future Higgs-factories (linear or circular) to providing a unique plasma acceleration R\&D facility. All these possibilities are described in Section~\ref{sec:ACCRnD}.
The full range of beams available to experiments, from injector to extracted beam from the SPS, are summarised in Table~\ref{tab:possible_beams}.

Most of the hardware, apart from the 3.5\,GeV linac and the 800 MHz SC RF cryomodule in the SPS, already exist. The main components, shown schematically in Fig.~\ref{fig:eSPS_layout}, are introduced below.

\vspace{\baselineskip}
\noindent \textbf{A 3.5\,GeV compact linac} would be built using the technology developed for CLIC at 12\,GHz, but with klystrons as power source instead of the two-beam acceleration method proposed for CLIC. An ideal location would be in transfer tunnels 4 and 5 (TT4 and TT5) at the entrance to the West Hall (building 180), and connected to the SPS by the TT60 tunnel complex. A 0.2\,GeV S-band photo-injector would provide a 200\,ns pulse with a very flexible bunch structure inside the pulse. Both of these elements are detailed in Section~\ref{sec:LINAC}.


\vspace{\baselineskip}
\noindent \textbf{Transfer and injection.} The beam from the linac is transferred via TT61, previously used to transport protons to the West Area, with the electrons being injected in the opposite direction to the protons. A new 3.5\,GeV kicker with a flat top of 200 ns and a 100 Hz repetition frequency would be installed at the SPS injection point in long straight section 6 (LSS6). The ring would be filled in about 700 milliseconds. Section~\ref{sec:SPS_LINACtoSPS} presents in detail the scheme.

\vspace{\baselineskip}
\noindent \textbf{Acceleration in the SPS.} A new vacuum sector in the SPS-LSS6 region was created to test superconducting crab cavities with proton beams for HL-LHC. This sector comprises two Y-shaped vacuum chambers articulated by mechanical bellows, the circulating proton beam line and the beam bypass consisting of the RF module. The mechanical bypass is equipped with a movable table to move the cryomodule in and out of the circulating beam. A dedicated RF system, cryogenic system and general infrastructure were put in place on the surface and in the tunnel, and then successfully operated with beam during 2018. The mechanically movable bypass allows for an installation of a RF module operating at 800\,MHz for a dedicated mode of operation with electrons. This configuration alleviates the strong constraints of impedance requirements for the high intensity proton beams and other modes of SPS operation. The changeover time to use the stage is around 10 min. Recent studies in the framework of the FCC study have led to design of several 800\,MHz cavities ranging from single-cell for high beam currents to five-cell structure for the high energy. Two five-cell cavities housed in a cryomodule will be suitable for acceleration of the electrons from 3.5\,GeV to 18\,GeV. 

\vspace{\baselineskip}
\noindent \textbf{Resonant extraction of the beam} from the SPS at 16\,GeV would be done by exciting the third integer resonance using existing sextupoles (see Section~\ref{sec:delivery_system}). Quantum excitation due to synchrotron radiation would provide a powerful method of pushing particles into resonances with no dynamic variation of the tune being required as is the case for the proton beam. The intensity and duration of the spill can be controlled down to very low currents by adjusting the distance in tune of the beam from the resonance. A new extraction channel using similar hardware to proton extraction must be installed in LSS1. Fast extraction through the same channel could be performed using existing SPS kickers if needed for a beam dump experiment. Preliminary simulation results supporting this scheme are shown in Section~\ref{sec:SPS_Extraction}. 

\vspace{\baselineskip}
\noindent \textbf{Beam transport to the detector.}
The beam would be extracted into TT10, the beamline which is also used to inject protons into the SPS. The TT10/TT2 switchyard magnets would not be powered during electron extraction so that the beam can be directed to a new experimental area. Quadrupoles would be used to blow up the beam to match the detector requirements. This is presented in Section~\ref{sec:delivery_system}.
The only civil construction required for the whole project would be a new short connection tunnel and an experimental hall housing one or two experiments, located just outside TT2 and shown in Section~\ref{sec:SPS_RF}.

%

\begin{table}[!hbt]
    \centering
    \caption{Summary of the parameters of available experimental beams after each accelerating section, the S-band linac, the X-band linac and the SPS. Note that single quantities listed here correspond to the upper limit and can be reduced for operation.}
    \label{tab:possible_beams}
    \begin{tabular}{lccc}
    \hline\hline
         \textbf{Parameter} &  &  \textbf{Accelerating section}  \\
         
         &  \textbf{S-band linac} (Section~\ref{sec:LINAC_GunInjector}) & \textbf{X-band linac} (Section~\ref{sec:LINAC_Design}) & \textbf{SPS} (Section~\ref{sec:SPS_RF}) \\
         
         \hline
        Energy [GeV] & 0.05--0.25 & 3.5 & 3.5--18 \\
        Electrons per bunch & \num{e6} -- \num{e10} & \num{e6} -- \num{e10} & \num{e9}\\
        Bunch length [ns] & \num{e-4} -- \num{4e-3} & \num{e-4} -- \num{2.5e-3} & 0.15 -- < 0.7\\
        Bunch spacing [ns] & Multiples of 0.33 & Multiples of 0.33 & 5 \\
        Bunches per cycle & 1 -- 200 & 1 -- 200 & \num{3e3} \\
        Cycle length  [s] & 0.02 & 0.02 & 0.02 \\
        \hline
        \hline
    \end{tabular}

\end{table}

\vspace{\baselineskip}
The CERN Council adopted an \textit{update of the European Strategy for Particle Physics}~\cite{TheCERNCouncil2020} in June 2020. The conceptual design report presented here, describes an infrastructure relevant for several of its key priorities, like an electron-positron Higgs factory, accelerator R\&D, dark sector physics, and neutrino physics. 
\begin{itemize}
\item The implementation of eSPS would make excellent use of the investment made in the CLIC programme and is the natural next step in the development of X-band \textit{high-gradient acceleration technology}, a key technology for compact and cost-effective electron/positron linacs.  

\item The multi-GeV electron beam from the linac would drive wakefields in the non-linear regime and would, with an independent electron witness bunch, demonstrate the applicability of \textit{plasma wakefields for high-gradient acceleration}. The facility would be unique in its ability to study collider related challenges, as the only facility with  multi-GeV  drive  bunch  and  truly independent electron witness bunch.  Addition of a positron witness bunch would make it a complete facility for collider studies. 

\item The 800\,MHz super-conducting cavities for the eSPS would be the same type as foreseen for a future \textit{electron-positron Higgs and electroweak factory} as the first stage of a next circular collider at CERN. The eSPS would be used for the development of and studies of a large number of components and phenomena for this circular collider. The operation of SPS with electrons would train a new generation of CERN staff on circular electron accelerators.

\item The electron beam would open a \textit{dark sector physics} programme and in particular provide sensitivity to \textit{light dark matter} production significantly beyond the targets predicted by a thermal dark matter origin, and for the nature of dark matter particles that are not accessible by direct detection experiments. 

\item The future \textit{neutrino physics programme} needs to precisely measure neutrino oscillation probabilities as a function of energy. This critically relies on the ability to model neutrino-nucleus interactions, and this in turn requires input data on electro-nuclear reactions; the beam from this facility would be excellent for such measurements. 
\end{itemize}
\vspace{\baselineskip}
eSPS could be made operational in about five years, and serve the programmes above. It could start already in LS3 and would operate in parallel and without interference with Run 4 of the LHC.

%% file: include/02-Physics_Goals/PhysicsGoals.tex
\section{Physics potential and requirements on the electron beam}
\label{sec:PhysicsGoals}
The following section is a summary of the physics case presented in the Expression of Interest~\cite{Torstenakesson2018} submitted to the SPS and PS Experiments Committee (SPSC) in October 2018.
\subsection{Reaching light dark matter with a primary electron beam}
\label{sec:LDM-at-eSPS}
Only 15\% of the observed matter is made of particles described by the Standard Model~\cite{ade:2015xua} (SM).
The evidence for dark matter (DM) does not give much direct guidance on the masses of its constituents, which could be anywhere from a tiny fraction of an eV up to many solar masses. More constraints can be obtained by focusing on likely scenarios for how the primordial DM was created. The most straightforward and simplest scenario is the thermal origin, in which DM arose as a thermal relic from the hot early Universe. This scenario only requires small non-gravitational interactions between dark and Standard Model matter, and is viable over the MeV to TeV mass range. The mass region $\sim$\,MeV to $\sim$\,GeV is largely unexplored. The fact that most stable forms of ordinary matter are found in this range, argues in favour of exploring this mass range.

\begin{figure}[!hbt]
\centering
\includegraphics[width=0.8\textwidth]{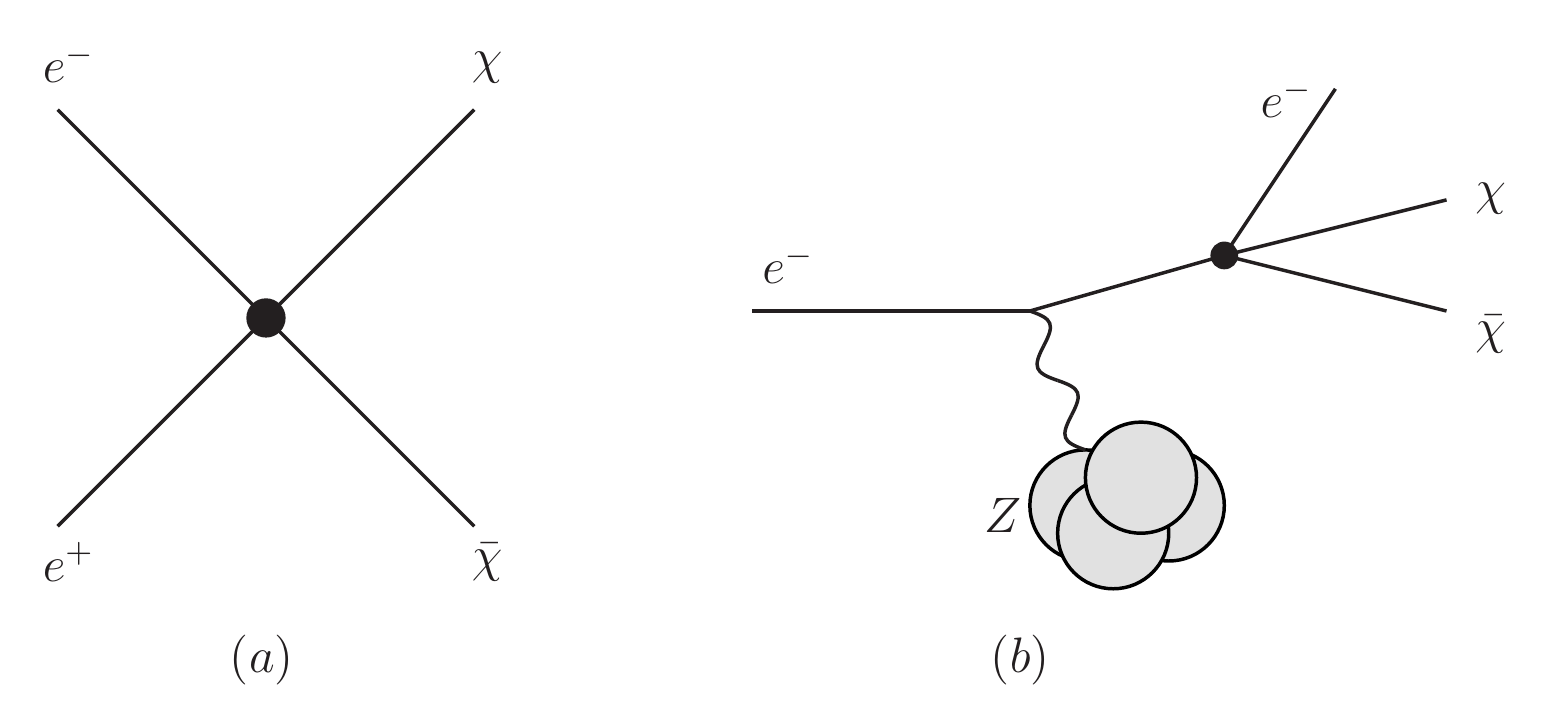}
\caption{\label{fig:introDiagrams}\textbf{Left:} Diagrammatic representation of a contact interaction between DM and electrons.  \textbf{Right:} Production of DM in electron-nucleus fixed target interactions, guaranteed by the existence of this contact interaction. As discussed below, models that resolve this contact interaction through the introduction of a new mediator at experimentally accessible energy scales can also be detected through related production modes.}
\end{figure}

A thermal origin for DM requires an interaction between DM and familiar matter, and if there is an interaction of light dark matter (LDM) with ordinary matter, 
(Fig.~\ref{fig:introDiagrams}(a))
then there necessarily is a production mechanism in accelerator-based experiments 
(Fig.~\ref{fig:introDiagrams}(b)), both in the minimal framework of a four-particle contact interaction and in realistic ultraviolet completions of this scenario by the addition of a new force carrier. The most sensitive way to search for this reaction is to use an electron beam to produce DM in fixed-target collisions, making use of missing energy and/or momentum to identify this process~\cite{izaguirre:2014bca,Izaguirre:2015yja}.
Dedicated searches for these production reactions thereby provide sensitivity to DM couplings to the Standard Model. 

The strength of this interaction determines when the DM froze out of equilibrium, therefore, the residual DM abundance. This production mechanism, together with the observed DM density thus  motivates a precise interaction strength for any given DM mass.  

\begin{figure}[!hbt]
\center
\includegraphics[width=6.5cm]{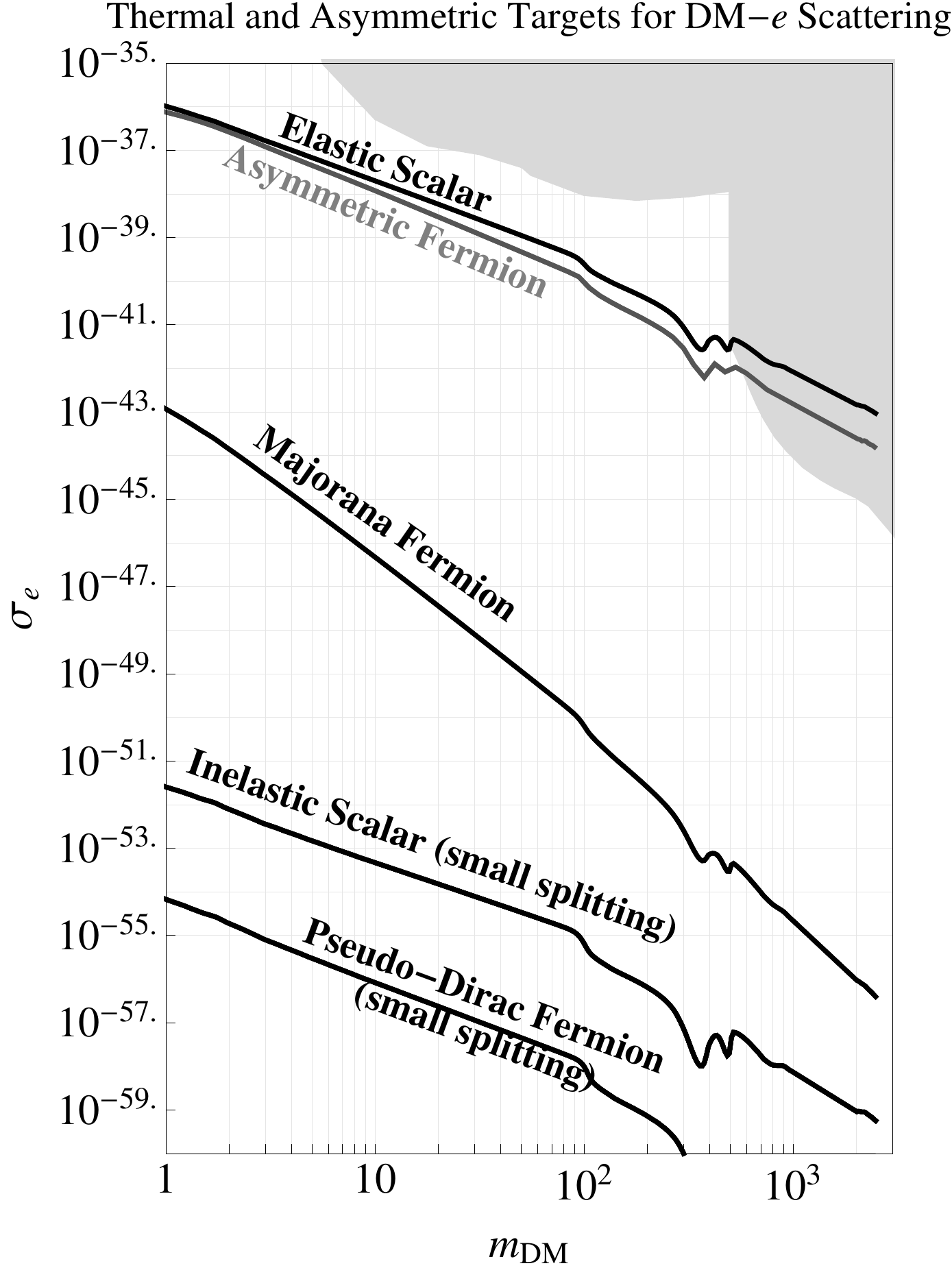}
\includegraphics[width=8.0cm]{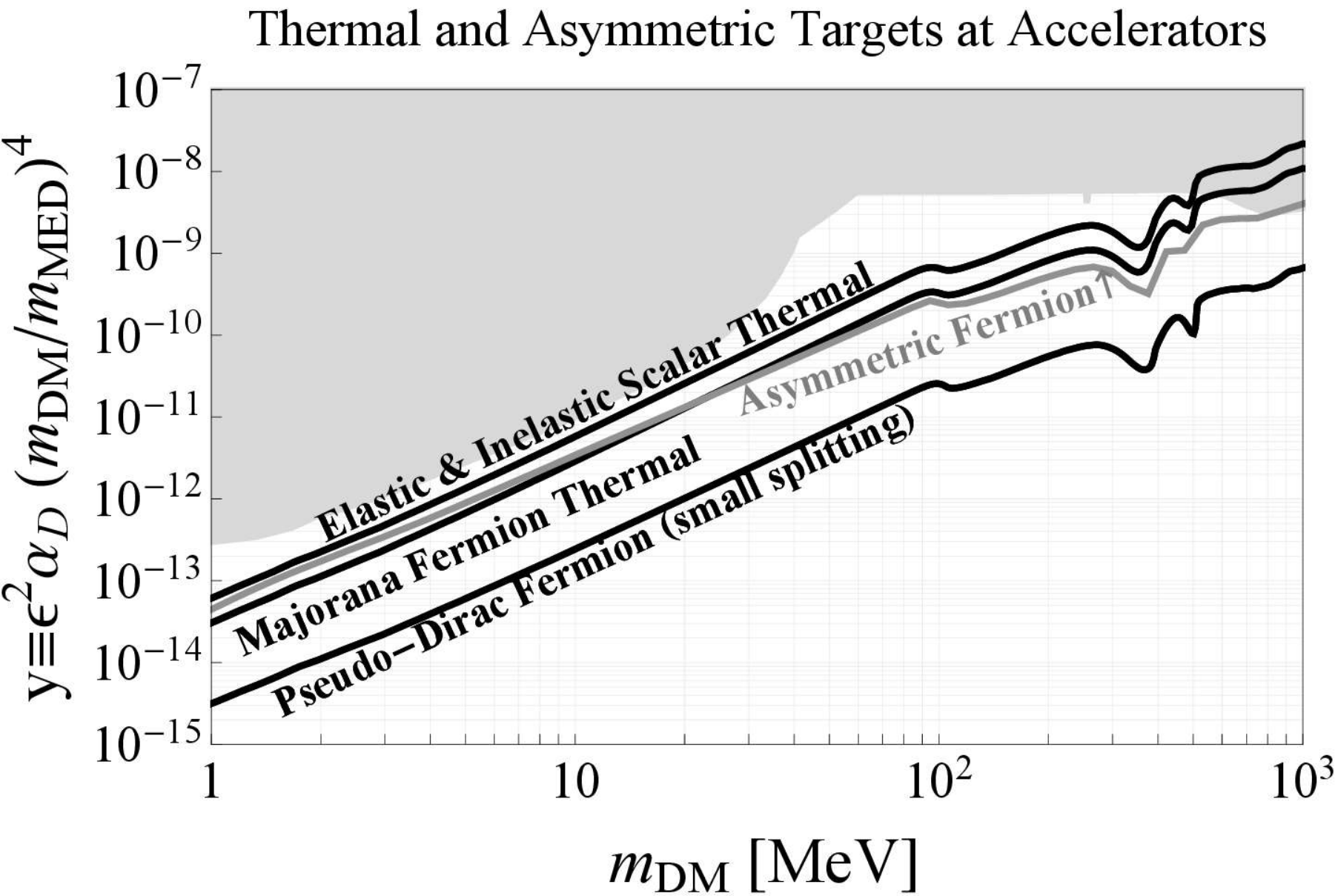}
\caption{\textbf{Left:} Thermal targets for DM plotted in terms of the electron-recoil direct detection cross section $\sigma_e$ vs. mass $m_{DM}$. The appropriate thermal freeze-out curve for each scenario differs by many orders of magnitude in the $\sigma_e$ plane due to velocity suppression factors, loop-level factors, or spin suppression, any of which are significant for non-relativistic scattering. 
\textbf{Right:}  By contrast, the dimensionless couplings (captured by $y$) motivated by thermal freeze-out do not differ by more than a couple of orders of magnitude from one another, as shown in the $y$ vs. $m_\chi$ plane. Probing couplings at this magnitude is readily achievable using accelerator techniques, which involve DM production and/or detection, as well as mediator production, all in a relativistic setting. Both plots above are taken from Ref.~\cite{Battaglieri:2017aum} and also show a target for asymmetric fermion DM, a commonly discussed variation on the thermal-origin framework. 
}
\label{fig:collapsing}
\end{figure}

Relative to other experimental techniques used to search for DM, such as direct detection scattering in underground experiments or indirect detection searches with satellite experiments, fixed-target accelerator experiments are the \textit{only} technique available that can probe the DM interaction at momentum scales comparable to those governing  freeze-out in the early Universe. This technique is, therefore, not hindered by some of the common challenges faced by, for example, direct detection experiments, where mass threshold or velocity suppression can severely inhibit signal rates in the non-relativistic limit as is shown in the left panel of Fig.~\ref{fig:collapsing}. This makes fixed-target probes of DM both complementary to other terrestrial techniques, and especially robust in exploring thermal freeze-out. 
Models of thermal DM in the MeV~--~GeV mass range require that the interaction governing freeze-out have a cutoff scale below the weak scale.  This is, essentially, a simple generalisation of the Lee-Weinberg bound~\cite{Lee:1977ua,Boehm:2003hm}, with two important consequences:
\begin{itemize}
   \item \textbf{Light Forces:} There must be new force carriers at the GeV-scale or below to mediate an efficient annihilation rate for thermal freeze-out.
   \item \textbf{Neutrality:} Both the DM and the mediator must be singlets under the full SM gauge group; otherwise they would have been produced and detected at LEP or at hadron colliders~\cite{ALEPH:2010aa}.  
\end{itemize}
These properties single out the hidden sector scenario highlighted in Refs.~\cite{alexander:2016aln, Battaglieri:2017aum}, which is the focus of considerable experimental activity. 

For the remainder of this chapter, we will use one of the simplest and most representative hidden sector models in the literature -- a DM particle charged under a $U(1)$ gauge field (i.e.\ ``dark QED''). We define the LDM particle to be $\chi$, the $U(1)$ gauge boson $A^\prime$ (popularly called a ``dark photon'' mediator), and $\epsilon$ as the kinetic mixing parameter.

This framework permits two qualitatively distinct annihilation scenarios in the early universe, depending on the $A^\prime$ and $\chi$ masses.  

\begin{itemize}
   \item  \textbf{Direct Annihilation:} A mediator with  $m_{A^\prime} >  m_{\chi}$ generates the effective contact interaction for non-relativistic DM particles.  In the resolved theory, annihilation proceeds via  $\chi \chi \to {A^\prime}^* \to ff$ to SM fermions $f$ through a virtual mediator. This scenario is quite predictive, because the SM-$A^\prime$ coupling $\epsilon$ must be large enough, and the $A^\prime$ mass small enough, in order to achieve the thermal relic cross-section.  Depending on the mass of the mediator, on-shell mediator production with decay to DM or production of DM through an off-shell mediator may be the dominant signal in a missing momentum experiment. In each case, the observed DM abundance implies a minimum DM production rate at accelerators. Constraints on this scenario can be extracted from CMB data, but are only relevant for some combinations of DM and mediator spin and couplings. This case will be the focus of the remaining discussion.  
   \item \textbf{Secluded Annihilation:} For $m_{A^\prime}  < m_{\chi}$, a new annihilation channel becomes kinematically allowed, and generically dominates.  In this case, DM annihilates predominantly into $A^\prime$ pairs~\cite{Pospelov:2007mp}.  This annihilation rate is independent of the SM-$A^\prime$ coupling $\epsilon$. The simplest version of this scenario is robustly constrained by CMB data~\cite{madhavacheril:2013cna}.  
\end{itemize}
Since the Feynman diagram that governs direct annihilation can be rotated to yield a scattering process of SM particles, the  direct detection cross section is uniquely predicted by the annihilation rate in the early universe for each choice of DM mass. Thus, direct annihilation models define thermal targets in the $\sigma_e$ vs. $m_{\chi}$ plane. The left panel of Fig.~\ref{fig:collapsing} shows how the non-relativistic direct detection cross sections can be loop or velocity suppressed in many models, and, therefore, these thermal targets vary by dozens of orders of magnitude in some cases.
%
%
However, these vast differences in the direct detection plane mask the underlying similarity of these models in relativistic contexts where both the scattering and annihilation cross sections differ only by order-one amounts. 
To study all direct annihilation models on an equal footing, we follow conventions in the literature (see Ref.~\cite{alexander:2016aln}), and introduce the dimensionless interaction strength $y$ as
\be
\sigma v (\chi \chi \to {A^\prime}^*  \to f f) \propto  \epsilon^2 \alpha_D \frac{m_\chi^2}{m_{A^\prime}^4} =  \frac{y}{m_\chi^2}~~~~,~~~~ y \equiv 
\epsilon^2 \alpha_D  \left(     \frac{m_\chi}{m_{A^\prime}  }    \right)^4~~.
\label{eq:ydef}
\ee

This is a convenient variable for quantifying sensitivity because for each choice of $m_\chi$ there is a unique value of $y$ compatible with thermal freeze-out independently of the individual  values of $\alpha_D, \epsilon$ and $m_\chi/m_{A^\prime}$. The right panel of Fig.~\ref{fig:collapsing} shows the thermal targets in the $y$ vs. $m_{\chi}$ plane. 
A measured (or upper limit on the) production cross section ($\sigma$) for the process shown in Fig.~\ref{fig:introDiagrams}(b), is translated to $y$ as, 

\be
y \propto \left(\frac{m_\chi}{m_{A^\prime}}\right)^2 \alpha_D m_\chi^2 \sigma 
\label{eq:sigma-to-y}
\ee

From~(\ref{eq:sigma-to-y}) it is seen that for a fixed values of $\alpha_D$ and of the $m_{A^\prime}$ to $m_\chi$ ratio, a measured (or limit on) $\sigma$ would translate to a parabola in the $y$ versus $m_\chi$ plane in the right panel of Fig.~\ref{fig:collapsing}. As also can be seen from~(\ref{eq:sigma-to-y}), small $m_\chi$ to $m_{A^\prime}$ ratios, and small $\alpha_D$ values, would result in stronger experimental reaches. We conservatively chose $m_{A^\prime} = 3 \times m_\chi$, and $\alpha_D = 0.5 $, when estimating the experimental reach described in this conceptual design report, and in the expression of interest (EoI)~\cite{Torstenakesson2018} that was submitted to the SPSC in 2018. 

More discussions on reach for such parameter settings, including off-shell production of $A^\prime$ and when approaching the resonance region of $m_{A^\prime} \approx2 \times m_\chi$, are in Ref.~\cite{LDMXSciencePaper}.

The thermal targets for various direct annihilation models shown on the right panel of Fig.~\ref{fig:collapsing} in the $y$ vs $m_{\chi}$ plane, are the same models as shown on the left panel, but the parameterisation in $y$ and $m_{\chi}$ reveals the underlying similarity of these targets and their relative proximity to existing accelerator bounds (shaded regions). 
Reaching experimental sensitivity to these benchmarks for masses between MeV and GeV would provide nearly decisive coverage of this class of models.
\newline
Reaching the sensitivity to find events as shown in Fig.~\ref{fig:introDiagrams}(b), or to establish the absence of such production, sets two contradictory requirements on the beam: 
\begin{enumerate}
   \item A large number of electrons on target (EOT) since the interaction strength is weak. More quantitatively, to reach the thermal targets for all possible natures of DM particles not excluded by CMB data~\cite{madhavacheril:2013cna} requires reaching at least $\sim$\,$10^{14}$ ($\sim$\,$10^{16}$) EOT for Scalar (Pseudo-Dirac) DM particles~\cite{whitepaper}. 
   \item A low current to be able to measure individual electrons entering the target, and to match them with potential signal electrons leaving the target. In addition, one has to be able to ensure the absence of any bremsstrahlung photons, but without rejecting signal events due to such photons produced by unrelated beam electrons.
\end{enumerate}
\textbf{Only a primary electron beam can deliver $\sim$\,$10^{14}$ to $\sim$\,$10^{16}$~EOT with low current and high duty cycle at 5--20\,GeV energy.} 
\vspace{\baselineskip}

\begin{figure}[!hbt]
\centering
\includegraphics[width=0.6\textwidth]{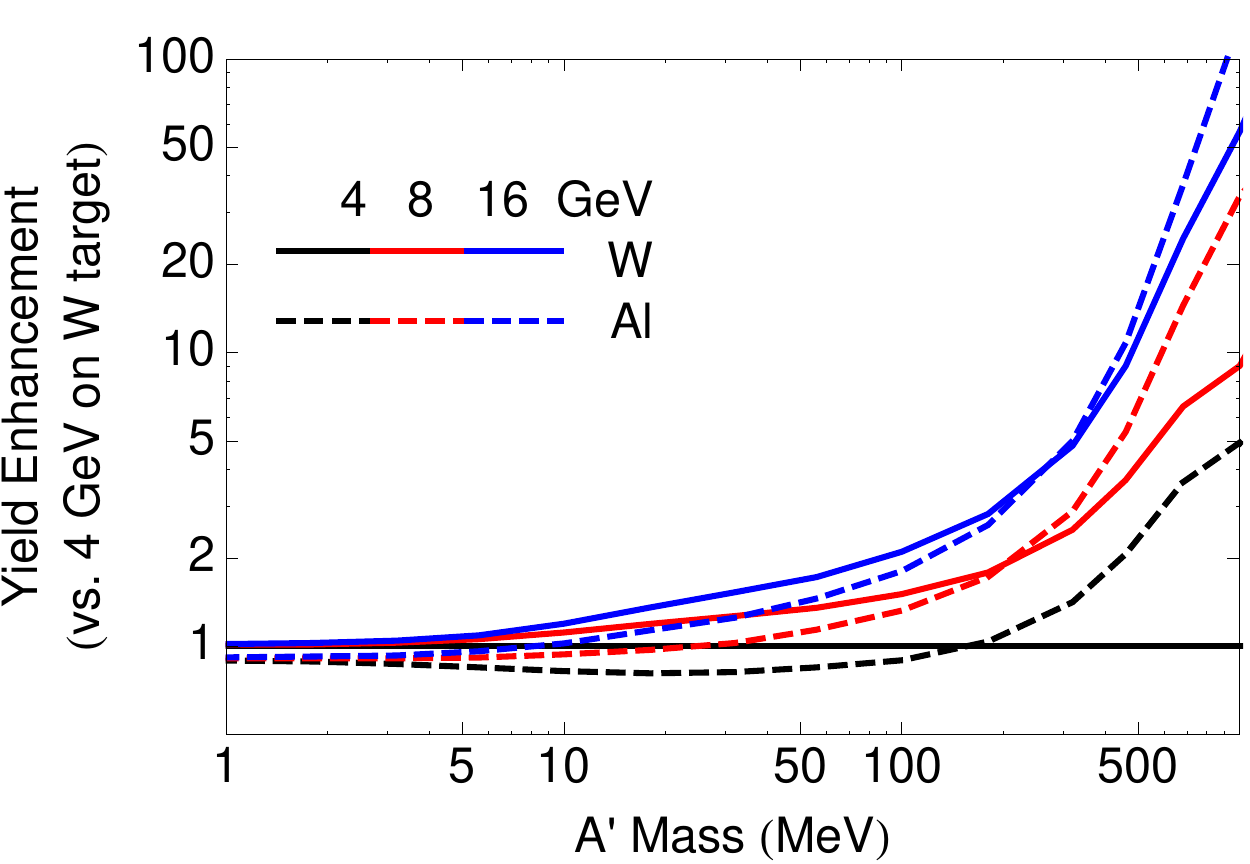}
\caption{\label{fig:xsecratio} Both beam energy and target material affect the dark photon production cross-section at the higher end of the mass range of interest.  This figure illustrates how increasing the beam energy to 8 or 16 GeV, and/or switching from a Tungsten to an Aluminium target (fixed at 0.1$X_0$), impacts the signal production cross-section for different dark photon masses.  We assume the kinematic selection $E_{recoil}<0.3 E_{beam}$. The plot is taken from Ref.~\cite{Torstenakesson2018}.}
\end{figure}

The higher end of this energy scale gives advantages on both signal production and background rejection (where the latter in particular requires discovering photo-nuclear events from bremsstrahlung photons). On the one hand, the signal cross sections increase with energy, improving the sensitivity, particularly in  the high mass region (several hundred MeV) as is shown in Fig.~\ref{fig:xsecratio}. On the other hand, the rates of certain backgrounds decrease with higher energy, e.g.\ that of the exclusive 2-body photo-nuclear reactions scales as $E_{\gamma}^{-3}$, and the products from these reactions carry more energy and are hence more visible in a detector. Similarly, in-flight decays within a detector, e.g.\ of charged kaons from photo-nuclear reactions, have a lower rate and more detectable products at higher energy.

\begin{figure}[!hbt]
\centering
\includegraphics[width=0.7\textwidth]{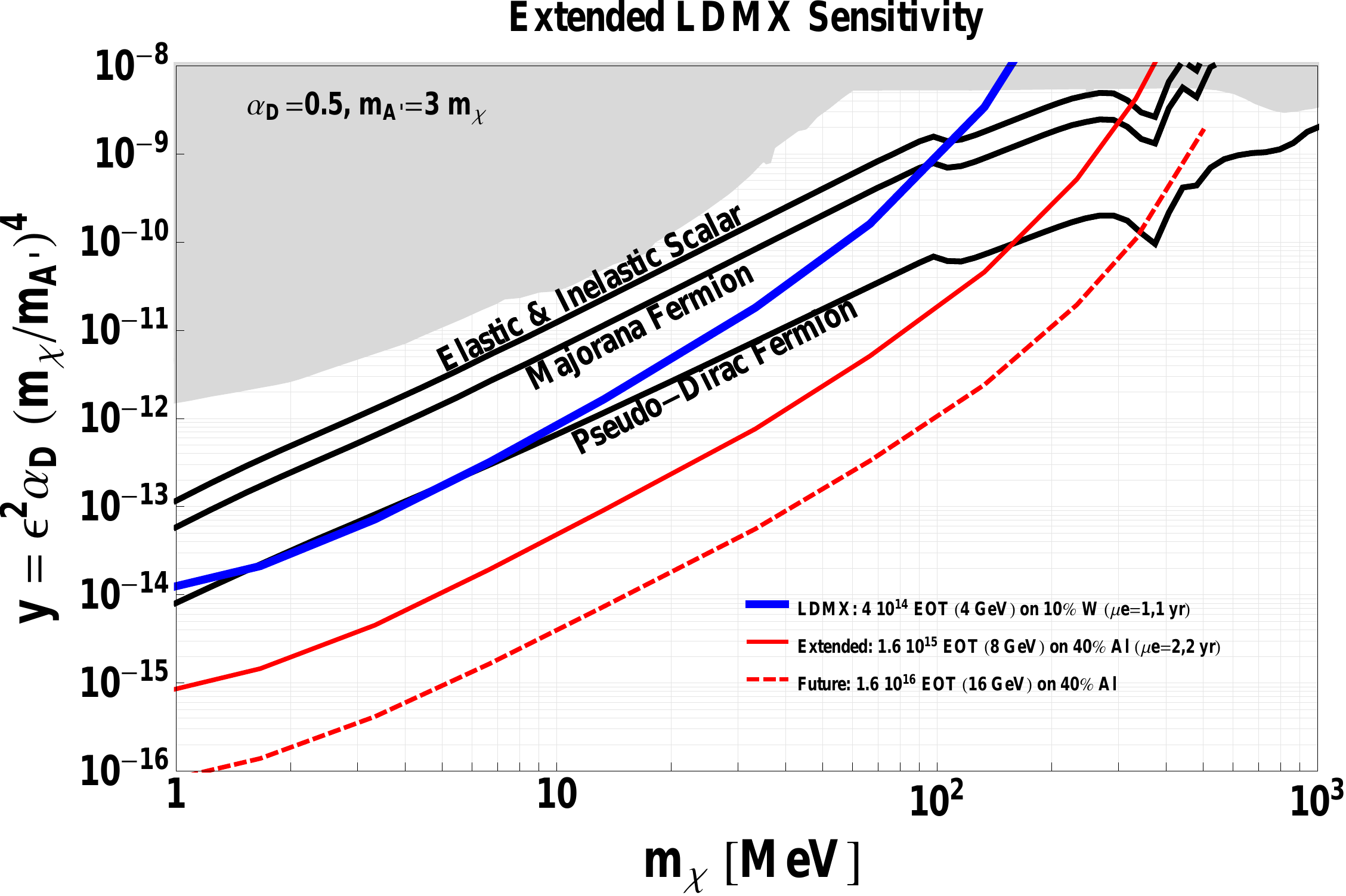}
\caption{\label{fig:extendedLDMX} The experimental reach compared to the thermal targets for various direct annihilation models shown on the right panel of Fig.~\ref{fig:collapsing}. The blue line is the sensitivity from the reference study discussed in Ref.~\cite{whitepaper}, that conservatively assumes 0.5 background events for $4\times10^{14}$~EOT at 4\,GeV.  A scaling estimate of the sensitivity of the configuration for the mass range $150 \mathrm{\,MeV} \le M_{\chi}<300 \mathrm{\,MeV}$ is illustrated by the solid red line.  The dashed red line represents a similar estimate of the projected reach for $\mu_e\sim$\,12 and roughly 3 years of running. For the latter two examples we have again assumed low background, consistent with reductions in yields of potential background sources, and better rejection, while increasing the effective luminosities to $1.6\times10^{15}$ and $1.6\times10^{16}$ EOT, respectively. The plot is taken from Ref.~\cite{Torstenakesson2018}.
}
\end{figure}

As mentioned above, an Expression of Interest~\cite{Torstenakesson2018} was submitted to the SPSC in 2018. It describes the potential of a missing-momentum experiment like the Light Dark Matter eXperiment (LDMX), using the primary electron beam delivered by the accelerator complex described in this Conceptual Design Report. The red dashed line in Fig.~\ref{fig:extendedLDMX} shows the expected performance compared with the thermal targets. The potential to cover a large fraction of sub-GeV mass range for all natures of DM particles, is clear. 

\begin{figure}[!hbt]
 \centering
   \includegraphics[width=0.45\textwidth]{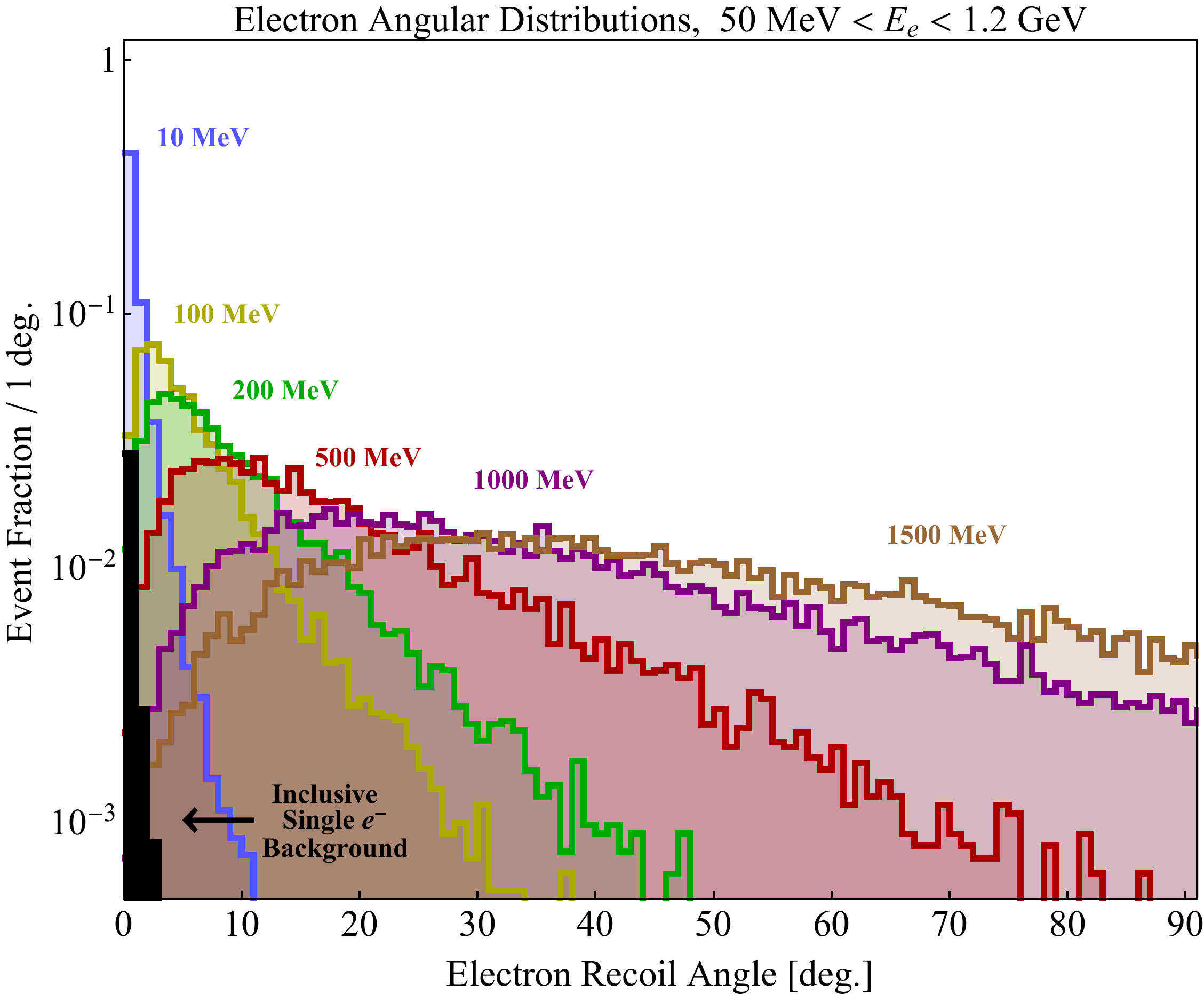}
   \includegraphics[width=0.45\textwidth]{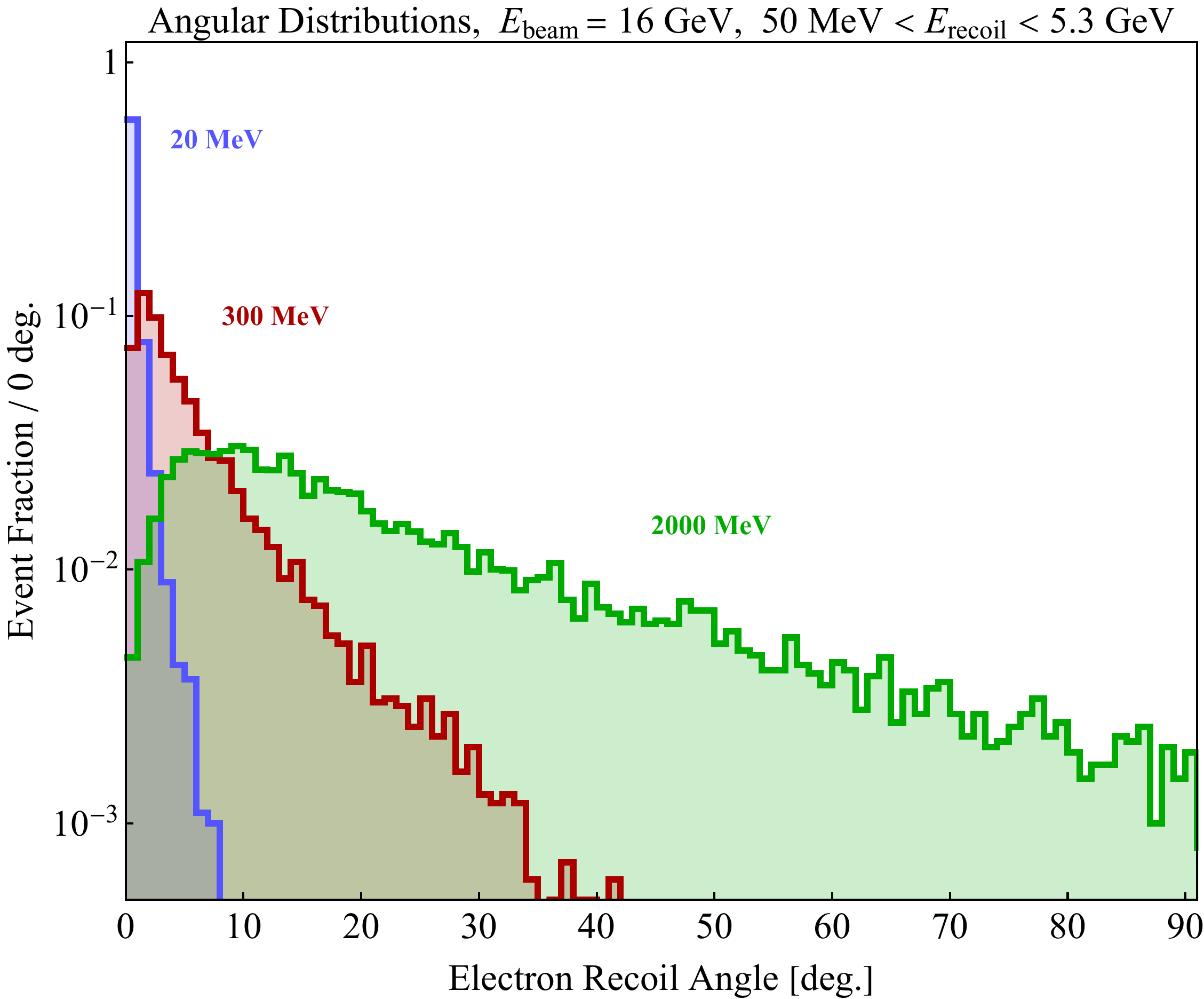}\\

   \caption{\label{fig:deflection-angle}  
   Electron deflection angle for DM pair radiation process, at various dark matter masses for electron beam energy of 4\,GeV (left) and 16\,GeV (right). The distributions are for electrons requiring $50~\mathrm{\,MeV} < E < 0.3 E_{\mathrm{beam}} $.  In both panels, the numbers next to each curve indicate $A'$ mass.The plots are taken from Ref.~\cite{Torstenakesson2018}.}
\end{figure}

As was discussed above, each coloured line in Fig.~\ref{fig:extendedLDMX}, corresponds to one value of a measured (or the limit on) $\sigma$. But, more information can be extracted if signal events are found. As demonstrated in Ref.~\cite{Akesson2019a}, the background to the signal, can be fully rejected without making use of the $p_T$ (or deflection angle) of the signal-electrons. The electron $p_T$ (or deflection angle) distribution can, therefore, be used as an independent signal hypothesis test, and as a DM mass estimator. Figure~\ref{fig:deflection-angle} shows how much these distributions change for different DM masses.    

Although the emphasis in this chapter has been on various models of DM with direct annihilation through a dark photon mediator $A^\prime$, the missing momentum technique can probe multiple other mediator scenarios with equally powerful sensitivity to the corresponding theoretical targets. For example, both dark and visible matter could be directly charged under a new $U(1)$  group which gauges an anomaly-free combination of SM quantum numbers (e.g.\ baryon minus lepton number). Such new forces can also mediate DM direct annihilation to SM particles with thermal targets analogous to those presented in Fig.~\ref{fig:collapsing}. Some of these are discussed in Ref.~\cite{Torstenakesson2018}. Furthermore, missing momentum techniques can also probe strongly interacting dark sectors~\cite{Berlin:2018tvf}, millicharged particles, minimal dark photons, minimal $U(1)$ gauge bosons, axion like particles, and light new leptophilic scalars~\cite{LDMXSciencePaper}.


\subsection{Mediators, millicharges, neutrino physics and nuclear physics}
\label{sec:More-at-eSPS}

The missing momentum search is sensitive to a range of other new-physics scenarios, potentially unrelated to dark matter. 

\begin{figure}[!hbt]
 \centering
\includegraphics[width=0.45\textwidth]{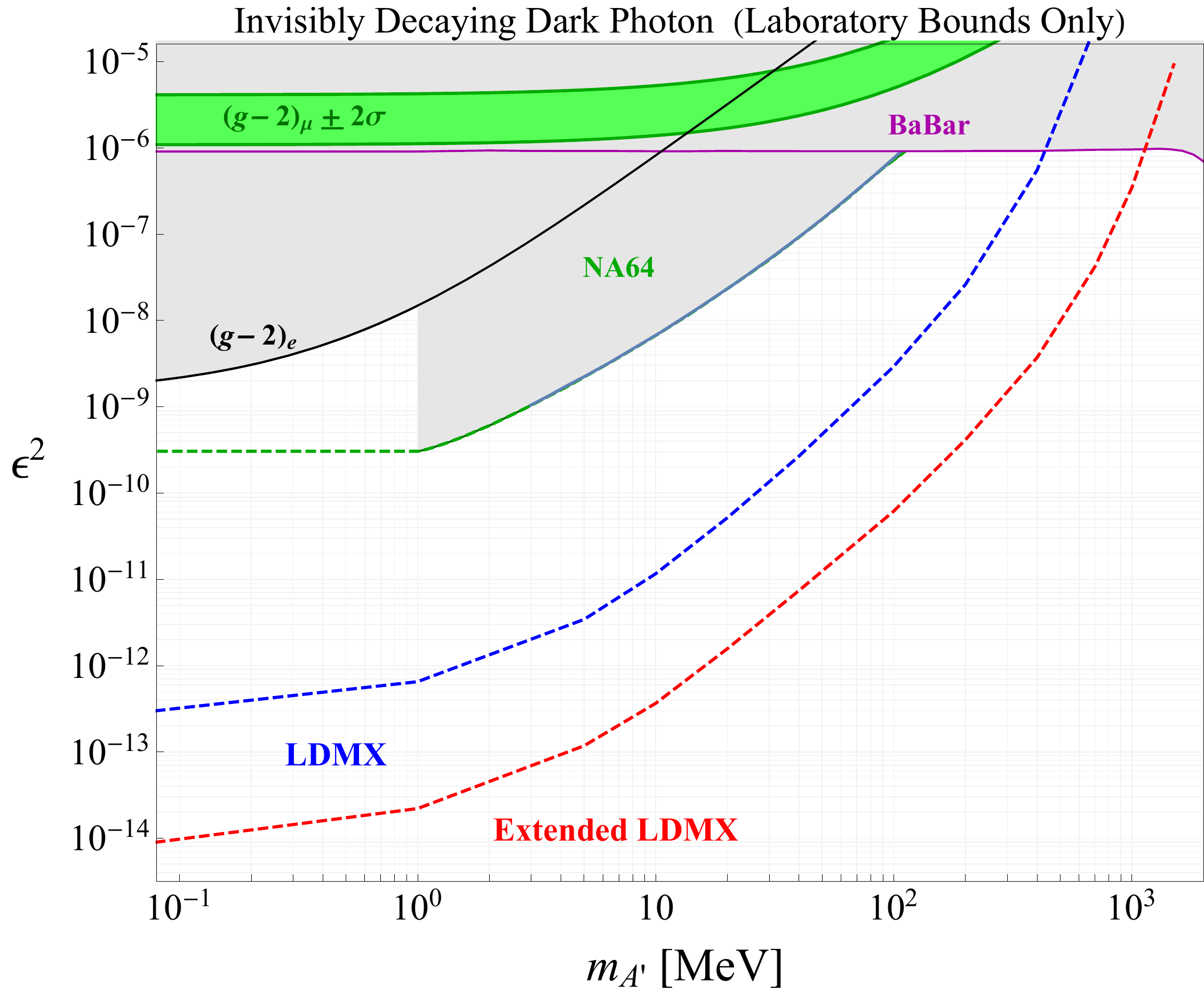}
\includegraphics[width=0.45\textwidth]{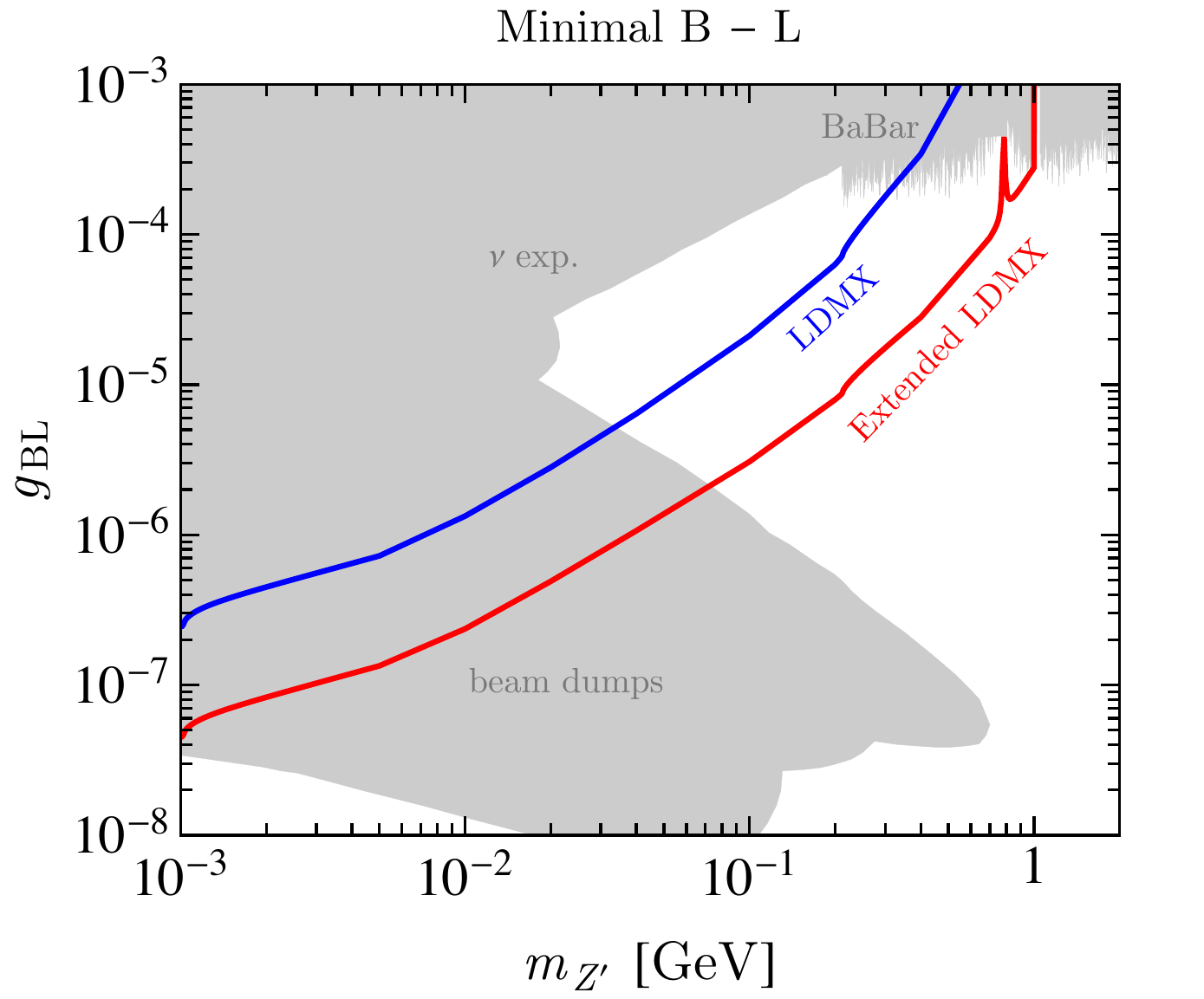}
\caption{\label{fig:InvisibleMediator} Sensitivity to invisibly decaying dark photons (left) and B-L gauge bosons (right). The blue and the red lines correspond to the full blue and red lines in Fig.~\ref{fig:extendedLDMX}. The plots are taken from Ref.~\cite{Torstenakesson2018}.}
\end{figure}

\begin{figure}[!hbt]
 \centering
\includegraphics[width=0.45\textwidth]{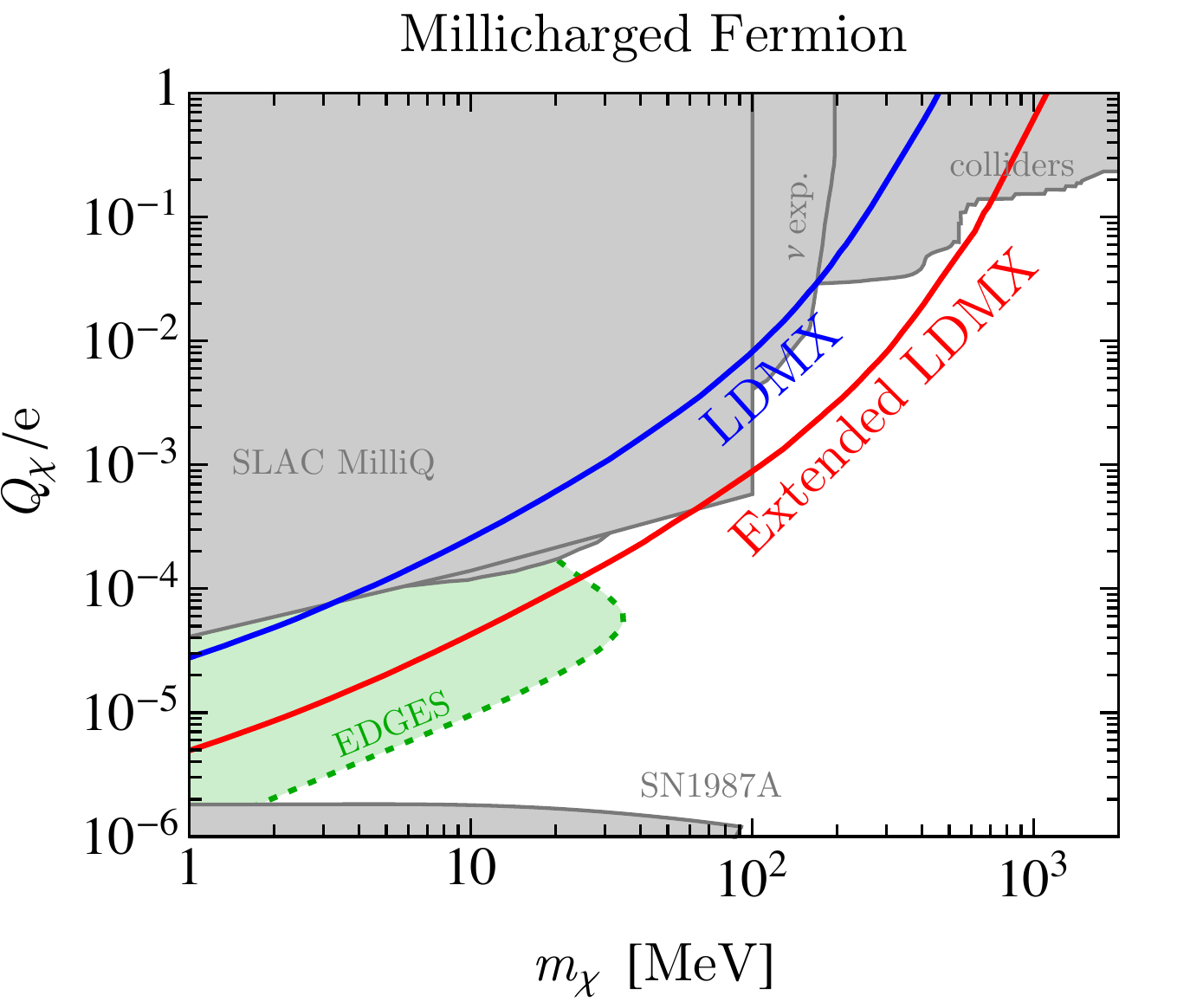}
\caption{\label{fig:millicharge} Sensitivity to millicharge fermion particles with charge $Q_{\chi}/e$ vs mass. Production occurs through an off-shell photon. Grey regions are existing constraints. The green shaded region represents parameter space where a millicharged dark matter subcomponent can accommodate the 21\,cm absorption anomaly reported by the EDGES collaboration~\cite{Bowman:2018yin,Barkana:2018lgd,Berlin:2018sjs,Kovetz:2018zan}.  The blue and the red lines correspond to the full blue and red lines in Fig.~\ref{fig:extendedLDMX}. The plot is taken from Ref.~\cite{Torstenakesson2018}.}
\end{figure}

Figure~\ref{fig:InvisibleMediator} illustrates the sensitivity to invisible dark photons and to minimal B-L $Z^\prime$ gauge bosons, via their invisible decays to neutrino final states. 
Figure~\ref{fig:millicharge} illustrates the sensitivity of a missing momentum search to production of millicharged particles.  Millicharge production occurs through off-shell photon exchange, and particles with sufficiently small millicharge $Q_{\chi}/e$ have no additional interactions in the detector. 

A primary electron beam facility described in this CDR, opens more possibilities than the missing momentum searches described in Section~\ref{sec:LDM-at-eSPS} and above in this section. Section~\ref{sec:ICE_ExperimentalArea}, therefore, outlines a scenario with two beamlines into the experimental hall, with space for two experiments.

The majority of the electrons remain in the SPS after the extraction of the long low current electron spill. These $1-5 \times 10^{12}$ electrons could be dumped in the other beamline in a 23\,$\mu$s spill. This could allow the accumulation of more than $10^{18}$ electrons in a year. More than that in fact, since if priority was given to such an operation, the cycle fill-accelerate-dump could be repeated every two seconds.

As can be seen in Ref.~\cite{Akesson2019a}, photo-nuclear and electro-nuclear reactions are major background sources for a dark matter missing momentum experiment. However, to measure such reactions is also important to understand neutrino-nuclear response. The future neutrino long baseline experiments need to precisely measure neutrino oscillation probabilities as a function of energy. This critically relies on the ability to model neutrino-nucleus interactions, and this in turn requires input data on electro-nuclear reactions; the beam from this facility would be excellent for for this purpose. Such measurements could maybe be done by a missing-momentum experiment, however, given the importance to understand such reactions for neutrino physics, there may be a case for a dedicated experiment. The physics of this is described in Ref.~\cite{Torstenakesson2018}.

There is a broad usage of electron beams for the study of hadrons and their underlying structures, like the momentum and spin distributions of sea quarks and gluons in the nucleons, the study of excitation spectra of nucleons and hyperons, and the prospect to produce mesons with exotic composition and/or exotic quantum numbers. The facility presented in this CDR would extend the energy range, but could not reach the beam intensity currently available, at Jefferson laboratory in the USA where the requests for beam go beyond what is available. 

%% file: include/03-LINAC/OverviewLINAC.tex
\section{Linac}
\label{sec:LINAC}


The electron linac produces the electron beam and accelerates it to an energy of to 3.5\,GeV, the energy required for injection into the SPS. The linac consists of two parts; the injector, that produces the electron bunches with the required time structure, emittance and charge and brings them at an energy of about 200 MeV, and the high-gradient X-band linac that further accelerates them to 3.5\,GeV. Each beam pulse consists of 40 bunches, separated by 5\,ns. Such bunch spacing corresponds to every 4\textsuperscript{th} RF bucket in the 800\,MHz SPS RF system and the number of bunches in turn depends by the optimised pulse length in the linac. The pulse structure and the 100\,Hz linac repetition rate allow to fill the SPS ring with 3000 bunches, the maximum given its diameter, within 1 second. 

After the injector, a low-energy beam line branches out from the main beam line, bending the beam by 180$^\circ$ to an experimental area.  This area can be used to perform independent experiments at a maximum beam energy of 250\,MeV, similar to what is presently ongoing in the CLEAR user facility~\cite{GAMBA2017}. At the end of the linac,  another independent beam line can be used for experiments requiring a higher beam energy. 

%% file: include/03-LINAC/BeamDynamics.tex
\subsection{Beam dynamics studies and start-to-end simulations}
\label{sec:LINAC_BeamDynamics}
The beam dynamics in the eSPS linac are dominated by the impact of wakefields, due to the small iris aperture of the X-band accelerating structures. The linac optics are based on a FODO lattice that uses the two quadrupoles of each pair of consecutive RF modules as focusing and defocusing magnets. The average $\beta$ function in the linac is less than 5 metres, and the phase advance per cell is 90 degrees. This setup is chosen to apply strong focusing in the transverse planes and to increase beam stability. At the start of the linac the beam energy is about 200\,MeV; the linac provides acceleration to the final energy of 3.5\,GeV. The beamline consists of 24 RF units and has a length of approximately 68\,m, with a filling factor of 86\%.

Start-to-end simulations of the linac have been performed in order to assess the beam quality performance in the presence of various imperfections. The simulations were carried out using the code PLACET, a particle tracking code developed at CERN for linear collider studies, that can simulate the transport of particle beams through a linear accelerator under the effects of various imperfections~\cite{Latina2013a}. PLACET enables the simulations of beam correction techniques, such as Beam-Based Alignment (BBA), to evaluate the beam quality after correction, and implements multi-bunch effects.

\subsubsection{Single-bunch effects}
\begin{figure}[!hbt]
\begin{center}
\includegraphics[width=0.6\textwidth]{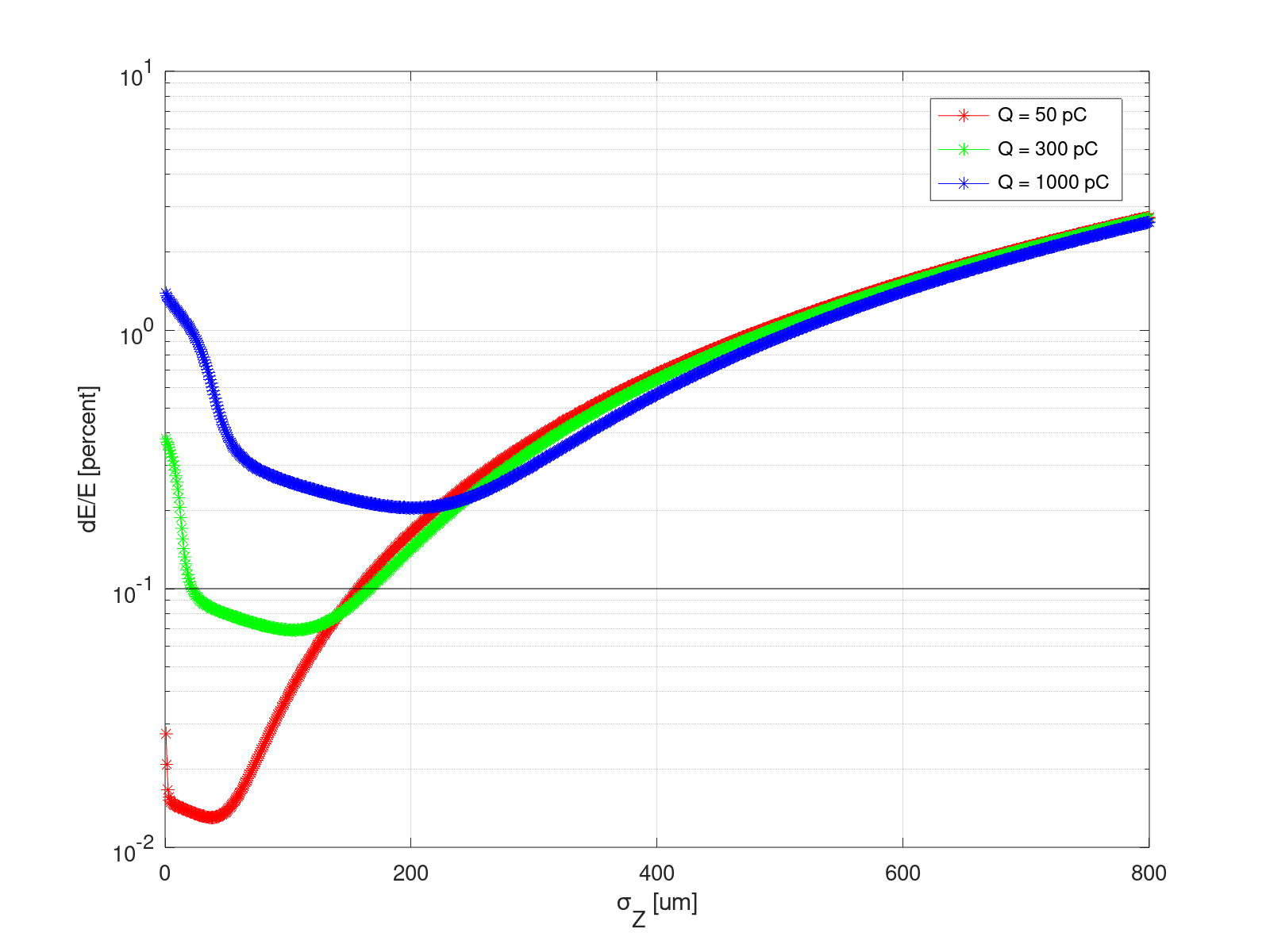}\\
\caption{Minimal energy spread achievable at the linac end as a function of the RMS bunch length, for different bunch charges. The black line indicates the required value of 0.1\% for injection in the SPS.}
\label{fig:espread_bunchlength}
\end{center}
\end{figure}

The wakefields generated in the X-band structures induce both emittance growth in the transverse plane and an increase in the correlated energy spread of the bunch. The effect on the energy spread can be partly compensated by running off crest in the phase of the accelerating structures with the optimal RF phase depending on both the charge and the bunch length. The requirement on the energy spread of 0.1\% at the end of the linac, for injection in the SPS, limits the maximum bunch length allowed for a given specific charge. Figure~\ref{fig:espread_bunchlength} shows the minimum energy spread achievable at the end of the linac as a function of the RMS bunch length, for different bunch charges. A bunch with charge 50\,pC and length 150\,$\mu$m reaches its minimum energy spread when the RF structures operate nearly on crest. A bunch charge of 1000\,pC can reach at best 0.2\% energy spread at a bunch length of approximately 200\,$\mu$m for an RF phase of $-7$ degrees.

Transverse misalignment of the elements also induce emittance growth, through two mechanisms: spurious dispersion due to off-axis quadrupoles, and transverse wakefield kicks due to the beam travelling off-axis through the accelerating structures. BNS damping can be considered to stabilise the beam, but it wasn't deemed necessary given the relatively large emittance requirements and the overall good robustness of the linac to such effects.

\subsubsection{Multi-bunch effects}
\begin{figure}[!hbt]
\begin{center}
    \includegraphics[width=0.6\textwidth]{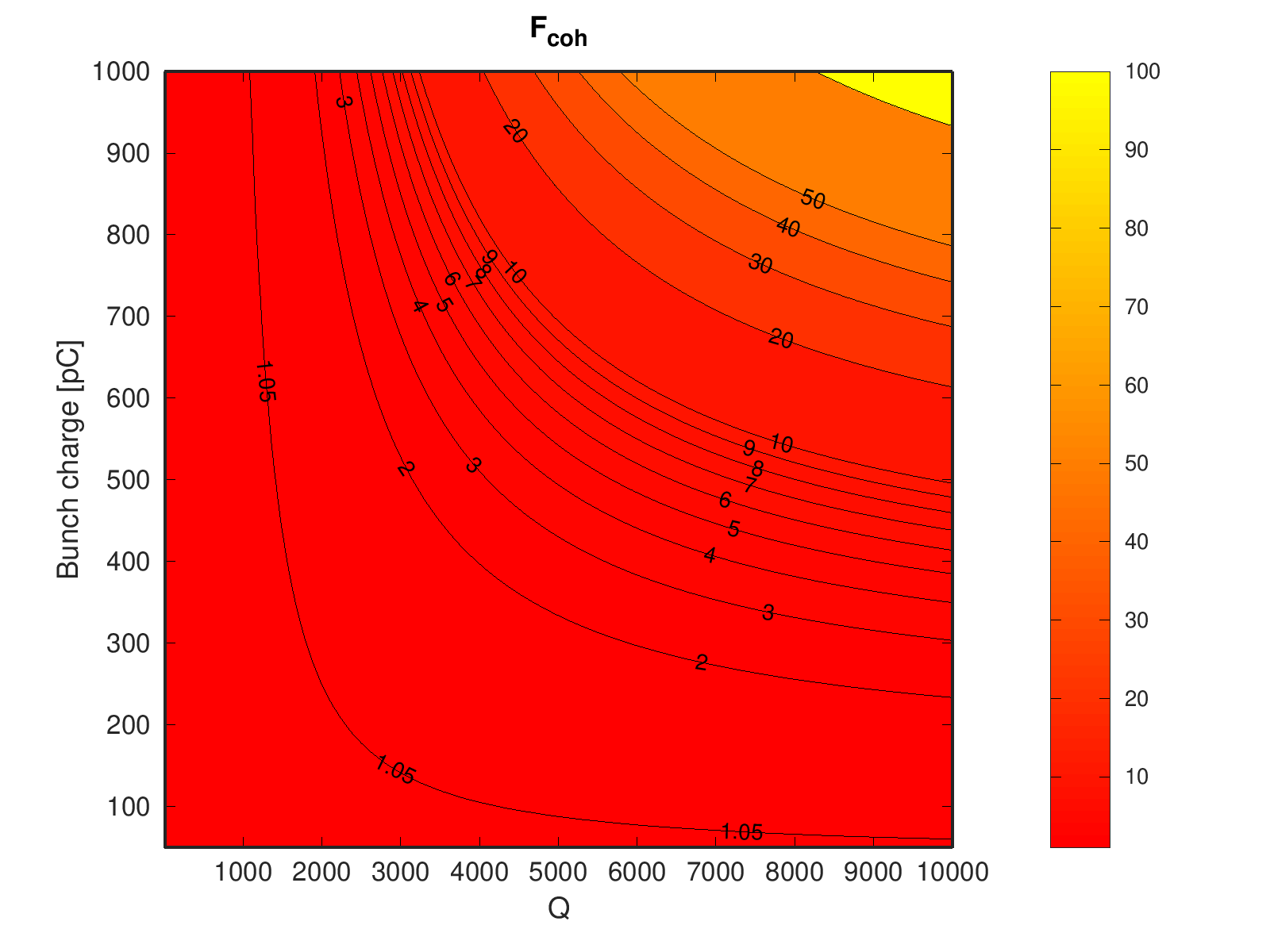}
    \caption{Multi-bunch operation: average beam offset amplification due to long-range wakefields  as a function of bunch charge and the structure's $Q$ factor, due to a coherent incoming offset of the bunch train. Bunch charges below 300\,pC show a factor 2 amplification even without HOM damping.}
 \label{fig:LRMB_coherent}
\end{center}
\end{figure}

Considerable advancement in the understanding and control of the effects of long-range wakefields on the beam stability has been made throughout the last few decades, for example within the context of the CLIC Study, supported by experimental verification at SLAC/FACET~\cite{Zha2016}. The tools developed for the CLIC Study have been used to evaluate the maximum bunch charge that can be transported through the 3.5\,GeV electron linac. The natural suppression of the long range wakefields, due to the tapering of the iris aperture, also guarantees beam stability in multi-bunch operation for all 40 bunches in the nominal 200\,ns-long train. A bunch charge of up to 300\,pC can be transported with negligible emittance growth through the linac without high-order mode damping. Semi-analytical estimations show that the average amplification factor of incoming beam offsets as a function both of the bunch charge and of the structure's $Q$ factor, in case of a coherent offset of all bunches in the train: even with no damping (that is, small $Q$), a charge of 300\,pC is transported through the linac with a small amplification factor of less than 2, as shown in Fig.~\ref{fig:LRMB_coherent}.

\subsubsection{Linac operational modes}\label{subSec:LinacOpModes}
Four operational modes have been studied:
\begin{itemize}
\item \emph{nominal:} for a missing momentum experiment, trains of 40 bunches of 50\,pC, spaced by 5\,ns are accelerated;
\item \emph{high-charge:} trains of 40 bunches with charge up to 300\,pC and bunch length about 200~$\mu$m are accelerated;
\item \emph{very-high-charge:} single bunches with 1\,nC charge and length about 750~$\mu$m are accelerated;
\item \emph{plasma R\&D:} for plasma acceleration R\&D, single bunches of 1.7\,nC and length less than 250\,$\mu$m are accelerated.
\end{itemize}
In the latter case the goal is to achieve the required high electron density at the plasma cell. Longitudinal bunch length compression would need to be implemented though a conventional 4-bend magnetic chicane. The initial and final bunch parameters, of all cases, are listed in Table~\ref{tab:linac_beam_params}. Other operational scenarios could be considered, like filling all buckets of 800 MHz RF system in the SPS: 160 bunches, 50\,pC (or 50/4), $\approx$\,1.25\,ns bunch spacing.
\begin{table}[!hbt]
	\centering
	\caption{Beam parameters in the linac.}
	\label{tab:linac_beam_params}
\begin{tabular}{lcccc}
\hline\hline
\textbf{Parameter}  & \textbf{Nominal}  & \textbf{High-charge}  & \textbf{Very-high-charge}  & \textbf{Plasma R\&D}\tabularnewline
\hline
Bunch charge {[}pC{]}  & 50  & 300  & 1000  & 1700\tabularnewline
Bunch length {[}$\mu$m{]}  & 150  & 200  & 750  & 250\tabularnewline
Bunches per train {[}\#{]}  & 40  & 40  & 1  & 1\tabularnewline

Normalised emittance {[}$\mu$m{]}  & \multicolumn{4}{c}{< 100 }\tabularnewline

Initial energy {[}MeV{]}  & \multicolumn{4}{c}{200}\tabularnewline
Initial relative energy spread {[}\%{]}  & \multicolumn{4}{c}{1 }\tabularnewline

Final energy {[}MeV{]}  & \multicolumn{4}{c}{3500}\tabularnewline
Final relative energy spread {[}\%{]}  & $<0.1\%$  & $0.2\%$ & 3\% & 0.5\%\tabularnewline
\hline\hline
\end{tabular}
\end{table}

\subsubsection{Injector simulations}
\par In order for the eSPS linac to operate in each of the four operational modes mentioned in Section~\ref{subSec:LinacOpModes}, the injector must be flexible enough to provide beams of several different types, the parameters of which are summarised in Table~\ref{tab:linac_beam_params}.  The beam dynamics in the injector are dominated by space-charge effects when the beam is at a low energy in the gun and bunching cavity. The beam parameters can be adjusted by altering the pulse length and spot size of the laser, the electric field strength and phase of the RF gun and bunching cavity and, the magnetic field strength of the gun and buncher solenoids.  The current instillation of the CLEAR injector uses a laser with a fixed pulse length of 4.7\,ps, and fixed electric field strengths of 80\,MV/m and 18\,MV/m in the gun and bunching cavity respectively.

\par The beam in the injector was simulated using the code ASTRA, to calculate the effect of space-charge forces~\cite{Floettmann2000AAlgorithm}.  Any wakefield effects in the S-band linac were considered negligible so were not included in simulation. Any misalignments and losses were also omitted. To verify the validity of the ASTRA model, the simulated beam was compared to experimental measurements of the beam in the CLEAR user facility. Different bunch charges of up to 2000\,pC with laser spot sizes of an RMS radius of up to 1.2\,mm were investigated.

\par
\begin{figure}[!hbt]
    \centering
    \includegraphics[width=0.8\linewidth]{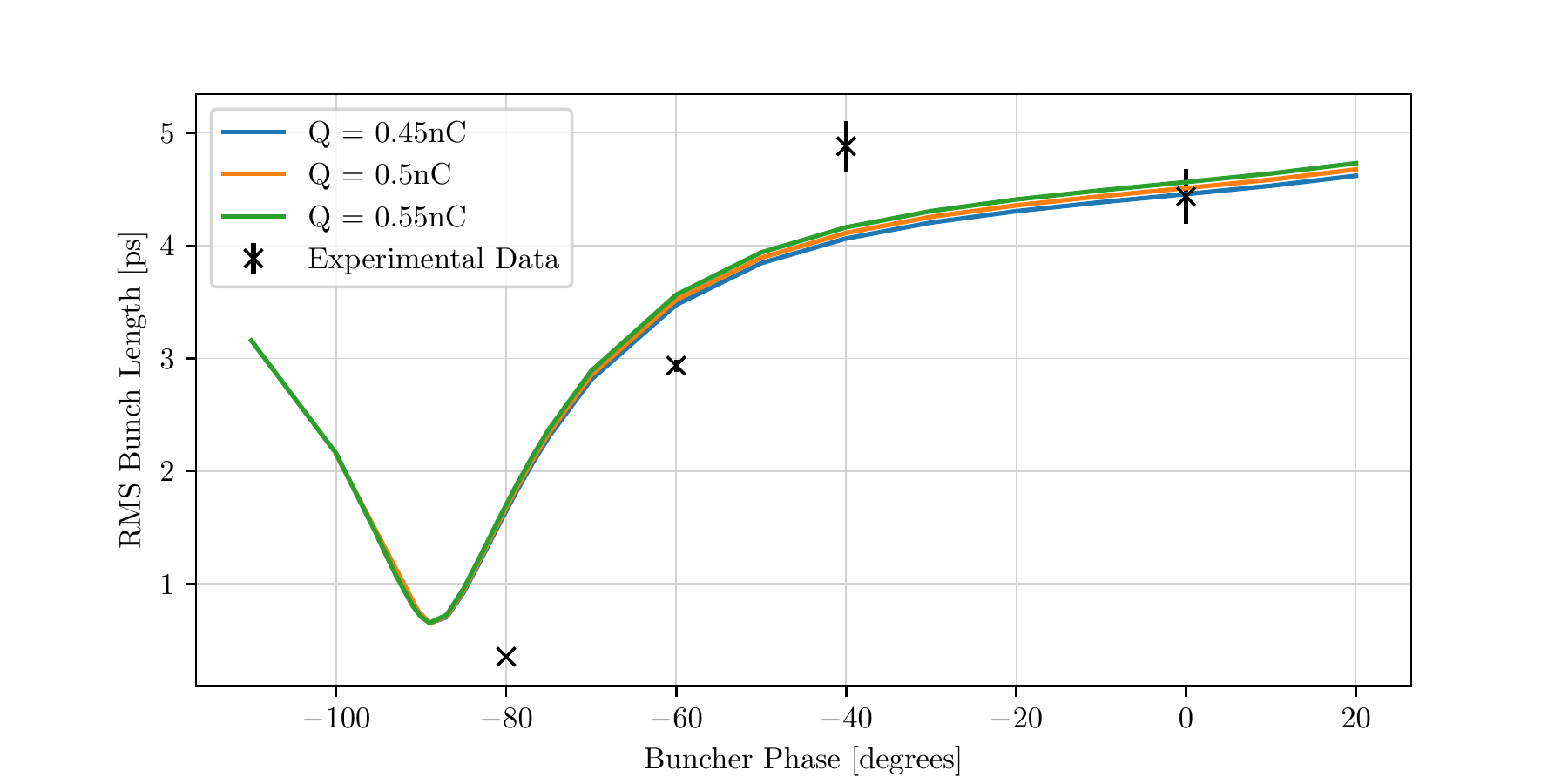}
    \caption{A comparison of simulated and experimentally measured bunch lengths at different phases of the bunching cavity on CLEAR for a laser spot size of 0.6\,mm.  The phase, $0^\circ$, is defined as the phase of maximum energy gain.}
    \label{fig:CLEARBunchLengthExperiment}
\end{figure}

\begin{figure}[!hbt]
    \centering
    \includegraphics[width=0.8\linewidth]{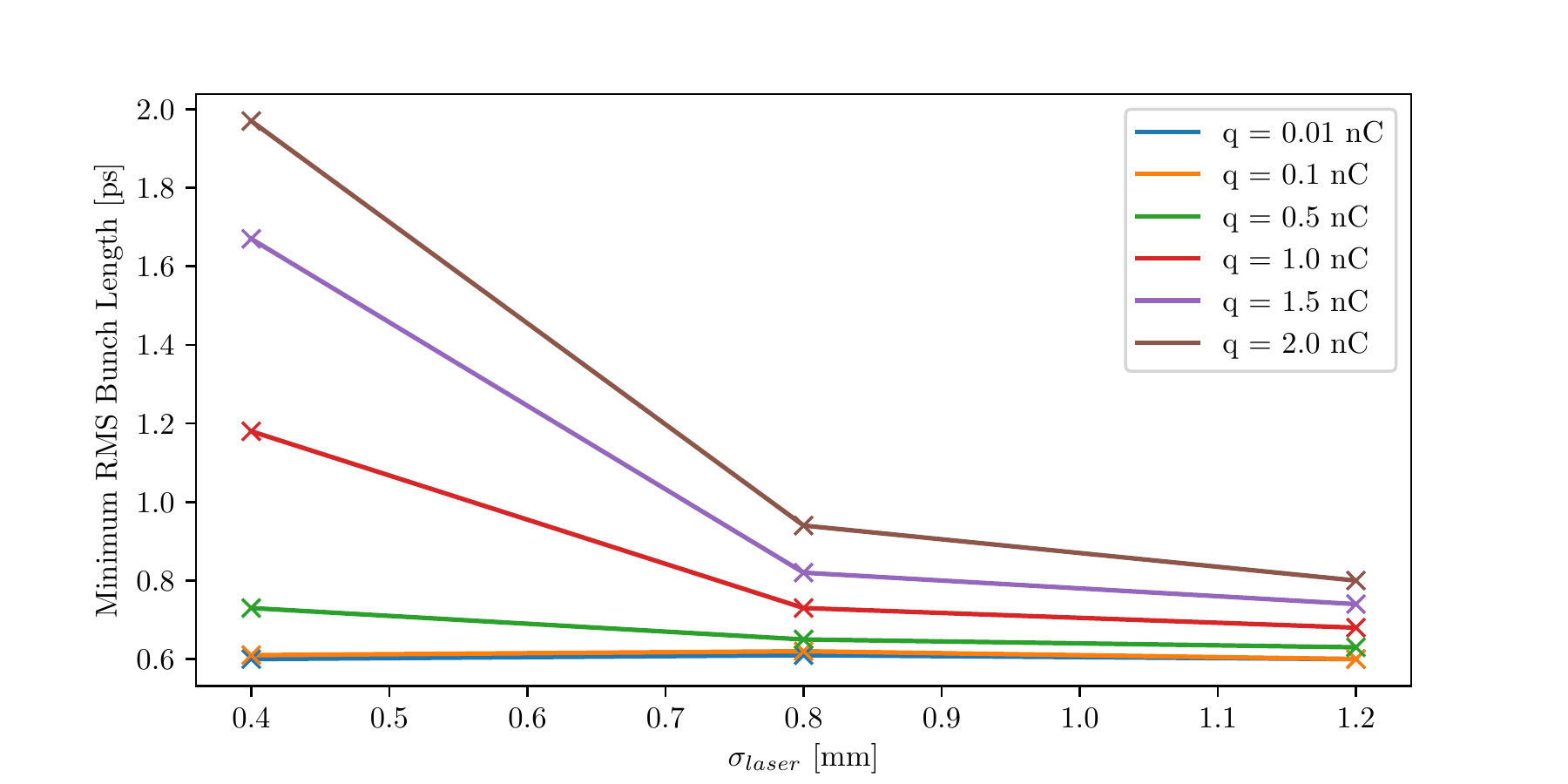}
    \caption{Simulated bunch lengths in CLEAR, at maximum compression for different laser spot sizes and different charges.  }
    \label{fig:CLEARBunchLengthSimulation}
\end{figure}

 When accelerated on the maximum energy phase in the bunching cavity the bunch length stays around the 4.7\,ps ($\sim$\,1400 $\mu$m) of the laser pulse.  To achieve shorter bunches the phase of the buncher is reduced, with maximum compression achieved around zero crossing, $-90^\circ$ below crest.  The simulated bunch length at different phases of the buncher gave reasonable agreement with experiment up to a charge of 500\,pC on CLEAR, shown in Fig.~\ref{fig:CLEARBunchLengthExperiment}.  
The minimum bunch length achievable depends on the bunch charge, due to the increased space-charge forces in higher charge bunches.  To reduce the effect of the space charge forces, thus to maximise compression, the size of the laser spot can be increased.  The minimum lengths achievable in simulation for different laser spot sizes and bunch charge are shown in Fig.~\ref{fig:CLEARBunchLengthSimulation}.
In these simulations it is assumed that the RF gun is operating at the peak energy phase.  By de-phasing the gun by up to $\sim$\,$20^\circ$, shorter bunch lengths can be achieved for bunches of charge less than $\sim$\,500\,pC. For the nominal 50\,pC case, a bunch length of less than 0.1\,ps (30 $\mu$m) can be achieved in both simulation and experiment.  Without any gun de-phasing the bunch lengths required for the high-charge and very-high-charge modes can be satisfied.  The bunch length for the Plasma R\&D beam may be able to be achieved with velocity bunching alone, but the tolerance will be very fine.  Increased compression with a magnetic chicane may be desirable for this operational mode. 

\par When beams are accelerated with the crest energy phase in the buncher the energy spread of each bunch simulated is less than the required 1\%.  When compressing, the energy spread grows.  When operating at maximum compression phase in the buncher, the energy spread of the beam after the buncher is less than 4\%, for each bunch up to 2000\,pC.  After the two following accelerating structures, the energy spread is reduced to less than 1\% for each bunch charge.  It must be noted that for compression phases in the buncher between the peak energy and maximum compression phases the energy spread is larger than 1\%, with a maximum of $\sim$\,4\% at the end of the injector.

\par The emittance tolerance for the eSPS linac in each of the operational modes is quite large at 100\,$\mu$m. When accelerating on crest in the bunching cavity, the emittance for bunches up to a charge of 2000\,pC remains below 20\,$\mu$m.  When operating at a compression phase in the buncher, the emittance grows relative to the peak energy phase by around a factor $\sim$\,3.  This growth can be suppressed by optimising the strength of the buncher solenoid.  After an optimisation, the emittance growth is reduced to a factor $\sim$\,1.7, with a maximum emittance of 35\,$\mu$m for a 2000\,pC beam produced with a laser spot of 1.2\,mm.

\par There are likely to be several small upgrades made to the CLEAR injector for use as the eSPS linac injector.   It is proposed to use two independent klystrons to power the gun and bunching cavity, instead of the current setup of one. Therefore, the electric fields in each would be able to be adjusted independently.  A higher field in the gun would reduce the emittance growth and bunch lengthening due to space-charge effects.  The gun would also be more able to operate at a lower phase for high charge bunches allowing compression in the gun.  The added klystron would also allow the optimisation of the buncher field to maximise compression. The use of a new laser, with a shorter laser pulse, would also enable the creation of shorter electron bunches.  The emittance of the beam could also be significantly reduced by optimising the drift length between the gun and the buncher.

\subsubsection{Start-to-end simulation results}
\begin{figure}[!hbt]
\begin{center}
Q = 50 pC \textit{(nominal)}\\
\includegraphics[width=0.8\textwidth]{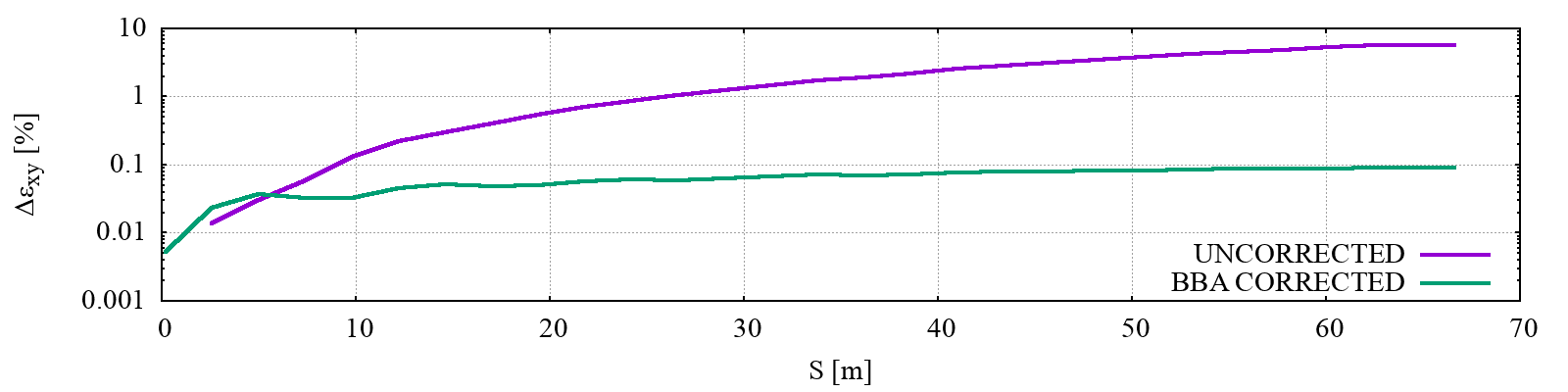}\\
Q = 300 pC \textit{(high-charge)}\\
\includegraphics[width=0.8\textwidth]{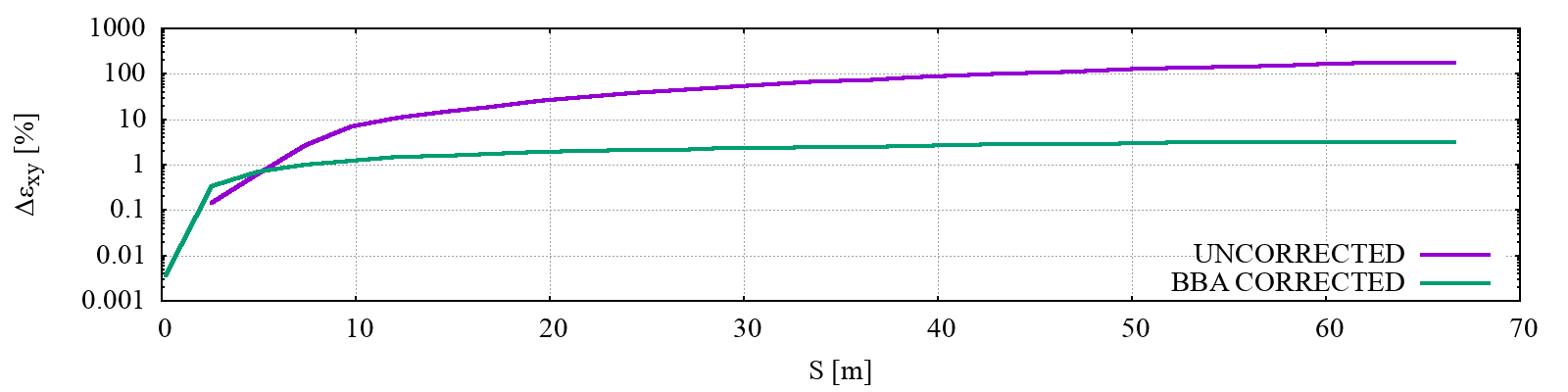}\\
Q = 1000 pC \textit{(very-high-charge)}\\
\includegraphics[width=0.8\textwidth]{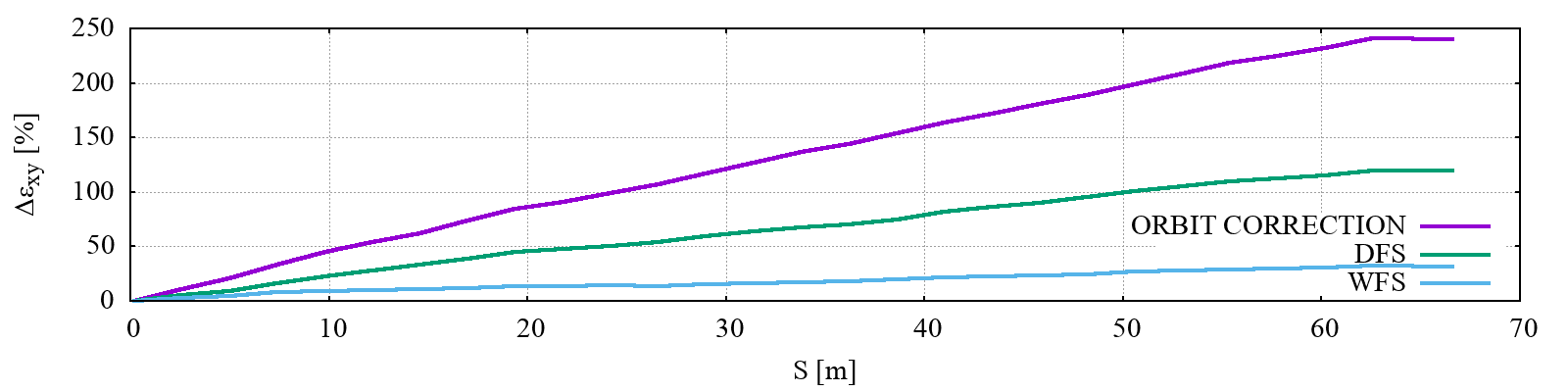}\\
Q = 1700 pC \textit{(plasma R\&D)}\\
\includegraphics[width=0.8\textwidth]{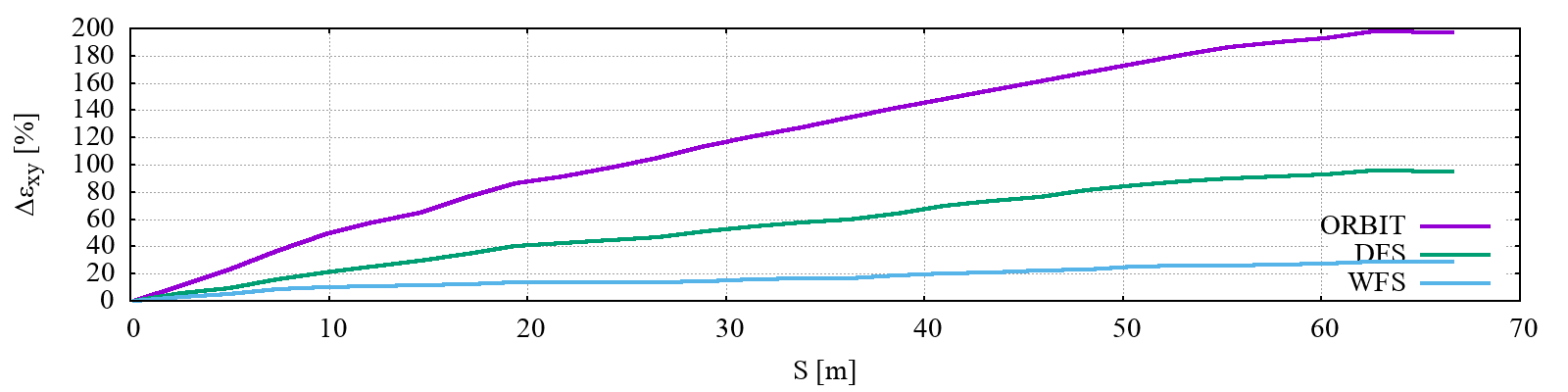}\\
\caption{Relative emittance growth along the linac in presence of static imperfections, before and after beam-based alignment. In the nominal and high-charge cases, the emittance growth remains below 10\%. In the case of very-high-charge and plasma-R\&D, WFS is needed to reduce the emittance growth to below 20\% and 40\%, respectively.}
\label{fig:emitt_bba}
\end{center}
\end{figure}

In the presence of imperfections beam-based alignment techniques must be applied in order to preserve beam quality. The alignment procedure follows the experience made in the context of linear collider studies, and relative experimental tests, and consist of three steps to be applied in cascade:
\begin{enumerate}
	\item Orbit correction, where dipole correctors are used to centre the beam through the BPMs;
	\item Dispersion-Free Steering (DFS), where the dipole correctors are used to remove residual dispersion introduced by misaligned quadrupoles;
	\item Wakefield-Free Steering (WFS), where the dipole correctors are used to remove wakefield kicks introduced by misaligned accelerating structures.
\end{enumerate}
In order to perform trajectory correction, it has been assumed that all quadrupoles are equipped with a beam position monitor, and corrector coils to deflect the beam transversely in both horizontal and vertical direction.
Table~\ref{tab:bba_imperfections} lists the RMS imperfections that have been considered. All simulation results shown are the average of 100 randomly misaligned machines, corrected with beam-based alignment. Figure~\ref{fig:emitt_bba} shows the relative emittance growth in presence of static imperfections for the four operational scenarios described. 

\begin{table}[!hbt]
    \begin{center}
    \caption{List of imperfections.}
    \label{tab:bba_imperfections}
        \begin{tabular}{lc}
        \hline\hline

        \textbf{Imperfection} & \textbf{RMS Value}  \\ 
        \hline
Quadrupole offset [$\mu$m] & 100  \\ 
Quadrupole pitch [$\mu$rad] & 100   \\ 
Structure offset [$\mu$m] & 100  \\ 
Structure pitch [$\mu$rad] & 100   \\ 
BPM offset [$\mu$m]  & 100  \\ 
BPM resolution [$\mu$m] & 10   \\ 
        \hline\hline 
        \end{tabular}
    \end{center}
\end{table}

Given the relatively large initial emittance, the impact of errors is small in both the nominal and high-charge cases; in the case of very-high-charge or plasma R\&D case the application of beam-based alignment techniques is necessary to reduce the emittance growth. 

%% file: include/03-LINAC/GunInjector.tex
\subsection{Gun and injector}
\label{sec:LINAC_GunInjector}


The CERN Linear Electron Accelerator for Research (CLEAR) injector operating at CERN can be used basically unchanged to produce the electron beam with the specifications required for missing momentum and beam dump experiments as well as for many other potential applications~\cite{GAMBA2017}. 

\subsubsection{The CLEAR injector linac}
\begin{figure}[!hbt]
\begin{center}
    \includegraphics[width=0.9\textwidth]{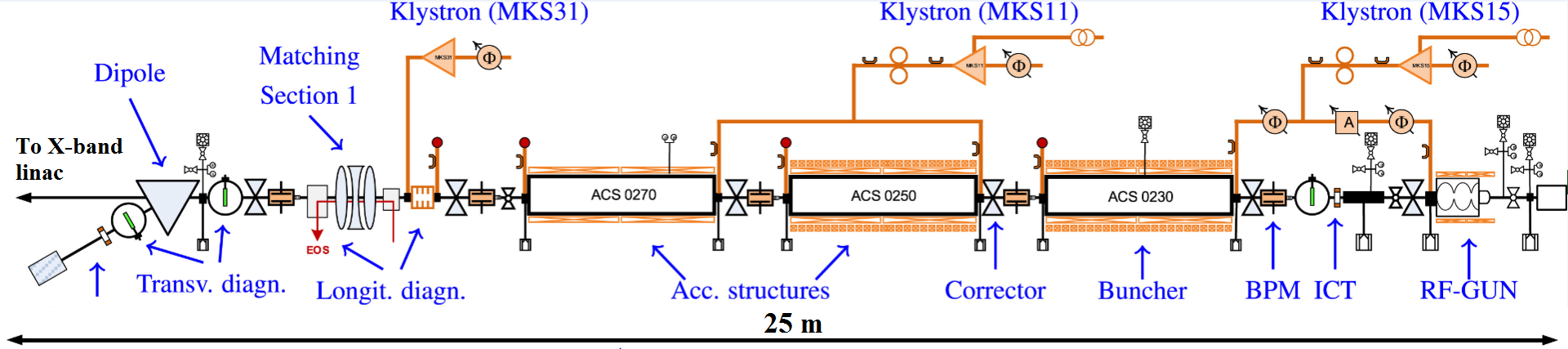}
    \caption{Layout of the CLEAR injector. The electron beam travels from right to left~\cite{GAMBA2017}.}
 \label{fig:CLEAR_injector_layout}
\end{center}
\end{figure}

The present installation is about 25\,m long and its schematic layout is shown in Fig.~\ref{fig:CLEAR_injector_layout}. The electron bunches are generated on a Cs$_{2}$Te photo-cathode by a pulsed UV laser.
This allows the generation of a beam with arbitrary time structure with the minimum bunch spacing of 0.33\,ns given by the 3\,GHz RF frequency of the gun (at present the spacing is limited to multiples of 0.66\,ns by the laser system, a modification will be needed to be able to get multiples of 0.33\,ns). The gun is followed by three LEP Injector Linac (LIL) 4.5\,m-long accelerating structures which are used for beam bunching and acceleration. The first structure can be used to vary the bunch length from 0.3 to 1.2\,mm rms by means of velocity bunching. The gun, buncher and first accelerating structure are inside solenoid magnets which provide tuneable focusing and space charge compensation. A matching section with three tuneable quadrupoles and a spectrometer line complete the injector. The range of beam parameters which can be obtained at the end of the CLEAR injector are summarised in Table~\ref{tab:CLEARinjectorPars}. 

\begin{table}[!hbt]
\begin{center}
\caption{Beam parameters at the end of the CLEAR injector.}
\label{tab:CLEARinjectorPars}
\begin{tabular}{p{6cm}cc}
\hline\hline
\textbf{Parameter}          & \textbf{Value range}   	& \textbf{Value for eSPS}\\
\hline
Energy [MeV]  				& 50 to 250           		& 200 \\
Bunch charge [nC]           & 0.001 to 1.5         		& 0.05 \\
Norm. emittance [$\mu$m]    & $\sim$\,3 for 0.05\,nC/bunch 		& 3 \\
							& $\sim$\,20 for 0.4\,nC/bunch   	&  \\
Bunch length rms [mm]     	& 0.3 to 1.2 				& 0.8 \\
Energy spread rms [\%]    	& below 0.2 				& 0.1 \\
Number of bunches     		& 1 to 200             		& 40 \\
Micro-bunch spacing [ns]	& multiple of 0.33			& 5 \\
\hline\hline
\end{tabular}
\end{center}
\end{table}

\subsubsection{Implementation of the CLEAR injector in the eSPS linac}

The present installation is located in B2010, with some technical equipment being hosted in adjacent building. All the equipment will be relocated in TT5 and in B183. A few upgrades and modifications of the CLEAR linac are also planned, the main being the implementation of a new Modulator/Klystron station, in order to provide more operational flexibility. At present, the RF gun and the first acceleration structure, used as a buncher, are fed by a single klystron (see Fig.~\ref{fig:CLEAR_injector_layout}). In the new configuration, they will be independently fed by two klystrons, giving the possibility of completely independent adjustment of the RF amplitude and phase, the use of compression only in the buncher, and slightly increasing the maximum obtainable beam energy. The present laser system should also be modified in order to provide pulses with a spacing of 0.33\,ns instead of the present value of 0.66\,ns. A likely better alternative is to substitute it with a new system, simplified and adapted to the eSPS requirements. 
Other minor upgrades and rearrangements will likely concern beam diagnostics and ancillary equipment. Some of these upgrades are under study for the next few years of CLEAR operation, therefore, the details will depend on their advancement at the time of installation in TT5.

\begin{figure}[!hbt]
\begin{center}
    \includegraphics[width=.85\linewidth]{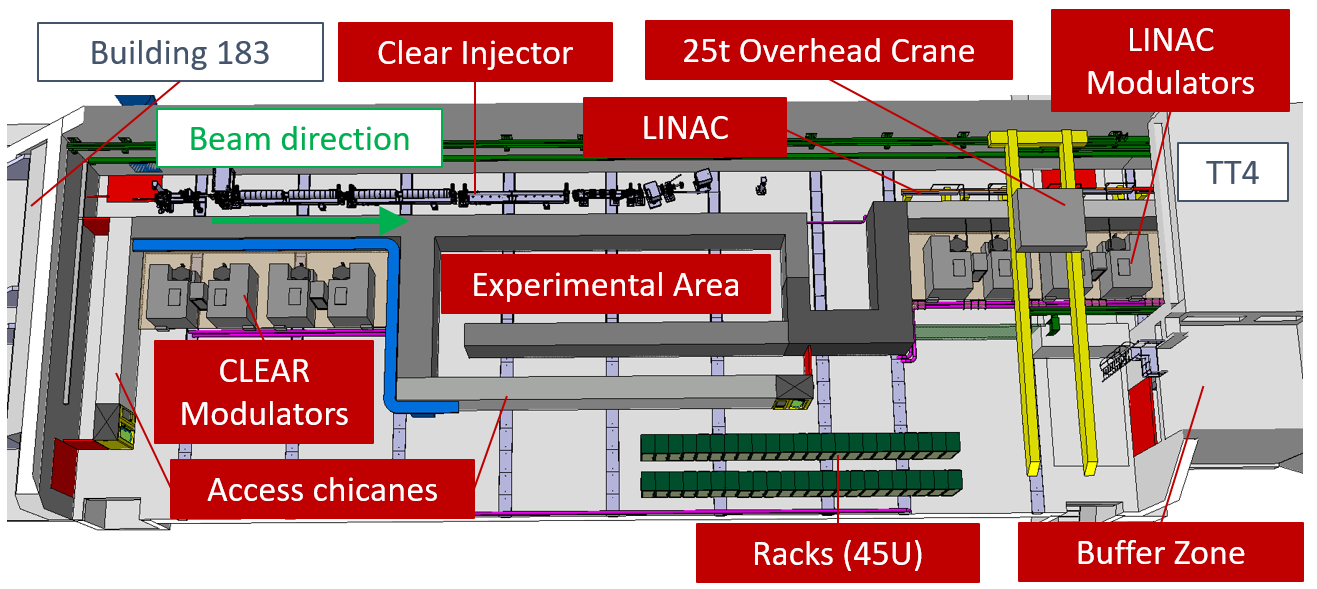}
    \caption{TT5 Layout.}
 \label{fig:TT5_Layout_Injector}
\end{center}
\end{figure}

In the new implementation layout (see Fig.~\ref{fig:TT5_Layout_Injector}) a 10\,m free space has been reserved after the CLEAR injector and before the linac, to host dedicated beam  diagnostics and possibly a three bends chicane, which will add flexibility for bunch compression especially at high bunch charge.
In this space, a beam line will branch off the main one and give the possibility to send the beam back, through a 180\,degrees achromatic arc cell, towards the low energy experimental beamline, hosted in an independently shielded zone within TT5. 

%% file: include/03-LINAC/X-BandLINAC.tex
\subsection{X-band linac design}
\label{sec:LINAC_Design}


Unlike the injector where S-band (3\,GHz) RF structures are used at a relatively low accelerating gradient of about 15\,MV/m in order to obtain the desired beam parameters, high gradient X-band (12\,GHz) RF accelerating structures are used in the linac in order to make it compact and accelerate the beam from 0.2\,GeV up to 3.5\,GeV within 70\,m. A high gradient is required due to limited space in the TT4/TT5 area at CERN. The X-band high-gradient RF technology has been developed in the framework of the CLIC study and is now being widely adopted. The high-gradient X-band RF systems have been extensively operated at CERN Xbox1, 2 and 3, as well as in several linac based light sources: SwissFEL, FERMI, and LCLS, where it is used for beam manipulation and beam diagnostic purposes. 

\begin{figure}[!hbt]
\begin{center}
    \includegraphics[width=7cm]{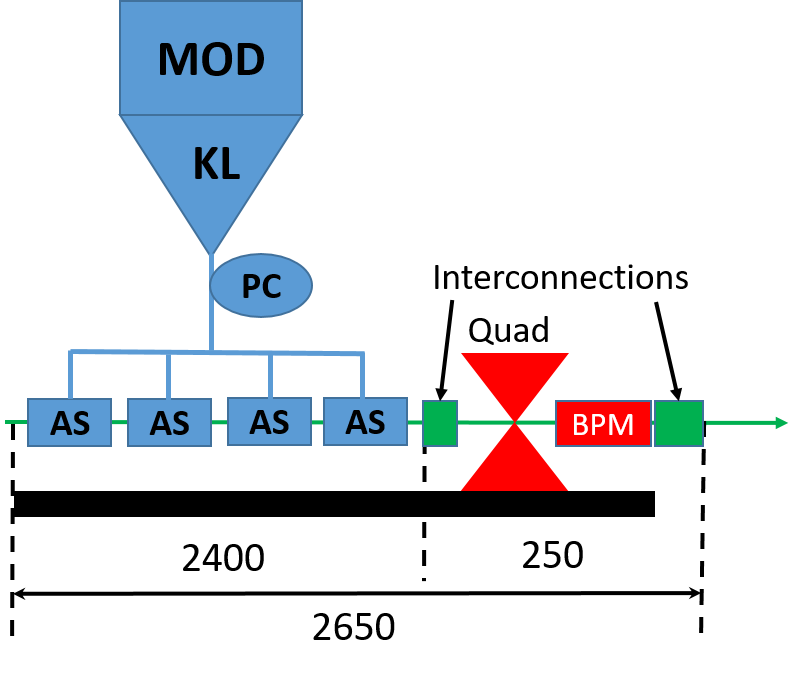}
    \caption{Schematic layout of one X-band RF unit. Dimensions are given in mm.}
 \label{fig:XbandRFunit}
\end{center}
\end{figure}

For the klystron-based option of the first stage of CLIC at 380\,GeV~\cite{Aicheler2019}, an average loaded acceleration gradient of 75\,MV/m (95\,MV/m unloaded) has been chosen as a compromise between making the main linac as short as possible and reducing the required peak power and associated number of klystrons. About one klystron per metre is necessary to feed the klystron-based CLIC main linac. This very ambitious specification requires development of a special compact modulator unit accommodating two high-efficiency klystrons. Going to slightly lower gradients for the eSPS reduces the peak power per metre and simplifies the integration in the available space in TT4/TT5 area. This makes it possible to use commercially available klystron/modulator units used, for example, in XBOX2 facility at CERN. The schematic layout of the RF unit is presented in Fig.~\ref{fig:XbandRFunit}.
The total length of one RF unit is 2650\,mm including 2400\,mm for four accelerating structures and 250\,mm for magnetic quadrupole, BPM and interconnections.  Section~\ref{sec:LINAC_Integration} provides more details on the integration of different components into an eSPS module and how the R\&D done for the klystron-based option of CLIC has been used. In particular, for the RF waveguide (WG) network connecting X-band klystron with four accelerating structures components designed for CLIC has been used. Based on this, total power loss from the klystron to the input of the accelerating structures has been estimated to add up to about 13\%. Table~\ref{tab:XbandRFloss} summarises losses in each component and also indicates whether single or double height is used.   
\begin{table}[!hbt]
\begin{center}
\caption{Power loss in the X-band RF WG network.}
\label{tab:XbandRFloss}
\begin{tabular}{p{6cm}cc}
\hline\hline
\textbf{Component}          	& \textbf{Double height [Y/N]}   	& \textbf{Power loss [\%]}\\
\hline
Long straight with bend			& Y           				& 4.5 \\
CCC                 			& Y             			& 2.3 \\
BOC             				& Y                         & 2.2 \\
3dB-Hybrid                   	& Y                			& 0.3 \\
Directional coupler             & Y 						& 0.3 \\
Straight with 2 bends       	& Y 						& 2.2 \\
3dB-Hybrid      	       	    & N         				& 0.5 \\
Straight with bend    	       	& N         				& 0.5 \\
\hline
Total               	       	&          				& 12.8 \\
\hline\hline
\end{tabular}
\end{center}
\end{table}

\begin{figure}[!hbt]
\begin{center}
    \includegraphics[width=8cm]{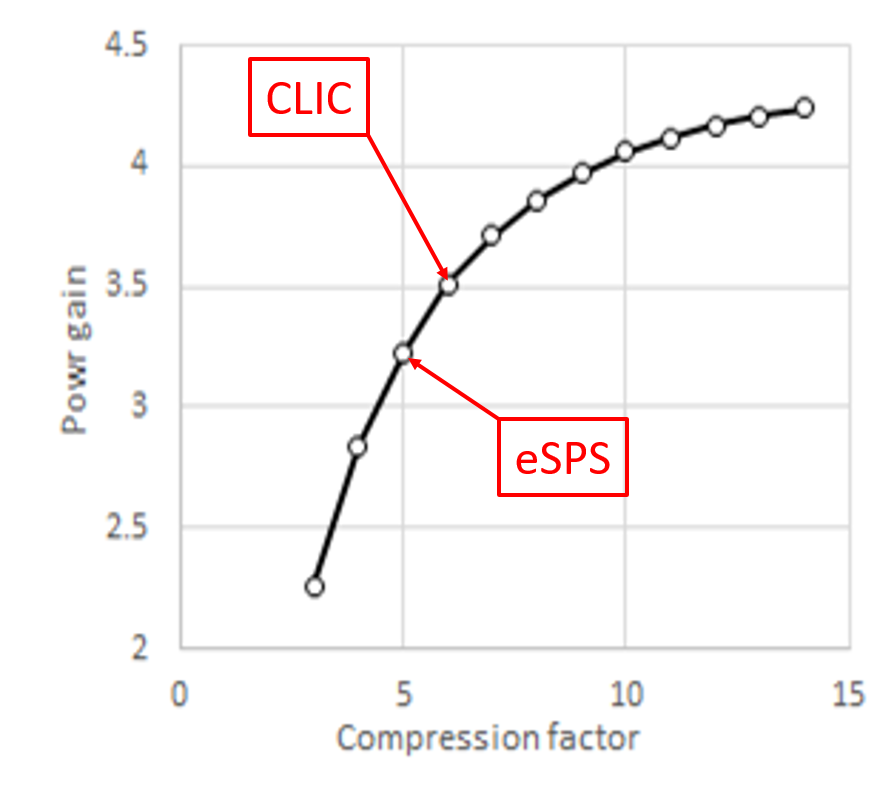}
    \caption{Power gain versus compression ratio for pulse compression system based on BOC + CCC is shown.}
 \label{fig:RFpulseCompression}
\end{center}
\end{figure}

The RF WG network includes an RF pulse compression system designed for klystron-based CLIC that includes SLED-type pulse compressors based on Barrel Open Cavity (BOC) with Correction Cavity Chain (CCC). Power gain versus compression ratio curve calculated for CLIC and a pulse shape with a flattop are shown in Fig.~\ref{fig:RFpulseCompression}.

For eSPS it is envisaged to use commercially available CPI VKX-8311A tubes. One tube can generate 50\,MW, 1500\,ns RF pulses at 100\,Hz repetition rate. Based on this klystron pulse length a compression ratio of five has been identified as good compromise between achievable power gain (3.2) and compressed pulse length (300\,ns) available for filling acceleration structure (AS) (100\,ns) and accelerating the train of bunches (200\,ns).   

\begin{figure}[!hbt]
\begin{center}
    \includegraphics[width=10cm]{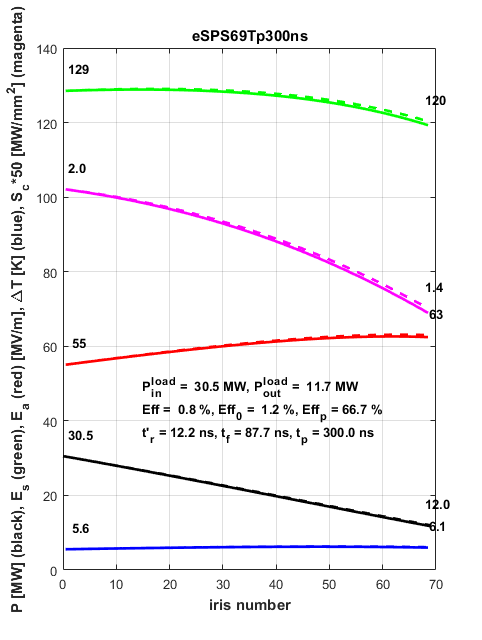}
    \caption{Distribution of accelerating gradient (red), surface electric field (green), $S_c$ (magenta), power (black) and pulse surface temperature rise (blue) along the accelerating structure is shown for 60\,MV/m average loaded accelerating gradient (solid lines). The unloaded case is also shown in dashed lines.}
 \label{fig:XbandASpars}
\end{center}
\end{figure}

The total flange to flange length of one accelerating structure is 600\,mm. Using the CLIC design for structure interconnection, an active length of 575\,mm per structure can be used for the RF design. With these two constraints on the filling time and the active length. The accelerating structure has been designed as a quasi constant gradient structure with linear tapering of the iris radius. In Fig.~\ref{fig:XbandASpars}, the distribution of surface field quantities, accelerating gradient, and power along the structure for a loaded gradient of 60\,MV/m averaged over the structure active length are shown (solid lines) for the nominal train of 40 bunches of 50\,pC every 5\,ns for a missing momentum experiment.
The dashed lines show the unloaded case, which is very close to the loaded one since the peak beam current in the train is very low (10\,mA) and the beam loading effect is very weak and hardly visible on the plot.

The parameters of the X-band linac for an energy gain of 3.3\,GeV are summarised in Table~\ref{tab:XbandLinacPars} for two cases: the nominal case with 24 RF units in operation and the sub-nominal case where one RF unit is defective and the corresponding energy gain is compensated by increasing the gradient in the remaining 23 RF units. 

\begin{table}[!hbt]
\begin{center}
\caption{Parameters of the X-band linac for energy gain of 3.3\,GeV.}
\label{tab:XbandLinacPars}
\begin{tabular}{p{7cm}cc}
\hline\hline
\textbf{Parameter}          	& \textbf{Nominal}   	& \textbf{Fault}\\
\hline
Klystron pulse length [ns]		& 1500           		&  \\
Power loss in WG [\%]			& 13           			&  \\
Compression factor              & 5           			&  \\
Power gain                   	& 3.2          			&  \\
\hline
Number of RF units              & 24 					& 23 \\
Energy gain per RF unit [MeV]	& 137.5					& 143.5 \\
Average acc. gradient [MV/m]   	& 59.8					& 62.4 \\
Required peak power per klystron [MW]   		& 43.5 					& 47.4\\
\hline\hline
\end{tabular}
\end{center}
\end{table}

In both cases there is a large enough operational margin in the required klystron peak power for an overall energy gain of 3.3\,GeV since it is significantly lower than the klystron maximum specified power of 50\,MW.   For operation in single bunch mode the compressed pulse is shorter (approx. 100\,ns) and higher compression factors can be used. However, since the RF pulse compression system and accelerating structure parameters (summarized in Table~\ref{tab:XbandASPars}) are optimised for multi-bunch operation and cannot be adjusted for the single bunch mode the improvement in the effective power gain is limited and estimated to go up to 3.5 only. 
\begin{table}[!hbt]
\begin{center}
\caption{Parameters of the X-band accelerating structure.}
\label{tab:XbandASPars}
\begin{tabular}{p{6cm}cc}
\hline\hline
\textbf{Parameter}          	&\textbf{First cell}    & \textbf{Last cell}\\
\hline
Aperture iris radius [mm]		& 3.7                   & 2.7 \\
Iris thickness [mm]		            & 1.35 & 1.35 \\
Q-factor 		            & 7090 & 7020 \\
Group velocity [\% of $c$] 		            & 3.6 & 1.34 \\
$R'/Q$ [k$\Omega$/m] 		            & 14.3 & 17.5 \\
\hline
Number of cells		            & 69 \\
Active length [mm]		            & 575 \\
Input power for <60MV/m> [MW]		            & 30.5 \\
Rise time (1/bandwidth) [ns]                         & 12 \\
Filling time [ns]                         & 88 \\
\hline\hline
\end{tabular}
\end{center}
\end{table}
This results in a rather small increase in the energy gain in the X-band linac for the same klystron input power. Keeping the same operational margin for nominal parameters in Table~\ref{tab:XbandLinacPars}, the beam energy goes up from 3.3\,GeV to 3.45\,GeV only.

\begin{figure}[!hbt]
\begin{center}
    \includegraphics[width=12cm]{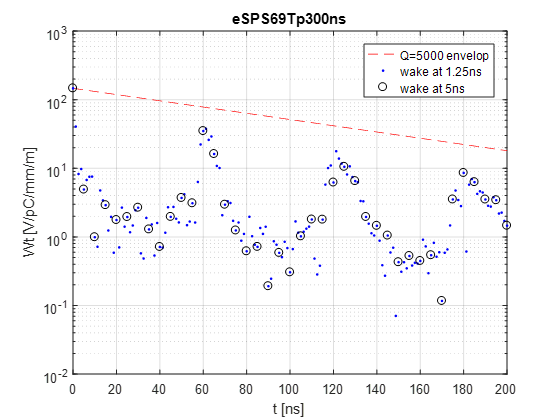}
    \caption{Magnitude of the transverse wakefields in AS is shown for the first dipole mode at the position of the bunches with 5\,ns spacing (circles) and 1.25\,ns (dots). Dashed line is the envelop of Q~=~5000 mode for comparison.}
 \label{fig:XbandAStransWake}
\end{center}
\end{figure}

The structure is designed without any higher order mode suppression features, except the detuning caused by the iris tapering. The frequency of the first dipolar Higher Order Mode (HOM) changes from 16\,GHz in the first cell to 17.2\,GHz in the last cell. The difference in frequency of 1.2\,GHz results in a suppression of the wakefield amplitude by an order of magnitude within 1\,ns, resulting in a very low value of the wakefields at the position of the second bunch. Since the structure has 69 cells, wakefield re-coherence takes place and the amplitude increases every 60\,ns. Nevertheless, ohmic losses in copper walls help to reduce the wakefield amplitude on this longer time scale. The amplitude of the transverse long range wakefield at the positions of all 40 bunches is presented in Fig.~\ref{fig:XbandAStransWake} demonstrating good wakefield suppression.


%% file: include/03-LINAC/LINACIntegration.tex
\subsection{Linac integration}
\label{sec:LINAC_Integration}

\begin{figure}[!hbt]
\begin{center}
    \includegraphics[width=0.8\textwidth]{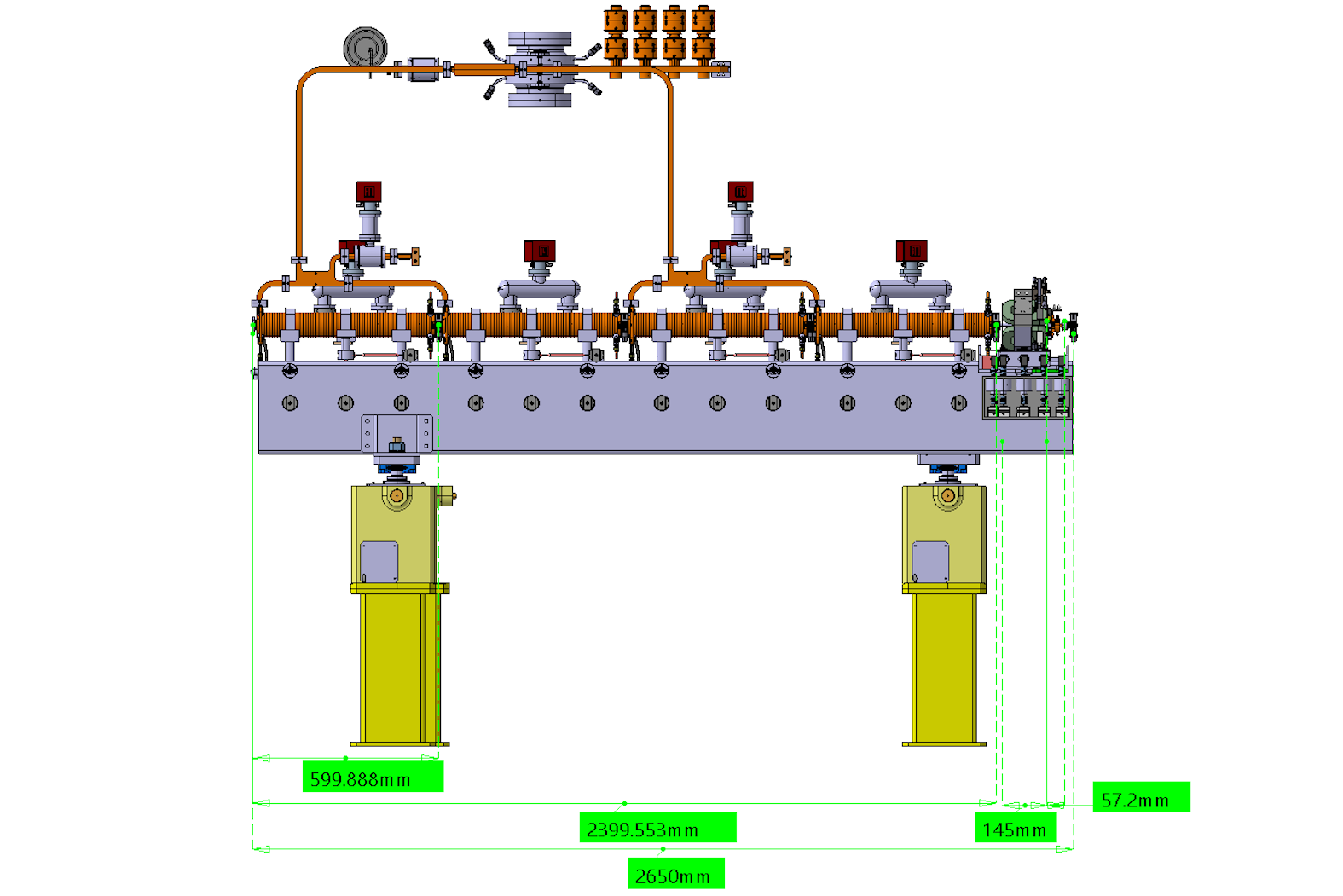}
    \caption{Side view of the eSPS Module, with longitudinal dimensions.}
 \label{fig:ModuleSize}
\end{center}
\end{figure}

As described in Section~\ref{sec:LINAC_Design}, four 0.6\,m long accelerating structures are assembled in one eSPS Module (see Fig.~\ref{fig:ModuleSize}).
Each of the structures is aligned by adjustable supports that are envisaged to provide an alignment precision of the accelerating structures within 10\,$\mu$m rms with respect to the beam axis. The Module is also supporting a quadrupole magnet and its associated Beam Position Monitor (BPM). In this case it has been chosen to adopt an adjustable support for the BPMs based on a platform developed by the Survey Group that can be equipped with motors, thus allowing realignment of the quadrupole during the accelerator operation or using it to introduce beam steering. The possibility to adopt integrated steering dipoles is also considered in order to avoid the complication introduced by the motorised displacement of the quadruple magnet.
The accelerating structures and BPMs are sitting on a steel girder, which is supported and aligned by means of jacks that have already been used for Linac4 installation.

\begin{figure}[!hbt]
\begin{center}
    \includegraphics[width=0.8\textwidth]{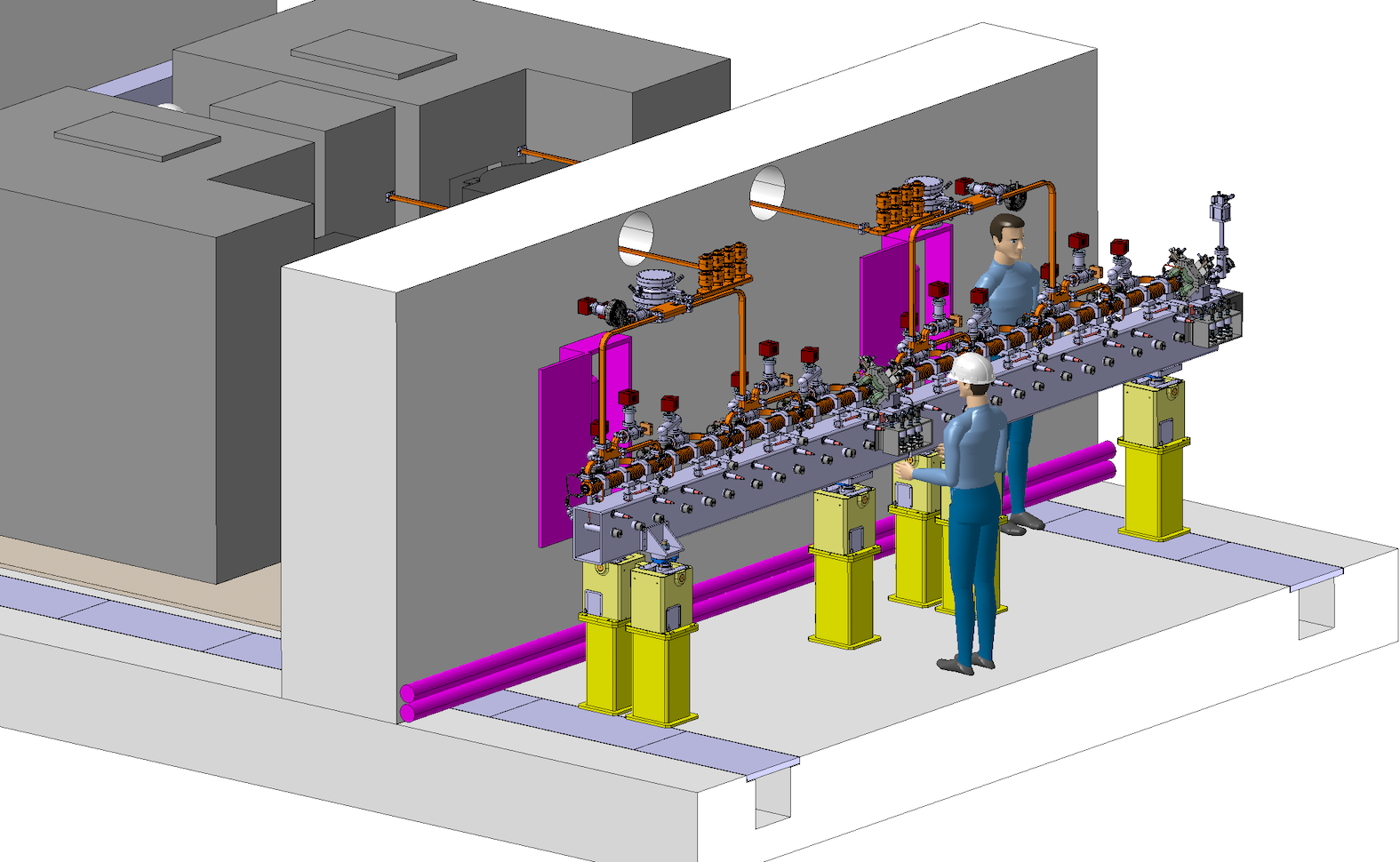}
    \caption{Two eSPS Modules in the accelerator tunnel.}
 \label{fig:ModuleInteg}
\end{center}
\end{figure}

A total of 24 Modules is required to achieve the final electron energy of 3.5\,GeV.  Due to the limited available space in the installation area an as compact as possible layout has been adopted (see Fig.~\ref{fig:ModuleInteg}).
Each Module is fed by a single 12\,GHz klystron delivering a 43.5\,MW, 1.5\,$\mu$s RF pulse to a pulse compression system. A compressed RF pulse 300\,ns long is delivered to each accelerating structure with 30.5\,MW peak power. The power delivery system adopts waveguide components that have been developed for the CLIC accelerator and have been built and tested at the Xbox test stations. We estimate that 13\% of the power is lost in the power distribution system. The average power dissipation is 0.9\,kW for each TW accelerating structure and 0.6\,kW in the output load, if the 100\,Hz operation is assumed. The accelerating structure temperature stabilisation is achieved by demineralised water circulation at the reference temperature of 28\,$^{\circ}$C. The water consumption is estimated in 6\,litres/min per accelerating structure.

%% file: include/03-LINAC/BeamInstrumentation.tex
\subsection{Linac beam instrumentation}
\label{sec:LINAC_BI}

The source and injector will be reusing the existing CLEAR facility along with its current beam instrumentation. The latter will benefit from the most recent development in beam instrumentation performed on CLEAR for low beam charge using inductive BPMs~\cite{Gasior2003}.
The beam position monitors along the X-band linac will be based on RF cavities also developed for the CLIC main linac~\cite{CLIC_BPM}. There will be one monitor per quadrupole, i.e.\ 24 BPMs in total for the full linac. 
The beam energy, emittance and bunch length will be measured at the entrance and the exit of the linac to monitor the evolution of beam properties before and after acceleration. Beam emittance and energy measurements will be based on Optical Transition Radiation (OTR) beam-imaging systems (BTV). Emittance monitoring is performed at the entrance and the exit of the linac using optical systems capable of providing resolution better than 10~$\mu$m~\cite{Bolzon2015}. The beam energy is measured in spectrometer lines equipped with optimised screen materials and shapes developed for the CLIC test facility 3 (CTF3)~\cite{Welsch2007,Olvegard:2013ska}. Non-invasive bunch length measurements will be performed non-invasively using coherent radiation~\cite{Lekomtsev:2011apa} or electro-optical techniques~\cite{Pan2016}.



%% file: include/03-LINAC/LINACPositrons.tex
\subsection{Possible upgrade for positron related studies and its impact on the electron linac}
\label{sec:LINAC_Positrons}

\begin{figure}[!hbt]
\begin{center}
    \includegraphics[width=10cm]{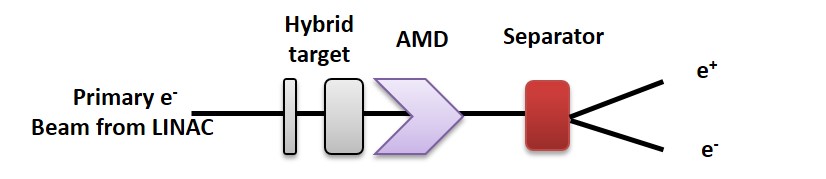}
    \caption{Schematic layout of a positron production target consisting of a tungsten crystal and an amorphous tungsten target followed by a adiabatic matching device.}
 \label{fig:positrontarget}
\end{center}
\end{figure}

There is an enormous interest in studying positron production and acceleration for future lepton collider projects. Worldwide only a few positron sources are still operated and usually part of facilities with little time for R{\&}D.
Efficient positron production typically requires a primary electron beam of a few GeV which is then converted into a positron beam using a dense target. The availability of a 3.5 GeV electron beam, therefore, presents an excellent opportunity to study positron production and possibly re-acceleration. 
A first step could be to implement a converter target area into the facility which allows the study of target technologies and their limitations, a very critical area in the design of a reliable positron source. The positron yield which can be achieved is a figure of merit for these sources. A second critical device is the so called adiabatic matching device, a strong pulsed magnet which focuses and collects the low energy positrons for re-acceleration.
A promising place for such a positron source R{\&}D area could be the first alcove within the TT61 transfer tunnel. The primary beam could be separated with a dogleg from the main SPS injection line allowing the installation of a positron target, a capturing device and some diagnostics including a beam dump. The location in the underground tunnel would not require excessive shielding to be able to perform such experiments. Figure~\ref{fig:positrontarget} shows a schematic of the CLIC positron source consisting of a hybrid target and an adiabatic matching device.

The CLIC positron source~\cite{Aicheler2019,Aicheler2012a} represents a typical example of such a source and is similar to the one proposed for FCC-ee.
In order to provide high quality, high energy positrons for experiments, the positrons have to be captured and re-accelerated to a few 100\,MeV and cooled in a damping ring.
After that they can be transported back to the beginning of the linac and finally accelerated to 3.5\,GeV. The yield of such a system would be of the order of one, which would result in positron bunches relevant for lepton collider studies. These beams could be used for plasma based acceleration studies with positrons, which is still an unsolved problem for future linear colliders based on plasma acceleration, see Section~\ref{sec:ACCRnD_Positrons}. Another possibility would be to inject the positrons into the SPS for further acceleration and for damping ring studies or Muon collider studies based on the LEMMA scheme.
It is, however, not straightforward to upgrade the proposed eSPS linac facility to a high energy positron source without additional civil engineering. Nevertheless such an integration could be feasible as shown by the proposed FACET II facility which includes a positron production target, damping ring and re-acceleration all within the moderately sized SLAC main tunnel\cite{facet2}.
To summarise, the implementation of an area for positron production studies seems straightforward and would require only minor modifications of the beam transfer line down to the SPS in TT61 while a full energy positron beam available for experiments represents a larger integration challenge. More details about the potential research program of such an upgrade of the facility can be found in Section~\ref{sec:ACCRnD_Positrons}.

%% file: include/03-LINAC/LINACPlasma.tex
\subsection{Possible upgrade for plasma related studies and its impact on the electron linac}
\label{sec:LINAC_Plasma}

\begin{figure}[!hbt]
\begin{center}
    \includegraphics[width=0.8\textwidth]{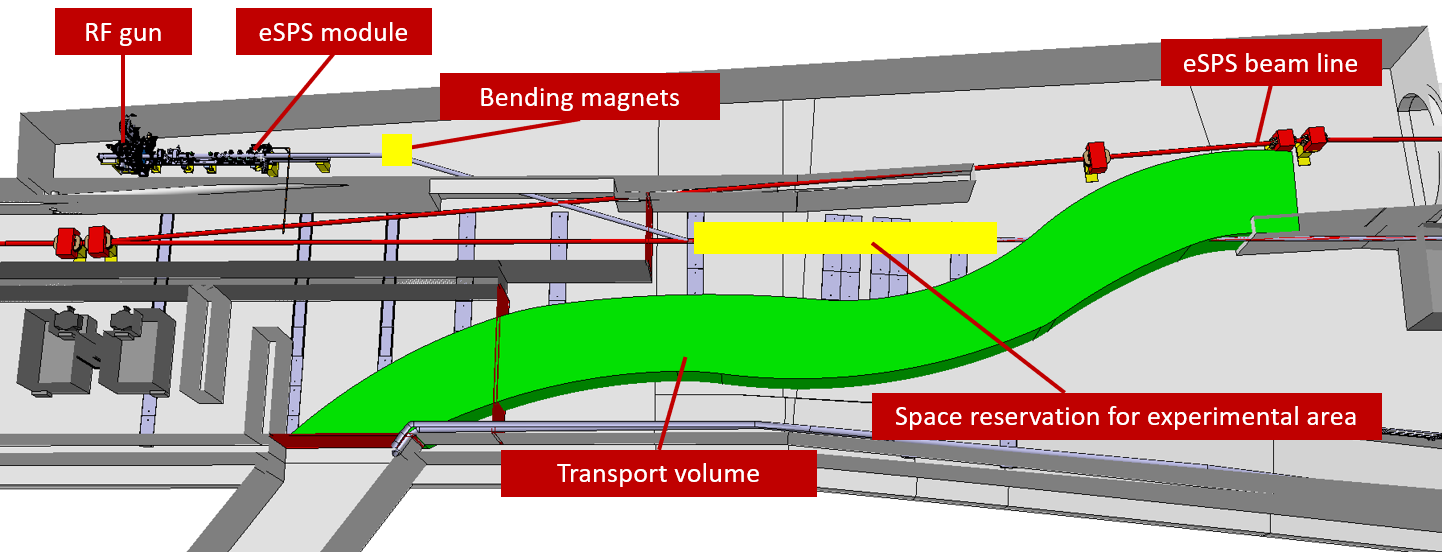}
    \caption{Schematic layout for a second injector dedicated for plasma wakefield acceleration research. The injector could be installed at the beginning of TT61 and connected via a dogleg to an experimental area at the end of the linac.}
 \label{fig:Second_injector_layout}
\end{center}
\end{figure}

The possibilities and the potential for plasma-based acceleration Research and Development taking advantage of the 3.5 GeV electron linac are described in Section~\ref{sec:ACCRnD_Plasma}. For these kind of experiments the high energy 3.5\,GeV linac with its excellent beam quality is used to accelerate the drive bunch. The witness bunch has to be truly independent to be able to probe the wakefields at variable distances without compromising the beam quality of the witness bunch. This can only be achieved with a second independent injector. Therefore, to fully exploit the potential of such experiments, a second injector could be added in an upgrade of the facility. An energy between 100 and 200\,MeV would be sufficient for this purpose. Further requirements as explained in Section~\ref{sec:ACCRnD_Plasma} are a very short bunch length of the order of 140\,fs (42~$\mu$m), a moderate charge of 100\,pC and a small emittance to be able to match the witness beam to the plasma wakefields.
Such an injector has been studied in detail for the AWAKE project\cite{Pepitone2018} and could be fitted into the facility at the beginning of the TT61 tunnel. The injector consists of an S-band RF-gun followed by X-band structures for velocity bunching and acceleration. The beam produced by the injector could be connected with a achromatic dogleg onto the main beam axis at the end of the linac. After the dogleg an approximately 2\,m long experimental area would be needed to accommodate a plasma cell and corresponding diagnostics. Finally a spectrometer capable of beam energies of up to 10\,GeV is needed at the end of the beam line. A schematic drawing of such an implementation can be seen in Fig.~\ref{fig:Second_injector_layout}. Space for two more modulators units can be found extending slightly in the klystron gallery.

The injector itself could use a slightly modified X-band acceleration module as described for the main linac. This would provide up to 160 MeV of acceleration when powered by a single X-band klystron. According to the AWAKE study the witness beam injector could produce the required beam parameters, see Table~\ref{tab:plasmainjectorpar}.

\begin{table}[!hbt]
\begin{center}
\caption{Tentative Beam parameters for a witness beam injector.}
\label{tab:plasmainjectorpar}
\begin{tabular}{p{9cm}cc}
\hline\hline
\textbf{Parameter}   	& \textbf{Value}\\
\hline
Energy [MeV]  	  		& 160 \\
Bunch charge [pC]    	& 100 \\
Norm. emittance [$\mu$m]  	& 1 \\
Bunch length rms [fs]     		& 140 \\
Energy spread rms [\%]    		& 0.1 \\
\hline\hline
\end{tabular}
\end{center}
\end{table}

Beside the space for the experimental area at the end of the X-band linac, no hardware modifications are necessary within the main linac and injector. However, the drive beam requires a high charge of 1.7\,nC which has to be accelerated and finally compressed to a bunch length of 0.8\,ps (240 $\mu$m). Beam dynamics studies have shown that the acceleration of such a bunch is feasible with the linac described in Section~\ref{sec:LINAC_BeamDynamics}.
Such a facility would represent unique opportunities to study beam driven plasma wakefield acceleration (PWFA). For the time being, there are no facilities in the world providing truly independent drive and witness bunches, which is essential to study emittance preservation, beam loading and energy spread compensation in detail and with high precision for collider applications.

%% file: include/04-SPS/OverviewSPS.tex
\section{SPS and transfer lines}
\label{sec:SPS_Transferlines}

From 1989, and for more than a decade, the SPS was used as injector for the LEP. It accelerated electrons and positrons coming from the PS from \SI{3.5}{GeV} to \SI{22}{GeV}~\cite{LepInjectorStudy:1983aa}. This clearly demonstrated capability was the genesis of this new proposal to once again accelerate leptons in the SPS ring. 

The concept draw on the LEP era but several aspects are adapted to the new experimental requirements. We discuss in particular the new the injection, acceleration and extraction systems proposed to produce a low-current but high frequency spill of electrons at \SI{16}{GeV}.

%% file: include/04-SPS/Linac-SPS.tex
\subsection{Linac to SPS transfer}
\label{sec:SPS_LINACtoSPS}
\begin{figure}[!hbt]
\begin{center}
   \includegraphics[width=1.0\linewidth]{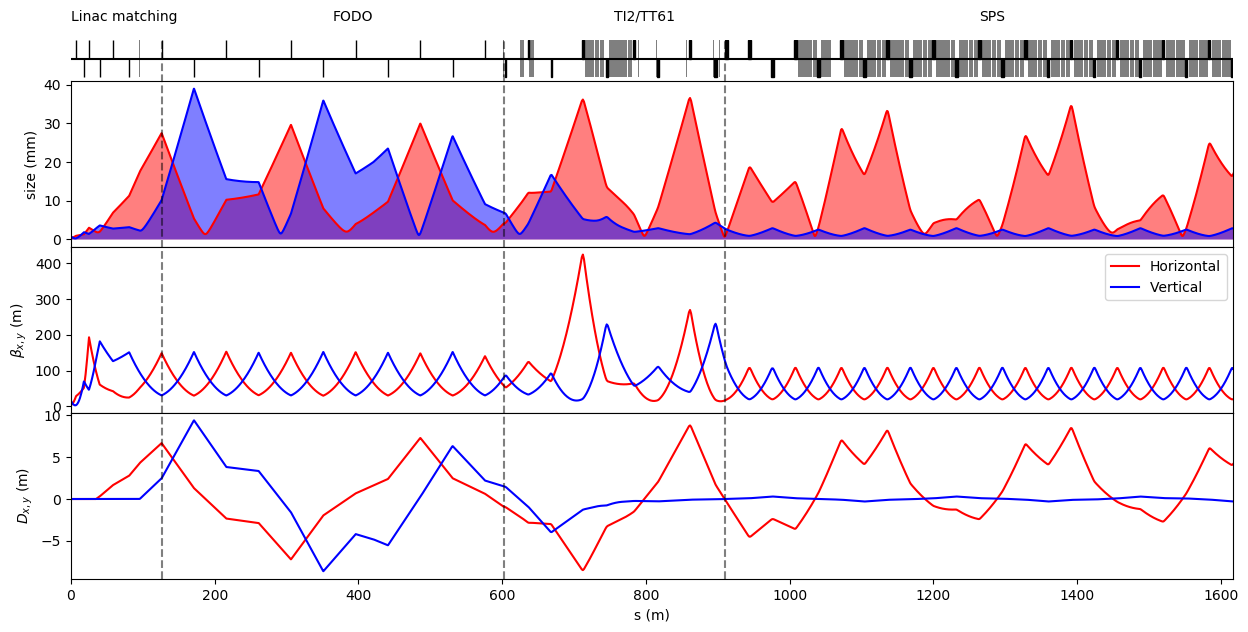}
   \caption{Beam envelopes and optical functions from the linac to the SPS.}
\label{fig:Linac_to_SPS}
\end{center}
\end{figure}

Electrons from the linac in transfer tunnel 4 (TT4) will be transported through TT61 to the injection system in SPS long straight section 6 (LSS6). Figure~\ref{fig:Linac_to_SPS} presents a conceptual design of these transfer line optics. Beam size is represented using a 4~$\sigma$ envelope with the beam parameters mentioned in Table~\ref{tab:CLEARinjectorPars} and includes maximum trajectory offsets of \SI{2}{mm}. 

\subsubsection{TT61 transport}
The linac matching section is composed of six independently powered quadrupoles. Those quadrupoles of type QTN allow matching of the beam envelope coming from the linac to larger sizes that are more convenient for the long and mostly straight transport in the TT61 tunnel.
Bending of the trajectory in both horizontal and vertical planes is needed to transfer the beam from the linac to the axis of the TT61. A total of four water-cooled dipoles of type BH2 type2 will be used in the linac matching section. 

A long FODO straight section uses QTR type quadrupoles in a regular arrangement of two families powered in series. Those low-field quadrupoles provide sufficient focusing strength, large aperture and present the advantage of being air-cooled. The use of air-cooled magnets in TT61 for around \SI{400}{m} minimises the costs by forgoing the need to supply cooling water. 

From the end of the FODO section the beam joins the TI2 and TT61 lines currently used for transfer of the proton beam from the SPS to the LHC. A further three dipole magnets are used to align the trajectory of the electron beam with the TI2 line. The existing magnets and powering scheme in TI2 are used to provide the optics and beam sizes shown in Fig.~\ref{fig:Linac_to_SPS}. However, this study does not consider power supply stability or remnant fields at the very low currents the magnets of the TI2 and TT61 lines will need to be operated. Table~\ref{tab:magnets_res_TT61} lists the existing magnets that are considered to be re-used in the design of the electron transport line, only needing at most a refurbishment before installation.

\begin{table}[!hbt]
\begin{center}
\caption{List of existing magnets used in the linac to SPS beam transport line design.}
\label{tab:magnets_res_TT61}
\begin{tabular}{llccc}
\hline\hline
\textbf{Name} & \textbf{Type}  & \textbf{Quantity}  & \textbf{Max. current [A]} & \textbf{Max. integrated strength}\\ 
\hline
QTN &  Water-cooled quadrupole & 6 & 150 & \SI{2.0520}{T} \\
QTR & Air-cooled quadrupole & 11 & 50 & \SI{0.4997}{T} \\
BH2 & Water-cooled dipole & 7 & 650 & \SI{0.7752}{T.m} \\
\hline\hline
\end{tabular}
\end{center}
\end{table}

\subsubsection{SPS injection}
Matching of the beam to the SPS lattice is imperfect despite the many independent quadrupoles available in TI2/TT61. In particular, a mismatch of the horizontal dispersion causes a large horizontal beam size in the SPS ring. A more careful design of the transfer line optics and FODO section may cure this mismatch but is not critical for our application. As long as the beam fits within the acceptance of the SPS, the synchrotron radiation damping will ensure that beam characteristics quickly converge towards their equilibrium values.

The injected bunch structure will depend on the linac injector and the SPS RF system frequency. We consider here that the linac will produce \SI{200}{ns} trains of bunches. A fast bunch-to-bucket fast injection scheme, very similar to the one in used during LEP era~\cite{LepInjectorStudy:1983aa}, will synchronise the injection of the bunches with the SPS buckets. The number of bunches per train will depend on the frequency of the SPS RF system and the experimental requirements. It could be as little as 40 bunches for a \SI{5}{ns} spacing between bunches or as much as 160 bunches to fill the SPS with an \SI{800}{MHz} RF system. 

\begin{figure}[!hbt]
\begin{center}
   \includegraphics[width=0.8\linewidth]{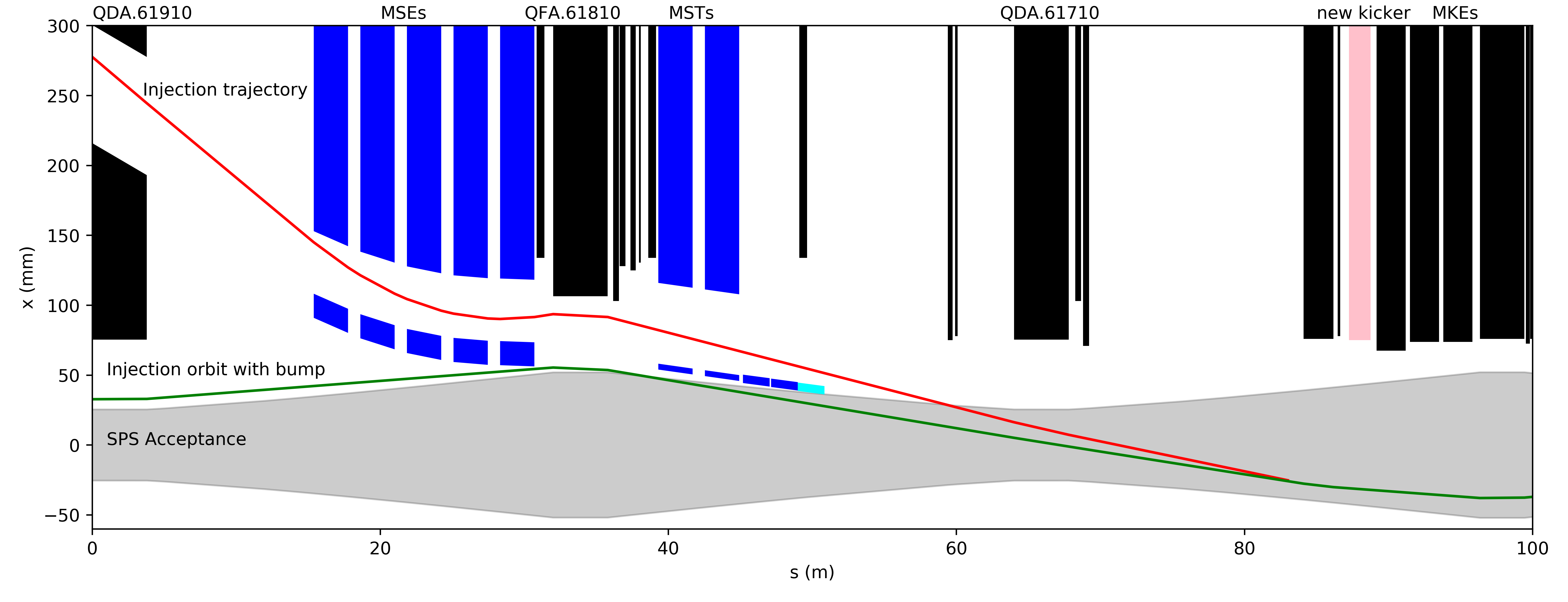}
   \caption{Horizontal electron injection trajectory in the SPS LSS6 region.}
\label{fig:inj_e_sps}
\end{center}
\end{figure}

Figure~\ref{fig:inj_e_sps} shows the injection trajectory through the side channel of the quadrupole QDA61910 followed by bending by the MSE septa and deflection onto the SPS orbit by a new fast kicker.
As during LEP operation in LSS1, the MSE will be powered at low field using a new low current power supply and the MST will be powered-off but demagnetised using another power supply.

Short trains of \SI{200}{ns} coming from the linac require a kicker system with fast rise or fall time, to allow maximum filling of the ring. A transmission line of magnets composed of two \SI{50}{cm} tanks could achieve a rise time of \SI{100}{ns} and the desired flat-top of \SI{200}{ns} to provide the required kick of \SI{500}{\micro rad}. The use of solid state switches will allow a \SI{100}{Hz} repetition rate. The SPS will be filled by up to 70 trains within less than a second. The new kicker system has to be designed and built but will largely make use of existing technologies and CERN competencies.

%% file: include/04-SPS/BI_TL.tex
\subsubsection{Beam instrumentation}
\label{sec:BI_TL}

Fourteen new inductive beam position monitors~\cite{Gasior2003} are required to measure the beam trajectory through the transfer line, with two new Optical Transition Radiation Beam TeleVision (OTR-BTV) systems~\cite{Bolzon2015} foreseen to measure the average beam size at the beginning and the end of the transfer line.

In addition, two new OTR-BTV systems~\cite{Bolzon2015} will also be required to measure the average beam size and position upstream and downstream of the new injection kickers to be installed in the SPS LSS6 section.

%% file: include/04-SPS/SPS_RF.tex
\subsection{SPS acceleration}
\label{sec:SPS_RF}
After injection into the SPS, the electron bunches should be accelerated from \SI{3.5}{GeV} to \SI{16}{GeV}.  In the LEP era, a dedicated RF system for lepton acceleration at \SI{200}{\mega\hertz} was installed in the SPS LSS3. If the same system is reinstalled in LSS3, a total of twelve cavities, providing \SI{1}{\mega\volt} RF voltage each, will be sufficient to store and accelerate electrons up to \SI{16}{GeV}. A frequency of \SI{200}{\mega\hertz} would be the minimum frequency to fulfil the time structure of the eSPS beams and also the 15$^\mathrm{th}$ harmonic of the injector frequency. Higher frequencies, up to \SI{800}{\mega\hertz}, open the possibility for increased number of bunches (x4) and give reduced beam loading for high current beams. Frequencies of 200, 400 and 800 MHz are already in use at the CERN accelerator complex with significant operational experience of both normal and superconducting RF systems. Although different options at these frequencies were considered, only the use of the normal conducting \SI{200}{\mega\hertz} cavities from the LEP era and a superconducting option at 800 MHz are discussed below.

\subsubsection{Normal conducting RF system}
 The LEP-era \SI{200}{\mega\hertz} cavity design and the RF system are described in~\cite{Herdrich1985,200MHzSPScavity:1987,Faugeras1989}.
 Although thirteen of LEP-era \SI{200}{\mega\hertz} RF cavities were stored in good conditions and are still available for re-installation into the SPS ring, only a few of the RF auxiliaries are still available. Most of the HOM dampers, fundamental power couplers and tuners have to be rebuilt. In addition, a new high power RF system including 60 kW power amplifiers for each cavity has to be built and installed in the SPS together with the new LLRF system. In general, the 200\,MHz RF system, accelerating leptons in the SPS used as a LEP injector, was very robust. It worked for twelve years and had a very little impact on the down time, also during the proton operation for fixed target experiments.

All RF systems for lepton acceleration were removed from the SPS, along with many other items, when the LEP was decommissioned. This allowed to reduce the impedance of the SPS ring which was necessary for the production of high intensity (nominal) proton beams for the LHC. To double the bunch intensity for HL-LHC, a broad impedance reduction campaign is underway in the framework of the LIU project~\cite{Damerau2014a}. Re-installation of twelve accelerating cavities for electron acceleration in the SPS ring will require a detailed study of the impact of their impedance on beam stability for proton operation including both their low-frequency inductive part ImZ/n and narrow-band HOMs. The so-called low-frequency ImZ/n, will be increased by \SI{0.55}{\ohm}, comparable with overall SPS impedance after LIU upgrades. This preliminary estimate was done using the values published in Ref.~\cite{Linnecar1996}.
A \SI{200}{\mega\hertz} normal conducting RF system designed for beam capture in the LHC~\cite{Boussard2000DesignSystem} could also be a potential alternative, but the compatibility with HL-LHC beams in the SPS have to be assessed, similar to the LEP-era cavities.

\subsubsection{Superconducting RF system}\label{sec:SPS:SRF}
\begin{figure}[!hbt]
\centering
\includegraphics[width=9cm]{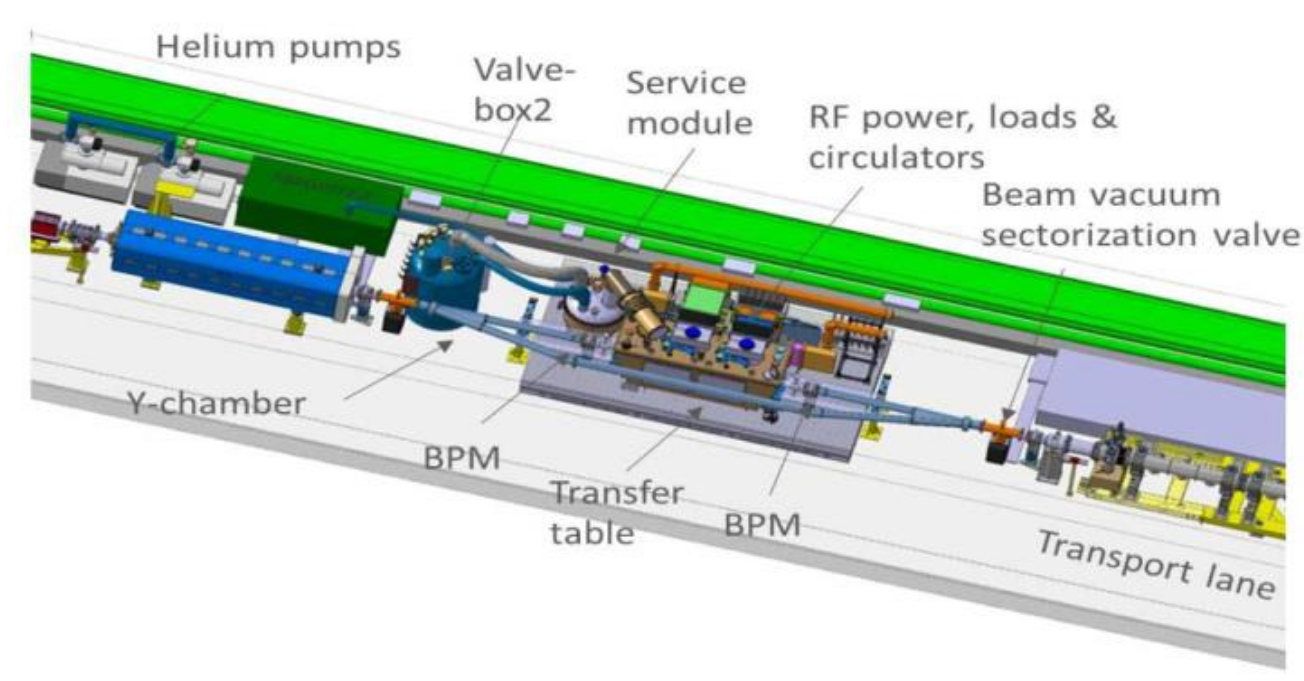}
\caption{Schematic view of the tunnel and surface installation of the cryogenic, mechanical and RF infrastructure in the SPS~\cite{Calaga2018}.}\label{fig:LAYOUT}
\end{figure}

In the framework of the HL-LHC project, a new  vacuum  sector in the SPS LSS6 region was created to test superconducting crab cavities with proton beams~\cite{Calaga2018}.
This sector comprises two Y-shaped vacuum chambers articulated by mechanical bellows, the circulating proton beam line and the beam bypass consisting of an RF module. The mechanical bypass is equipped with a movable table to move the cryomodule in and out of the circulating beam path. A dedicated RF system, cryogenic system and general infrastructure were put in place on the surface (BA6) and in the tunnel (see Fig.~\ref{fig:LAYOUT}),  and then successfully operated with beam during 2018. 
The mechanically movable bypass allows the installation of an RF module operating at \SI{800}{\mega\hertz} for a dedicated mode of operation with electrons. This configuration alleviates the strong constraints of impedance requirements for the high intensity proton beams and other modes of SPS operation. The remotely operated bypass with \SI{2}{\kelvin} helium, allows a rapid exchange between the dedicated mode and regular operation in about 10 minutes, already demonstrated in 2018. An alternative method could foresee a fixed dog-leg at the same location using the SPS magnets (or additional elements) to allow the electrons to circulate through the bypass with the RF cryomodule while the protons circulate through the empty vacuum chamber. Feasibility of such a configuration requires a detailed study on magnetic requirements, integration and aspects related to synchrotron radiation in the dog-leg region.

\begin{figure}[!hbt]
\centering
\includegraphics[height=2cm]{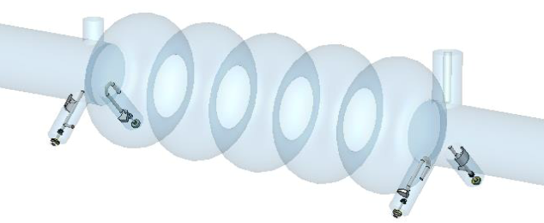}
\includegraphics[height=2cm]{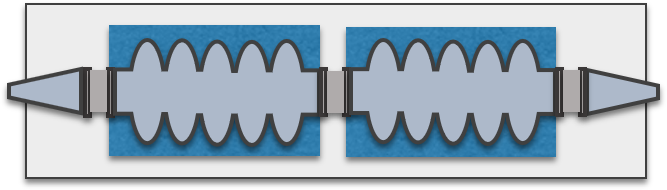}
\caption{Two-cavity concept for a cryomodule using five-cell \SI{800}{\mega\hertz} cavities for the bypass option.}
\label{fig:Multicell800}
\end{figure}

Recent studies in the framework of the FCC study have led to design of the \SI{800}{\mega\hertz} cavities ranging from single-cell for high beam currents to five-cell structures for the high energy and lower beam currents~\cite{Calaga2015,Zadeh2019}. A schematic of two five-cell cavities housed in a cryomodule and suitable for acceleration of the electrons from \SI{3.5}{\GeV} to \SI{18}{\GeV} is shown in Fig.~\ref{fig:Multicell800} and some relevant parameters are listed in Table~\ref{TAB:RFPARAM}. 
\begin{table}[!hbt]
\begin{center}
\caption{Relevant parameters of the RF systems in the SPS for electrons and, where applicable, for protons in parentheses.}
\label{TAB:RFPARAM}
\begin{tabular}{lcccc} 
        \hline\hline
      &  \textbf{Unit}    &  \textbf{LSS6 bypass}  & \textbf{Inline}  & \textbf{Inline}\\ \hline

Cavity type &   & \multicolumn{2}{c}{SC} & NC \\
Frequency & [\si{\mega\hertz}]  & \multicolumn{2}{c}{801.58} & 200.39 \\
Number of cells &  & 5 & 1 & 1\\ 
Number of cavities & & 2 & 4 & 12 \\
Voltage/cavity   & [\si{\mega\volt}]  & 5.0   & 2.5 ($<$1.0) & 1.0 \\
R/Q  (circuit def)     & [$\Omega$]   & 196  & 45 & $\approx$\,170 \\
RF Power  &  [\si{\kilo\watt}] & $\le$ 20 & $\le$ 20 ($\approx$\,300) & 60\\ 
\hline\hline
\end{tabular}
\end{center}
\end{table}
The two-cavity configuration would only require a moderate \SI{5}{\mega\volt/\meter} per cavity to keep the RF power to moderate level and be compatible with the existing crab cavity RF chain. A five-cell cavity in bulk niobium was built in collaboration with Jefferson lab and obtained an accelerating gradient of \SI{30}{\mega\volt/\meter}~\cite{Marhauser2018b}. 

The length of such a cryomodule is approximately the same as the crab cavity module presently installed. The two-cavity option is also compatible with the rest of the infrastructure needs. Due to the moderate electron beam current, both fundamental mode beam loading and higher-order mode impedance requirements are largely within the design capabilities of the FCC-cavity study outcome. The aperture of the \SI{800}{\mega\hertz} cavities is similar to the \SI{150}{\milli\meter} aperture in the SPS LSS6 region. Therefore, machine developments with protons beams could be performed to understand better the intensity limitations and eventually progress towards operating the high intensity proton beam with these cavities in the beam line.

\begin{figure}[!hbt]
\centering
\includegraphics[width=0.99\linewidth,trim={0.0cm 1.1cm 0.0cm 0.0cm},clip]{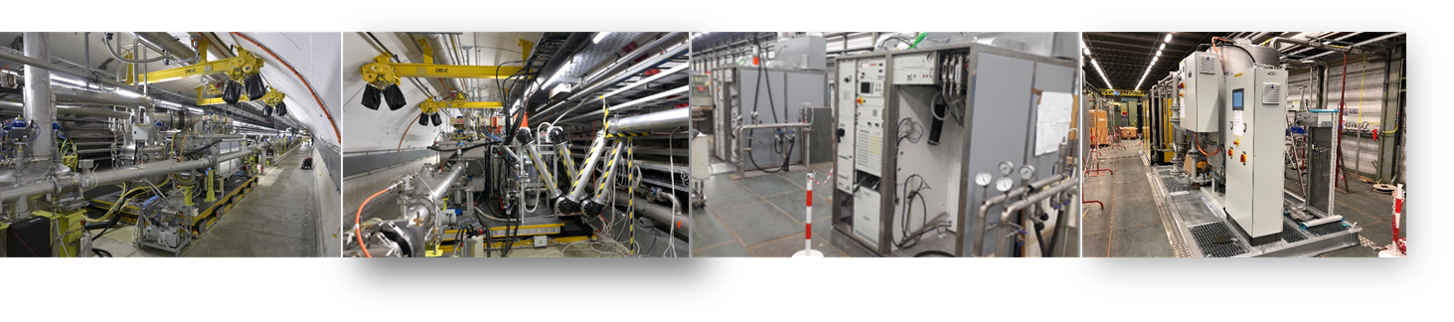}
\caption{Surface and tunnel installation of the RF and cryogenic system in the SPS for crab cavity tests.}
\label{fig:BA6INFRA}
\end{figure}

Two RF amplifiers of Inductive Output Tube (IOT) type are installed in the surface building (BA6) as shown in Fig.~\ref{fig:BA6INFRA}. 
They could provide up to \SI{60}{\kilo\watt}-CW power at \SI{800}{\mega\hertz} at the amplifier output to power each cavity independently for maximum flexibility. They are similar to the IOTs that are already operational in the SPS for the 4\textsuperscript{th} harmonic (\SI{800}{\mega\hertz}) RF system. The two IOTs were modified to provide power at \SI{400}{\mega\hertz} for the crab cavity tests and can be readily transformed back to their original configuration at \SI{800}{\mega\hertz}. First estimates of the RF power requirements are listed in Table~\ref{TAB:RFPARAM}, but exact power requirements and compatibility with the existing IOTs will need further studies. Any additional power will have to be compensated with an extra IOT per cavity. The power is transferred to the cavities in LSS6 zone via RF coaxial lines of approximately \SI{150}{\meter} in length. Taking into account the RF losses in the coaxial lines, it is estimated that only 40\,kW-CW might be available at the cavity input. Two specially designed high-power V-shaped RF lines with rotating connections provide the required movement to follow the table movement of \SI{510}{\milli\meter}.  The cavity control (aka LLRF) is located in BA6 next to the IOTs with its own Faraday cage, electromagnetically shielding it from neighbouring equipment. 

A new cryogenic system, including a mobile refrigerator, provides the required cooling capacity in the SPS.
The warm compressor unit, located at the surface (see Fig.~\ref{fig:BA6INFRA}), is connected to the cold-box underground by two warm lines for high and low pressure gas. 
The cryogenics system is capable of providing a total cooling capacity of \SI{3.5}{\gram/\second} at \SI{30}{\milli\bar} (saturation at \SI{2}{\kelvin}). This translates to approximately a total heat load (static and dynamic) of \SI{48}{\watt} at  \SI{2}{\kelvin} or equivalent to \SI{2}{\gram/\second} including a safety factor of \num{1.5}.

\begin{figure}[!hbt]
\centering
\includegraphics[width=14cm]{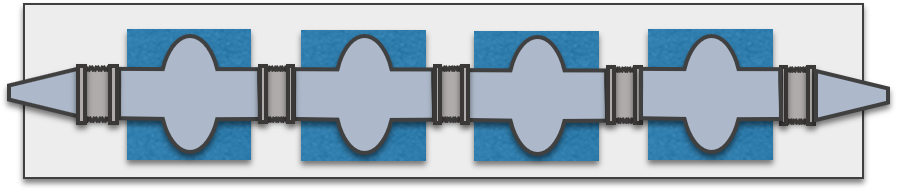}
\caption{Four-cavity concept for a cryomodule using one-cell \SI{800}{\mega\hertz} cavities for an in-line option.}
\label{fig:Singlecell800}
\end{figure}

An alternative option would be to install the \SI{800}{\mega\hertz} SC cavities directly in the SPS beam line (aka in-line option). This option, like the NC \SI{200}{\mega\hertz} option, requires strong beam loading compensation and very strong damping of HOMs using single-cell cavities. A schematic of four single-cell cavities comprising a cryomodule to provide the \SI{10}{\mega\volt} voltage is shown in Fig.~\ref{fig:Singlecell800}.
Table~\ref{TAB:RFPARAM} includes some relevant RF parameters for both five-cell and single-cell cavities. The present infrastructure in BA6 has to be increased by a factor of two or more including the change to higher power RF amplifiers and RF waveguides to be able to operate in this configuration. The long-delay loop for the RF control will be very challenging to maintain strong feedback. The cavities will have to be operated at fixed frequency due to the slow tuning capabilities of SC cavities. The compatibility of a such a system with the SPS high intensity proton operations requires a detailed study for the multitude of the SPS beams.

\subsubsection{Impedance considerations}
In the inline option, the eSPS will operate in parallel with the proton operation, supplying beam to both the fixed target experiments and to the HL-LHC. In order to provide the increased intensity required by the HL-LHC (compared to the nominal LHC) the LIU project has been executed~\cite{Damerau2014a}. One major component of this was an impedance reduction campaign in the SPS encompassing improved HOM damping in the \SI{200}{\mega\hertz} RF cavities, vacuum flange shielding and optimization of the electrostatic septa. 
The \SI{200}{\mega\hertz} RF cavities themselves were shortened and rearranged to cope, together with upgraded LLRF system, with high beam loading of HL-LHC beam.  
In addition, equipment exposed to the LIU beams must be able to withstand the high intensities if parasitic energy loss is caused by the device. This type of loss often manifests itself in the form of reduced peak accelerating voltage or damage to the equipment itself.


In addition to the RF system, equipment required for injection, extraction and specific beam instrumentation must also be installed. For injection there will be two additional kickers and for extraction one electrostatic and one magnetic septa. Experience during LHC runs I and II shows that exposing these types of equipment to the proton beam can result in excessive heating~\cite{Barnes2009a} and possible sparking~\cite{Balhan2016a}. 
In order to mitigate these effects the broadband impedance of any kicker magnet installed must be reduced either through minimizing the use of ferrites or shielding with serigraphy as has been done successfully for the SPS extraction kickers. For the septa the low frequency resonant impedance must be minimized to reduce the magnitude of beam induced fields that can cause sparking. An implementation using the detailed knowledge gained from the ZS replacement will achieve this~\cite{Balhan2016a}.
%

Aperture changes which are introduced will require tapered transitions where steps in aperture cannot be avoided. In keeping with the other tapers in the SPS, a maximum angle of 15 degrees is enforced. Further, the installation of kickers, septa and RF cavities may require additional sectorisation in the SPS beam line. Additional sector valves should match the neighbouring apertures as well as possible and have RF shields installed~\cite{Kaltenbacher2017a}.


%% file: include/04-SPS/SPS_Acceleration.tex


\subsubsection{Beam dynamics and stability}
\label{subsec:SPS_stability}

\begin{figure}[!hbt]
\begin{center}
   \begin{subfigure}{0.48\linewidth}
        \begin{center}
            \includegraphics[width=\textwidth]{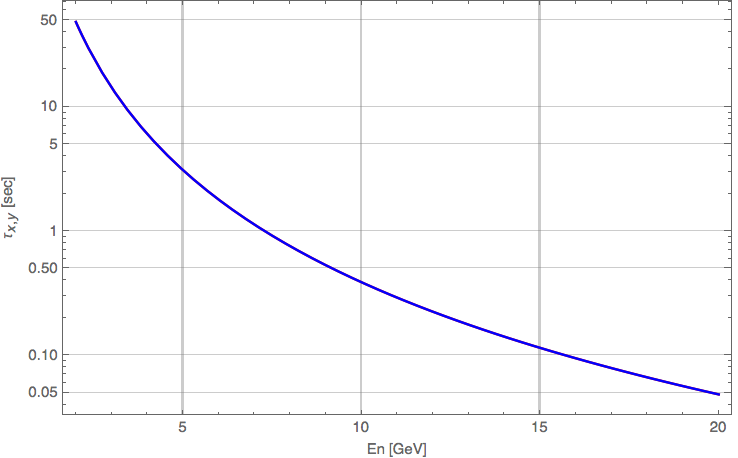}
            \caption{}
        \label{fig:damp_ener:sub1}
        \end{center}
    \end{subfigure}
    \begin{subfigure}{0.48\linewidth}
        \begin{center}
            \includegraphics[width=\textwidth]{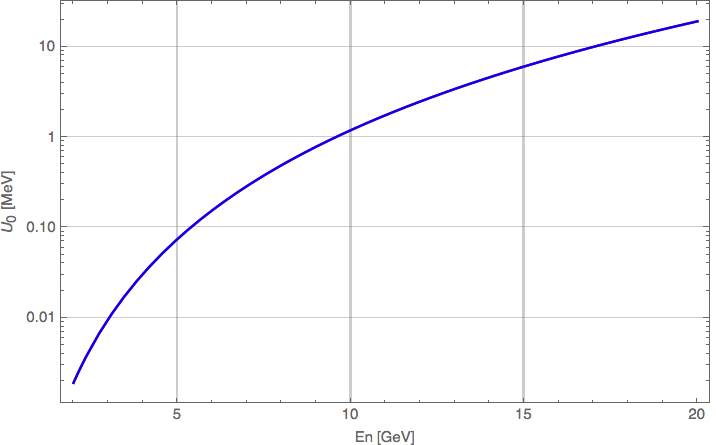}
            \caption{}
        \label{fig:damp_ener:sub2}
        \end{center}
    \end{subfigure}
    \caption{Transverse damping times \subref{fig:damp_ener:sub1} and energy loss per turn \subref{fig:damp_ener:sub2} versus electron energy in the SPS.}
\label{fig:damp_ener} 
\end{center}
\end{figure}

The electron beam is injected at \SI{3.5}{GeV}, the same energy that the SPS operated as an injector of positrons and electrons for LEP~\cite{LepInjectorStudy:1983aa}. The injection time of around \SI{0.8}{s} is very short compared to the long damping time which corresponds to 9\,s at \SI{3.5}{GeV}, as can be observed in Fig.~\ref{fig:damp_ener:sub1}, where the dependence of damping time versus energy is shown. This means that all emittances will essentially preserve their injected characteristics coming from the linac during this short injection plateau.

At this energy, particular attention has to be given for the first turn beam threading and establishing the closed orbit. In fact, the SPS dipoles have to operate at a very low current of around \SI{40}{A}, as compared to their flat top current which reaches \SI{6}{kA} at \SI{450}{GeV}. During the LEP era, special control loops had to be included in the SPS power converters in order to guarantee the required current stability~\cite{Cornelis:188928,Cornelis:217377}.

\begin{figure}[!hbt]
\begin{center}
   \includegraphics[width=10cm]{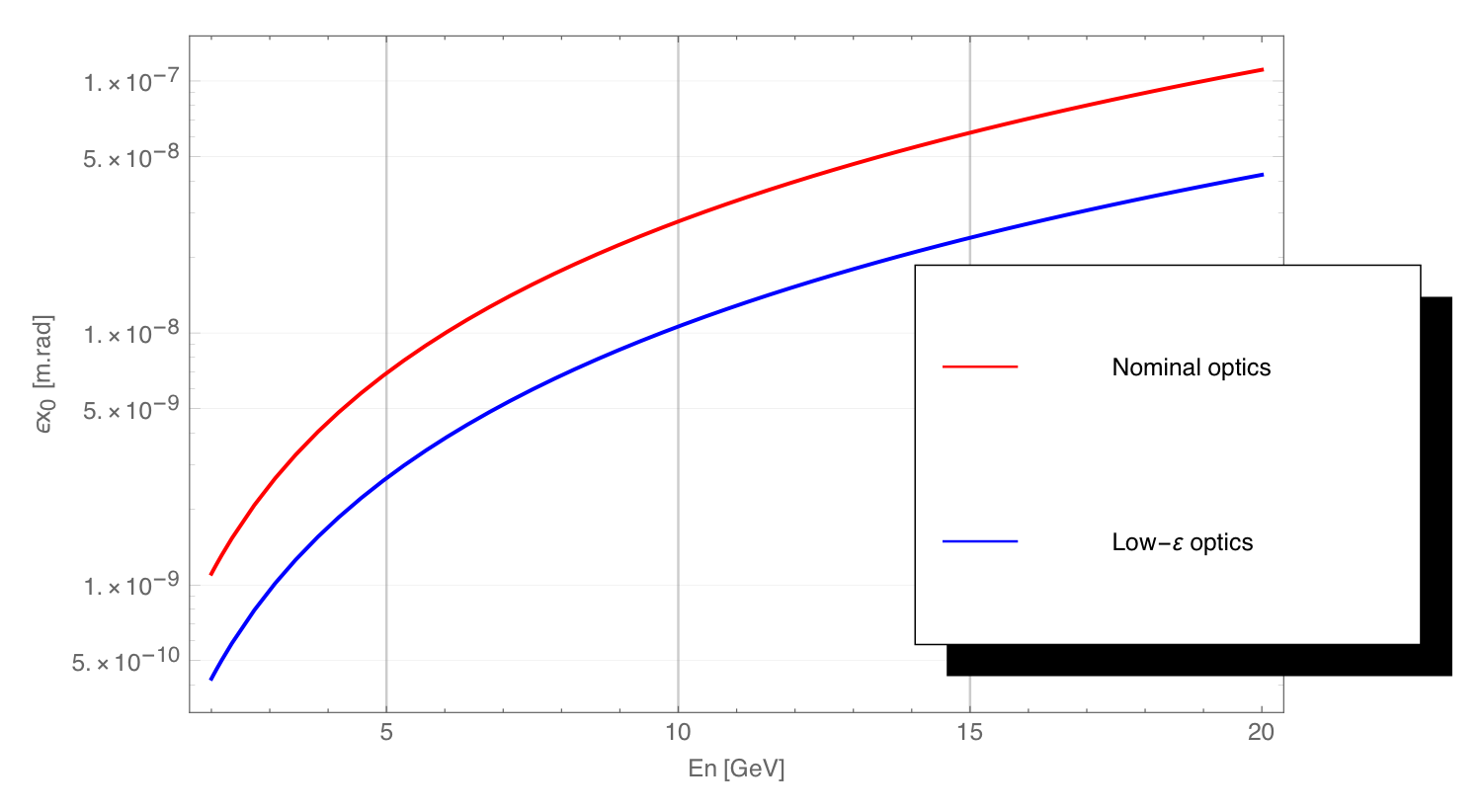}
   \caption{Equilibrium emittance versus energy for the nominal SPS working point (red) and for low emittance optics (blue).}
\label{fig:low_emittance}
\end{center}
\end{figure}

The SPS is a 6-fold symmetric ring, based on a FODO lattice.
It was traditionally tuned to phase advances of $\pi/2$, which provide an integer tune of 26 (Q26-optics) and a total arc phase advance multiple to 2$\pi$ for dispersion suppression in the straight sections. This was the working point used for injecting leptons in the SPS for LEP. The quadrupoles at the injection energy of \SI{3.5}{GeV} may have similar low current issues for controlling the optics of the ring. Their strength could be increased by moving the cell horizontal phase advance to around \SI{135}{\degree} (i.e.\ $3\pi/4$), which approaches the optimal phase advance for emittance minimisation in FODO cells (see Fig.~\ref{fig:low_emittance}), while keeping the total arc phase advance a multiple of 2$\pi$. This is not strictly necessary for the present application but can open the way to low emittance ring R\&D~\cite{Papaphilippou:1588122}.

In particular at injection but also through the cycle (although less critical due to the increase of radiation damping),  it is necessary to mitigate collective instabilities. The classical "head-tail" instability can be suppressed by imposing positive chromaticity, primarily using the SPS sextupoles (two families), throughout the cycle. For establishing a good chromaticity control, effects such as remnant fields in the main dipoles and sextupoles and eddy currents have to be taken into account. Regarding the Transverse Mode Coupling Instability (TMCI), and based on a broad-band resonator model for the impedance and typical SPS parameters, the threshold at injection is found at \SI{2e10}{electrons/bunch},  considering a bunch length of \SI{1}{mm} and RF voltage of \SI{1}{MV}. This provides a comfortable margin with respect to the injected bunch population. Some emittance blow-up may be observed due to intrabeam scattering at injection, but this will heavily depend on the injected beam parameters, whose brightness can be tuned to be relatively low, in particular by injecting large transverse and longitudinal emittances. Ion instabilities can be mitigated using trains with gaps and standard multi-bunch transverse feedback systems, as in modern electron storage rings.

The acceleration cycle of \SI{0.2}{s}, although quite short, is below the maximum ramp rate of \SI{90}{GeV/s} achieved for the lepton cycles in the SPS. At \SI{16}{GeV}, the dynamics are dominated by synchrotron radiation damping, with transverse damping times of around 94~ms and energy loss per turn of \SI{7.8}{MeV} (see~Fig.~\ref{fig:damp_ener}). Although the synchrotron radiation power becomes important at that energy, it has an average value of below \SI{1}{W/m} for \num{e11} electrons in the ring.

\paragraph{Collective Effects}

Collective effects can limit the ultimate performance of any accelerator. In this respect, an analytical estimation of intensity thresholds and impedance budgets have been performed for the eSPS, studying the impact of:  space charge (SC), longitudinal microwave instability (LMI), transverse mode coupling instability (TMCI), ion effects, intra-beam scattering (IBS) and coherent synchrotron radiation (CSR)~\cite{Chao1993PhysicsAccelerators, Zimmermann2005CollectiveRings, Ferrario2016WakefieldsAccelerators, Rumolo2008, Wolski2014BeamAccelerators, Koukovini-Platia2011, Nagaoka2017Ions, Byrd1997, FerrarioSpaceEffects, Zimmermann2010EstimatesRings, Bane2010ThresholdRings, Chao2013HandbookEdition, Lonza2017Multi-bunchSystems, Brandt1989BeamInjector}. Two optics options (Q26 and Q40) for the SPS are investigated and the results are summarised in Table~\ref{tab:collectiveeffects}.

\begin{table}[!hbt]
    \begin{center}
    \caption{Collective effects estimations for the eSPS.}
    \label{tab:collectiveeffects}
        \begin{tabular}{lcc}
        \hline\hline
        \textbf{Parameters and thresholds} &  \textbf{Q40}   & \textbf{Q26}  \\ 
        \hline
Phase advance (h/v) & $135^\circ/90^\circ$ & $90^\circ/90^\circ$  \\
Eq. hor. emittance @ injection. [nm.rad] &1.62 & 3.56 \\
Eq. hor. emittance @ extraction [nm.rad] &34.7 & 74 \\
Eq. Bunch length~[mm]  & 10 & 14 \\
Injected hor. emittance [nm.rad] & \multicolumn{2}{c}{0.43} \\
$e^{-}$ per bunch & \multicolumn{2}{c}{$10^{8}$} \\
Bending radius~[m] & \multicolumn{2}{c}{70.1}  \\
Average chamber radius~[m] & \multicolumn{2}{c}{0.04}  \\
Longitudinal impedance~$[\Omega]$ & \multicolumn{2}{c}{6.4} \\
Transverse impedance~[M$\Omega$/m] & \multicolumn{2}{c}{9.77} \\
\hline
SC tune shift @ injection [$10^{-4}$]& 4 & 4 \\
SC tune shift @ equilibrium [$10^{-4}$]& 8 & 3 \\
SC tune shift @ extraction [$10^{-7}$]& $1.2$ &  $0.6$\\
\hline
LMI impedance threshold @ injection~$[\Omega]$  & 17912  & 33988 \\
LMI impedance threshold  @ equilibrium~$[\Omega]$ & 21.67 & 41.12 \\
LMI impedance threshold  @ extraction~$[\Omega]$ & 682 & 1294 \\
\hline
TMCI impedance threshold @ injection~[M$\Omega$/m] & 5060  & 6900 \\
TMCI impedance threshold @ equilibrium~[M$\Omega$/m] & 506 & 690 \\
TMCI impedance threshold @ extraction~[M$\Omega$/m] & 1121 & 1612 \\
\hline
Tune shift due to ions [$10^{-5}$] & $4$ & $4$ \\
FII rise time ~$[t_{\textrm{rev}}]$ & 6.7 & 15 \\
\hline
CSR LMI bunch length threshold @ injection~[m]& 0.41 & 1.76  \\
        \hline\hline 
        \end{tabular}
    \end{center}
\end{table}%

Based on the analytical estimations, no major limitations are expected from SC, LMI, CSR, TMCI and IBS. On the other hand, the only critical point would be the control of the  fast ion instability (FII) with rise times of 15\,$t_{\textrm {rev}}$ and 6.7\,$t_{\textrm {rev}}$ for the aforementioned options. These rise times can be mitigated with a feedback system,  provided that the SPS achieves average vacuum pressures of \SI{e-8}{mbar}. The instability rise times can be also enhanced by large clearing gaps in the bunch train structure.

%% file: include/04-SPS/BI_SPS.tex
\subsubsection{Beam instrumentation}
\label{sec:BI_SPS}

\paragraph{Position Measurement} The SPS orbit and trajectory system will be upgraded during LS2 with new electronics read-out over optical fibres~\cite{Wendt:2018mrh}. No changes are foreseen for the pickups, consisting of 204 single plane electrostatic shoebox monitors and 12 dual plane electromagnetic strip-line monitors. The strip-line monitors are directional and hence additional cabling and electronics will be necessary to observe the counter-rotating electron bunches (with respect to proton bunches) using these 12 monitors. The nominal linac bunch intensity of \num{3e8} electrons is at the lower operational limit of the electronics, but the orbit system should still function for a single bunch train containing the nominal 40 bunches spaced by 5\,ns. A lower number of bunches or increased bunch separation may prove problematic for orbit measurement. The 100\,$\mu$m turn-by-turn position resolution requested should be achievable when the SPS is filled with the nominal 3080 bunches but may be significantly worse when operating with a lower number of bunches or with reduced bunch intensity. 

\paragraph{Intensity Measurement} The SPS Direct Current Current-Transformer (DCCT) has a resolution of $\sim$\,5\,$\mu$A which limits the detectable intensity change to \num{7e8}  electrons. The nominal requested resolution of 0.1\% should, therefore, be achievable for total beam intensities above \num{7e11} electrons. 
The wall current transformer~\cite{Krupa:2017qgh}, used to compute bunch-by-bunch intensity in the SPS, is adapted to measure LHC type beams separated by 25\,ns. Its bandwidth is, therefore, intentionally limited to $\sim$\,100\,MHz. Hence it will not be possible to measure individual electron bunches, with the monitor instead providing the integral of two to three bunches. The resolution of this system is \num{5e7} charges, implying that averaging over many turns would be required to obtain any meaningful data from this system. It should, nevertheless, be able to provide a qualitative intensity profile along the injected trains that could help with commissioning and optimisation. If this is considered not sufficient, a new monitor with an adequate electronic read-out system can be envisaged to provide better sensitivity. 

\paragraph{Profile Measurement} A new technique will be required to measure the beam profile of electrons in the SPS, as the existing instrumentation is not designed to work with such small beam sizes ($<$50\,$\mu$m in the horizontal plane and $<$10\,$\mu$m in the vertical plane). Equipping the existing wire-scanners with additional detectors (to deal with the change of direction compared to the proton beams) may allow comparative, qualitative measurements to be made in the horizontal plane, but would certainly not allow any measurement in the vertical plane. The existing beam gas ionisation profile monitors also do not presently offer the required resolution to measure such small beam sizes.
The synchrotron radiation emitted by the highly relativistic electron beam provides a way to measure such small beam sizes, with several techniques developed over the last twenty years for such purposes using x-ray optic imaging systems~\cite{Takano2006} or interferometric techniques in the visible range~\cite{Naito:2006zs}. Recent studies using randomly distributed interferometric targets indicate that such techniques could also be considered for use in the eSPS~\cite{Siano2017,mazzoni:2018inst}. At least one such system should therefore be envisaged in the SPS ring if the electron beam transverse profile is to be monitored. If fitted with a streak camera, such a system would also be able to measure the longitudinal profile of the beam. The existing SPS synchrotron monitor in LSS5 is designed for a clockwise rotating beam and a new light extraction system in a suitable location, i.e.\ with minimum dispersion in both planes, would need to be designed, installed and equipped in order to use synchrotron light monitors for electrons.

%% file: include/04-SPS/Extraction.tex
\subsection{SPS beam extraction and delivery}
\label{sec:SPS_Extraction}

Following acceleration in the SPS, the \SI{16}{GeV} electron beam extraction and transfer to the experimental area are discussed in this section. The concept also proposes a novel slow extraction scheme to reach the required experimental beam parameters.

\subsubsection{SPS internal dump}
\label{subsec:SPSdump}
An internal dump of the electron beam needs to be considered, both for machine development and for machine protection purposes. For instance, a beam instability may cause losses that necessitate to quickly dispose of the circulating beam. 

The existing SPS internal dump (the so called TIDVG) makes use of fast kicker magnets followed by absorption blocks to quickly channel and dilute the circulating beam into an internal absorber, below the circulating beam path. The rise time of the system is in the order of \SI{1}{\micro s}, faster than the revolution period of around \SI{23}{\micro s}. Therefore, the absorbing material has to be downstream of the fast kicker magnets, which is no longer the case with electrons circulating counter-clockwise in the SPS~\cite{Heler2018}.

An internal target limiting the SPS horizontal aperture in a dispersive region can make use of the natural energy loss of the electrons and serve as dump system. The existing TIDP collimator is already meant to intercept off momentum trajectories. The \SI{4}{m} long device certainly has the stopping power to easily absorb \SI{16}{GeV} electrons. The average maximum electron beam power on the dump with \SI{2}{s} cycle period and up to \num{e12} electrons in the ring will be around \SI{1.2}{kW}. Those will need to be considered for this device but also for any future consolidation of the TIDP.

At this location the horizontal dispersion is around \SI{2.5}{m} with the SPS Q26 optics. The electrons relative energy loss per turn is between \num{e-6} at injection and \num{0.5e-3} at \SI{16}{GeV}. We consider that the beam will be completely intercepted by the collimator when drifting horizontally inwards by \SI{1}{cm}. Without RF acceleration the beam will drift at this location by \SI{1}{cm} within around 4000 turns or \SI{90}{ms} at \SI{3.5}{GeV} and 8 turns or \SI{180}{\micro s} at \SI{16}{GeV}. The time necessary to switch off the acceleration, or considerably reduce the field of the cavity, will depend on the detailed design of the cavity and its ancillary systems. However, we can reasonably consider that the acceleration provided by an SRF cavity can be interrupted within less than \SI{1}{ms}. Therefore, such internal dump system can also be triggered by a machine interlock to safely discard the circulating beam.

The existing TIDP is a \SI{4}{m} long device which absorbing material is made of water-cooled aluminium alloy and copper blocs. In the case of counter-clockwise circulating beam, the electrons would connect first with the copper part of the device. Both the electron energy of \SI{16}{GeV} and average power of \SI{1.2}{kW} are below the nominal capabilities of the device. However, the very small minimum electron beam size (see Section~\ref{subsec:SPS_stability}) may result in large maximum local energy density reached in the absorber material during the dump event. Careful calculations of the beam dynamics, energy deposition and resulting thermo-mechanical stress during a dump event will be required to validate this concept.

Mitigation solutions such as foreseeing a movable beam diluter using lighter material could reduce the peak local energy density on the TIDP absorber. Alternatively, an upgrade or redesign of the device for proton operation foreseen during LS3 could be the opportunity to include the electron beam absorption requirement. 

\subsubsection{Slow extraction}
Resonant slow extraction allows the delivery of beam over long periods of time and using low fluence, which is particularly close to the experimental requirement discussed in Section~\ref{sec:LDM-at-eSPS}. Control of the extraction rate and the method used to bring particles into resonance are critical. Experimental requirements and specificities of the electron beam at \SI{16}{GeV} suggest two promising methods to drive the extraction process.
\begin{itemize}
\item Using RF noise and chromatic extraction. Band-limited noise forces particles to exit the bucket on one side and start drifting in energy along a longitudinal separatrix. The advantage is that the control of RF noise power controls the diffusion rate of particles out of the bucket. This can provide a very low extraction rate and may reduce the effects of machine ripples on the spill structure~\cite{Stockhorst:1997nk};
\item Using quantum excitation and amplitude extraction. Zero amplitude transverse tune is set at a distance $\delta$ from the resonant tune. Amplitude detuning makes all particles beyond a certain amplitude resonant. Quantum excitation diffuses particles from the core to the resonant and unstable region. The extraction rate can be controlled by changing the distance to the resonant tune $\delta$.
\end{itemize}

Regardless of the method used for slow extraction, the feasibility of reaching the very low rates required for the detector has already been demonstrated in other facilities~\cite{Frommberger:2016ieb}.

Of the two methods presented above, the latter was investigated in more detail. At \SI{16}{GeV} the electron beam dynamics in the SPS are dominated by synchrotron radiation effects with characteristic damping times below \SI{100}{ms} in all planes. This may prove problematic for conventional slow extraction methods as particles can spontaneously jump on and off resonance. We take advantage of the quantum excitation to propose an extraction method that will use the existing SPS lattice while providing the capability for an arbitrarily low spill rate.

\begin{figure}[!hbt]
\begin{center}
   \begin{subfigure}{0.4\linewidth}
        \begin{center}
            \includegraphics[width=.95\textwidth]{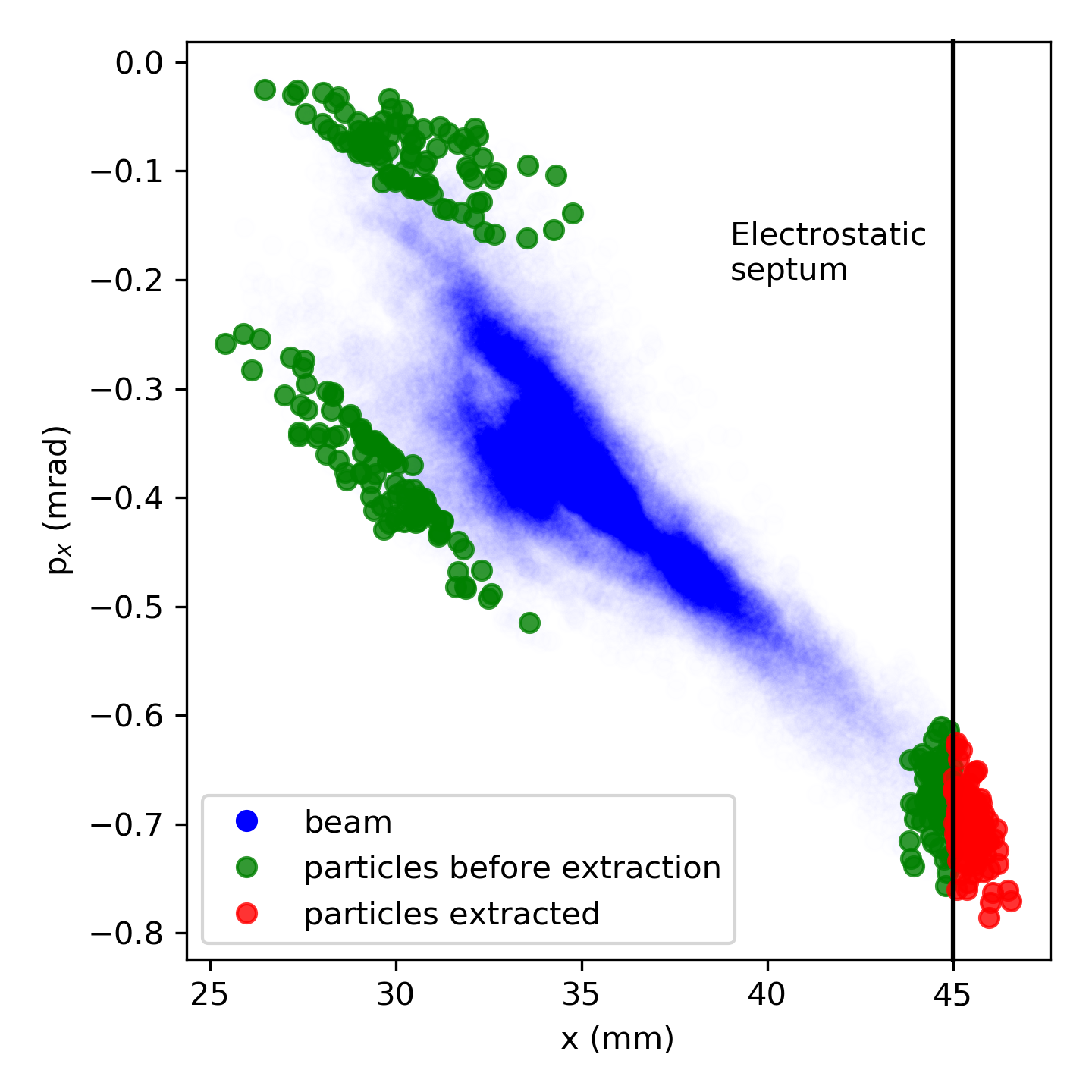}
            \caption{}
        \label{fig:extraction:PS}
        \end{center}
    \end{subfigure}
    \begin{subfigure}{0.59\linewidth}
        \begin{center}
            \includegraphics[width=.95\textwidth]{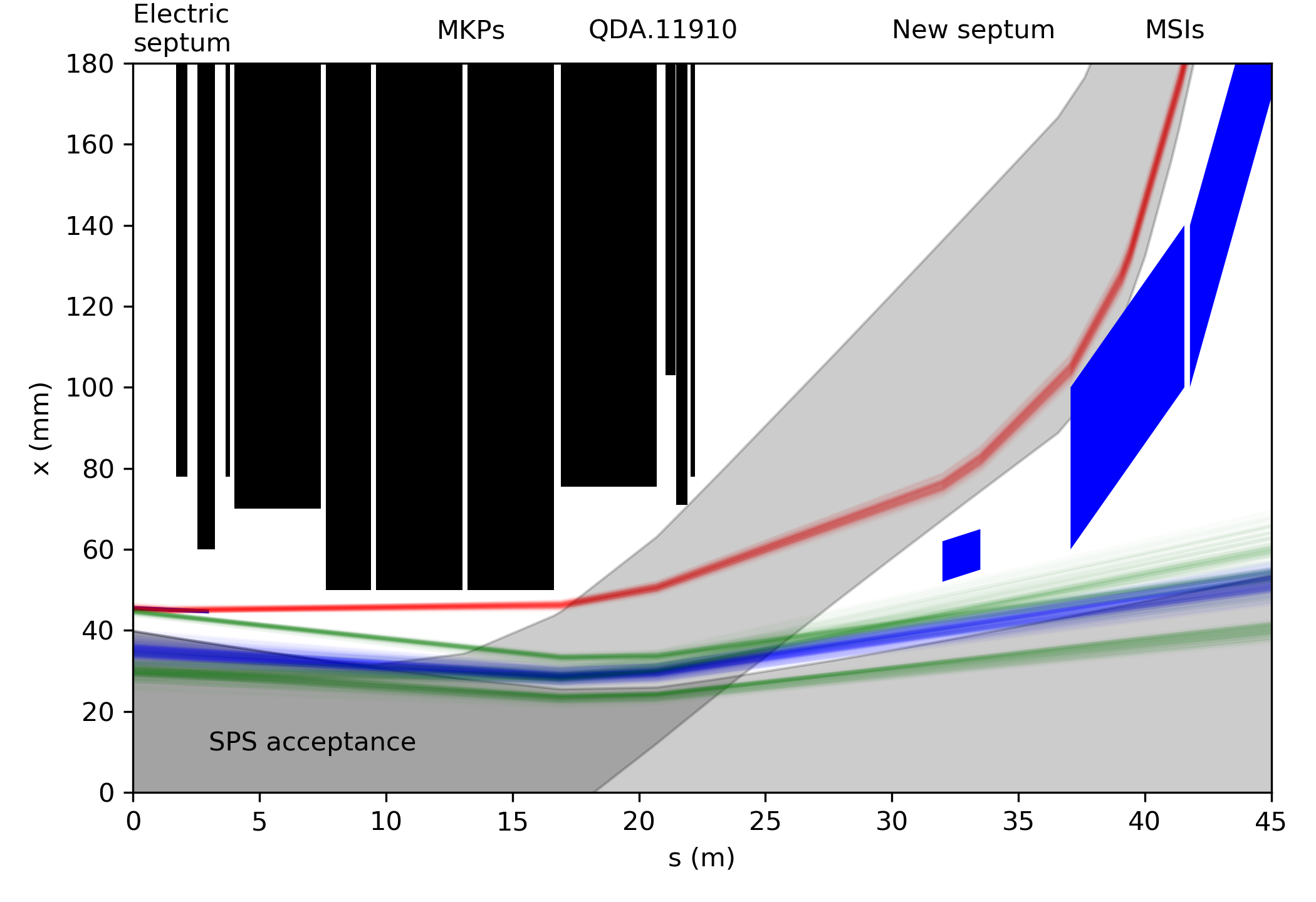}
            \caption{}
        \label{fig:extraction:ap}
        \end{center}
    \end{subfigure}
    \caption{Extraction process in the horizontal phase space at the electric septum \subref{fig:extraction:PS}. Particle trajectories in the transverse horizontal space in the extraction region with apertures and the injected \SI{14}{GeV/c} proton beam envelope in grey \subref{fig:extraction:ap}.}
\label{fig:extraction} 
\end{center}
\end{figure}

Figure~\ref{fig:extraction:PS} shows the horizontal phase space during the extraction process simulated using the MADX thin tracking with synchrotron radiation.
The core of the beam is stable while large amplitude particles are trapped on the third order resonance, as the triangle shape of the beam shows. A new electric septum placed at a specific distance from the beam core channels particles reaching an excursion of more than \SI{45}{mm} towards the extraction channel. This simulation allowed a first estimation of the parameters of the extracted beam shown in red on Fig.~\ref{fig:extraction:PS} and in particular a geometrical emittance of \SI{e-8}{m.rad}.

\paragraph{Extraction Rate Control} 
The main experimental program considered here requires a very low but highly constant flux of particles, making the control of the extraction rate critical. The scheme presented above addresses the control of the spill rate but machine and in particular power supplies stability is critical to ensure constant spill rate.

Regulation of the SPS dipole magnets at low field relied on a switch of the gain settings of the Analog Control board during the LEP era. This system has since been removed but the current regulation is now implemented in a virtual FGC within the SPS Mugef Control system. The Current Loop is now fully digital hence the resolution problem does no longer exist and stability at low field should not be an issue.

The quadrupole converter hardware was modified during LS2 in order to improve performances for the proton mode. The stability of these converters at very low currents is no longer obvious, hence requests for machine developments time in the SPS at low energy shall be booked in 2021.

In view of the LEP era, all other SPS auxiliary power converters were modified to allow operation at low energy. However, during the SPS consolidation campaign at the end of the 1990’s, these converters were modified or replaced: operation of the SPS at low energy is not possible anymore today. Considering the ageing state of these converters and the ongoing consolidation program that will last until LS4, the requirements for the operation at low energy should considered within the framework of this program and a budget should be granted accordingly.

\paragraph{Extraction Trajectories} 
Slowly extracted particle trajectories were investigated in LSS1 of the SPS towards the Meyrin site. LSS1 is used for injection of the proton beam coming from the Proton Synchrotron (PS). The scheme shown in Fig.~\ref{fig:extraction:ap} requires two new elements.
\begin{itemize}
    \item An electric septum of length \SI{2.5}{m} and field \SI{5}{MV/m} to provide a deflection of the extracted particles of around \SI{780}{\micro rad}; 
    \item A thin magnetic septum with a blade of \SI{10}{mm} and a field of \SI{150}{mT} will provide an additional deviation of \SI{4}{mrad}, to reach the existing MSI septa and enter the TT10 line. 
\end{itemize}
Further synergies with possible upgrades of the proton injection scheme are also considered.

\subsubsection{Fast extraction scheme}
A fast extraction scheme was not studied in detail here. However, the slow extraction orbit bump can be combined with a non-local fast kicker to achieve complete extraction of the beam within a single turn. A preliminary study indicates that non-local extraction using an MKE kicker in LSS4 would be capable of quickly extracting the beam in the TT10 channel with a Q53, or double tune, SPS optics. Similar schemes have already been studied for high energy protons~\cite{Velotti:2013noa}. A fast extraction scheme for electrons will be feasible, should not present any specific technical challenge and will come at little to no additional costs.

\subsubsection{Transfer and delivery to experimental hall}
\label{sec:delivery_system}
The TT10 line is used to transport protons from the PS to the SPS. The transport in the opposite direction of the extracted electron beam comes naturally. The polarity of magnetic elements will remain the same and the required strengths will be similar to proton operation, as the extracted beam rigidity is in the same range as the injected beam coming from the PS.

Of particular interest is the powering scheme and optics of this line. The first part of the line is powered by only two power supplies in a regular FODO scheme. The FODO lattice was designed to match the injected proton beam into the SPS lattice. In the case of slow extraction the beam parameters are very different from the ring ones. Therefore, the extracted beam is not matched to the FODO lattice of the TT10 line. However, transporting the beam in this mismatched transfer line is still possible due mainly to its small emittance.

\begin{figure}[!hbt]
\begin{center}
   \includegraphics[width=16cm]{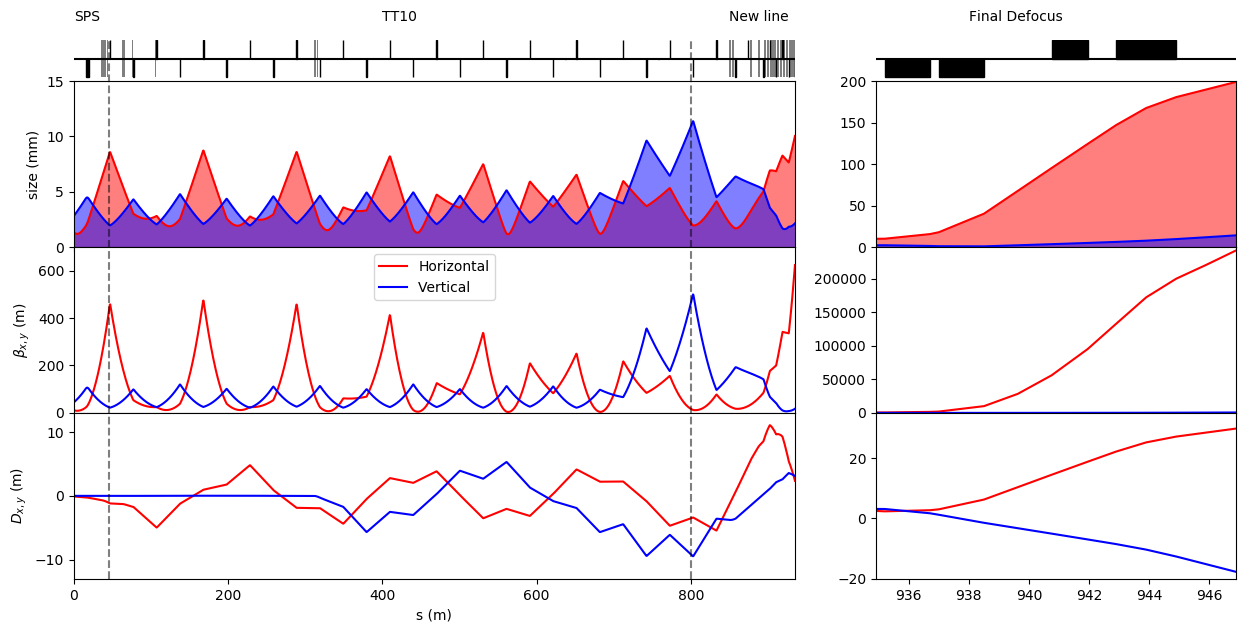}
   \caption{Beam sizes and optical functions from the SPS to the experimental target.}
\label{fig:SPS_to_LDMX}
\end{center}
\end{figure}

Figure~\ref{fig:SPS_to_LDMX} shows the evolution of the 4~$\sigma$ beam envelope with the beam parameters estimated earlier, a momentum spread of \num{e-4} and a maximum trajectory offset of \SI{2}{mm}.
It is clear from the evolution of the beta functions in TT10 that the beam is not correctly matched. However, due to the particularly low geometrical emittance of the extracted beam, the beam size along the line remains reasonable and well within the \SI{108}{mm} aperture of the quadrupoles.

Starting $s=\SI{650}{m}$ the powering scheme is composed of two independent FODO cells followed by five independent quadrupoles. In the conceptual design presented here, this section is used to match the incoming beam to the new line. The dispersion created by both horizontal and vertical bends in TT10 is also maintained below around \SI{10}{m}.

\begin{figure}[htpb]
\begin{center}
    \begin{subfigure}{0.20\linewidth}
        \begin{center}
           \includegraphics[angle=90, width=.95\textwidth]{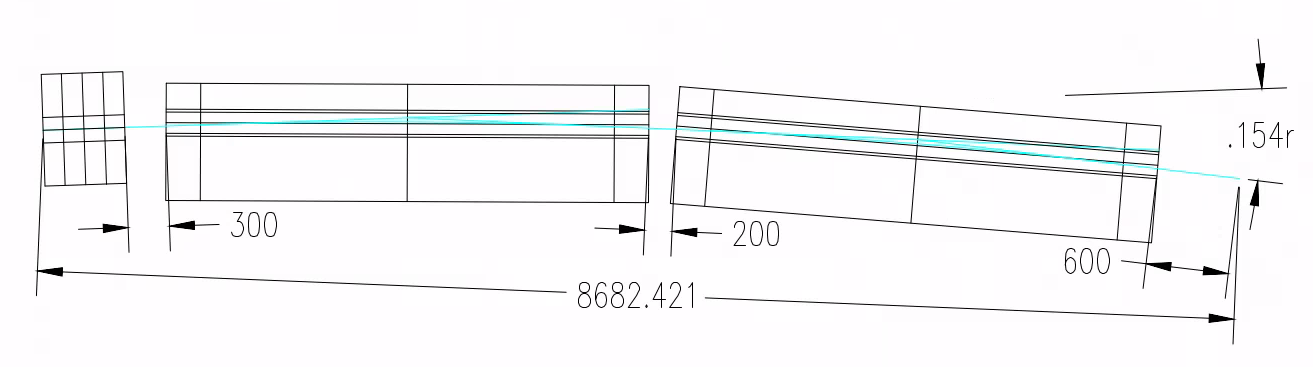}
           \caption{}
        \label{fig:Line_cell}
        \end{center}
    \end{subfigure}
   \begin{subfigure}{0.70\linewidth}
        \begin{center}
            \includegraphics[width=.95\textwidth]{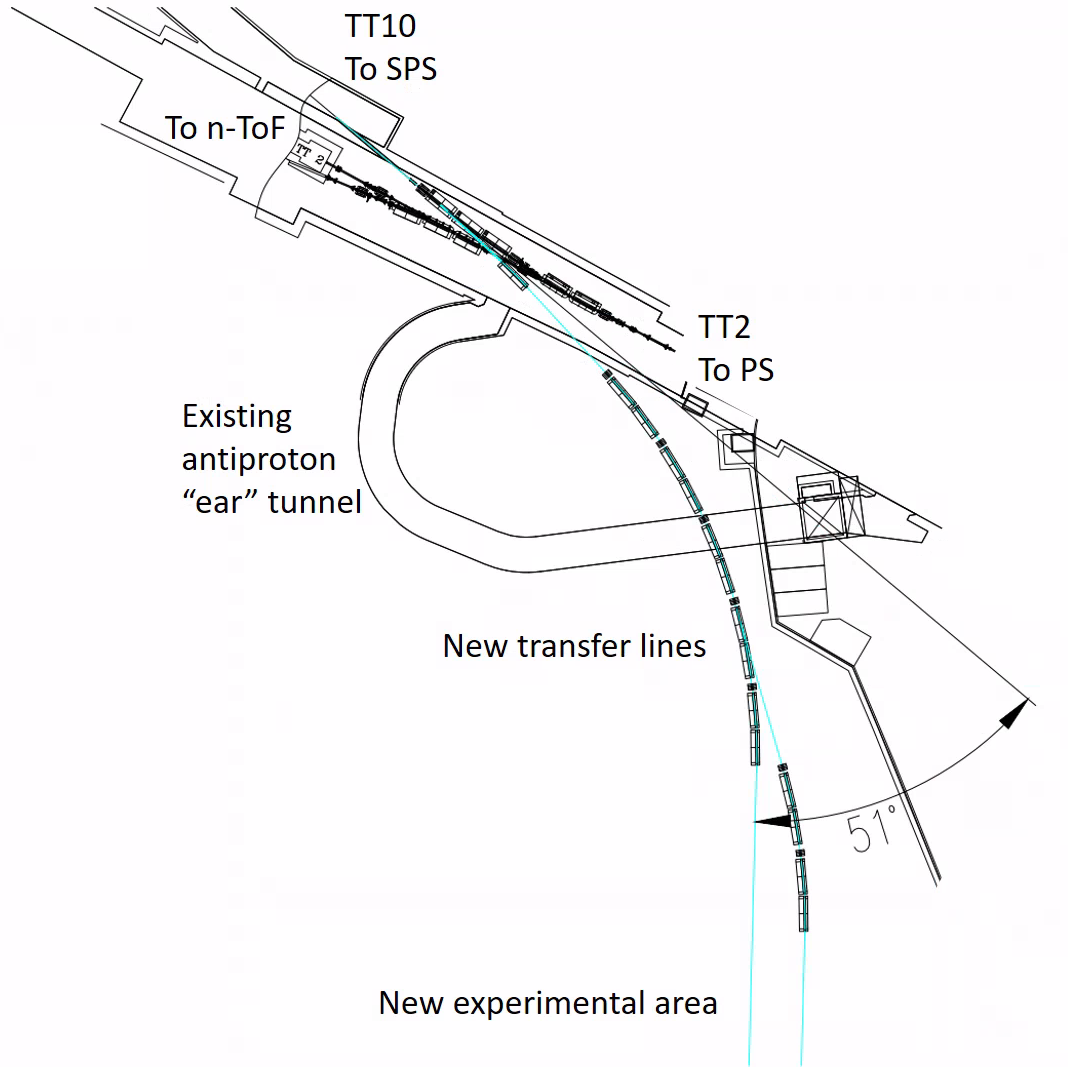}
            \caption{}
        \label{fig:Layout_Junction:all}
        \end{center}
    \end{subfigure}
    \caption{Schematic view of the base cell used in the new line with important dimensions in \si{mm} or \si{rad} \subref{fig:Line_cell} and layout of the new line in the junction region \subref{fig:Layout_Junction:all}, with electron trajectory in cyan.}
\label{fig:Layout_Junction} 
\end{center}
\end{figure}

Figure \ref{fig:Layout_Junction:all} shows the switchyard region at the end of the TT10 line. 
\begin{figure}[!hbt]
\begin{center}
   \includegraphics[width=0.60\linewidth]{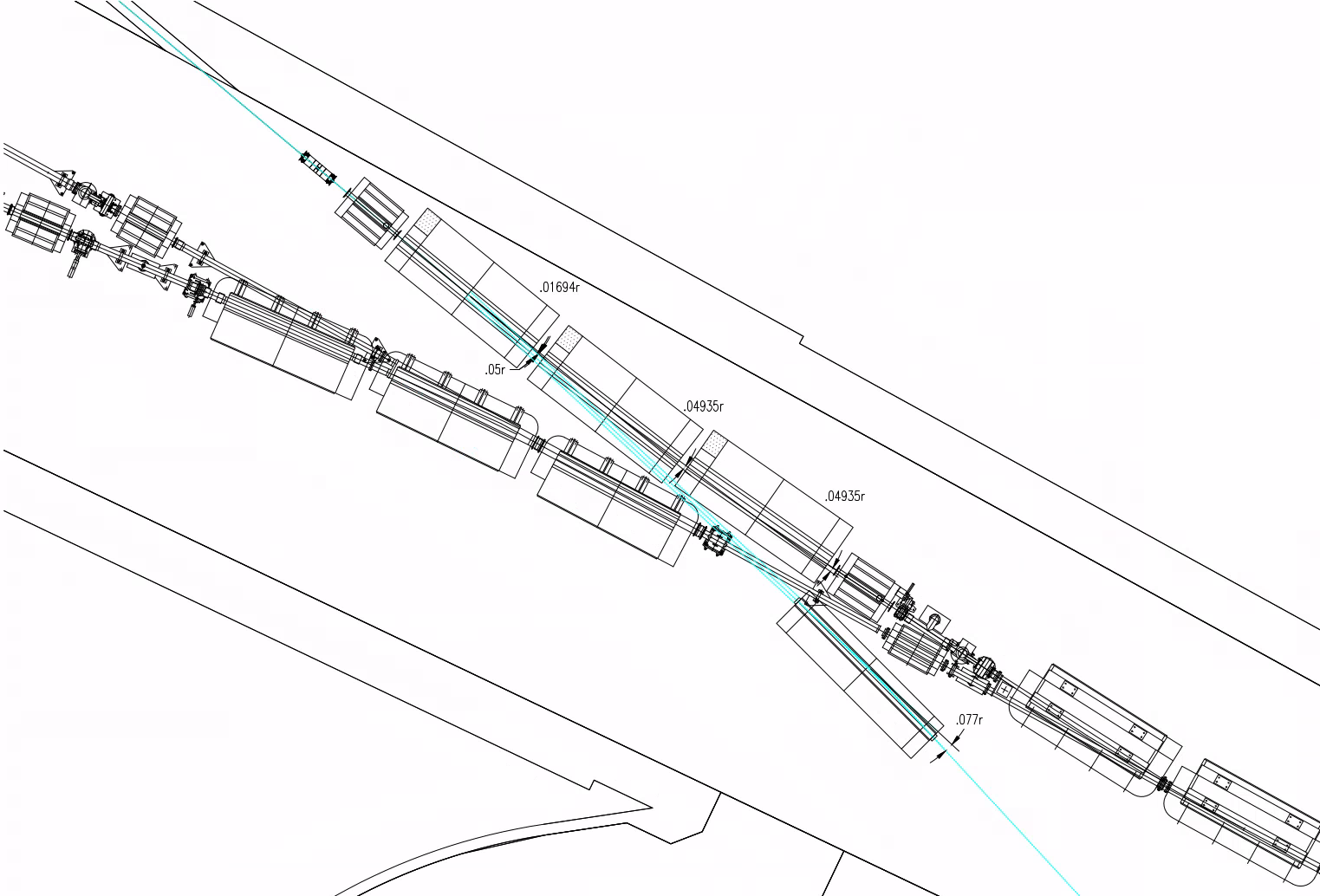}
   \caption{Layout of the junction region between the TT2, TT10 and n\_ToF lines with the proposed electron trajectory in cyan.}
\label{fig:Layout_Junction:zoom}
\end{center}
\end{figure}
This region is already filled by magnetic elements due to the split of the TT2 line coming from the CPS into the TT10 and n\_ToF lines. The design proposed here aims at maintaining the proton trajectories while allowing the electron beam to branch off the existing line towards a new line. Many factors where included in the design and integration of all the elements and ancillary systems was decisive (see Section~\ref{sec:ICE:extract:integration}).

Currently two MAL-type dipoles bend the proton beam towards the TT10 line. In Fig.~\ref{fig:Layout_Junction:zoom} those dipoles are replaced by three HB2 C-type dipoles. The deviation of the proton beam coming from the PS is around \SI{50}{mrad} for the first two dipoles and around \SI{17}{mrad} for the last one. This peculiar choice of strengths preserves the proton trajectories outside of these three dipoles, in TT2 and TT10.  
The quadrupole immediately before the wall on the TT10 side and its downstream beam monitor will have to be shifted downstream by \SI{30}{cm} (see Section~\ref{sec:ICE:extract:integration}), with very small effect on the optics. Two new power supplies will have to be wired to those three dipoles.
This new layout for the transport of the proton beam will be able to transport protons up to \SI{26}{GeV} but further studies on the HB2 magnets at higher fields could allow operation at higher rigidity.

Electrons coming from TT10 will experience a deviation of \SI{50}{mrad} in the first BH2 dipole. The beam will travel through the opening of the following two dipoles before crossing the n\_ToF line. The beam pipe and vacuum chamber in this region will become particularly complex, but the careful integration study ensured that this concept is within the capabilities of CERN experts. 
This solution is only one among several explored. A complete bypass of this area with a new beamline above those magnets could be studied in case a technical design of the concept presented here identifies show stoppers. However the concept discussed here was developed in close coordination with 3D integration and other groups to ensure its feasibility (see Sections~\ref{sec:ICE:extract:integration} and \ref{sec:ICE:extract:transport}).

The new beam line starts with a single MCW C-type dipole as close to the region of crossing with the n\_ToF line as possible. The following drift space up to the wall allows for passage of personnel and equipment, if the beam pipe is removed. There follows a penetration through the existing tunnel wall towards the new tunnel. The concept presented here for the new tunnel was directed by several constraints:
\begin{itemize}
    \item The two experimental areas cannot overlap with existing structures;
    \item The two experimental areas should be capable of receiving beam successively;
    \item The beamline should make use of existing and available magnets due to the high cost of new magnets manufacturing;
    \item The design should be capable of transporting \SI{16}{GeV} electrons with possible upgrade to at least \SI{18}{GeV};
    \item The beam delivered to the primary experimental area should be manipulated to reach large transverse sizes of up to a few tens of centimetres.
\end{itemize}

The bending section of the new line is built around a simple FODO arrangement, shown in Fig.~\ref{fig:Line_cell}. 
It is designed around two MCW dipoles of \SI{3.5}{m} that each impart an angle of \SI{77}{mrad} to the electron beam. Each bending cell starts with a water-cooled quadrupole of type QTN and ends with a \SI{600}{mm} straight section. The design of this bending cell provides the space needed to install ancillary systems not included in this study such as beam position monitors, vertical correctors or vacuum ports.

Five bending cells plus the other dipoles in the junction region provide the total angle of \SI{51}{\degree} and align the beam trajectory with the experimental area. The beam is directed towards the secondary experimental area by switching the powering of the last two bending cells to the secondary line, around \SI{5}{m} besides the primary one. This is possible due to the opening on the C-type dipoles used for the bending cells.

The very large total angle of the beam trajectory from the SPS extraction is a concern since the high energy electrons may radiate a non- negligible fraction of their energy. From the extraction of the SPS to the end of the TT10 line the cumulative angle is quite small and the average relative energy lost by electrons at \SI{16}{GeV} is in the order of \num{e-4}. The large angle of \SI{51}{\degree} between the TT10 line and the experimental area (see Fig.~\ref{fig:Layout_Junction:all}) causes a relative average loss of energy of \num{1.3e-3} for electrons of at \SI{16}{GeV}. The change of the average beam energy along the new line can be compensated by splitting the powering of the cells in different circuits. A technical design will also need to account for the increase in momentum spread associated with stochastic energy loss.

The beam delivery system to a missing momentum experiment must be capable of provide a particularly large beam size. A \SI{12}{m} straight section after the last dipole is dedicated to the manipulation of the beam size. The design studied here is very simple and limited to the use of quadrupoles that are available and stored at CERN. The synoptic of this Final Defocus scheme is shown in Fig.~\ref{fig:SPS_to_LDMX}. The beam sizes achieved here range from \SI[product-units=power]{10x1.9}{cm} to \SI[product-units=power]{30x0.4}{cm}. Figure~\ref{fig:SPS_to_LDMX} presents the optics leading to a beam size on the experimental target of \SI[product-units=power]{20x1.4}{cm}. This concept demonstrates an approximate range of beam sizes that could be provided. However, a more detailed design can be established together with precise experimental requirements with an experiment design optimised for large beam sizes.

The design makes use of existing and available magnets to minimise the cost. Table~\ref{tab:magnets_res_newline} lists those magnets with some of their most important characteristics.

\begin{table}[!hbt]
\begin{center}
\caption{List of existing magnets used in the junction area and the new line.}
\label{tab:magnets_res_newline}
\begin{tabular}{lccrrr}
\hline\hline
\textbf{Name} & \textbf{Type}  & \textbf{Quantity}  & \textbf{Max. current [A]} & \textbf{Max. integrated strength}\\ 
\hline
QTN &  Water-cooled quadrupole & 9 & 150 & \SI{2.052}{T} \\
QTS & Water-cooled quadrupole & 2 & 416 & \SI{37}{T} \\
Q200 & Water-cooled quadrupole & 1 & 750 & \SI{22.05}{T} \\
MCW & Water-cooled dipole & 15 & 1000 & \SI{4.63}{T.m} \\
MCB & Water-cooled dipole & 3 & 880 & \SI{4.44}{T.m} \\
\hline\hline
\end{tabular}
\end{center}
\end{table}

In this section we discussed how to produce and deliver a low intensity beam to a missing momentum experiment. The intensity considered will be around \num{e8} to \num{e9} electrons on target per second, considerably below the maximum number of electrons in the SPS ring, in the order of \num{e12}. A dump-type experiment making use of that beam could be housed at the end of the secondary new line and beside the missing momentum experiment, as shown in Fig.~\ref{fig:Layout_Junction:all}. 

%% file: include/04-SPS/BI_Extr.tex
\subsubsection{Beam instrumentation}
\label{sec:BI_Extr}

The instrumentation for the SPS extraction system and TT10 transfer line to the experimental area will need to be able to deal with two modes of operation. The first mode is based on a fast extraction for setting-up the line, where the full beam intensity is extracted in one SPS turn. The second one is a slow extraction performed while running the experiment, where only a fraction of electron beam circulating in the SPS is extracted over 10 seconds giving, on average, a few electrons per 5 ns period.

For the fast extraction, the 10 existing OTR-BTV systems can be used to monitor the beam, provided some modifications are made to their light extraction systems. The 10 existing strip-line beam position monitors in TT10 would also be capable of giving the trajectory of such beams, but would require additional cabling and electronic systems to be installed due to the directionality of these monitors. In addition, some 6 additional BPMs will need to be installed in the new transfer line that will send the beam to the experiment hall.

The case of the slow extraction is much more challenging for beam instrumentation as it would require extremely high sensitivity to be able to detect the very few particles sent to the experiment. A new type of monitor will need to be developed to cope with the extremely low beam fluence. One candidate to measure the position of the extracted beam is to detect the Cherenkov light emitted by the electrons as they propagate close to the surface of a long dielectric~\cite{Cherenkov_diffraction2}. This method, recently pioneered in the framework of the CLIC study on the Cornell Electron Synchrotron Ring (CESR)~\cite{Kieffer2018}, would nevertheless require an exhaustive R\&D phase to assess the sensitivity limit of such monitors. One candidate to measure transverse beam profile is an ultra-high vacuum (UHV) version of the scintillating fibre monitor developed for the CERN Neutrino Platform, capable of detecting particle rates as low as 100 Hz~\cite{ruiz:2018inst}. Significant development work will nevertheless need to be carried out to adapt the existing monitor and make it UHV compatible, probably requiring replacement of the scintillating fibres by Cherenkov fibres.
If successful, one can assume that 16 Cherenkov beam position monitors and 10 optical fibre profile monitors would then be built to monitor the extracted low charge beam while running the experiment.

%% file: include/05-ICE/OverviewICE.tex
\section{Infrastructure and civil engineering}
\label{sec:ICE}

\begin{figure}[!hbt]
\begin{center}
    \includegraphics[width=\linewidth,trim={0.0cm 0.0cm 0.0cm 0.5cm},clip]{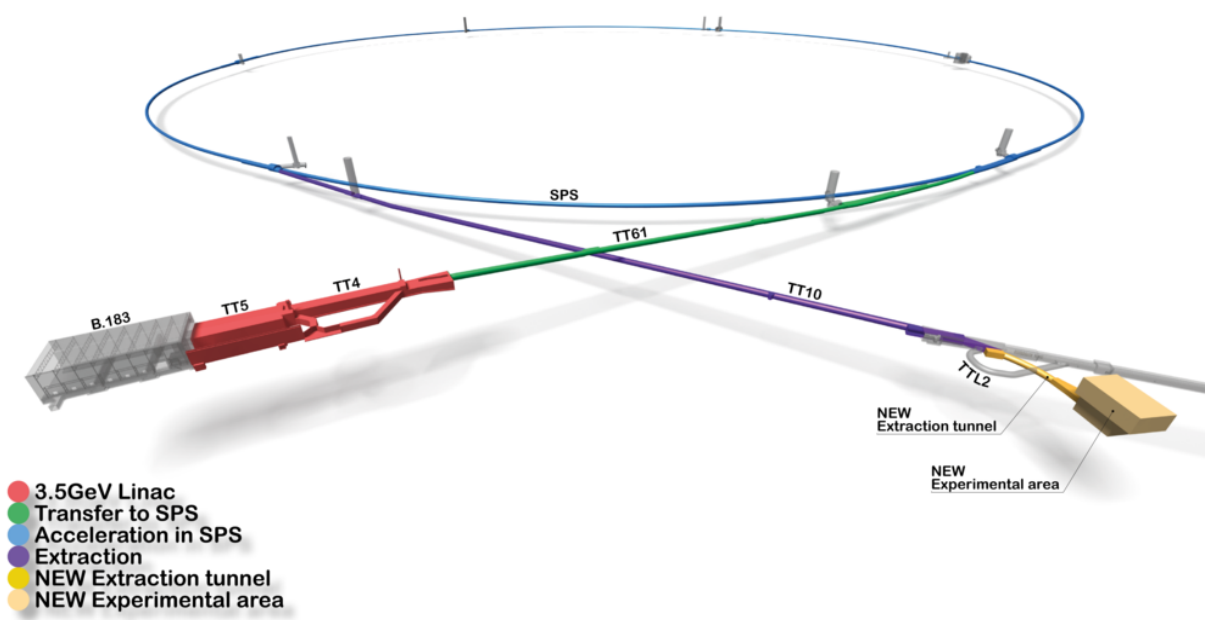}
    \caption{Overview of main elements of project.}
 \label{fig:Overview}
\end{center}
\end{figure}

Civil engineering (CE) and infrastructure generally represents a significant proportion of the total budget for projects at CERN.  On that basis, a significant amount of attention has been given to the CE and other infrastructure required to implement eSPS. This chapter will discuss the infrastructure and civil engineering considerations for the project, the optioneering process undertaken and the key factors in choosing the selected design options. Figure~\ref{fig:Overview} shows the main elements of the scheme. The scope of the study in this area includes the infrastructure and civil engineering required for the following:

\begin{itemize} 
\item Beam injection in TT5,TT4 (shown in red);
\item Transfer of beam via TT61, TCC6 and TT60 into the SPS (shown in green);
\item Extraction of the beam via TT2, TTL2 and a new extraction tunnel (shown in dark yellow);
\item Construction of a new experimental hall (shown in light yellow).
\end{itemize}

The section breaks the study down into general considerations (applicable to some or all parts of the scheme) and the specific considerations divided into the areas noted above.

The scope of this chapter also specifically excludes any changes required as part of the following which will be covered elsewhere in the CDR:

\begin{itemize}
\item Acceleration of the beam in the SPS (shown in blue);
\item Extraction of the beam in TT10 (shown in purple).
\end{itemize}
\newpage

%% file: include/05-ICE/GeneralConsiderations.tex
\subsection{General considerations}
\label{sec:ICE_General}

\subsubsection{Civil engineering}

\paragraph{Location}

\begin{figure}[!hbt]
\begin{center}
   \includegraphics[width=0.9\linewidth]{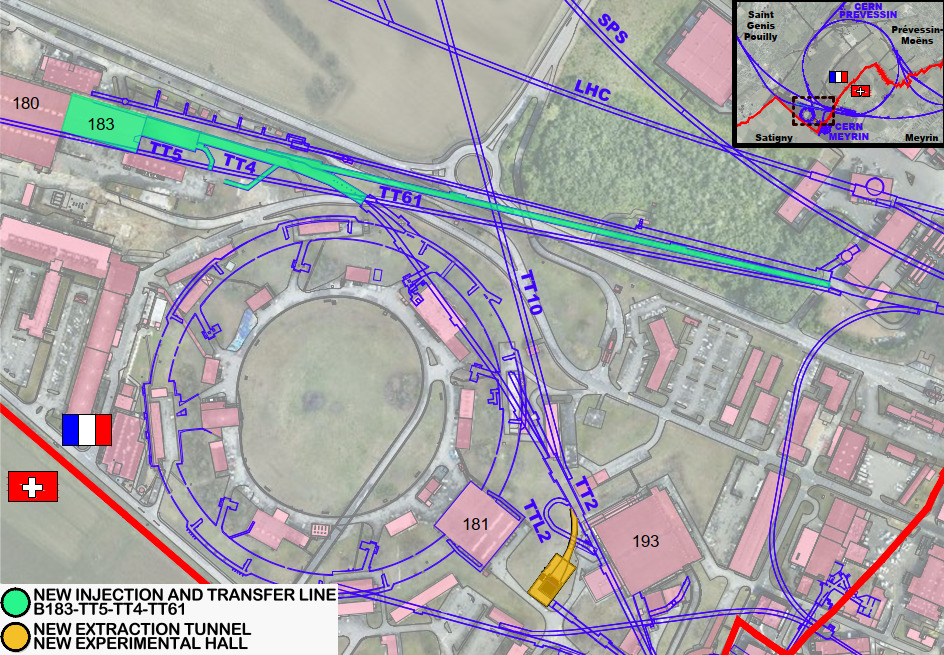}
   \caption{Location of the planned facilities at the CERN Meyrin site showing Franco-Swiss border.}
\label{fig:Fig_1}
\end{center}
\end{figure}

The eSPS scheme would span the CERN site, housed within a mixture of new and existing infrastructure as depicted in Fig.~\ref{fig:Overview}. The areas within the CE scope of this document are located entirely within CERN land in France as shown in Fig.~\ref{fig:Fig_1}. 

\paragraph{Geology}

The location is well suited from a CE point of view since the ground conditions are relatively stable and well understood with detailed geological records in and around the area, reducing the geotechnical risk. New construction would generally be founded within the molasse rock, which is very close to the surface in the relevant areas. The molasse is composed of an alternating sequence of marls and sandstones (and formations of intermediate compositions) and is generally considered good rock for tunnelling. This rock is overlain by Quaternary glacial moraines related to the Würm and Riss glaciations and in many places on the Meyrin site by made ground. 

\paragraph{Civil engineering layout}
The layout required for each part of the project and the CE works needed to achieve it are detailed in the appropriate sections of this chapter. The main elements of CE could be summarised as follows:

\begin{itemize}
    \item Refurbishment, minor structural modifications and provision of shielding block arrangement in B183, TT5 and TT4;
    \item Monitoring and possible maintenance of TT61 to allow re-installation of a transfer beam line;
    \item Construction of a new 55\,m long, 3.57\,m wide (varying to 9.09\,m) extraction tunnel;
    \item Construction of a 20\,m wide by 40\,m long experimental hall and detector pit as well as associated  infrastructure.
\end{itemize}

\subsubsection{Integration}

Integration studies have been performed to evaluate the feasibility of siting the eSPS facility within CERN's Meyrin site. The infrastructure requirements were defined and integrated within the civil engineering layout. The overall layout was optimised in terms of radiation protection, general safety, accessibility, and practicality. Figure~\ref{fig:Overview} shows the layout of the overall facility.

The integration of the facility is divided into the following four areas:

\begin{itemize}
\item Beam injection: Building 183, Transfer tunnels TT5 and TT4: this houses the CLEAR injector, the linac beam line and services;
\item SPS transfer: Transfer tunnel TT61 and junction cavern TCC6: this houses the SPS transfer beam line;
\item SPS extraction: Transfer tunnels TT2, TTL2 and the new extraction tunnel: this houses the SPS extraction beam line;
\item Experimental area: Surface and underground experimental halls: to house and install the experiment and to provide the services required for the operation of the detector.
\end{itemize}

The following sections describe in detail the integration studies performed for each of the four areas listed above. Table \ref{tab:SmarTeam references of the integration 3D models} contains the SmarTeam numbers of the integration models (ENOVIA SmarTeam is a product data management tool that enables organisations to manage and collaborate on component information).

\begin{table}[h!]
\begin{center}
\caption{SmarTeam references of the integration 3D models.}
\label{tab:SmarTeam references of the integration 3D models}
\begin{tabular}{p{4cm}ccc}
\hline\hline
\textbf{Description}   	& \textbf{SmarTeam number}\\ 
\hline
Injection                     & ST1153166 \\
SPS transfer                  & ST1283509 \\
SPS extraction                & ST1222240 \\
Experimental area             & ST1222240 \\

\hline\hline
\end{tabular}
\end{center}
\end{table}

\subsubsection{Electrical engineering and infrastructure}

\begin{figure}[!hbt]
\begin{center}
   \includegraphics[width=0.9\linewidth]{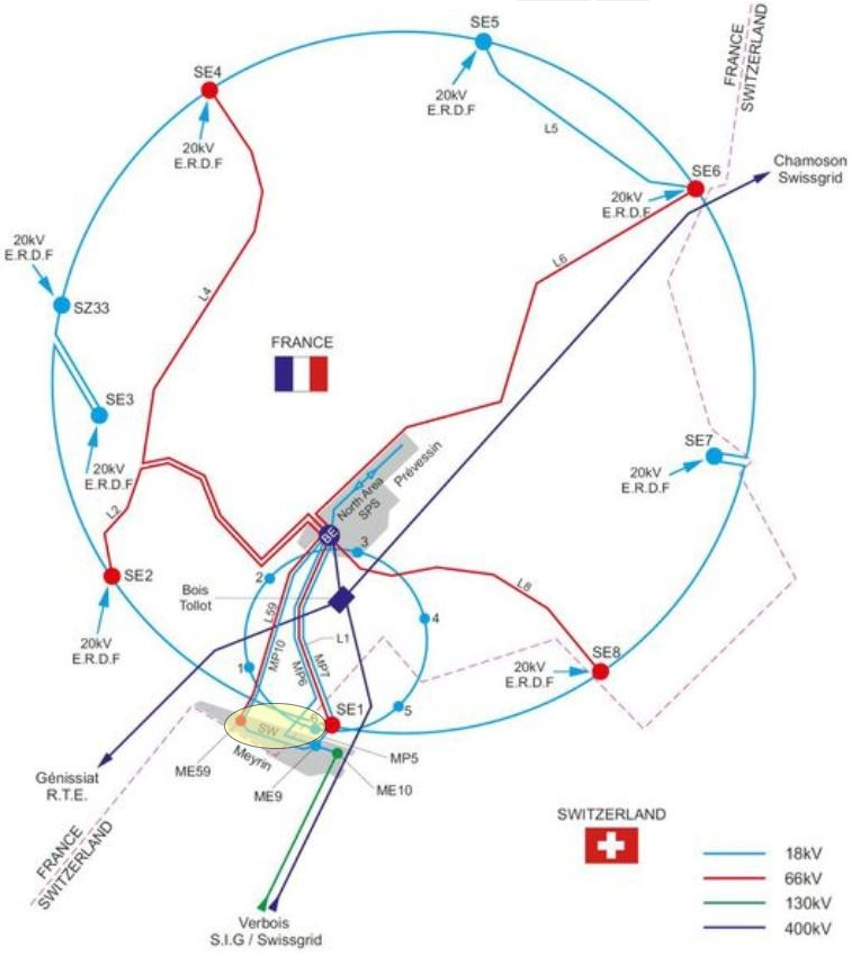}
   \caption{Schematic view of the CERN transmission and distribution network. Approximate location of eSPS according to the existing CERN electrical infrastructure is shown in yellow.}
\label{fig:EL1}
\end{center}
\end{figure}

The CERN electrical network is composed of a transmission and a distribution level. The transmission level transmits the power from the existing source of the European Grid to the different CERN locations including the Meyrin campus and the SPS complex where the eSPS is planned to be constructed. The transmission network operates at high voltage levels of 400\,kV, 66\,kV and 18\,kV. The distribution level distributes the power from the transmission level to the end users at medium and low voltage levels comprised of between 400\,V and 18\,kV. A schematic view of the CERN transmission and distribution network is shown in Fig.~\ref{fig:EL1}. The yellow ellipse in Fig.~\ref{fig:EL1} shows the approximate geographical location of the eSPS according to the existing CERN electrical infrastructure.

\newpage 
The concept for the design of the eSPS electrical network is driven by four factors:

\begin{enumerate}
    \item The estimated electrical power requirements;
    \item The status existing electrical network infrastructure;
    \item The location and type of equipment to be supplied;
    \item The electrical load class and load types.
\end{enumerate}
  
  \paragraph{eSPS power requirements}  
    
Table~\ref{tab:ELT1} summarises the electrical power loads for each of the three eSPS electrical areas. Power loads include values received as of April 2020 by concerned stakeholders. In the case of unknown values such as for the new experimental area, an estimate has been derived from equivalent and comparable CERN accelerators and utilities infrastructures. A total electrical power need of 4800 kW is considered at conceptual level. The exact values of the electrical power requirements will have to be confirmed during the technical design phase.

\begin{table}[!hbt]
\begin{center}
\caption{eSPS power requirements shown by electrical area.}
\label{tab:ELT1}
\begin{tabular}{p{6cm}ccc}
\hline\hline
\textbf{Electrical area}   	& \textbf{Nominal power (kW)}\\ 
\hline
Linac                     & 3600 \\
SPS including injection and extraction                  & 400 \\
Experimental area                & 800 \\

\hline\hline
\end{tabular}
\end{center}
\end{table}

\paragraph{Electrical network infrastructure status and requirements}

From an electrical supply point of view the eSPS accelerator is in areas where several electrical substations are available, operational, and equipped to cover the additional eSPS demand in terms of power and energy consumption. Three electrical zones, which correspond to the accelerator description in this chapter are defined for the eSPS. The three zones can be considered independent from an electrical point of view and are : The linac (B183, TT4 and TT5), the SPS ring including the injection and extraction of the SPS (TT61, TT10, TT2 and the new extraction tunnel) and the new experimental area. 

\subparagraph{Linac}

\begin{figure}[!hbt]
\begin{center}
   \includegraphics[width=\textwidth,height=7cm,keepaspectratio]{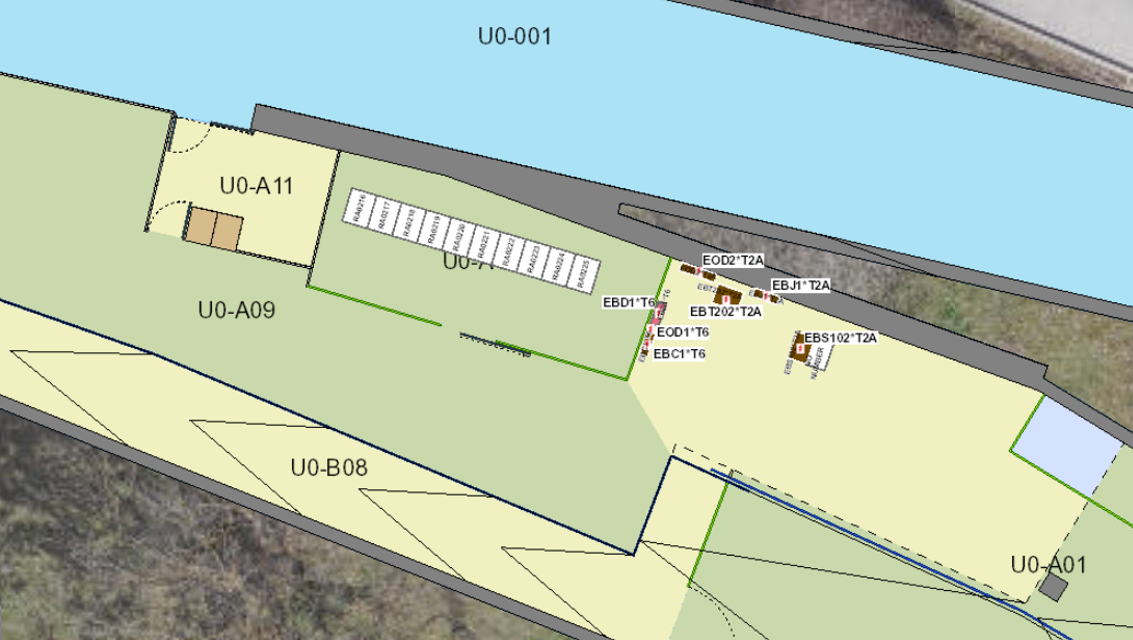}
   \caption{U0-A01 area hosting UPS infrastructure for the n\_TOF experiment.}
\label{fig:EL2}
\end{center}
\end{figure}

The linac is located nearby Building 112 where a primary electrical substation, hereafter called ME59, is located. Considering the expected power requirements given in Table~\ref{tab:ELT1}, the power required for the linac can be supplied form the substation ME59. The existing electrical infrastructure of the buildings where the linac will be installed dates from the seventies. A full refurbishment of the general services is necessary to replace obsolescent electrical equipment, comply with applicable safety rules and engineering standards. In TT4 the construction of the shielding wall and the installation of the beam dump will require the displacement of the UPS electrical installation in the area which supplies, among other users, the n\_TOF experiment. The concerned area U0-A01A is shown in Fig.~\ref{fig:EL2}. A new suitable location will have to be identified.

\subparagraph{The SPS ring including the injection and extraction}

The electrical infrastructure of the SPS complex has recently been renovated and sufficient power is available for the needs of the eSPS including the needs for the new extraction tunnel. The general services network of the existing SPS tunnels are in a good state. Changes or modifications might be necessary according to eSPS requirements. The additional loads for the injection and extraction of the SPS will be supplied from existing substations located in  points SPS-6 and SPS-7. The additional loads for the new extraction tunnel can be supplied from either the SPS infrastructure or from the new experimental area.

\subparagraph{The new experimental area}

The new experimental area will be constructed on the Meyrin campus as shown in Fig.~\ref{fig:Fig_1}. The new experimental area will be supplied from the electrical substation ME59, which is also the source for the linac zone.

\paragraph{Electrical load class and load types}

The electrical supply concept for eSPS is based on the requirement to keep essential parts of the accelerator infrastructure operational if the normal power source fails. Emphasis is put on loads related to personnel and machine safety during degraded situations. The various load classes and types can be characterised as shown in Table~\ref{tab:ELT2}. The main ranking parameters are the acceptable duration of the power interruption and whether the load is part of a personnel or accelerator safety system.

\begin{table}[!hbt]
\begin{center}
\caption{Load classes and main characteristics.}
\label{tab:ELT2}
\begin{tabular}{  p{2.8cm} p{6.8cm}  p{4.5cm} }
\hline\hline
\textbf{Load class}  & \textbf{Load type (non-exhaustive list)} & \textbf{Maximum duration of power unavailability }\\
\hline
\textbf{Machine} & Power converters, cooling and ventilation motors, radio frequency, klystrons & Until return of main supply\\
\hline
\textbf{General Services}  & Lighting, pumps, vacuum, wall plugs        & Until return of main or secondary supply\\
\hline
\textbf{Secured}  & \textbf{Personnel safety:} Lighting, pumps, wall plugs, lifts &  10--30 seconds\\
\hline
\textbf{Uninterruptible}  &  \textbf{Personnel safety:}  evacuation and anti-panic lighting, fire-fighting system, oxygen deficiency, evacuation  &  Interruptions not allowed, continuous service mandatory\\
 & \textbf{Machine safety:} sensitive processing and monitoring, beam loss, beam monitoring, machine protection & \\

\hline\hline
\end{tabular}
\end{center}
\end{table}

Machine loads do not have a second source of supply, therefore, in case of an upstream electrical power cut the equipment will be cut. The general service loads typically accept power cuts between several minutes and several hours, sufficiently long to commute to the second source or to wait until the main source is restored. Both the machine and general services loads do not include personnel or machine safety equipment or systems. Secured loads include personnel and machine safety equipment or systems that can sustain short power cuts up to a duration of 30 seconds. In a degraded situation or upstream electrical power cut a level backup is provided by the diesel power station, which typically starts up within 10~seconds. Uninterruptible loads include personnel and machine safety equipment or systems that require continuous and stable power supply. Wherever necessary, the uninterruptible network is created locally by installing an uninterruptible power supply (UPS) powered from the secured network.

\paragraph{Distribution network proposal}

\begin{figure}[!hbt]
\begin{center}
   \includegraphics[width=0.9\linewidth]{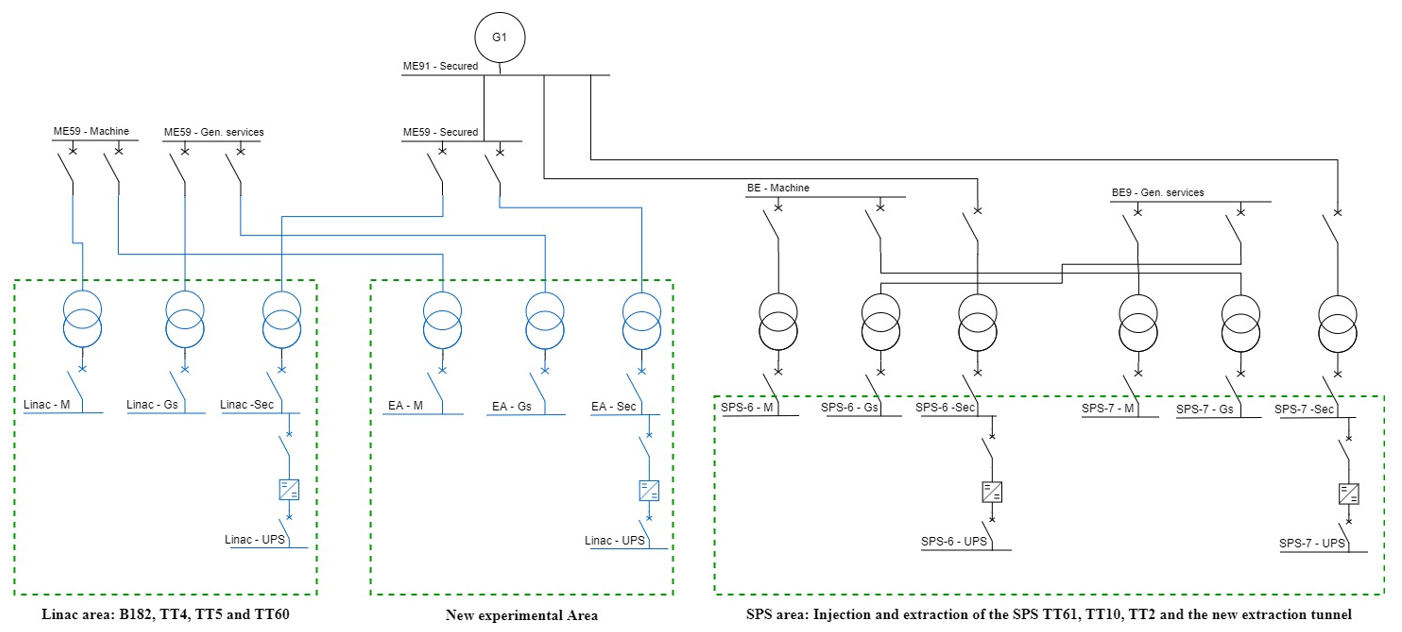}
   \caption{eSPS simplified electrical supply schematic of the distribution network.}
\label{fig:EL3}
\end{center}
\end{figure}

Considering the information gathered at the time of designing the eSPS layout and writing the electrical requirements, a simplified conceptual electrical supply scheme of the distribution network is proposed and shown in Fig.~\ref{fig:EL3}, see Ref.~\cite{EDMS}. The scheme describes how the four types of load classes are made available at the level of medium voltage and low voltage to the three eSPS electrical areas. For simplicity of the scheme, the 400\,kV and 66\,kV transmission network are not shown but are available in the drawing GENEM0033~\cite{EDMS}.The part of the diagram shown in black corresponds to the existing infrastructure while the part in blue represents the new infrastructure to be installed. The proposed scheme allows margin for optimisation as soon as the technical details become available during the technical design phase.

\subsubsection{Cooling and ventilation}

The Cooling and Ventilation section is composed of three parts: the first presents the piped utilities, the second the air-related systems and the third addresses safety and environmental protection aspects.

The design herewith presented is based on the user requirements available in February 2020. A safety margin of approximately 20\% was applied to the heat loads, given the early stage of the project. Additionally, the infrastructure was designed to be flexible and allow for future upgrades and extensions. Energy efficiency is considered a priority and different strategies are adopted to enhance it.

\paragraph{Piped utilities} \label{sec:ICE_General_Considerations_PipedUtilities}

The piped utilities are mainly dedicated to the cooling of the accelerator equipment and the related infrastructure such as power converters, electronic racks, cables, etc., as well as to the ventilation and air-conditioning of underground and surface premises. In addition, specific systems are foreseen to cover other general needs such as fire extinguishing plant, drainage and sumps (for both surface and below-ground areas), drinking water and compressed air for ancillary equipment (vacuum pumps, valves, dampers, etc.). The main circuit typologies are:

\begin{itemize}
\item Demineralised and industrial water: for the cooling of accelerator equipment (excluding: magnets, power converters, etc.) and infrastructure if needed;
\item Chilled water: at present, used exclusively in ventilation and air-conditioning plant;
\item Firefighting systems;
\item Wastewater: water from underground and surface premises that is to be rejected;
\item Drinking water: for sanitary purposes and make up of industrial water circuits;
\item Compressed air: mainly for ancillary equipment such as pneumatic valves or vacuum pumps.
\end{itemize}

\begin{figure}[!hbt]
\begin{center}
    \includegraphics[width=10cm]{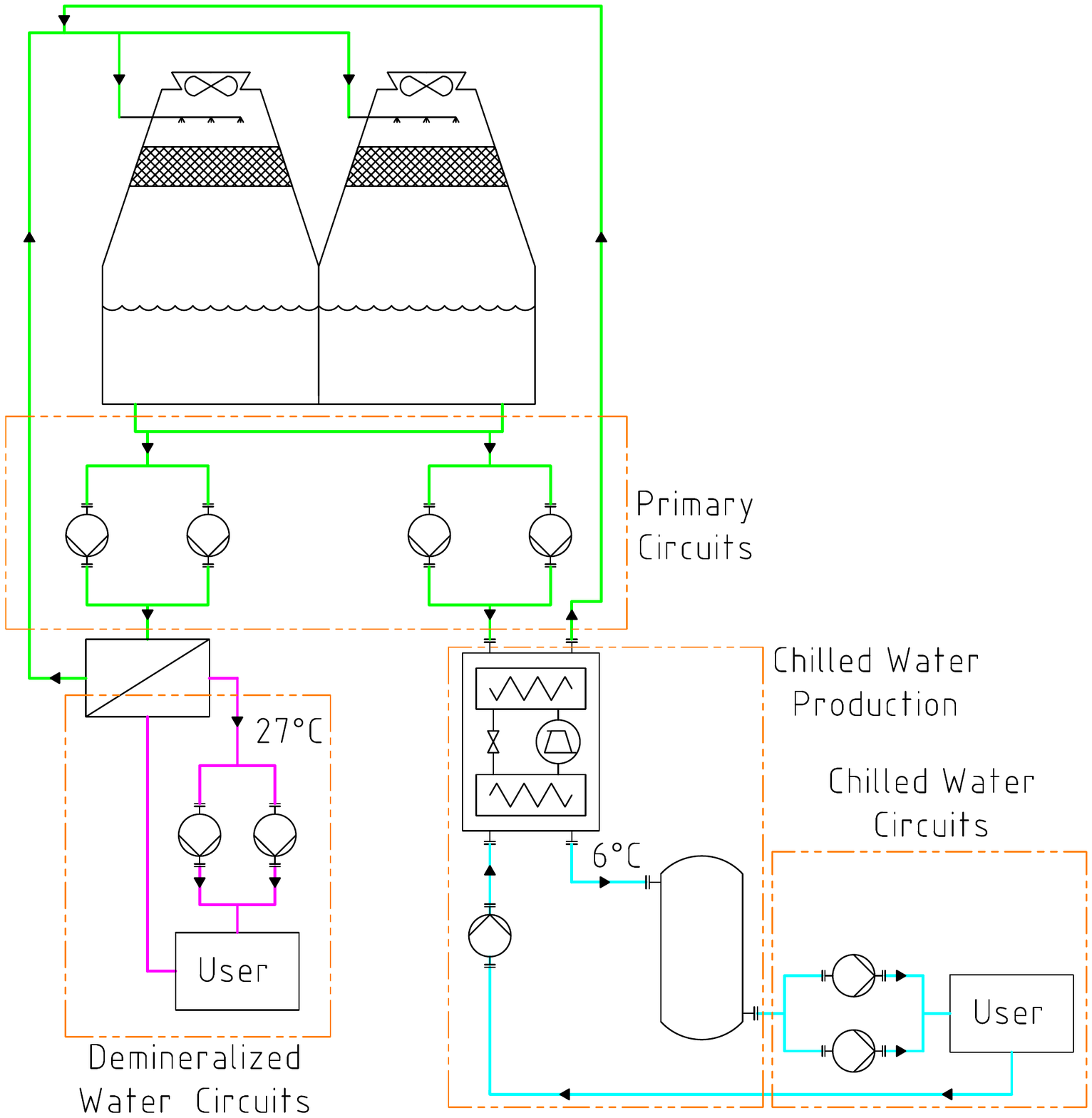}
    \caption{General diagram for cooling circuits.}
 \label{fig:CV_PID_General}
\end{center}
\end{figure}

Two different typologies of water cooling circuits are defined according to their equipment and working temperatures: demineralised water circuits and chilled water ones. The first are cooled by primary circuits (industrial water) and the latter by chillers, that are able to produce water at lower temperatures. Figure~\ref{fig:CV_PID_General} illustrates a simplified layout for the two types of circuits, although they can assume different configurations, depending on the requirements. 

The primary water circuits, cooled by open wet cooling towers, are configured in closed loops to minimise water consumption. Drinking water make-up is used to compensate for evaporation, leaks and blowdown. A continuous water treatment against Legionella, scaling and proliferation of algae is foreseen.

Generally, demineralised water circuits are used to refrigerate equipment and have a maximum conductivity of 0.5 $\mu$S/cm. A set of demineralisation cartridges is foreseen for the cooling circuits, in order to locally control the conductivity. The temperature of the water leaving the cooling station is set to 27\si{\degree}C~($\pm0.5$\si{\degree}C).

Chilled water is used to condition air in air-handling units. Where dehumidification is required, the leaving water temperature is approximately 6\si{\degree}C, otherwise it is usually raised to 14\si{\degree}C.

Most electromechanical cooling components, such as cooling towers, pumps and chillers are foreseen with N+1 redundancy. However, no redundancy is foreseen for electrical cubicles, control cubicles and heat exchangers. Under present conditions, there is no need to provide a secure power supply for the majority of the cooling plant. In case of power failure, all accelerator related equipment would stop and would not require any cooling.

The drinking water network is used for sanitary purposes and as make-up water for the cooling towers. The existing drinking water network at CERN is extended and upgraded to satisfy the requirements in the new premises.

The existing firefighting water is upgraded to cover areas that require such systems. For the underground infrastructure, such as TT4 and TT5, rigid pipelines are installed. They are kept dry to avoid corrosion and water stagnation. In case of fire, the fire brigade opens manual valves at the surface to supply water to the concerned area.

The reject water is split in two distinct networks - one for clear water and one for sewage; these networks are connected to the corresponding existing systems in each area. For safety reasons, each sump is equipped with sensors providing alarms for high and too high level.
Thorough control mechanisms are implemented before releasing reject water to existing natural water lines. Relevant parameters such as pH and temperature are monitored.

The user requirements for compressed air are not defined at this stage of the project. However, existing networks are present in CERN premises. If their capacity is not sufficient, or if they cannot be used (for any other reason), new production stations will be installed. An N+1 redundancy is foreseen for the compressors.

\paragraph{Heating, ventilation and air-conditioning} \label{sec:ICE_General_Considerations_Heating}

The heating, ventilation and air conditioning (HVAC) plants are designed to provide fresh filtered air to people and to purge (where required) contaminants and pollutants that might be harmful, allowing for safe access. For certain facilities, the exhaust air is filtered before being rejected to the atmosphere. Moreover, HVAC plants provide the desired indoor temperature and/or humidity in underground and surface premises. Finally, these plants provide smoke extraction in places where it is required, such as for TT4 and TT5.

\paragraph{Safety and environmental protection}

The safety of people and environmental protection are priorities in the early design herewith described. Collaboration with CERN’s Occupational Health \& Safety and Environmental Protection (HSE) Unit is established to ensure the satisfaction of the regulations and application of best practice in these domains.

Two safety aspects were considered in the context of the cooling and ventilation infrastructure - fire protection and radiation protection. For the first, firefighting water networks are put in place, as described in Section~\ref{sec:ICE_General_Considerations_PipedUtilities}. Additionally, hot smoke extraction systems are foreseen for certain selected areas, as mentioned in Section~\ref{sec:ICE_General_Considerations_Heating}. Other safety related mechanisms like pressurisation of safe zones, particularly in the new detector region, are likely to be required. However, this will be defined in the future, during a more advanced stage of the study. For radiation protection, pressure cascades are evaluated, as well as requirements concerning different operational modes, such as purge and access.

Environmental protection is addressed from two different perspectives - mitigation of impacts related to waste fluids and energy/water efficiency of cooling and ventilation plant. Strategies are put in place to design systems that treat and prepare fluids that are to be rejected. For instance, filters are foreseen at the air outlets of certain premises and water treatment is put in place for the cooling towers blowdown.

Water conservation is considered central in the design of the cooling and ventilation infrastructure. For instance, a new system to treat and reuse cooling tower waste water is currently being developed at CERN and is foreseen to be implemented. This allows for large savings in drinking water used as make-up. Energy efficient measures are taken to minimise electricity consumption. For instance, air is recycled as much as possible, avoiding the energy costs associated with treating fresh air. For TT4 and TT5, this requires long ducts; however, the additional fan power is largely offset by the savings in ventilation loads. A detailed study to optimise energy efficiency is required for future more detailed stages of the present study.


\subsubsection{Radiation protection}

To mitigate risks associated with ionising radiation, CERN radiation protection rules and procedures must be applied~\cite{CERN2006}. Risks resulting from ionising radiation must be analysed to develop mitigation approaches. Design constraints ensure that the doses received by personnel working next to the beam installations as well as the public will remain below regulatory limits under all operational conditions. A~reliable radiation monitoring system coupled with an effective early warning and emergency stop systems are important parts of the radiation protection infrastructure.

Radiation protection is concerned with two aspects: protection of personnel operating and maintaining the installations and the potential radiological environmental impact. The facilities were optimised based on general radiation protection guidelines and specific studies on prompt and residual dose rates and air activation. To assess the radiation protection aspects, extensive simulations were performed with the FLUKA Monte Carlo radiation transport code~\cite{Battistoni2015,Boehlen2014,Vlachoudis2009}.

The radiological hazards mainly arise from the accelerated electron beam with energies of up to \SI{16}{GeV} and a beam power of up to \SI{1.28}{kW} and the prompt stray radiation generated by beam losses. This hazard represents the major challenge for the shielding design given the construction constraints in the existing infrastructure. The basic parameters used for the radiation protection study are listed in Table~\ref{tab:intens}. They provide sufficient margin compared to the projected range of parameters as shown in Table~\ref{tab:linac_beam_params}, except for the high charge mode, for which the repetition rate would need to be lowered accordingly.

By the nature of electron beams, which have a considerably lower activation potential compared to proton beams of similar power, the activation of materials, liquids and air represent a lesser concern. Radiation from induced activity will be limited by actual admissible beam losses driven by the shielding in place. The radiological environmental impact of the eSPS installations is negligible given the proposed design and beam parameters. 

The operation of klystrons to produce high-power RF and generating parasitic X-rays is a further source of radiation. Standard prescriptive methods to mitigate these risks are well established and allow an efficient control.

A more detailed radiation protection study was conducted for the linac part to demonstrate feasibility regarding the envisaged beam parameters and operational and shielding constraints. The injection, acceleration and extraction from the SPS have no significant radiological impact. The design of the new experimental hall and cavern have been covered in less detail but no limiting constraints have been identified for their foreseen implementation. More details on the radiological study presented here can be found in Ref.~\cite{Widorski2020}.

\begin{table}[!hbt]
\centering
\caption{Operational scenarios and beam parameters used for the radiation protection study. These scenarios basically cover the intended Linac beam modes as shown in Table~\ref{tab:linac_beam_params}.}
\begin{tabular}{lcccc}
\hline\hline
	& \multicolumn{2}{c}{\textbf{S-Band Linac}}	& \multicolumn{2}{c}{\textbf{X-Band Linac}}	\\
	& \textbf{multi bunch} 	& \textbf{single bunch}	& \textbf{multi bunch} 	& \textbf{single bunch}	\\
\midrule
Max. energy [GeV]	& \SI{0.250}{}	& \SI{0.250}{}	& \SI{3.55}{}	& \SI{3.65}{}	\\
Avg. beam intensity [\SI{}{e^-\per\second}]	& \SI{1.875E13}{}	& \SI{3.125E12}{}	& \SI{1.25E12}{}	& \SI{1.25E12}{}	\\
Avg. beam current [\SI{}{nA}]	& \SI{3000}{}	& \SI{500}{}	& \SI{200}{}	& \SI{200}{}	\\
Avg. beam power [\SI{}{W}]	& \SI{750}{}	& \SI{125}{}	& \SI{710}{}	& \SI{730}{}	\\
Repetition rate [\SI{}{Hz}]	& \SI{10}{}	& \SI{100}{}	& \SI{100}{}	& \SI{100}{}	\\
Pulse charge [\SI{}{nC}]	& \SI{300}{}	& \SI{5}{}	& \SI{2}{}	& \SI{2}{}	\\
Pulse charge [\SI{}{e^-}]	& \SI{1.875E12}{}	& \SI{3.125E10}{}	& \SI{1.25E10}{}	& \SI{1.25E10}{}	\\
Bunch charge [\SI{}{nC}]	& \SI{2}{}	& \SI{5}{}	& \SI{0.05}{}	& \SI{2}{}	\\
Bunch charge [\SI{}{e^-}]	& \SI{1.25E10}{}	& \SI{3.125E10}{}	& \SI{3.125E8}{}	& \SI{1.25E12}{}	\\
Bunches per pulse	& \SI{150}{}	& \SI{1}{}	& \SI{40}{}	& \SI{1}{}	\\
dE per module [MeV]	&	&	& \SI{137.5}{ }	& \SI{141.67}{ }	\\
Accelerating gradient [MV/m]	&	&	& \SI{59.8}{}	& \SI{61.6}{}	\\
No. of modules	&	&	& \SI{24}{}	& \SI{24}{}	\\
Module length [m]	&	&	& \SI{2.65}{}	& \SI{2.65}{}	\\
Accelerating structures	&	&	& 96 (4/mod.)	& 96 (4/mod.)	\\
\midrule
	& \multicolumn{2}{c}{\textbf{SPS}}	& \multicolumn{2}{c}{\textbf{Missing Momentum Experiment}}	\\
\midrule
Max. energy [GeV]	& \multicolumn{2}{c}{\SI{16}{}}	& \multicolumn{2}{c}{\SI{16}{}}\\
Avg. beam intensity [\SI{}{e^-\per\second}]	& \multicolumn{2}{c}{\SI{5.0E11}{}}	& \multicolumn{2}{c}{\SI{3.25E9}{}}\\
Avg. beam power [\SI{}{W}]	& \multicolumn{2}{c}{\SI{1280}{}}	& \multicolumn{2}{c}{\SI{8.33}{}}	\\
\hline\hline
\end{tabular}
\label{tab:intens}
\end{table}

All generally accessible areas inside buildings which are dedicated to the exclusive operation of the therein installed accelerators shall not be classified higher than Supervised Radiation Area during nominal beam operation. The areas outside, which are generally classified as non-designated public domain areas shall be protected such to comply with the applicable limits. The design target values were chosen a factor 3 lower than the actual limits to include a margin for uncertainties in the models, calculations and imperfections in the construction of the infrastructure. The corresponding dose limits and design values are shown in Table~\ref{tab:limits}.

\begin{table}[!hbt]
\centering
\caption{Applicable ambient or effective dose and dose rate limits for on-site areas~\cite{Forkel-Wirth2006} and for the reference group for the environmental impact (all sources combined).}
\begin{tabular}{llcccc}
\hline\hline
\textbf{Area classification} 	&\textbf{Permanent stay}	&\textbf{Low occupancy} 	&\textbf{Optimisation} 	&\textbf{Design target} 	\\
	&	&	&\textbf{threshold}	&	\\
\midrule
Non-designated	& \SI{0.5}{\micro\sievert\per\hour}	& \SI{2.5}{\micro\sievert\per\hour}	&\SI{100}{\micro\sievert\per\year} 	& \SI{0.05}{\micro\sievert\per\hour} 	\\
Supervised	& \SI{3}{\micro\sievert\per\hour}	& \SI{15}{\micro\sievert\per\hour} 	& - 	& \SI{1}{\micro\sievert\per\hour}	\\
Simple	& \SI{10}{\micro\sievert\per\hour}	& \SI{50}{\micro\sievert\per\hour} 	& - 	& \SI{3}{\micro\sievert\per\hour} 	\\
Limited Stay	& -	& \SI{<2}{\milli\sievert\per\hour} 	& - 	& -	\\
High Radiation	& -	& \SI{<100}{\milli\sievert\per\hour} 	& - 	& - 	\\
Prohibited 	& - 	& \SI{>100}{\milli\sievert\per\hour} 	& - 	& - 	\\
\midrule
Reference group	& \SI{0.3}{\milli\sievert\per\year}	&	&\SI{10}{\micro\sievert\per\year} 	&\SI{<0.5}{\micro\sievert\per\year}	\\
\hline\hline
\end{tabular}
\label{tab:limits}
\end{table}

\subsubsection{Safety engineering} 
\label{sec:ICE-Gen-safety}

\paragraph{Workplace safety}
Many standard industrial hazards will need to be considered for working in the eSPS, such as noise, lighting, air quality, and working in confined spaces, which can be satisfied by following existing CERN safety practices and the Host State’s regulations for workplaces. 

The ambient temperature in the tunnels will be of particular interest for access and work; following a guidance document provided by the French INRS~\cite{TravailINRS} a limit of 28$^{\circ}$C is recommended for manual work within the tunnels (across the full range of relative humidity values) during extended access periods. At~temperatures above this where manual work is required, care must be taken to monitor the condition of workers, ensuring reduced working time and increased rest and hydration in proportion to any increased heat induced stress.

\paragraph{Fire safety}
All buildings, experimental facilities, equipment and experiments installed at CERN shall comply with CERN Safety Code E and other fire safety related instructions and notes listed in Ref.~\cite{CERN-HSE}. In view of the special nature of the use of certain areas, in particular underground, and the associated fire risks, CERN’s HSE unit is to be considered the authority for approving and stipulating special provisions.  

As the project moves to the technical design report stage, finalising layouts and interconnecting ventilation systems, detailed fire risk assessments will have to be made for the eSPS complex. At this stage, general fire safety considerations have been outlined, and implemented in the civil engineering, integration and ventilation designs, based upon the latest fire safety strategies employed at CERN. As such, the complex has been considered to apply an extension of the fire safety concept being implemented in the SPS during CERN’s LS2 period and beyond.

The most efficient protection strategy is one that uses complementary “safety barriers”, with a bottom-up structure, to limit fires at the earliest stages with the lowest consequences, thus considerably limiting the probability and impact of the largest events. 

In order to ensure that large adverse events are possible only in very unlikely cases of failure of many barriers, measures at every possible level of functional design need to be implemented: 

\begin{itemize}
    \item In the conception of every piece of equipment (e.g.\ materials used in electrical components, circuit breakers, etc.); 
    
    \item In the grouping of equipment in racks or boxes (e.g.\ generous cooling of racks, use of fire-retardant cables, and fire detection with power cut-off within each rack, etc.); 
    
    \item In the creation and organisation of internal rooms (e.g.\ fire detection, power cut-off and fire suppression inside a room with equipment);   
        
    \item In the definition of fire compartments; 
    
    \item In the definition of firefighting measures.  
\end{itemize}

The key fire safety strategy concepts are set out below.

\subparagraph{Access and egress}
\begin{itemize}
    \item Ensure that the evacuation distance and path width is within the acceptable range for underground facilities;
    
    \item Ensure the evacuation routes are practicable;
    
    \item Ensure two ways in and out for the fire service intervention.  
\end{itemize}

\subparagraph{Reasonable fuel load and fire ignition sources}
\begin{itemize}
    \item Reduce additional fuel load, beyond that of the oil in the klystrons, especially due to storage of material or equipment;
    
    \item Minimise the number of possible fire ignition sources, avoiding workshops within the same fire compartment.
\end{itemize}

\subparagraph{Compartmentalisation}
Compartmentalisation impedes the propagation of fire and potentially activated smoke through a facility, allowing occupants to escape to a comparatively safe area much more quickly than otherwise, as well as facilitating the effective fighting of the fire, and evacuation of victims by the CERN Fire and Rescue Service. The following requirements have been set:

\begin{itemize}
    \item All ventilation doors must be fire doors EI90;
    
    \item Isolate communicating galleries with fire doors EI90; 
        
    \item Isolate neighbouring surface facilities with fire doors EI120;
        
    \item Normally opened fire doors to be equipped with remote action release mechanism, monitoring position and self-action thermal fuse. 
\end{itemize}

\subparagraph{Fire detection, fire alarm, safety action integration}
An early fire detection system, integrated into the safety action system is a crucial component of the SPS fire safety WP2 technical solution, which shall also be implemented in the eSPS. Early detection is such that it allows evacuation (last occupant out) before untenable conditions are reached. The system must be capable of transmitting an alarm, triggered upon fire detection, action on evacuation push buttons, CERN Fire Brigade action out of CERN FB SCR/CCC or BIW (Beam Imminent Warning) situations. Evacuation push buttons shall cover all premises. The fire detection and evacuation push buttons must also be integrated with safety actions such as compartmentalisation, ventilation stop and other machine functions according to a predefined fire protection logic. The existing fire safety system shall be dismantled. The key features of the system are:

\begin{itemize}
    \item Fire detection and fire alarms throughout facility: smoke detection by air sampling, and manual call points;
    
    \item Standard ventilation interlocked with fire detection: command \& control of fire dampers, fire doors and ventilation stops;
    
    \item Voice alarm system : broadcast of audible signals with loudspeakers (evacuation signal and beam imminent Warning signal);    
        
    \item Triggering of an evacuation alarm in the fire compartment of origin and the adjacent fire compartments; 
    
    \item Triggering of a Level 3 alarm in the CSAM system, which alerts the CERN Fire Brigade control room, and results in crews being dispatched immediately.   
\end{itemize}

The eSPS fire safety system will be connected to the SPS fire safety system by including the control cabinets in the optical fibre loops of the SPS system. Space allocation shall be made in an accessible area (not subject to radiation) in the TT5 Hall, for the evacuation and fire detection control panels. The eSPS system shall have the same functionalities as the SPS system: manual command of safety functions, broadcast of instructions with safety microphones for the CERN Fire and Rescue Service use, and microphones for SPS OP use.

\subparagraph{Smoke extraction}
Smoke extraction facilitates both the safe evacuation of occupants, and the effective intervention of the CERN Fire and Rescue Service to locate victims and prevent the further spread of a fire. A buildup of smoke can also result in lasting damage to the sensitive and valuable equipment present, an effect that can be limited through early extraction. This will be addressed in detail for each area of the eSPS complex.

\subparagraph{Fire suppression}
The CERN Fire Brigade need adequate means of fighting a fire on arrival. This will be addressed in detail for each area of the eSPS complex.

\paragraph{Electrical safety}
The electrical hazards present for this project are considered standard for such an installation. There will be a number of high voltage systems, including the klystrons, modulators and magnets. The supply infrastructure represents additional hazards frequently found at CERN, such as uninterruptible power supplies, transformers and power converters; these shall be mitigated through sound design practice and execution. The CERN Electrical Safety Rules, alongside NF C 18-510, shall be followed throughout the design process; where exceptions are required, this shall be subject to an appropriate level of risk assessment to evaluate the residual risk, and determine the mitigation strategies required. NF C 18-510 compliant covers, interlocks preventing access to high voltage equipment, and restriction of access to the electrical rooms to those with the appropriate level of CERN electrical training and certification shall be used to protect personnel from any electrical hazards present. 

For all electromagnets, appropriate grounding measures shall be implemented for the magnet yokes, and all live parts protected to a minimum of IP2X for low voltage and IP3X for high voltage circuits or locked-out for any intervention in their vicinity. Interventions may only be carried out by personnel with the necessary training and certification, after following the work organisation procedures and authorisation of the facility coordinator (VICs, IMPACT etc.). 


    
    




    



\paragraph{Electromagnetic safety}
Radiofrequency (RF) components purchased from industry are required to be CE marked and to comply with the EU emission norms for industrial environments. RF equipment built or installed at CERN shall be “leak-tight”, and tested against EU industrial norms for RF emissions in situ. In case of a RF-related accident (for example the breaking of a waveguide), the mismatch in reflected power shall be detected, and trigger a cut in the electrical supply. 

The magnetic fields from the proposed electromagnets in the current eSPS design represent a hazard similar to that found in many of the facilities at CERN, and shall be handled with standard mitigation strategies. The facility shall follow the Directive 2013/35/EU on the occupational exposure of workers, alongside CERN Safety Instruction IS 36 and its Amendment. Any activity inside the static magnetic field shall be subject to risk assessment and ALARA. Personnel shall be informed about the hazards and appropriately trained. Areas with magnetic flux densities exceeding 0.5 mT shall be delimited (use pacemaker warning signs), while areas with magnetic flux densities exceeding 10 mT shall be rendered inaccessible to the public.

\paragraph{Mechanical safety}
Currently the TT4 and TT5 galleries are equipped with electrical overhead travelling (EOT) cranes, installed in the 1970s; these will be replaced with modern EOT cranes. The new experimental building will be equipped with an EOT crane and a lift for equipment and personnel. Special lifting beams will also be required for the installation of the modulators and accelerators components.






As the project moves into the Technical Design Report stage, the mechanical Safety aspects shall be reviewed in greater detail.

\paragraph{Civil engineering safety}
The installation of the eSPS facility will involve civil works, including tunnel modifications, new concrete structures, and a new building. At CERN, all infrastructure shall be designed in accordance with the applicable Eurocodes to withstand the expected loads during construction and operation, but shall also consider accidental actions, such as seismic activity, fire, release of cryogens (if any – not currently foreseen) and the effect of radiation on the concrete matrix and other tunnel construction fabric.

\subparagraph{Fire resistance}
New structures and infrastructure shall be designed and executed to guarantee a mechanical resistance for 120 minutes of exposure to the design fire. Eventual passive protection systems, e.g.\ intumescent paintings and plasters, will be foreseen only for those elements that are unable to respect such a requirement. The structural assessment will need to be carried out in accordance with EN~1991-1-2, EN~1992-1-2 and EN~1993-1-2. 

\subparagraph{False floors}
A number of considerations must be taken into account when deciding the type of floor to install and the space required to meet the above criteria. Each phase of the floor’s life must also be designed for and must allow for safe installation, utilisation and finally decommissioning of the floor. Key design criteria include:

\begin{itemize}
    \item Minimise the area of floor that is freely removable (this reduces the risks of damage and deterioration of the floor from multiple interventions);
    
    \item Where possible provide a full height access to the cable trays and pipework;    
    
    \item Identify cable pulling routes for the current layout and where possible future modifications and try to fix smaller access points for pulling the cables;
    
    \item Identify points of access and egress from the floor for maintenance purposes that have the minimum impact on the normal walkways through the building;
    
    \item Detail the pulling points and access points such that there is a rigid barrier integrated into the access trap (access via a trap door which in the open position is supported by barriers that protect the opening); 
    
    \item Provide a suitable ladder or steps to access under the floor;
    
    \item Consult all groups who may need to pull cables to ensure the layout chosen is suitable and if necessary pre-equip the cable trays with pulleys and pulling wires;
    
    \item Consult with transport to identify the method for transport of foreseen equipment into the building and verify the floor loadings; 
    
    \item Develop a plan of the building showing all access and egress points for daily use, access under the floor for works, fire escape routes and transport routes. This can be used to quickly validate the loadings and mark the floor areas with the permissible loads;
    
    \item Once the basic layout of the floor and structures is made, the removable parts can be either designed in-house or bought in as a system. In either case it is highly recommended to have a system that fixes the position of the tiles with a solid frame to prevent creep of the tiles when repeatedly lifted and replaced. It is also worth considering a numbering system or pattern to help ensure tiles are replaced with the correct orientation and position.
\end{itemize}

\paragraph{Environmental safety}
Many efforts have been made in the conception of the eSPS design to economise on the use of energy and water, particularly in the design of the cooling and ventilation system for TT4 and TT5. With the refurbishment of this linac area, along with the construction of the new experimental area, waste, and particularly the excavation of soil are foreseen to be significant considerations in the delivery of this project.  As the design of the experimental area is still at a preliminary stage, the general requirements for environmental protection have been set out here, with a full study to follow as the project moves into detailed design. The new and refurbished facilities for this project are located entirely within French territory, and French regulations are, therefore, to be applied.

\subparagraph{Air}
Atmospheric emissions shall be limited at the source and shall comply with the relevant technical provisions of the following regulations:

\begin{itemize}
    \item \textit{Arrêté du 02 février 1998 relatif aux prélèvements et à la consommation d'eau ainsi qu'aux émissions de toute nature des installations classées pour la protection de l'environnement soumises à autorisation, Articles 26, 27, 28, 29, 30.}
\end{itemize}

The design of exhaust air discharge points shall comply with the requirements of the Section~5.1.3 of the CERN Safety Guideline C-1-0-3 - Practical guide for users of local exhaust ventilation (LEV) systems. 

\subparagraph{Water}
The project leader shall ensure the rational use of water. The discharge of effluent water into the CERN clean and sewage water networks shall comply with the relevant technical provisions contained in the following regulations:

\begin{itemize}
    \item \textit{Loi no 2006-1772 du 30 décembre 2006 sur l'eau et les milieux aquatiques;}
    
    \item \textit{Arrêté du 02 février 1998 relatif aux prélèvements et à la consommation d'eau ainsi qu'aux émissions de toute nature des installations classées pour la protection de l'environnement soumises à autorisation.}
\end{itemize}

The direct or indirect introduction of potentially polluting substances into water, including their infiltration into ground is prohibited. Applicable emission limit values for effluent water discharged in the Host States' territory are defined in the following regulations:

\begin{itemize}
    \item \textit{Arrêté du 02 février 1998 relatif aux prélèvements et à la consommation d'eau ainsi qu'aux émissions de toute nature des installations classées pour la protection de l'environnement soumises à autorisation Art. 31 and Art. 32.}
\end{itemize}

Retention measures for fire extinguishing water are required for any CERN project in which large quantities of hazardous, or potentially polluting substances are used or stored. As the project moves to the technical design report stage, through discussions with the environmental protection specialists from HSE, it will be determined whether the following guidance document shall be applied (in accordance with the French \textit{Code de l'Environnement}):

\begin{itemize}
    \item \textit{Référentiel APSAD D9 : Dimensionnement des besoins en eau pour la défense contre d’incendie and Référentiel APSAD D9A : Dimensionnement des rétentions des eaux d’extinction available from the Centre National de Prévention et de Protection (CNNP) (http://www.cnpp.com).}
\end{itemize}

\subparagraph{Energy}
The use of energy shall be done as efficiently as possible. For the entire facility, adequate measures shall be taken to comply with the relevant technical provisions contained in the following regulations:

\begin{itemize}
    \item \textit{Loi no 2010-788 du 12 juillet 2010 portant engagement national pour l’environnement (Grenelle II).}
\end{itemize}

In addition, construction of new buildings sited in France shall comply with the relevant technical provisions relating to thermal efficiency contained in the following regulation:

\begin{itemize}
    \item \textit{Décret no 2012-1530 du 28 décembre 2012 relatif aux caractéristiques thermiques et à la performance énergétique des constructions de bâtiments;}
    
    \item \textit{Arrêté du 26 octobre 2010 relatif aux caractéristiques thermiques et aux exigences de performance énergétique des bâtiments nouveaux et des parties nouvelles de bâtiments and the French Réglementation Thermique 2012 (RT 2012);}
    
    \item \textit{NF EN 15232 Performance énergétique des bâtiments - Impact de l'automatisation, de la régulation et de la gestion technique.}
\end{itemize}

\subparagraph{Soil}
The natural physical and chemical properties of the soil must be preserved. All the relevant technical provisions related to the usage and/or storage of hazardous substances to the environment shall be fulfilled to avoid any chemical damage to the soil. Furthermore, the excavated material shall be handled adequately and prevent further site contamination.  All excavated material must be disposed of appropriately in accordance with the associated waste regulations.

\subparagraph{Waste}
The selection of construction materials, design and fabrication methods shall be such that the generation of waste is both minimised and limited at the source. Waste shall be handled from its collection to its recovery or disposal according to: 

\begin{itemize}
    \item \textit{Code de l’environnement,  Livre V: Titre IV -Déchets;}
    
    \item \textit{LOI no 2009-967 du 3 août 2009 de programmation relative à la mise en œuvre du Grenelle de l'environnement (1), Art. 46.}
\end{itemize}

The traceability of the waste shall be guaranteed at any time.

\subsubsection{Personnel protection system and access control}

\paragraph{Personnel protection system}

The eSPS access process will be based on building access control System and a personnel protection system (PPS) with dedicated access points that will regulate the personnel access to the “controlled” Machine areas. The access point will be made of one personnel access booth and an optional material booth, and will regulate with the highest safety level the access according to the operating modes managed remotely by the CERN control centre (CCC). It will perform several checks: such as user’s identification, biometric authentication, access and training validity, proper work authorisation (IMPACT) as well as the use of Operational dosimeters (DMC).

The PPS controls a number of independent beam zones divided into access sectors equipped with various safety elements such as sector doors, end-of-zone doors, mobile shielding, crinoline, patrol boxes, etc. The access sectors are particularly important to organise the patrol of the machine and to minimise the radiation exposure by means of radiation veto handled locally or remotely by the RP responsible person. Several access modes are foreseen: `General', `Restricted' with a `Safety token' and `Equipment Test', automatically managed by the PPS or controlled by human operator, locally or remotely from the CCC. The eSPS being a facility installed both in the PS and the SPS complex, it would probably be operated by the SPS operators from the CCC for the most part.

In the machine, the safety is ensured by access or beam “Important Element for Safety (EIS)”. Each beam zone has its own independent access conditions. The absence of beam in each independent beam zone is guaranteed by at least two beam safety elements, with at least one passive element (e.g.\ a movable stopper) and one active element (e.g.\ magnet power converter interlock). These safety measures are activated and interlocked by the PPS: the access status can make a zone unsafe for operation with beam, or forbid the access to the Machine if the status of a safety element is unsafe. 
The proposed zoning has been designed in accordance with the RP classification. 
The following beam areas have been considered:

\begin{itemize}
    \item Electron linac;
    \item CLEARER experimental area;
    \item TT5 Hall;
    \item TT4 experimental area and the upper part of the TT61;
    \item TT2-TT10;
    \item eSPS experimental area.
\end{itemize}

\subsubsection{Survey and alignment}

\paragraph{Hypotheses}

The following systems would be implemented for survey and alignment:

\begin{itemize}
\item 	24 modules to be aligned along a straight line of 70\,m;
\item On each module, 4 accelerating structures (length   $\sim$\,0.5\,m) and 1 quadrupole over an independent adjustment platform;
\item The overall tolerance for alignment of each component axis (RF electromechanical of AS and magnetic axis of quadrupole): would be $\pm$ 0.1\,mm in transverse position ($1\sigma$), for roll, pitch and yaw with respect to a straight line.

\end{itemize}
Schedule to be based on preparation during a run with installation during long shutdown. Due to uncertainties, a schedule is provided later in the document which is from any given start point and not rooted in the CERN LHC and injector timetable. 

\paragraph{Survey and alignment strategy proposed}

\subparagraph{Introduction}
The tolerances of alignment requested for the eSPS are tighter than for current accelerators at CERN but larger than in lepton colliders (CLIC or FCC) under study. As a matter of fact, in the LHC, each fiducial, on top of any cryostat has to be aligned within $\pm$ 0.15\,mm over 150\,m ($1\sigma$), while for CLIC, each reference axis of component has to be included within a cylinder with a radius of 14 $\mu$m over a sliding length of 200\,m. The survey and alignment strategy proposed for eSPS will integrate specific steps to cover these tighter tolerances.

The configuration of the tunnel in which eSPS components will be installed is very particular: 
\begin{itemize}
\item Narrow and straight: 2.10\,m in width (including the components to align) and 2.5\,m in height, once the shielding wall and roof are in place;
\item The main components in the “tunnel” will be linked to modulators and klystrons located on the other side of the shielding wall by RF waveguides;
\item The shielding roof will be put in place once all the components are installed. Components will be pre-aligned before shielding walls are installed and their alignment will  be controlled again once the shielding has been added;
\item TT4 and TT5 halls should be a very stable in terms of ground movement due to their age, however, this is to be confirmed by regular measurements before the installation of components.
\end{itemize}

\subparagraph{Steps of survey and alignment strategy proposed}
First, a geodetic network will be installed in all the transfer tunnels required for eSPS and the position of points w.r.t.\ to the existing underground network and components (located in TCC6) re-determined. The new points installed in TT4 and TT5, will be located preferably on the ground floor (to limit the impact of the shielding roof), close to the existing wall, in the “transport” area. Additional targets will be installed on the permanent walls.

\par
The fiducialisation and initial alignment of components on the same module will be a very important step in the overall alignment. If such an “assembly” alignment on each module is performed within $\pm$~40~$\mu$m~($1\sigma$), this will allow use of standard means of alignment later in the tunnel for each module. Each accelerating structure will have to be fiducialised independently, i.e.\ its mechanical axis will be determined w.r.t.\ external targets, at the metrology lab or using laser tracker. In both cases, this means we will align them w.r.t.\ to their mechanical axis, which does not correspond to their RF axis. For a more accurate determination, it is recommended to use a stretched wire setup (see Ref.~\cite{GalindoMunoz2017}): the stretched wire will be put at the mean RF axis and then the position of the wire will be determined w.r.t.\ external fiducials. The same for the determination of the quadrupole magnetic axis (see Ref.~\cite{Caiazza2017a}). Once all the components are fiducialised, they will be pre-aligned on a common support, using universal adjustment platforms. The mean axis of components will be determined in the support coordinate system, materialised by fiducials or sensors interfaces, depending on the solution chosen for their smoothing. Some components coming from the CLEAR injector will need to be re-fiducialised as well, by a combination of laser tracker and Romer arm measurements.

\par
Once all the MAD-X files and all element drawings are available, the marking can take place in the tunnels, including the beam line, and the projection of the jacks’ position on the ground floor. 

\par
Once the modules and the components are ready, they can be transferred and pre-aligned w.r.t.\ the geodetic network using laser tracker measurements, within $\pm$ 0.3 mm ($1\sigma$).
Different options are possible concerning the smoothing (final relative alignment) of the modules within $\pm$ 0.1 mm ($1\sigma$):

\begin{itemize}
\item	\textbf{Option 1:} Use of standard means of measurements to position them in vertical and in radial position (offsets measurements w.r.t.\ a stretched wire for radial alignment, levelling measurements for vertical alignment), combined with inclinometer measurements. Such a smoothing operation will have to be carried out at least twice: before the installation of the shielding and after the installation of the shielding; 
\item	\textbf{Option 2:} Use of temporary WPS sensors, measuring w.r.t.\ a permanent wire stretched between two metrological platforms (located at each extremity) with an absolute position determined in the geodetic network of the tunnels. The two metrological platforms will be equipped with permanent WPS sensors and HLS sensors, while the modules will be equipped with permanent sensors interfaces, determined in the coordinate system of the module. The configuration of temporary sensors per module would consist of two wires, four WPS sensors, providing a redundant and controlled determination of the position of each module. Each module would be aligned one after the other w.r.t.\ the extremity reference platforms, using four sensors plugged temporary. Such a solution provides the possibility to upgrade later on each module with permanent sensors if needed;
\item	\textbf{Option 3:} Use of WPS sensors installed permanently, with the same options concerning their configuration as before.

\end{itemize}
\par
In all cases, the standard fiducials (or the sensors interfaces) shall be located above the adjustment means to avoid the level arms effect during the adjustment process and to both facilitate and quicken the process of alignment. Adjustment solutions shall be studied and installed  below each module as well, with the possibility of motorisation (if needed).

%% file: include/05-ICE/InjectionTT4_TT5.tex
\subsection{Linac in B183, TT4 and TT5}
\label{sec:ICE_Injection}

\subsubsection{Civil engineering}

\paragraph{Location}

\begin{figure}[!hbt]
\begin{center}
    \includegraphics[width=.8\linewidth]{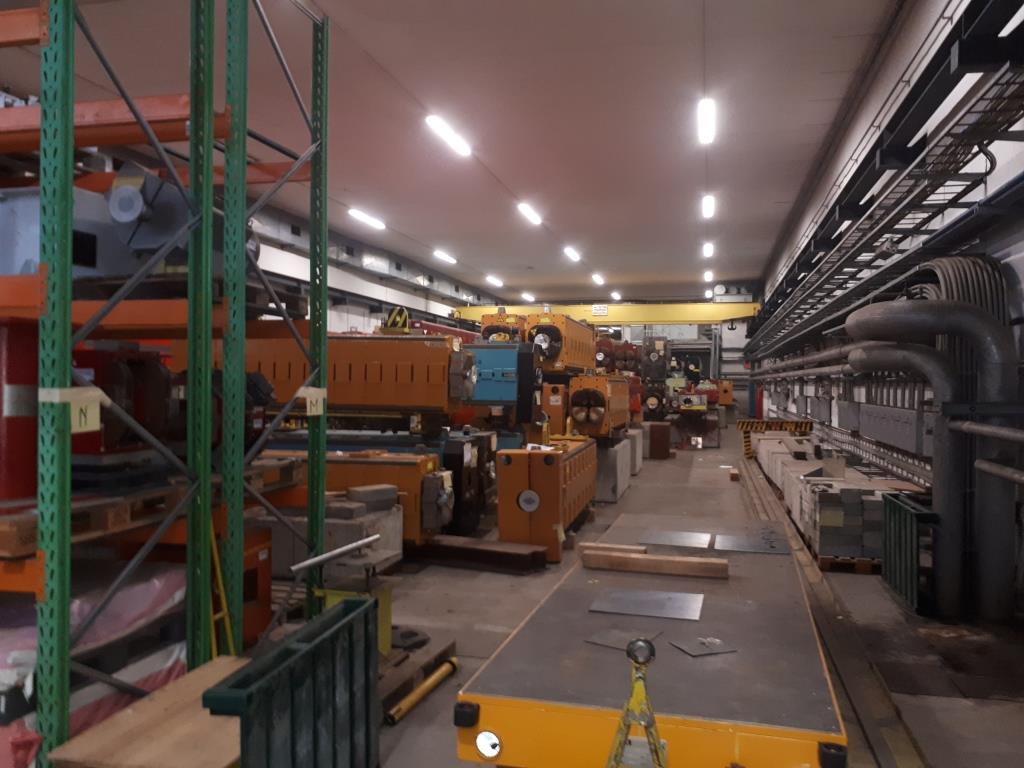}
    \caption{View within existing TT5 tunnel looking east towards TT4.}
 \label{fig:TT5photo}
\end{center}
\end{figure}

The injection infrastructure is to be housed in the existing B183 (B183), transfer tunnel 5 (TT5) and transfer tunnel 4 (TT4). The area which would be used to house eSPS is currently used as low-level radioactive material storage as can be seen from Fig.~\ref{fig:TT5photo}. The existing stored materials would need to be relocated in advance of re-purposing for eSPS. This is not covered as part of this feasibility study although the subject is discussed further in Section~\ref{sec:Widercerncontext}. 

\paragraph{Existing structures}

\subparagraph{B183}

\begin{figure}[!hbt]
\begin{center}
    \includegraphics[width=.85\linewidth]{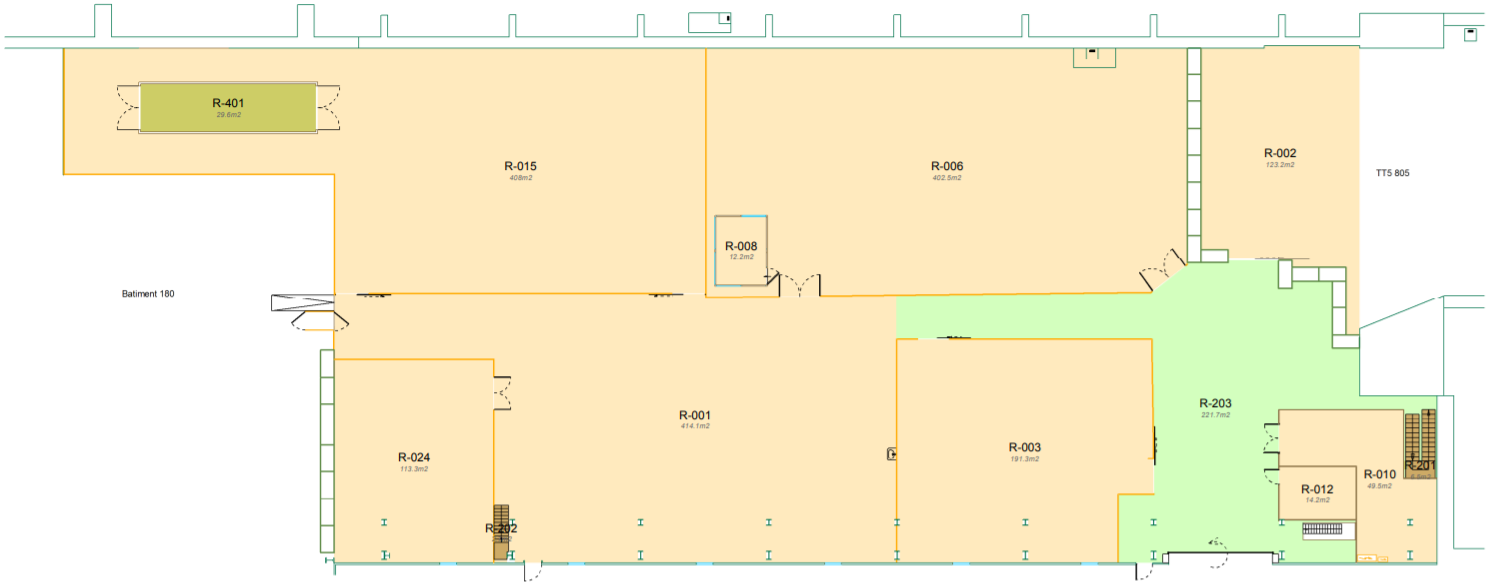}
    \caption{B183 floor plan.}
 \label{fig:B183floorplan}
\end{center}
\end{figure}

B183 is an existing structure built in 1971 as an experimental hall forming part of the 'West area'. The building is constructed in a mixture of reinforced concrete and steel frame construction with a truss roof structure. The building is predominately used as part of the magnet assembly building but the area (designated R-002) abutting TT5 is a continuation of the low-level radioactive material storage area as shown on Fig.~\ref{fig:B183floorplan}. This area is surrounded by and delineated by shielding blocks and requires no civil engineering works to be used for the scheme. 

\subparagraph{TT5}

\begin{figure}[!hbt]
\begin{center}
    \includegraphics[width=.8\linewidth]{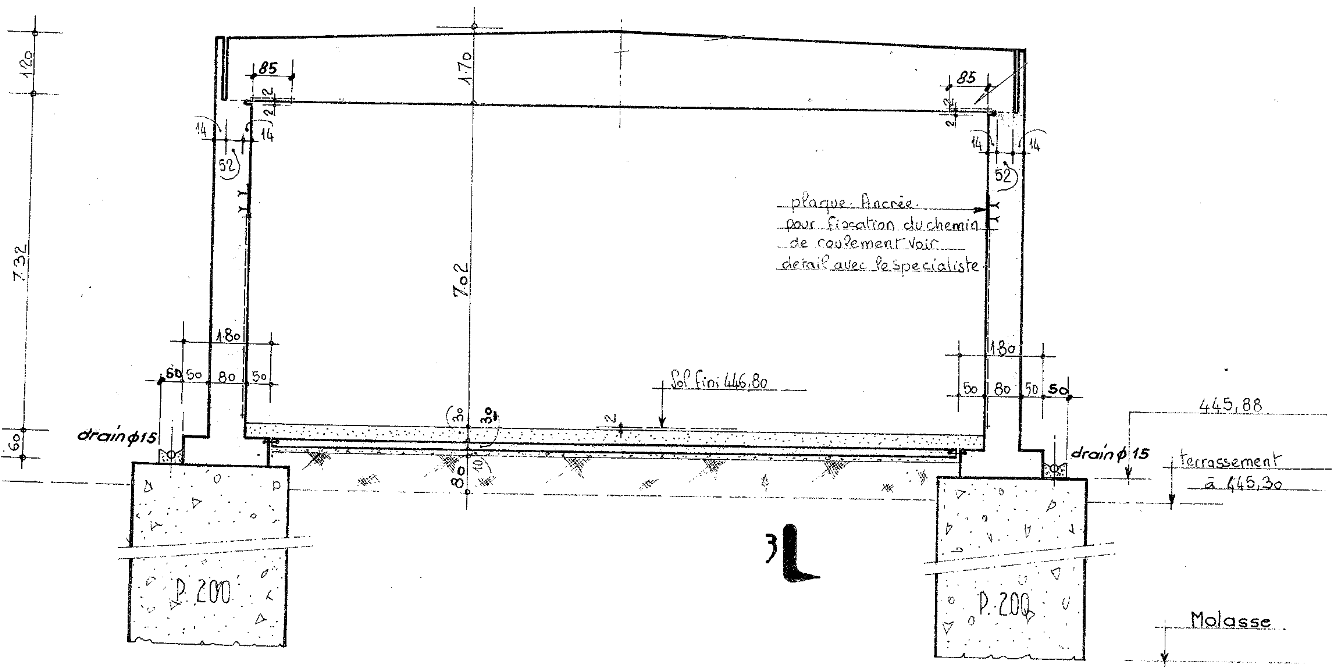}
    \caption{Existing TT5 typical section.}
 \label{fig:TT5xsec}
\end{center}
\end{figure}

TT5 is a buried reinforced concrete structure built in 1971 to house three transfer beamlines delivering beam from the SPS to the West area. The main part of the structure comprises deep strip foundations of variable depth sat on the molasse rockhead with a 620~mm deep reinforced concrete slab above a further 600-800\,mm thick layer of improved structural fill spanning between founds. The slab is made up of a 300\,mm lean concrete base below a 300\,mm reinforced concrete slab finished with a layer of PVC waterproofing and a 2\,mm layer of screed. The walls are 7.32\,m tall, 800\,mm thick reinforced concrete columns with a 1.8\,m wide, 600\,mm deep footing supported on the strip foundations. The reinforced concrete roof slab is cast integrally with the columns in nine bays and varies in depth from 1.2\,m at the outside of columns to 1.7\,m at the centre (see Fig.~\ref{fig:TT5photo}). The typical section is shown in Fig.~\ref{fig:TT5xsec}. 

The building in plan measures 47.53\,m long by 16\,m wide. There are a number of channels which run the full width of the hall with metal plate covers. Channels are 400\,mm wide and at irregular spacing varying between 2.98\,m to 7.9\,m but typically at around 4.5\,m spacing. The channels have drains at the mid-point of the hall with a carrier drain below running along the centre line of the building.

\begin{figure}[!hbt]
\begin{center}
    \includegraphics[width=.85\linewidth]{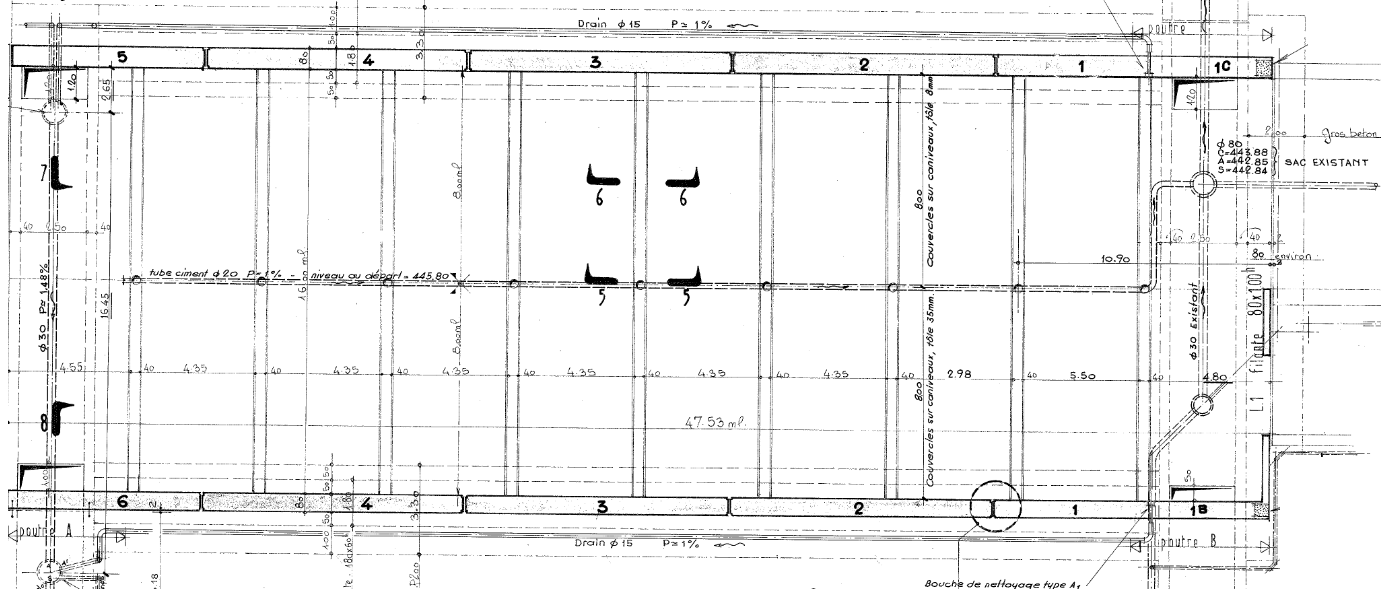}
    \caption{Existing TT5 plan view showing channels and drains in floor slab.}
 \label{fig:TT5floorplan}
\end{center}
\end{figure}

\begin{figure}[!hbt]
\begin{center}
    \includegraphics[width=.85\linewidth]{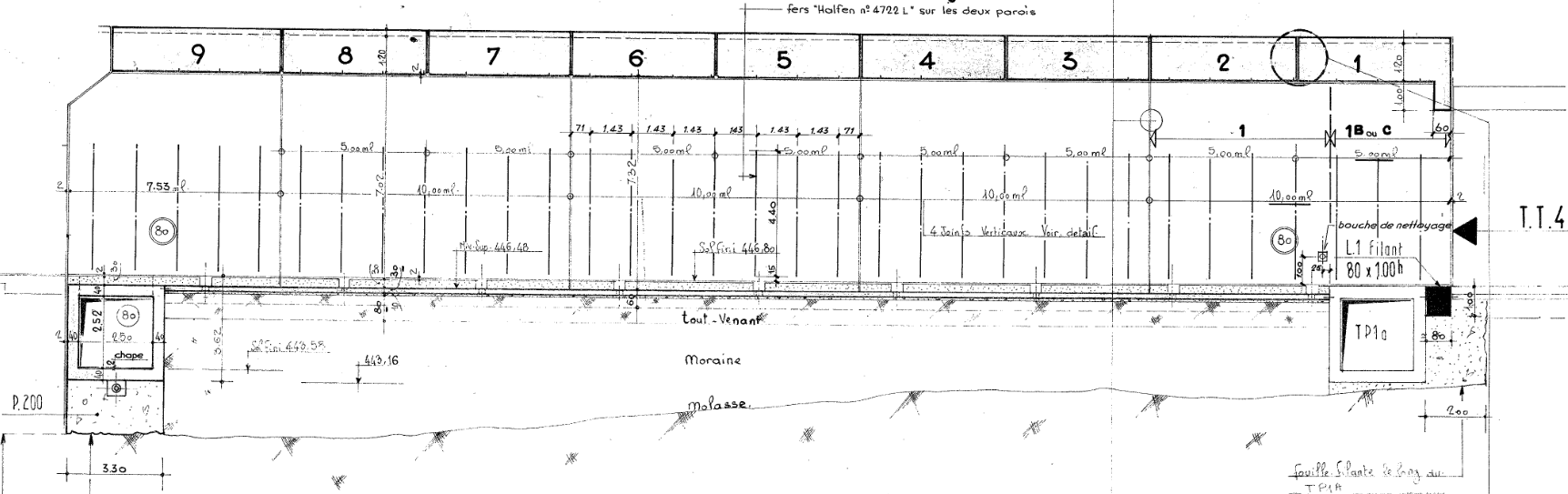}
    \caption{Existing TT5 long-section.}
 \label{fig:TT5longsec}
\end{center}
\end{figure}

As-built drawings for the structure exist and are quite detailed, including reinforcement and detailing drawings. Details of the existing structure are shown in Figs.~\ref{fig:TT5xsec},~\ref{fig:TT5floorplan}~and~\ref{fig:TT5longsec}. 

\par
TT5 also has several technical galleries beneath it to carry services and provide access for maintenance which are shown in the tunnel long-section Fig. \ref{fig:TT5longsec} . Additional tunnels provide level access to the main building from the surrounding hard-standing.

\begin{figure}[!hbt]
\begin{center}
    \includegraphics[width=.85\linewidth]{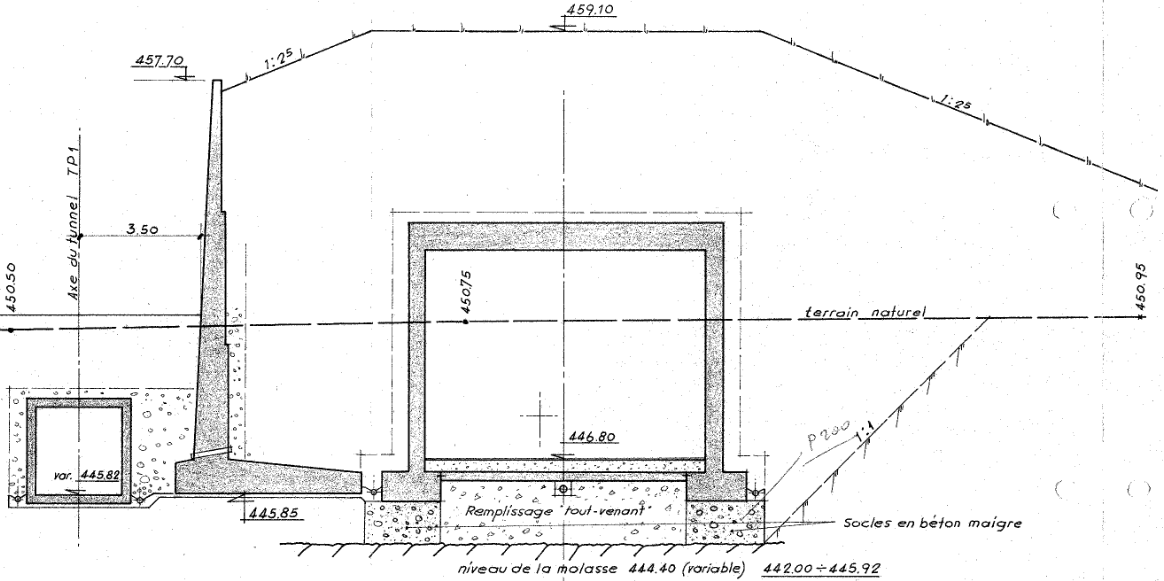}
    \caption{Existing TT4 typical cross section.}
 \label{fig:TT4XSec}
\end{center}
\end{figure}

\begin{figure}[!hbt]
\begin{center}
    \includegraphics[width=.85\linewidth]{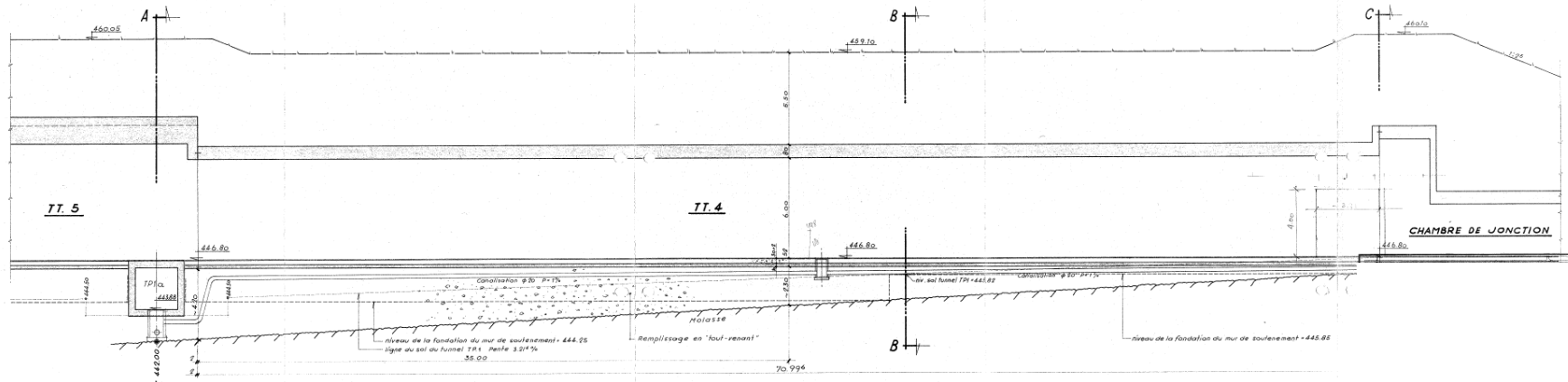}
    \caption{Existing TT4 long-section showing junction with TT5.}
 \label{fig:TT4LongSec}
\end{center}
\end{figure}

\newpage
\subparagraph{TT4}
TT4 is another buried reinforced concrete structure built in 1971 as a continuation of the structures housing transfer beamlines bringing beam from the SPS to the West area. TT4 abuts the eastern end of TT5 as shown in Fig.~\ref{fig:TT4LongSec}, connecting it with TT61 and TT3 to the east. TT4 comprises variable depth lean mass concrete strip foundations which extend from the molasse rockhead to the base of the box section footings as shown in Fig.~\ref{fig:TT4XSec}. TT4 has a concrete slab \SI{500}{mm} deep with a PVC wasterproofing layer and 2mm screed on top. Figure~\ref{fig:TT4XSec} also shows the significant depth of earth above the structure as well as the retaining wall running alongside TT4 retaining the higher ground above \emph{Route Nord}, a general purpose road.

The reinforced concrete tunnel structure above spans between 1\,m deep, 1.8\,m wide footings. The  walls are 600\,mm wide and 6m tall while the integral roof slab is 800\,mm thick with a span of 8\,m over the majority of the structure. The layout in plan and in terms of floor levels is more complex at its junction with TT61 and TT3/TT2A. TT4 has channels in every way similar to those found in TT5, again at regular spacings. Some channels have been filled in with concrete. 

Drawings for TT4 are less detailed, potentially making future works to modify the structure more challenging but still achievable. 

\begin{figure}[!hbt]
\begin{center}
    \includegraphics[width=.85\linewidth]{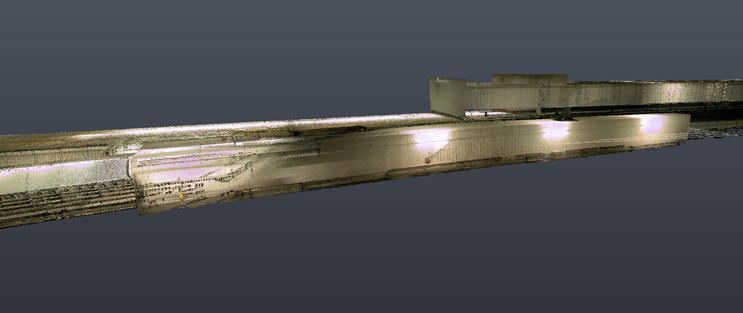}
    \caption{Interim scan output during processing.}
 \label{fig:3Dscan}
\end{center}
\end{figure}

As part of the study, 3D scans of TT5 and TT4 were carried out by the SMB-SE-DOP section to allow an accurate model to be produced as shown in Fig.~\ref{fig:3Dscan}. This has been used as the basis of a new integration layout. In general the scans showed good agreement with the as-built drawings, giving a fair degree of confidence in the drawings' quality, although there were some areas which required adjustments.

\par
All of the structures were found to be in generally good condition with no major issues apparent from a civil engineering perspective. 

\newpage 
\paragraph{Civil engineering enabling works}

Limited civil engineering works are required in order to facilitate the implementation of the eSPS scheme. The integration study carried out focused on fitting the required injection and associated infrastructure into the existing space within B183, TT5 and TT4 and through careful planning and optimisation between all disciplines, this has been achieved. Some limited civil engineering enabling works are still required and these are summarised here.

\begin{figure}[!hbt]
\begin{center}
    \includegraphics[width=9cm]{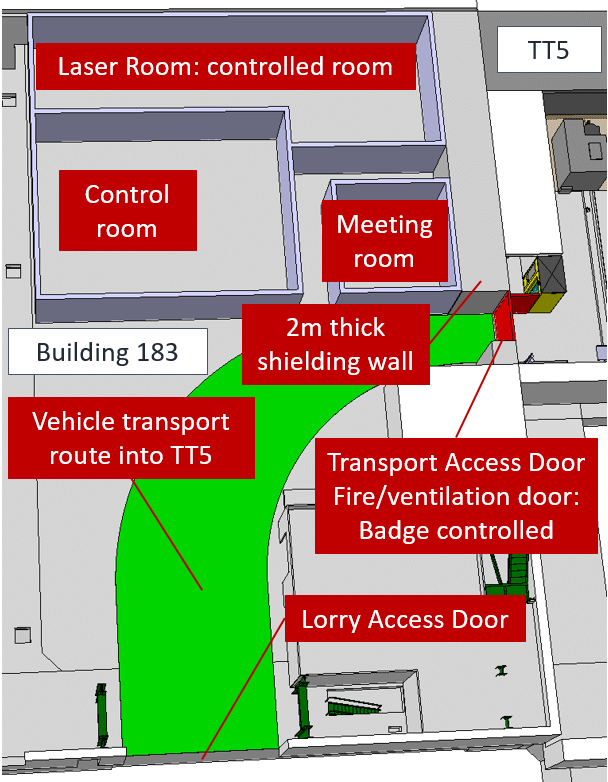}
    \caption{B183 layout.}
 \label{fig:Building_183_Layout}
\end{center}
\end{figure}

\subparagraph{B183}
In B183 there are minimal enabling works required to implement the scheme. The main elements are as follows:

\begin{itemize}
   
   \item A dividing wall between B183 and TT5 will be needed to provide compartmentalisation for fire safety and to provide some additional shielding to equipment located in B183. The requirements for the wall are set by applicable fire safety standards and shielding requirements detailed in other chapters. The wall must be 2\,m thick, 4.5\,m high constructed in normal density concrete;
 \item Although the form of construction is not yet finalised, rooms will need to be constructed to accommodate ancillary infrastructure as shown in Fig.~\ref{fig:Building_183_Layout}:
 \begin{itemize}
     \item 9\,m $\times$ 10\,m Control room;
     \item 4\,m $\times$ 3\,m Laser room;
     \item 5\,m $\times$ 5\,m Meeting Room.
 \end{itemize}
\end{itemize}

\begin{figure}[!hbt]
\begin{center}
    \includegraphics[width=.85\linewidth]{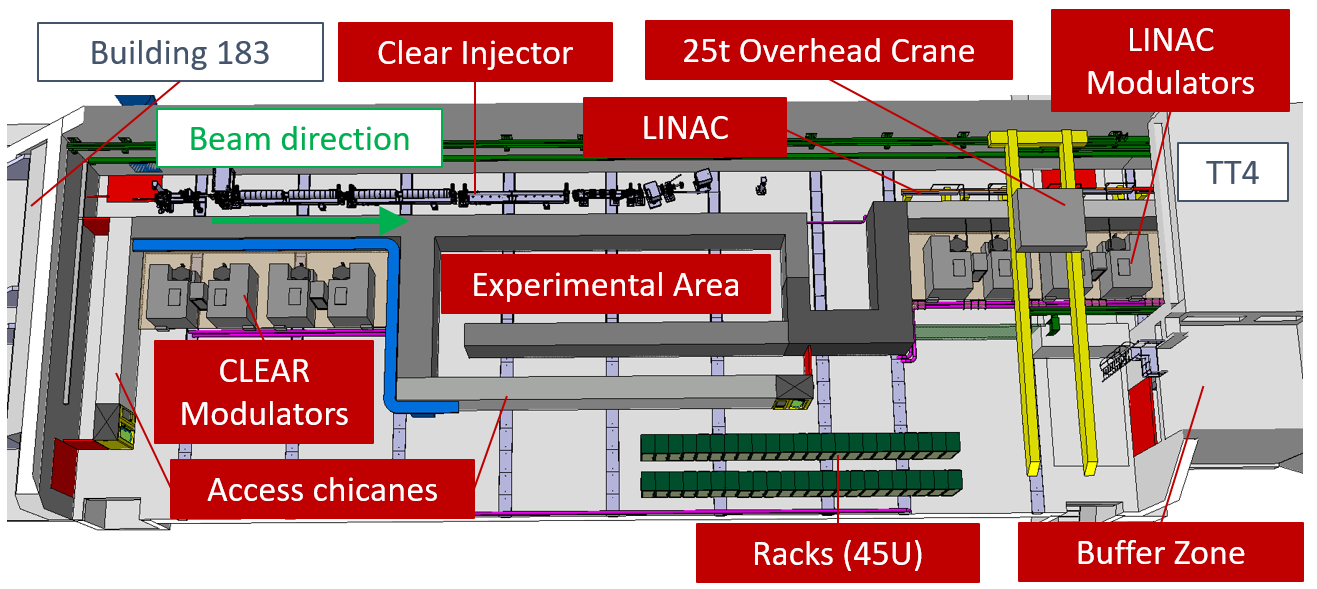}
    \caption{TT5 layout.}
 \label{fig:TT5_Layout}
\end{center}
\end{figure}

\subparagraph{TT5}
In TT5, the layout required is shown in Fig.~\ref{fig:TT5_Layout}. The main civil engineering contribution is the required shielding walls and roof slabs over the injector and linac. The shielding has been defined by the RP study which has identified the thickness of shielding needed along with the grade:

\begin{itemize}
\item Normal density concrete 2.4\,g/cm$^{3}$;
\item High density concrete 3.9\,g/cm$^{3}$;
\item Iron 7.9\,g/cm$^{3}$.  
 \end{itemize}
 
The concrete shielding walls will be formed from a number of temporary blocks to allow greater flexibility for any future changes in layout and greater speed of construction. The blocks will be cast offsite and transported into position at a suitable time during refurbishment. The arrangement of temporary shielding blocks will be designed to allow blocks to be standardised as far as possible. Blocks will, however, need to interlock and ensure there is no shine path for radiation to pass meaning there will be slightly reduced opportunity for modularisation. 

The concrete mix for normal density concrete is flexible and needs only to support itself and any blocks above. The concrete mix for high density concrete has been selected based on a density high enough to be suitable for radiation protection criteria while being practical for standard concrete suppliers without being cost prohibitive. The concrete mix is based on using high density magnetite, an ore of iron, as aggregate in the concrete:

Formula per cubic meter given as an indication for a density of 3.9\,g/cm$^{3}$,
 \begin{itemize}
    
 \item MD8S 0/6\,mm: 1180\,kg;
 \item MD20S 0/16\,mm: 2360\,kg;
 \item CEM II type cement: 310\,kg;
 \item Water (including moisture from aggregates): 160/170\,l;
 \item High water reducing super-plasticiser: 1.5--2\% of the weight of the cement.
 \end{itemize}
 
 In addition to the shielding walls, the higher levels of radiation from the injector and linac necessitate a shielding roof. This is to be removable to allow access to equipment by bridge cranes. The roof will be formed of a number of separate slabs. Each slab unit must be lifted and transported via the crane so the length of has been selected to ensure they remain within the crane's safe lifting capacity. For radiation protection, slabs must be 800\,mm thick and have been sized (in terms of width) to span between the central shielding wall and a support beam to be fixed to the wall of TT5 (and also TT4). An outline design of the support beam has been carried out by the EN-AC-INT section using conservative assumptions which confirmed this would be feasible. The initial design was based on a new steel beam 200mm wide such as HEB 200 with HEB 140 supports every metre attached to the existing walls with heavy duty anchorages such as Hilti HSL-3-G. 

An initial assessment of floor loading for the shielding block arrangement has been carried out by the SMB-SE-DOP section which found the existing slab and foundations will be sufficient. 

In several places in TT5, cores or breakouts in the existing concrete will be required to allow CV ducts to enter and leave the structure. Initial study shows there should be no issue with feasibility, however, further study will be needed to design and detail these penetrations. Experience has shown reinforcement drawings cannot be absolutely relied upon so ground penetrating radar (GPR) scans will be needed as well as assessment and design to ensure there is no detrimental reduction on the structural capacity of walls, roof slabs or floor slabs. To ensure the water tightness of the structure is not impacted, a suitable waterproofing joint detail will be used with hydrophillic sealant to prevent water ingress around service penetrations. 

\subparagraph{TT4}

\begin{figure}[!hbt]
\begin{center}
    \includegraphics[width=16cm]{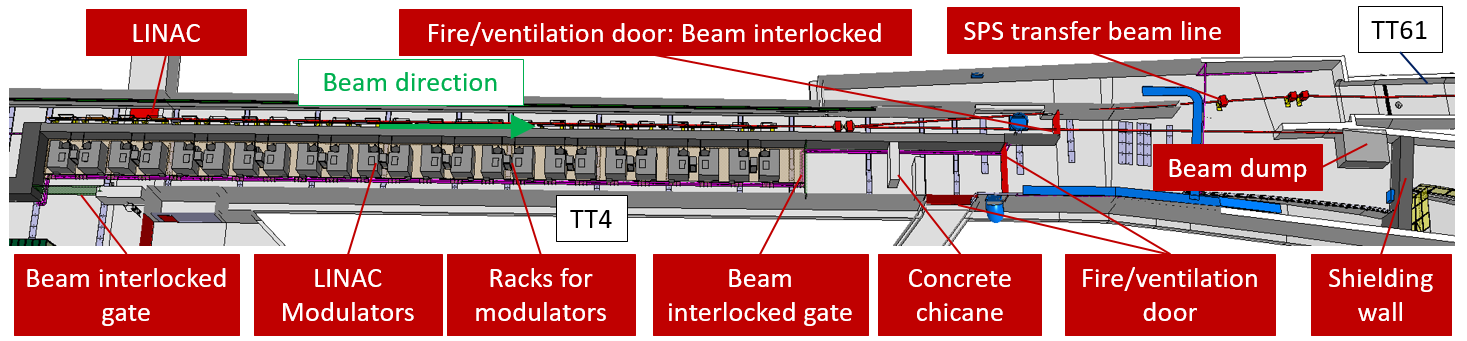}
    \caption{Layout of TT4.}
 \label{fig:Layout of TT4}
\end{center}
\end{figure}

In TT4, the layout required is shown in Fig.~\ref{fig:Layout of TT4}. The enabling works will practically be the same as those required in TT5 with the addition of some minor works to modify access arrangements to n\_TOF at the junction with TT3. Equipment transport and access will not be affected, other than a shielding wall will need to be moved to allow access. Options for this include a movable wall on rails or a wall composed of blocks which can be moved and replaced when infrequent access is required. For personnel access which is far more frequent, a set of steel stairs or access ladder can be provided beyond the shielding wall which will allow access via TT3 and the ISR. 

Significant enabling is also required for CV services. Above TT4 and TT5, several ducts and a service building must be sited as shown in Fig.~\ref{fig:External cooling and ventilation layout for TT4 and TT5}. Although this layout shows the ducts and building sited immediately above the building, it is likely all ducts and building will be sited at or just below existing ground level. Some deep trench-box excavations will be needed to enable duct connections to be formed. Lagged and trace heated duct-work can be installed at ground level. As there is a considerable earth mound above the tunnel here, this will be more cost effective than removing all earthworks and will, in addition, leave the considerable earth shielding in place for the future in case it is needed for radiation protection.

The overall load of the building and any applied loads when considered at the level of the tunnel should not be problematic, although a more detailed study will be required to confirm this at the next stage of development. 

\subsubsection{Integration}
\label{sec:linac:integration}
\paragraph{B183} 

B183 is an old experimental hall connected to transfer tunnel TT5. The control room, meeting room and laser room for the eSPS facility are located at the western end of B183 next to transfer tunnel TT5, in which this area is currently used for storing shielding blocks. A 2\,m thick shielding wall will separate B183 from TT5.  The size of the control room and meeting room is like that of CERN's existing CTF3 control and meeting room in B2008. The laser room is a controlled room with an area like like that of CERN's existing laser room in B2013. There is an entrance and a laser transport pipe with a diameter of 50\,mm goes through the shielding wall to the CLEAR RF gun in TT5 approx. 5\,m away. As the total distance from the laser output to the cathode needs to be matched to the focal length of the lenses in use, a revised layout will be studied at the technical design stage of the project. The vehicle access into TT5 is via B183 in which vehicles enter through the existing lorry access door and into TT5 through the fire/ventilation badge-controlled access door as shown in Fig.~\ref{fig:Building_183_Layout}.

\paragraph{Transfer tunnel TT5}

Transfer tunnel TT5 is an underground tunnel connected to TT4. Historically, three beamlines coming from TT4 were focused on primary targets located in TT5 after which secondary beams were sent to experiments in B180. The tunnel is currently used to store low-level radioactive magnets and shielding blocks.

TT5 will house the CLEAR injector, modulators, the experimental beamline and the first 10\,m of the linac as well as some racks for the different infrastructure as shown in Fig.~\ref{fig:TT5_Layout}.

The CLEAR injector is an existing injector at CERN housed in the B2010. The start of the injector will be located 2~m from the shielding wall separating B183 from TT5. This allows personnel to easily access the injector and it is also positioned away from an existing technical gallery access shaft of which will be permanently closed. For radiation protection the injector is surrounded by concrete shielding with an clear width of 2.8~m to allow for the installation and maintenance of the injector. With respect to the beam direction starting on the left-hand side the breakdown is as follows:

\begin{itemize}

\item \SI{200}{mm} allowance for safety lighting and services; 
\item  \SI{900}{mm} allowance for personnel access; 
\item \SI{700}{mm} allowance for the width of the injector; 
\item \SI{1000}{mm} allowance for personnel access. 

\end{itemize}

The layout of the concrete shielding is such that it allows the CLEAR injector to continue straight into TT4 and an experimental beamline to change direction by 180\si{\degree} into an experimental area. The layout of which was specified by the equipment owners. Personnel access to the injector and experimental area is via the access chicanes as shown in Fig.~\ref{fig:TT5_Layout}. The width of the chicanes is 1.4\,m to ensure that fire fighters can access the area with a stretcher and have sufficient room to turn. 

\begin{table}[!hbt]
\begin{center}
\caption{Required number of racks for each sub-systems.}
\label{tab:Racks in TT5}
\begin{tabular}{p{9cm}ccc}
\hline\hline
\textbf{Equipment}   	& \textbf{Number of racks}\\ 
\hline
Access control                     & 3 \\
Fire detection                  & 3 \\
Experimental beamline vacuum control                & 3 \\
Experimental beamline beam instrumentation             & 2 \\
Experimental beamline power converters             & 7 \\
Experimental beamline spare             & 3 \\
Linac vacuum control                & 6 \\
Linac beam instrumentation             & 6 \\
Linac power converters             & 5 \\
Linac spare             & 3 \\
Transfer line beam instrumentation and vacuum control             & 1 \\
Transfer line power converters             & 3 \\
Transfer line spare             & 1 \\
\hline 
Total             & 46 \\

\hline\hline
\end{tabular}
\end{center}
\end{table}

The control racks for the access control, fire detection, beam instrumentation, vacuum and the power converters for the facility were specified by the equipment owners. The racks are standard SPS 45U racks used at CERN. The required number of racks for each sub-system is listed in Table~\ref{tab:Racks in TT5}.

\paragraph{Transfer tunnel TT4}

Transfer tunnel TT4 is an underground tunnel connected to transfer tunnel TT61. Historically, the SPS beamline transferred via TT61 was split into three proton beamlines in TT4 and continued into TT5. The tunnel is currently used to store low-level radioactive magnets and shielding blocks.

TT4 will house the linac beamline and modulators as well as an experimental beam dump area at the downstream end of TT4. The linac is approx. 65\,m in length of which is composed of 24 modules each 2.65~m long with a gap of 200\,mm every 4$^{\textrm{th}}$ module for the installation of the sector valve, vacuum pumping port and for the connection of the primary pumping. The beamline splits at the downstream end of TT4 with the SPS transfer beamline bending into TT61 towards the SPS and the beam dump beamline continuing straight as shown in Fig.~\ref{fig:Layout of TT4}.

Personnel access to the linac is via the fire/ventilation door at the downstream end or via the CLEAR injector chicane entrance in TT5. Access to the modulators at the downstream end is via a concrete chicane and access gate or via the access gate in TT5. Because of the restricted geometry of transfer tunnel TT4, a detailed breakdown of the cross-section was undertaken to ensure that the linac and modulators could fit within the existing tunnel whilst adhering to the radiation protection and general safety requirements specified by CERN experts. The dimensions required for equipment and personnel/transport access, as shown in Fig.~\ref{fig:Cross-section of TT4} are:

\begin{figure}[!hbt]
\begin{center}
    \includegraphics[width=16cm]{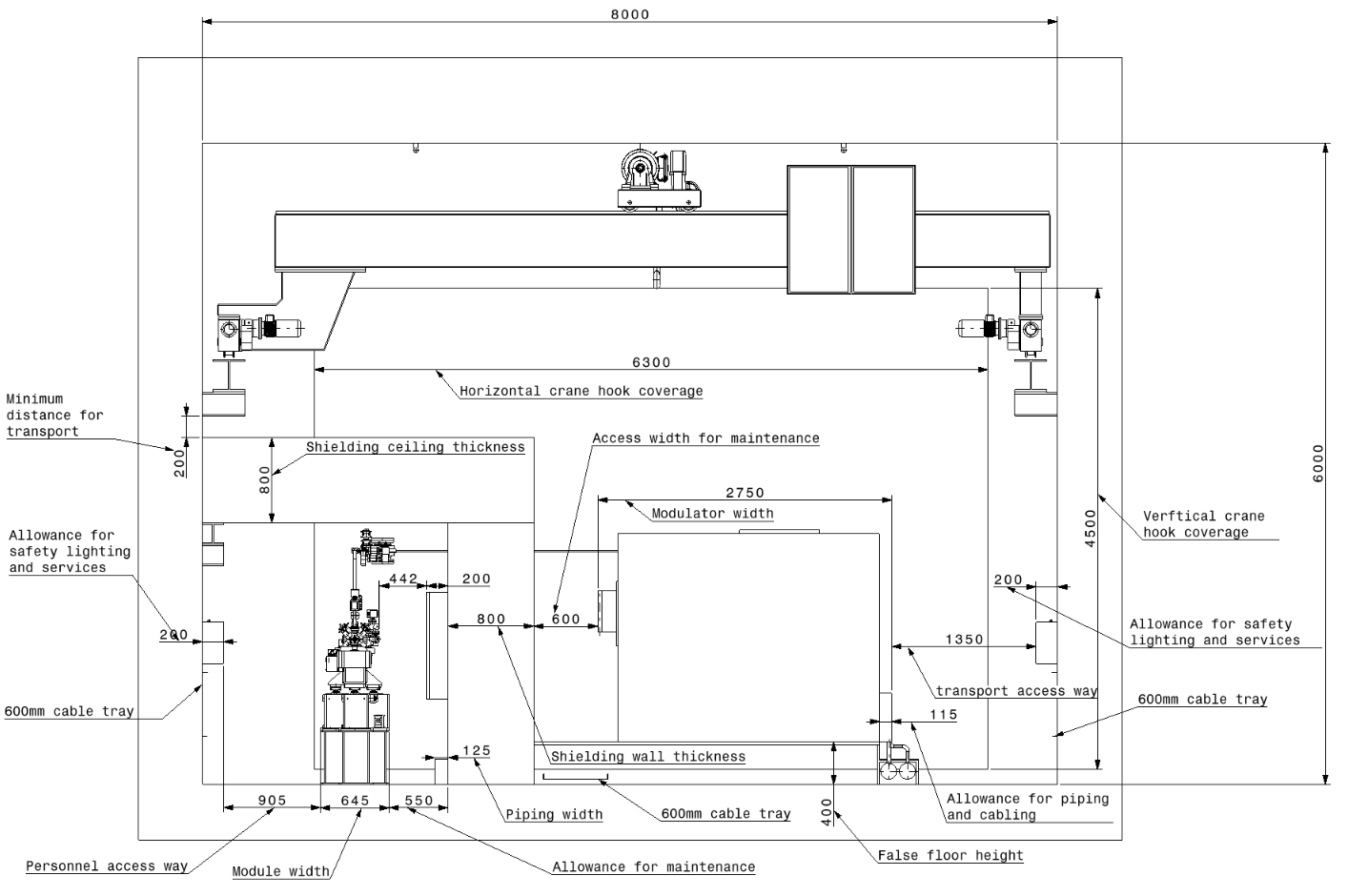}
    \caption{Cross-section of TT4.}
 \label{fig:Cross-section of TT4}
\end{center}
\end{figure}

\begin{itemize}

\item \SI{200}{mm} allowance for safety lighting and services; 
\item \SI{905}{mm} allowance for personnel access; 
\item \SI{645}{mm} width for the linac; 
\item \SI{550}{mm} allowance for personnel access for maintenance of the linac and cooling equipment;
\item \SI{800}{mm} width of shielding wall;
\item \SI{600}{mm} allowance for personnel access; 
\item \SI{2750}{mm} width of the modulator;
\item \SI{1350}{mm} allowance for the overhead crane to transport the linac;
\item \SI{200}{mm} allowance for safety lighting and services.

\end{itemize}

\begin{figure}[!hbt]
\begin{center}
    \includegraphics[width=0.7\linewidth]{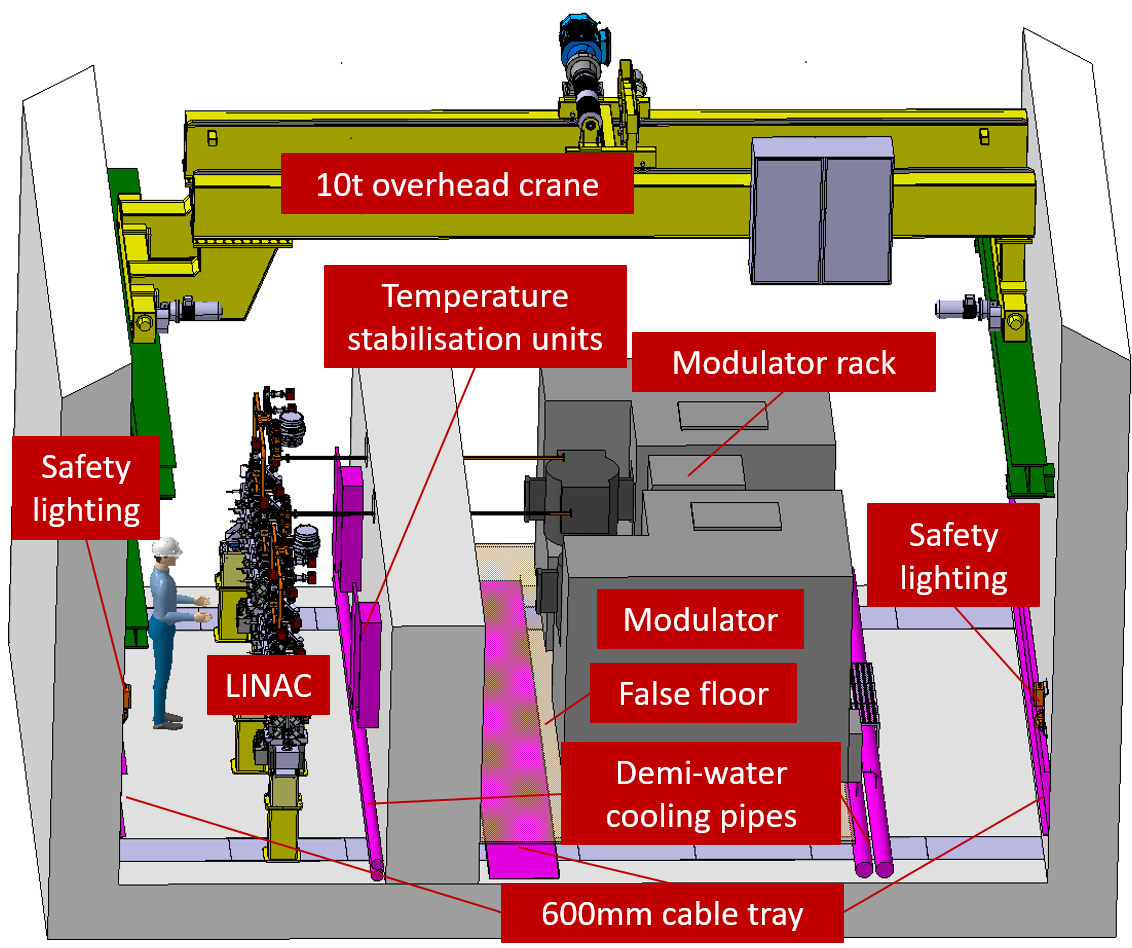}
    \caption{Integration of linac and modulators in TT4}
 \label{fig:Integration of LINAC and modulators in TT4}
\end{center}
\end{figure}

As the modulators are quite large, the access around the modulators is restricted because the space between them is controlled by the spacing of the linac beamline as detailed above. The longitudinal distance between each unit is 900\,mm, apart from every 4$^{\textrm{th}}$ unit in which this distance is increased to 1100\,mm. There is a movable rack between every two units as shown in Fig.~\ref{fig:Integration of LINAC and modulators in TT4}. Due to the restricted cross-sectional geometry, the access on the side close to the shielding wall is only 600\,mm. There is a false floor surrounding the modulators to allow the cabling and piping to be distributed underneath the modulators allowing unobstructed access all around the modulators.    

\paragraph{Cooling and ventilation layout}

\begin{figure}[!hbt]
\begin{center}
    \includegraphics[width=0.85\linewidth]{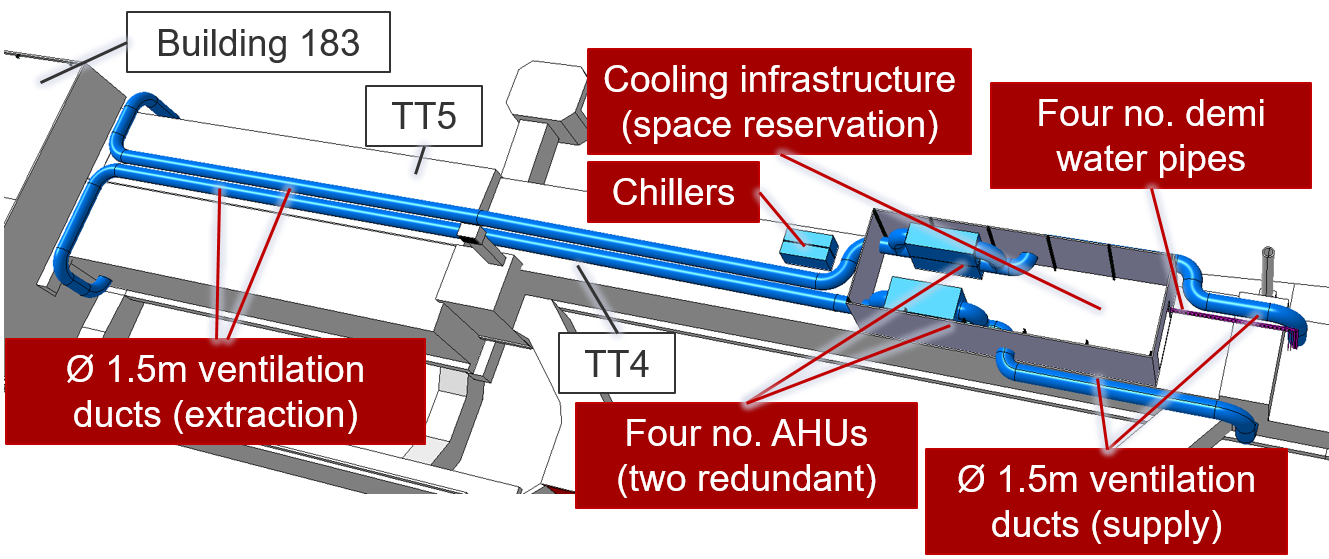}
    \caption{External cooling and ventilation layout for TT4 and TT5.}
 \label{fig:External cooling and ventilation layout for TT4 and TT5}
\end{center}
\end{figure}

\begin{figure}[!hbt]
\begin{center}
    \includegraphics[width=0.85\linewidth]{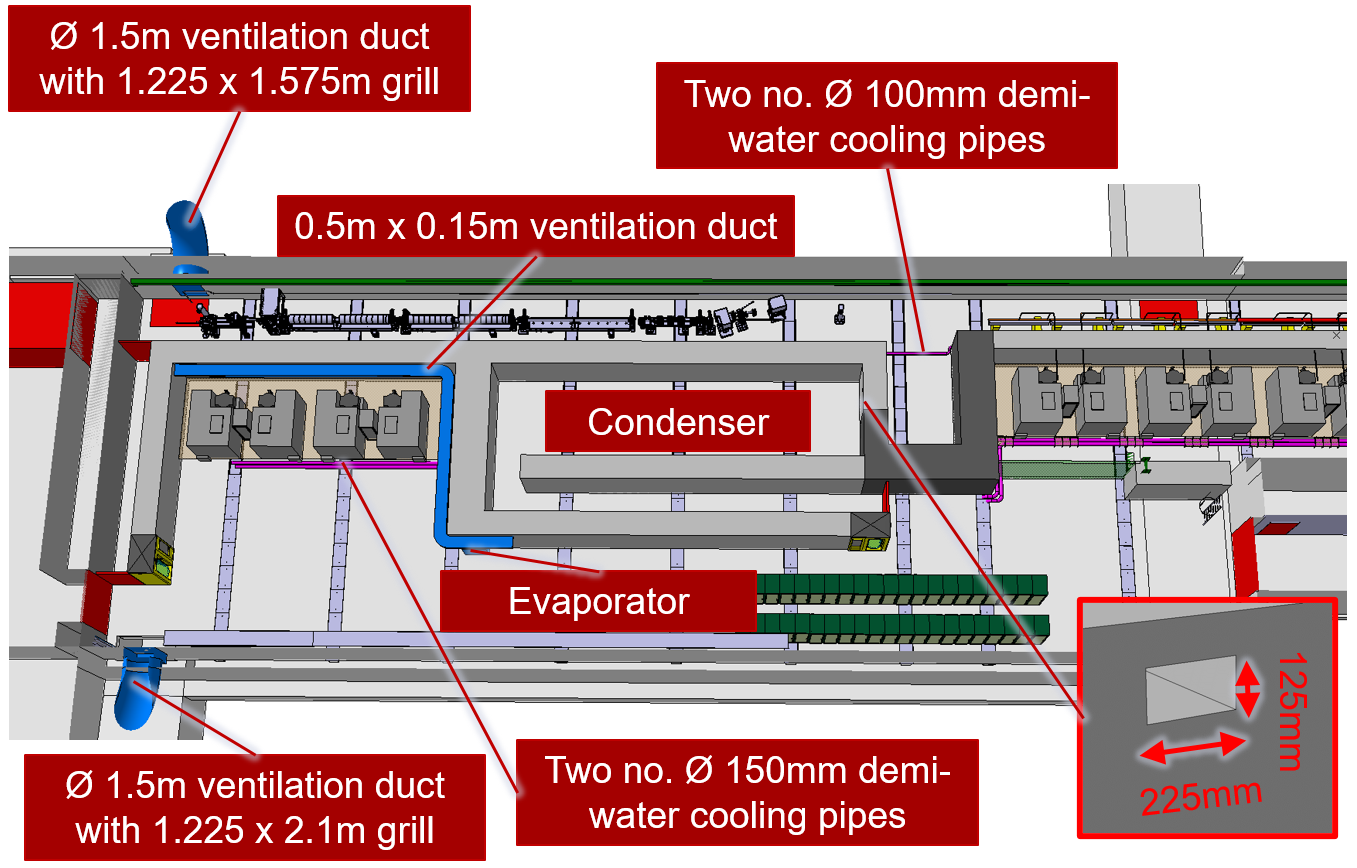}
    \caption{Internal cooling and ventilation layout for TT5.}
 \label{fig:Internal cooling and ventilation layout for TT5}
\end{center}
\end{figure}

\begin{figure}[!hbt]
\begin{center}
    \includegraphics[width=0.85\linewidth]{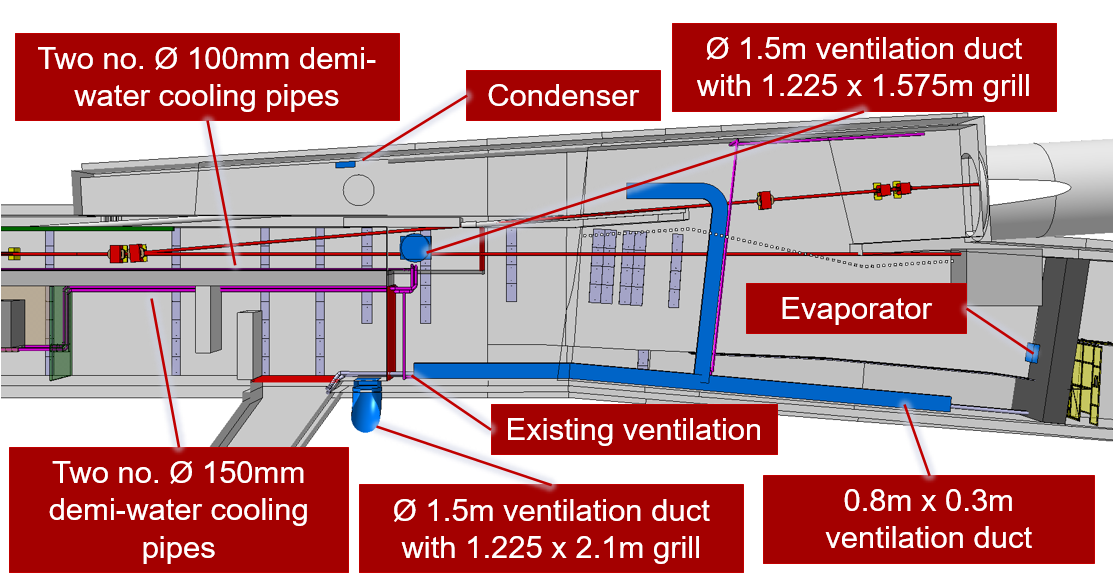}
    \caption{Internal cooling and ventilation layout for TT4.}
 \label{fig:Internal cooling and ventilation layout for TT4}
\end{center}
\end{figure}

A CV study was undertaken for the facility and the infrastructure requirements integrated into the existing layout. A new CV building will be located on top of TT4 ideally located such that the length of cooling piping and ventilation ducts was optimised throughout the facility. The CV layout of the facility is shown in Figs.~\ref{fig:External cooling and ventilation layout for TT4 and TT5}, \ref{fig:Internal cooling and ventilation layout for TT5} and \ref{fig:Internal cooling and ventilation layout for TT4}. For further details on the CV study undertaken see Section~\ref{sec:CV_TT4}.

\newpage 
\paragraph{Radiation protection layout}

\begin{figure}[!hbt]
\begin{center}
    \includegraphics[width=0.8\linewidth]{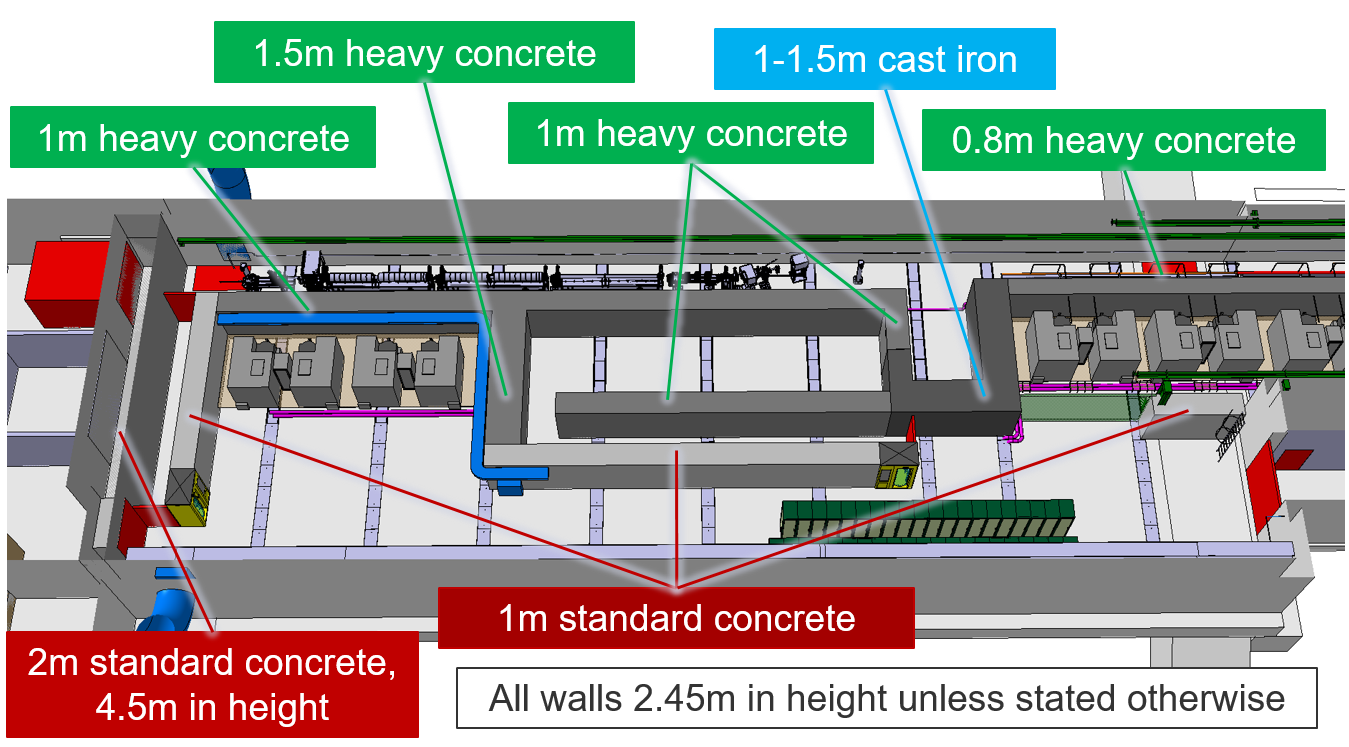}
    \caption{Shielding wall layout in TT5.}
 \label{fig:Shielding_wall_layout_in_TT5}
\end{center}
\end{figure}

\begin{figure}[!hbt]
\begin{center}
    \includegraphics[width=0.79\linewidth]{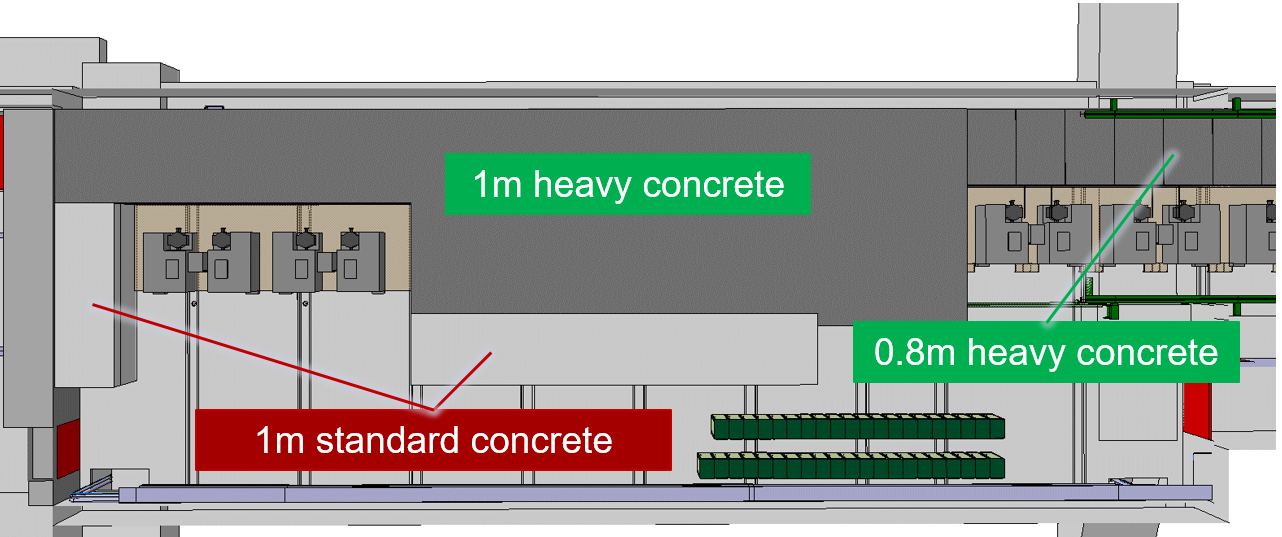}
    \caption{Shielding roof layout in TT5.}
 \label{fig:Shielding_roof_layout_in_TT5}
\end{center}
\end{figure}

\begin{figure}[!hbt]
\begin{center}
    \includegraphics[width=0.79\linewidth]{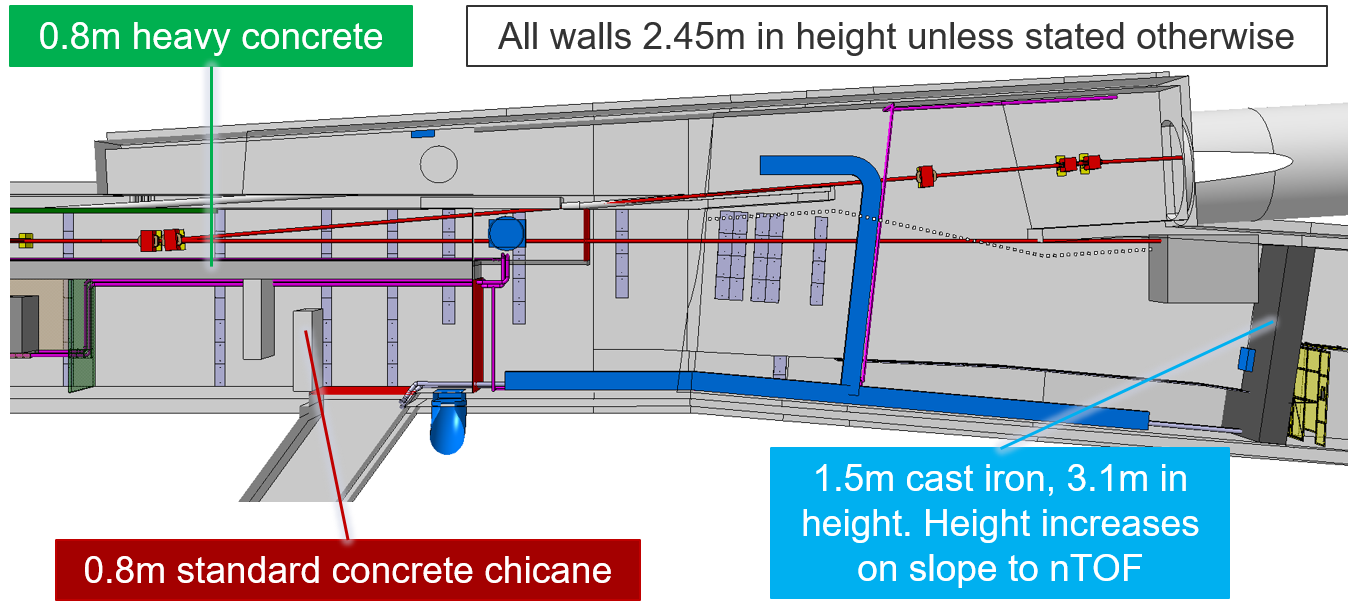}
    \caption{Shielding wall layout in TT4.}
 \label{fig:Shielding_wall_layout_in_TT4}
\end{center}
\end{figure}

Radiation protection was one of the key aspects in the overall layout of the facility. Detailed radiation protection studies have been undertaken to determine the material and thickness of the shielding walls and roofs in the facility (see Section~\ref{sec:ICE:injection:radprotection}). Different densities of shielding material have been utilised to fit within the existing geometry of facility as shown in Figs.~\ref{fig:Shielding_wall_layout_in_TT5}, \ref{fig:Shielding_roof_layout_in_TT5} and \ref{fig:Shielding_wall_layout_in_TT4}.

\newpage 
\paragraph{Access control layout}

\begin{figure}[!hbt]
\begin{center}
    \includegraphics[width=0.79\linewidth]{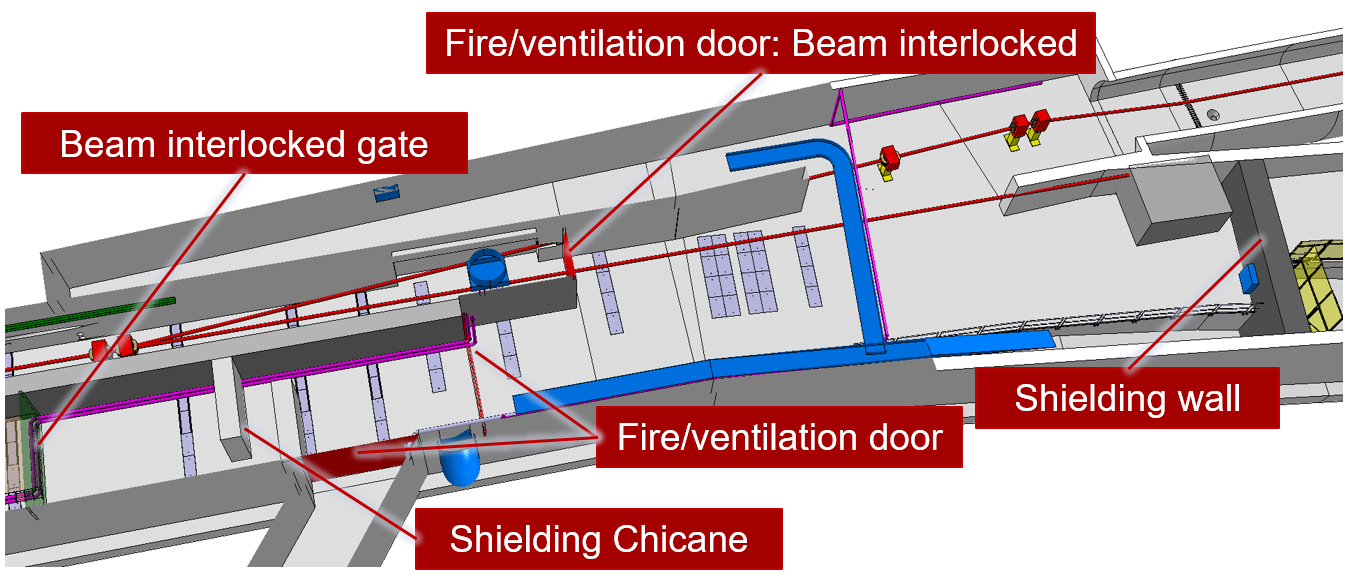}
    \caption{Access control layout in TT4.}
 \label{fig:Access control layout in TT4}
\end{center}
\end{figure}

\begin{figure}[!hbt]
\begin{center}
    \includegraphics[width=0.8\linewidth]{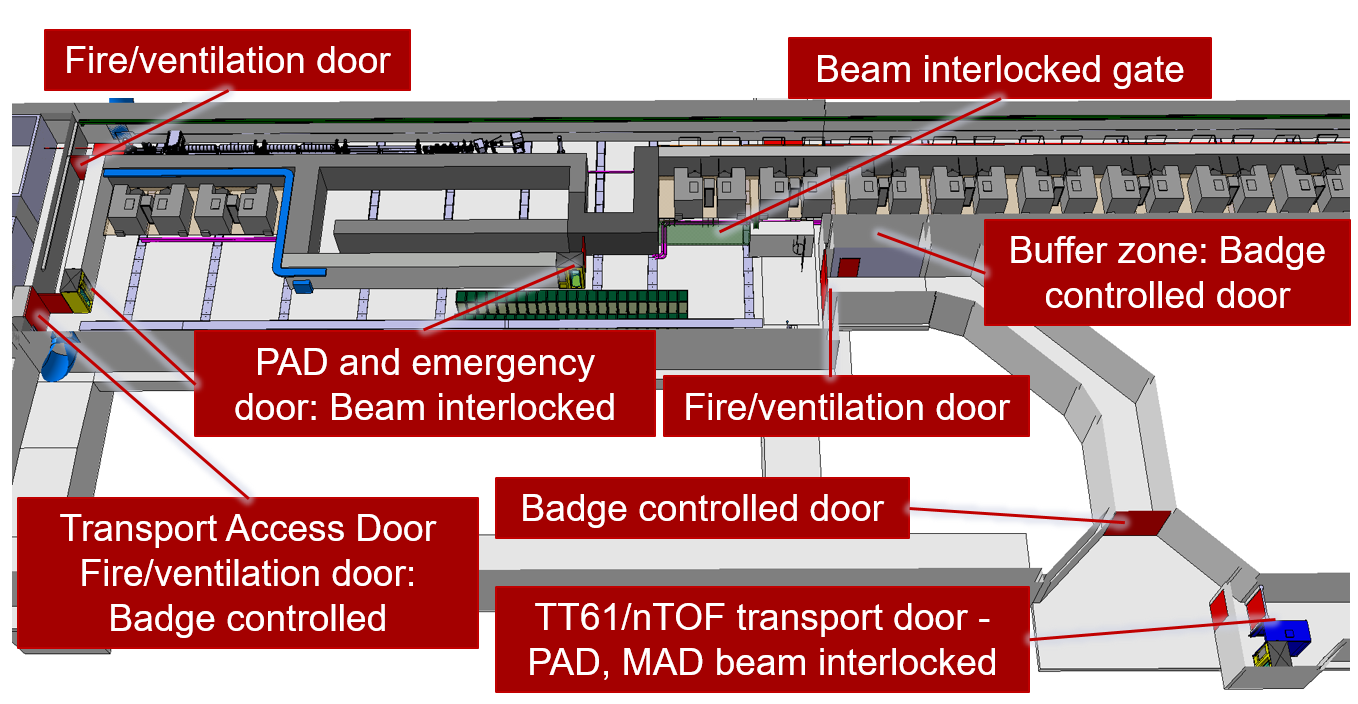}
    \caption{Access control layout in TT5.}
 \label{fig:Access control layout in TT5}
\end{center}
\end{figure}

The access control systems have been defined in accordance with the access requirements of the different equipment owners. The access control layout for transfer tunnels TT4 and TT5 are shown in Figs.~\ref{fig:Access control layout in TT4} and~\ref{fig:Access control layout in TT5}. Access rules to the different areas was studied in detail, in particular to the n\_ToF target~\cite{Dougherty}.

\subsubsection{Cooling and ventilation}
\label{sec:CV_TT4}
The cooling and ventilation conceptual design for TT4 and TT5 is herewith presented. It is based on heat loads provided by machine experts in February 2020 and is strongly dependent on these numbers. A~safety factor of 1.2 is applied to the heat loads to account for uncertainties and future facility expansions. 

\begin{figure}[!hbt]
\begin{center}
    \includegraphics[width=0.9\linewidth]{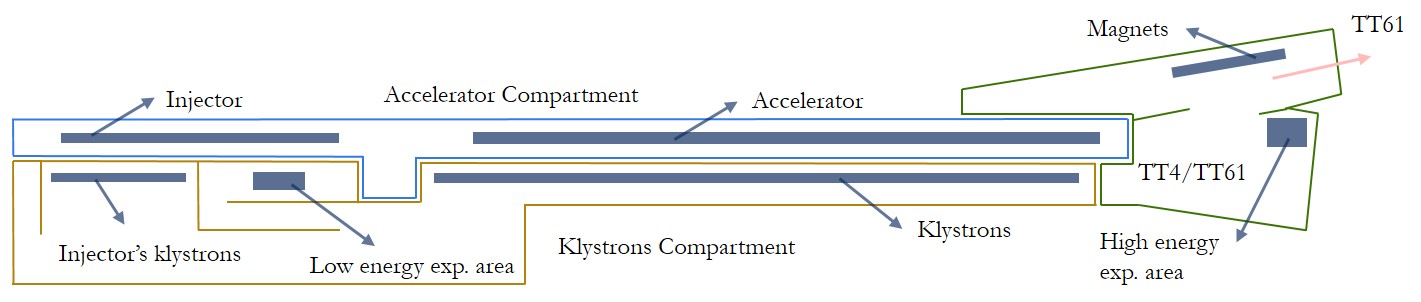}
    \caption{Top view of the studied compartments and heat load sources.}
 \label{fig:CV_TT4_TT5}
\end{center}
\end{figure}

The premises are divided in three regions according to operational requirements. The first encompasses the accelerator and injector, the second includes the klystrons and the low energy experimental area and the third comprises the area TT4/TT61. These regions, or compartments, are presented in Fig.~\ref{fig:CV_TT4_TT5}, along with the heat load sources.

\paragraph{Piped utilities}

\begin{figure}[!hbt]
\begin{center}
    \includegraphics[width=16cm]{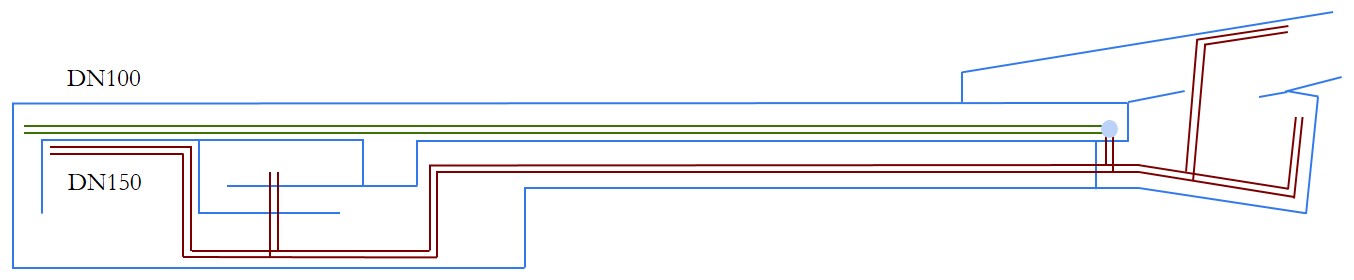}
    \caption{Top view of demineralised water circuits.}
 \label{fig:CV_DEMI_TT4_TT5}
\end{center}
\end{figure}

\begin{figure}[!hbt]
\begin{center}
    \includegraphics[width=8cm]{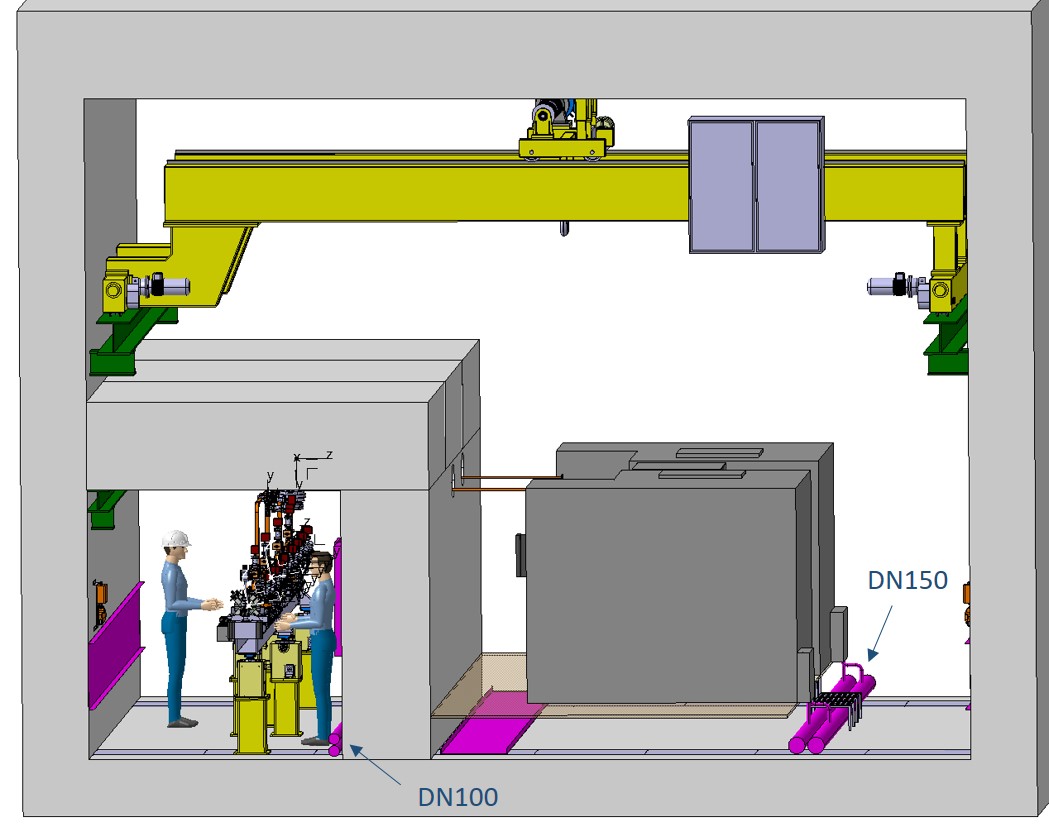}
    \caption{Integration of the demineralised water pipes in the accelerator and klystron compartments.}
 \label{fig:CV_Cross_Section_Piping}
\end{center}
\end{figure}

Demineralised water circuits are foreseen to cool equipment installed in TT4, TT5 and TT4/TT61, as portrayed in Figs.~\ref{fig:CV_DEMI_TT4_TT5} and~\ref{fig:CV_Cross_Section_Piping}. The list of corresponding heat loads is presented in Table~\ref{tab:Inj_WaterHeatLoads}. Local fine temperature stabilisation is provided for the pulse compressors (in the accelerator compartment), using electrical resistances, placed on the shielding wall.

\begin{figure}[!hbt]
\begin{center}
    \includegraphics[width=16cm]{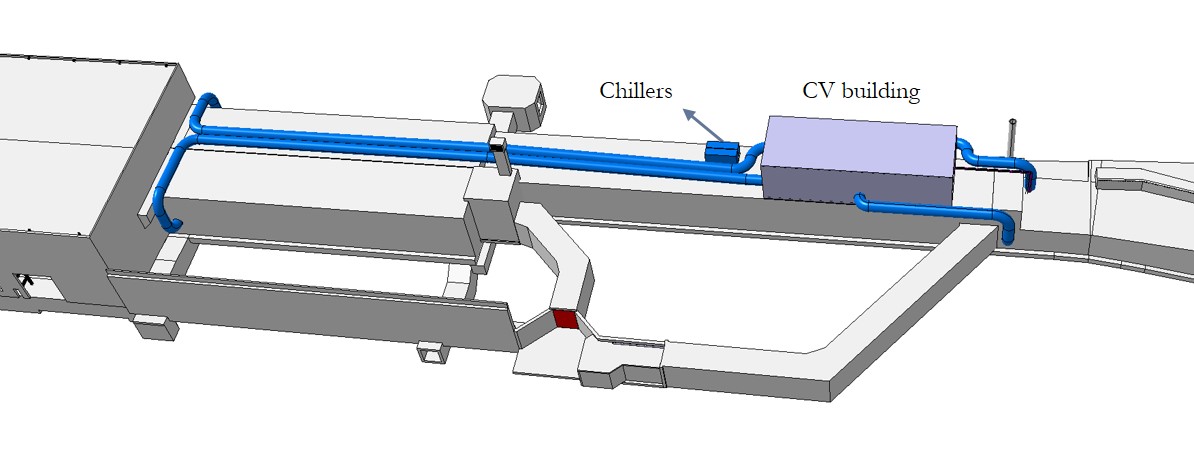}
    \caption{Chillers and CV building positioning.}
 \label{fig:CV_Chillers}
\end{center}
\end{figure}

\begin{figure}[!hbt]
\begin{center}
    \includegraphics[width=8cm]{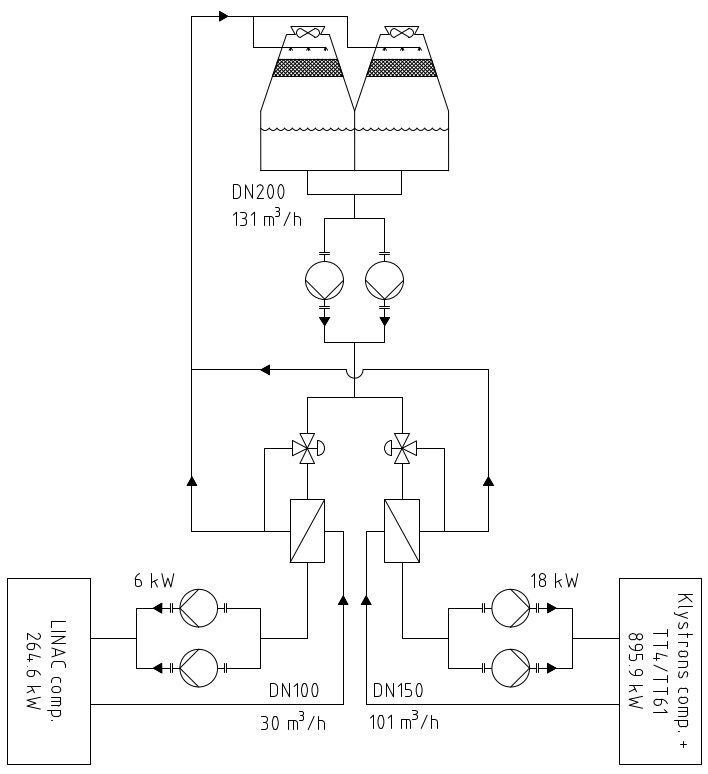}
    \caption{Simplified diagram of the primary and demineralised water circuits.}
 \label{fig:CV_PID_DEMI}
\end{center}
\end{figure}

Existing cooling towers (B280) are used in the primary circuit to cool the demineralised water loop. Pipes extend from the cooling towers to the CV building, sited on top of TT4, as shown in Fig.~\ref{fig:CV_Chillers}. Two cooling stations are foreseen to be installed there - one for the linac compartment (lines 1 and 2 in Table~\ref{tab:Inj_WaterHeatLoads}) and one for the klystrons compartment and TT4/TT61 region (lines 3 to 5 in Table~\ref{tab:Inj_WaterHeatLoads}). The pumps in cooling stations have an N+1 redundancy, contrary to heat exchangers, for which no redundancy is considered. Details concerning the primary and demineralised water circuits can be found in Table~\ref{tab:Primary and demineralized water circuits} and in the simplified diagram presented in Fig.~\ref{fig:CV_PID_DEMI}.

\begin{table}[!hbt]
\begin{center}
\caption{Water heat loads.}
\label{tab:Inj_WaterHeatLoads}
\begin{tabular}{p{4cm}ccc}
\hline\hline
\textbf{Component}          	& \textbf{Heat Load (kW)}   	& \textbf{Safety Margin}  & \textbf{Final Load (kW)}\\ 
\hline
Accelerator                     & 178.4           				& 1.2                 & 214.1 \\
Injector                        & 42.1               			& 1.2                 & 50.5 \\
Klystrons                       & 615.5           				& 1.2                 & 737.5 \\
Injector's klystrons            & 102.4           				& 1.2                 & 123.0 \\
Low energy exp. area            & 10.0            				& 1.2                 & 12.0 \\
High energy exp. area           & 10.0            				& 1.2                 & 12.0 \\
Magnets in TT4/TT61             & 9.5             				& 1.2                 & 11.4 \\
\hline\hline
\end{tabular}
\end{center}
\end{table}

\begin{table}[!hbt]
\begin{center}
\caption{Primary and demineralised water circuits.}
\label{tab:Primary and demineralized water circuits}
\begin{tabular}{p{3cm}ccccc}
\hline\hline
\textbf{Circuit}             	& \textbf{Heat (kW)}   & \textbf{Design flow (m\textsuperscript{3}/h)}   & \textbf{Piping (mm)}   & \textbf{dP (bar)}   & \textbf{Pumps (kW\textsubscript{e})}\\
\hline
Primary                         & 1160.5              & 131              	                    & DN200                 & -                        & -\\
Accelerator Comp.               & 264.6               & 30              	                    & DN100                 & 5                        & 2 x 6\\
Klystrons Comp.                 & 895.9               & 101              	                    & DN150                 & 5                        & 2 x 18\\

\hline\hline
\end{tabular}
\end{center}
\end{table}

\newpage 

\begin{figure}[!hbt]
\begin{center}
    \includegraphics[width=6cm]{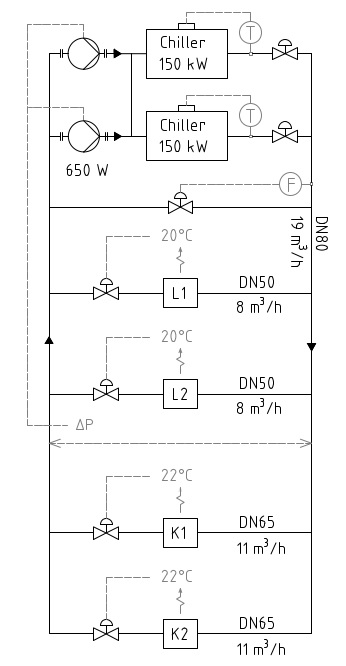}
    \caption{Simplified diagram of the chilled water plant.}
 \label{fig:CV_PID_CHW}
\end{center}
\end{figure}

Two air cooled chillers (one is redundant) with a maximum refrigeration capacity of 150\,kW (each) are foreseen to generate chilled water for air handling units (AHU). The two chillers are installed outdoors, close to the CV building, where the AHUs are placed, as displayed in Fig.~\ref{fig:CV_Chillers}. In the future, during a more detailed stage of the study, one should investigate whether water cooled chillers, that tend to be more efficient, are to be used instead. The details concerning the chilled water plant and circuits can be found in Tables~\ref{tab:Chilled water plant} and~\ref{tab:Chilled water circuits}, as well as in the diagram presented in Fig.~\ref{fig:CV_PID_CHW}. The chilled water pumps have an N+1 redundancy, to ensure proper operation in case of technical failure of a pump.

It is often energy efficient to vary the chilled water temperature as a function of the loads. In particular, when they are reduced, the water temperature is increased, resulting in lower chiller electricity consumption. This strategy has a negative impact on pump consumption (in variable flow systems), as the flow rate is higher. However, for short distribution systems, as in this case, the gains greatly offset the losses. A detailed study on this topic would be of interest for a future technical design report.

\begin{table}[!hbt]
\begin{center}
\caption{Chilled water plant.}
\label{tab:Chilled water plant}
\begin{tabular}{p{3cm}cccc}
\hline\hline
\textbf{Equipment}          	& \textbf{Refrigeration Capacity (kW)}   	& \textbf{Design water flow (m\textsuperscript{3}/h)}    & \textbf{Piping (mm)} \\ 
\hline
Chiller 1                   & 150                                       & 19                                                    & DN80 \\
Chiller 2                   & 150                                       & 19                                                    & DN80 \\

\hline\hline
\end{tabular}
\end{center}
\end{table}

\begin{table}[!hbt]
\begin{center}
\caption{Chilled water circuits.}
\label{tab:Chilled water circuits}
\begin{tabular}{p{3cm}ccccc}
\hline\hline
\textbf{Circuit}          	& \textbf{Heat (kW)}   	& \textbf{Design flow (m\textsuperscript{3}/h)}   & \textbf{Piping (mm)}  & \textbf{dP (bar)}   & \textbf{Pumps (kW\textsubscript{e})}\\
\hline

Main branch               & 150             	                & 19                                             & DN80                 & 1                        & 2 x 650 \\ 
AHU L1                      & 55                                & 8                                              & DN50                 & -                        & - \\ 
AHU L2                      & 55        	                    & 8                                              & DN50                 & -                        & - \\ 
AHU K1                      & 75                                & 11                                             & DN65                 & -                        & - \\ 
AHU K2                      & 75                                & 11                                             & DN65                 & -                        & - \\ 

\hline\hline
\end{tabular}
\end{center}
\end{table}

\newpage 

\begin{figure}[!hbt]
\begin{center}
    \includegraphics[width=0.9\linewidth]{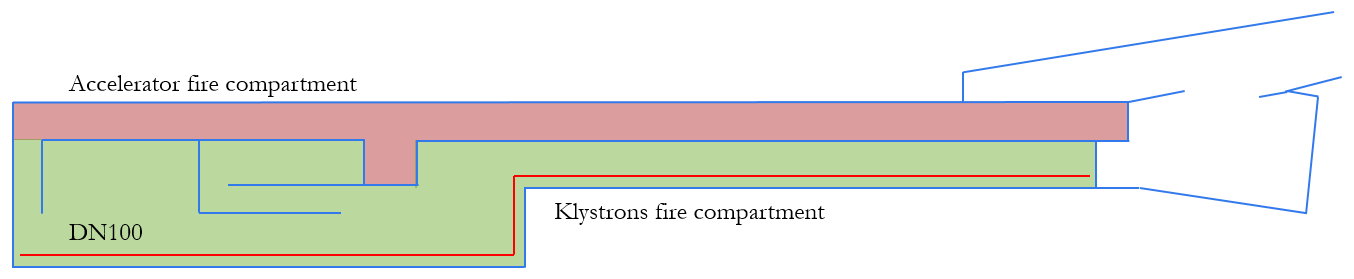}
    \caption{Fire compartments and firefighting water pipe.}
 \label{fig:CV_Fire_Compartments}
\end{center}
\end{figure}

A dry line for firefighting is put in place in the klystron fire compartment, where large fire heat loads are present. A DN100 pipe is foreseen to transport 60\,m$^{3}$/h (enough to supply two firefighting teams) of industrial water in the compartment. It has an outlet for connection every 20\,m. In case of fire, valves can be opened at the surface in order to provide the desired flow. Figure~\ref{fig:CV_Fire_Compartments} shows the two fire compartments and the DN100 pipe.

\paragraph{Heating ventilation and air-conditioning}

The HVAC plant is designed to guarantee the required indoor conditions for the most demanding scenario. Three operational modes are foreseen - run mode, access mode and smoke extraction. In run mode, the linac is running and people are not allowed in the accelerator compartment. On the other hand, they are allowed in the klystron compartment. Access mode is activated when the machine is switched off and people have to work within the accelerator compartment. Smoke extraction is engaged by the fire brigade in case of need. The system is designed to handle both cold and hot smoke.

A push-and-pull ventilation system is adopted for the three operational modes. During run and access modes, air is supplied at 20\si{\degree}C and 22\si{\degree}C to the klystron and accelerator compartments respectively. The maximum supply dew point temperature is 11\si{\degree}C for both regions and modes so that condensation is avoided on surfaces with temperatures lower than 12\si{\degree}C. Air can be partially recycled in each compartment, as ducts are installed to drive the extracted flow back to the AHUs, sited in the CV building, as illustrated in Fig.~\ref{fig:CV_Chillers}. This allows for great energy savings, as treating fresh air is minimised. During smoke extraction, fresh air is supplied to the fire compartments, displayed in Fig.~\ref{fig:CV_Fire_Compartments} without any treatment.

The heat loads for run mode are presented in Table~\ref{tab:Air heat loads during run mode}, and correspond to values provided by machine experts for the desired ambient temperature. In access mode, equipment is switched off and the loads are based on the maximum foreseen occupancy: 10 people in the accelerator compartment and 20 people in the klystron region undertaking hard physical work. The respective loads are presented in Table~\ref{tab:Air heat loads during access mode}.

\begin{table}[!hbt]
\begin{center}
\caption{Air heat loads during run mode.}
\label{tab:Air heat loads during run mode}
\begin{tabular}{p{4cm}cccc}
\hline\hline
\textbf{Component}          	& \textbf{Sensible (kW)}   	& \textbf{Margin}  & \textbf{Final Sens. (kW)}    & \textbf{Latent (kW)}\\
\hline
Accelerator                     & 17.3           				& 1.2                 & 21.0                  & 0.0 \\
Injector                        & 4.1               			& 1.2                 & 5.0                   & 0.0 \\
Klystrons                       & 44.9           				& 1.2                 & 54.0                  & 0.0 \\
Injector's klystrons            & 7.5           				& 1.2                 & 9.0                   & 0.0 \\
Low energy exp. area            & 3.0            				& 1.2                 & 4.0                   & 0.0 \\
High energy exp. area           & 3.0            				& 1.2                 & 4.0                   & 0.0 \\
Magnets in TT4/TT61             & 0.5             	    		& 1.2                 & 0.6                   & 0.0 \\

\hline\hline
\end{tabular}
\end{center}
\end{table}

\begin{table}[!hbt]
\begin{center}
\caption{Air heat loads during access mode.}
\label{tab:Air heat loads during access mode}
\begin{tabular}{p{4cm}cc}
\hline\hline
\textbf{Compartment}          	& \textbf{Sensible (kW)}   	 & \textbf{Latent (kW)}\\
\hline
Accelerator Comp.               & 1.5           			 & 2.7\\
Klystron Comp.                 & 3.1           			 & 5.4\\

\hline\hline
\end{tabular}
\end{center}
\end{table}

\begin{figure}[!hbt]
\begin{center}
    \includegraphics[width=16cm]{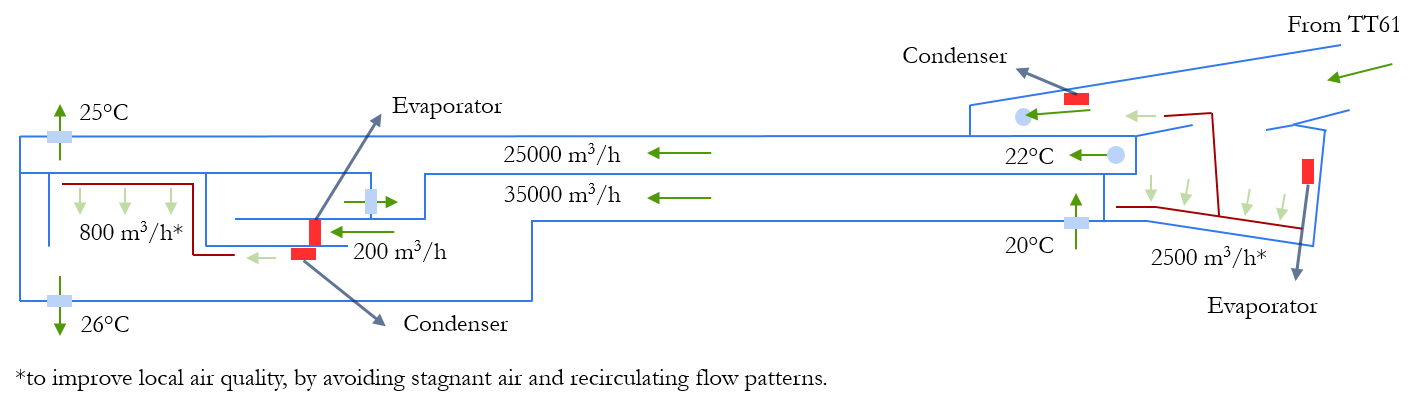}
    \caption{Ventilation layout during run mode.}
 \label{fig:CV_Ventilation_RunMode}
\end{center}
\end{figure}

The ventilation layout during run mode is shown in Fig.~\ref{fig:CV_Ventilation_RunMode}. During this mode, air is mostly recirculated to avoid the introduction of fresh air heat loads. The flow rates are 25000\,m$^{3}$/h and 35000\,m$^{3}$/h in the accelerator and klystron compartments respectively. They are selected according to the desired inlet-outlet temperature difference, whilst providing reasonable longitudinal velocities. The inlet-outlet temperature difference in the accelerator compartment is smaller than the standard values. This is to ensure a relatively small temperature gradient in the area, which is relevant for the beam alignment systems. On the other hand, a higher difference can be taken for the klystron region, where there are no constraints on this parameter. The low energy experimental area is ventilated (200\,m$^{3}$/h, corresponding to one air change per hour) with air transferred from the klystron compartment to the accelerator one (from an area with lower to higher radiation). This implies a minimum fresh air flow of 200\,m$^{3}$/h in the klystron gallery during run mode, which is desired, given that people can access this area during run mode (nonetheless, more people are expected to work in the space during access mode). The loads in the low and high energy experimental areas are treated by local direct expansion units. Moreover, two systems are added to avoid stagnant air or recirculating flow patterns in the injector's klystron area and TT4/TT61. The flow rates are defined to ensure at least three air changes per hour in these regions.

\begin{figure}[!hbt]
\begin{center}
    \includegraphics[width=16cm]{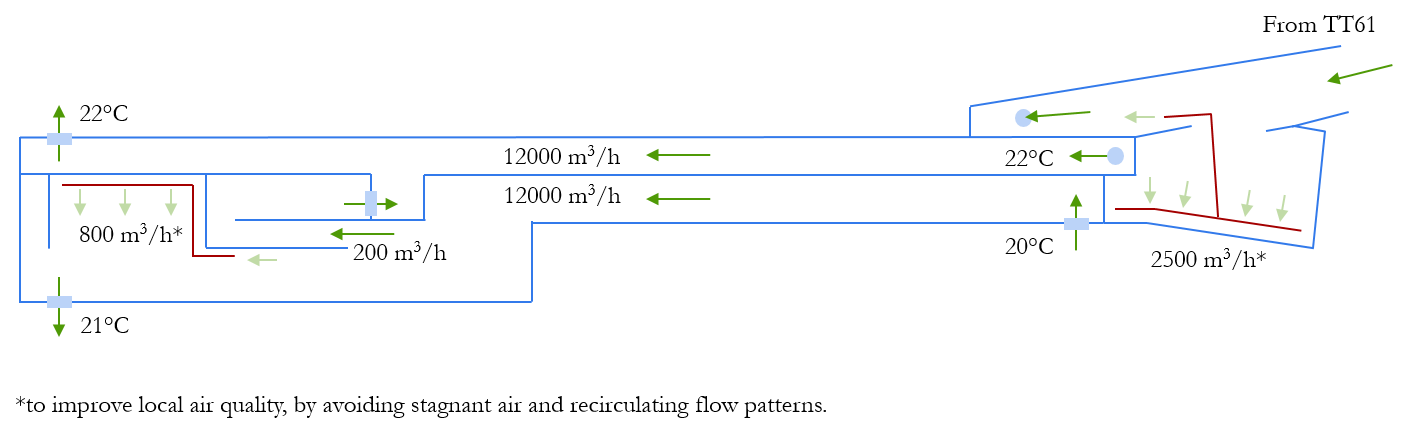}
    \caption{Ventilation layout during access mode.}
 \label{fig:CV_Ventilation_AccessMode}
\end{center}
\end{figure}

The ventilation layout during access mode is presented in Fig.~\ref{fig:CV_Ventilation_AccessMode}. The design flow rate is 12000\,m$^{3}$/h for both the accelerator and klystron regions. However, they are calculated based on different parameters. For the accelerator compartment, a minimum longitudinal velocity of 0.7\,m/s is taken, whilst for the latter, two air changes per hour are ensured. Most of the air is recirculated; however, minimum flows of 400\,m$^{3}$/h and 750\,m$^{3}$/h of fresh air are guaranteed in the accelerator and klystron compartments respectively, to provide proper air quality for the maximum number of people foreseen in these regions. The systems aiming at enhancing air mixing in the injector's klystron and TT4/TT61 regions run during access mode, as portrayed in Fig.~\ref{fig:CV_Ventilation_AccessMode}.

\begin{figure}[!hbt]
\begin{center}
    \includegraphics[width=16cm]{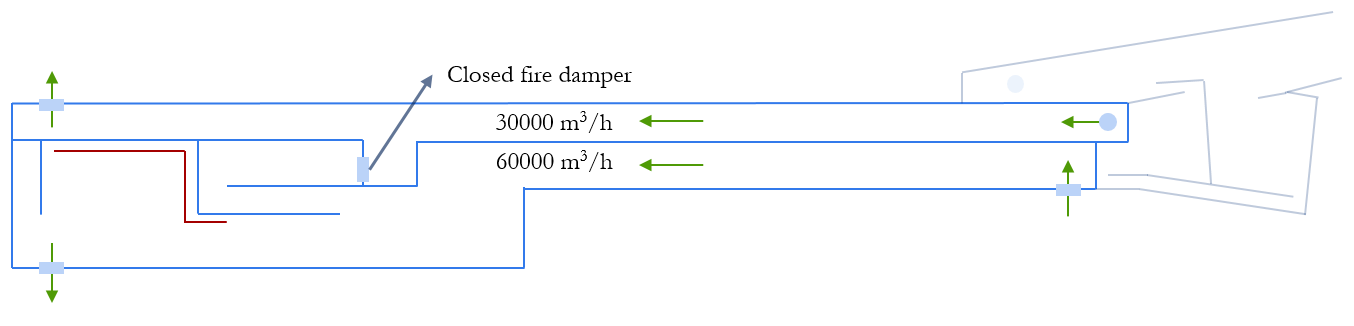}
    \caption{Ventilation layout during smoke extraction.}
 \label{fig:CV_Ventilation_SmokeExtraction}
\end{center}
\end{figure}

During smoke extraction, a fully fresh air ventilation is adopted. The layout is illustrated in Fig.~\ref{fig:CV_Ventilation_SmokeExtraction}, where the selected flow rates are shown: 30000\,m$^{3}$/h for the accelerator region and 60000\,m$^{3}$/h for the klystron compartment. The flow rates were defined by analogy, using French regulations for buildings. The system is able to extract both cold and hot smoke. Hence, the extraction ducts and the extraction units (EXUs) are fire rated. The EXUs are not part of the integration model displayed in Fig.~\ref{fig:CV_Chillers}, but they are foreseen to be installed close to the extraction points, to avoid the installation of long fire rated ducts. According to radiation protection studies, the smoke can be freely released to the atmosphere without any sort of treatment or filtration.

The HVAC infrastructure, presented in Table~\ref{tab:HVAC Infrastructure}, is designed to withstand the most demanding operational conditions. Moreover, most of the equipment has an N+1 redundancy to ensure normal operation in case of failure. The air handling units dedicated to the accelerator compartment, AHU L1 and L2 (one is redundant), have to provide a maximum flow of 30000\,m$^{3}$/h during smoke extraction mode. Similarly, the maximum flow rate for the AHUs treating the klystrons compartment, AHU K1 and K2 (one is redundant), is 60000\,m$^{3}$/h. The smoke extraction units match the flows of the supply ones: EXU L1 and EXU L2 (one is redundant) are able to extract 30000\,m$^{3}$/h whilst EXU K1 and EXU K2 (one is redundant) extract 60000\,m$^{3}$/h. The maximum cooling and heating capacity of AHU L1 and L2 are is required for access mode, due to the dehumidification and reheating needed to treat 400\,m$^{3}$/h of fresh air. For AHU K1 and K2, the maximum heating is also defined during access mode, particularly for the reheating stage after dehumidification of 750\,m$^{3}$/h of fresh air. However, the maximum cooling happens during run mode, to deal with the technical sensible heat loads, as well as to condition 200\,m$^{3}$/h of fresh air. 

At the moment, the heating is foreseen to be provided by electrical resistances. However, the energy released by the chiller's condenser might well be recovered for reheating after dehumidification. This would greatly reduce the energy consumption. A possible alternative is the use of solar panels to heat the water for reheating air. A detailed study on these and other energy efficiency aspects is needed at the next stage of design.

\begin{table}[!hbt]
\begin{center}
\caption{HVAC Infrastructure.}
\label{tab:HVAC Infrastructure}
\begin{tabular}{p{2cm}ccc}
\hline\hline
\textbf{Equipment}         & \textbf{Cooling (kW) @ flow (m\textsuperscript{3}/h)}   	& \textbf{Heating (kW) @ flow (m\textsuperscript{3}/h)}   & \textbf{Max flow (m\textsuperscript{3}/h)} \\
\hline
AHU L1                   & 55 @ 12000                                               & 50 @ 12000                                              & 30000 \\
AHU L2                   & 55 @ 12000                                               & 50 @ 12000                                              & 30000 \\
AHU K1                   & 75 @ 35000                                               & 40 @ 12000                                              & 60000 \\
AHU K2                   & 75 @ 35000                                               & 40 @ 12000                                              & 60000 \\
EXU L1                   & -                                                        & -                                                       & 30000 \\
EXU L2                   & -                                                        & -                                                       & 30000 \\
EXU K1                   & -                                                        & -                                                       & 60000 \\
EXU K2                   & -                                                        & -                                                       & 60000 \\

\hline\hline
\end{tabular}
\end{center}
\end{table}

\subsubsection{Radiation protection} 
\label{Radiation Protection}
\label{sec:ICE:injection:radprotection}


\paragraph{Area classification and access conditions}

The linac, including the two CLEARER experimental areas, will be installed in the existing infrastructure of TT5 and TT4. The linac will operate with varying beam power depending on the beam destination. Areas in TT4 and TT5 will be separated into different access zones to ensure personnel safety during the operation of the accelerator and experimental areas. Figure \ref{fig:Accesssafetyplan} gives an overview on the zoning and sectorisation concept.

\begin{figure}[!hbt]
\centering
\includegraphics[width=\textwidth]{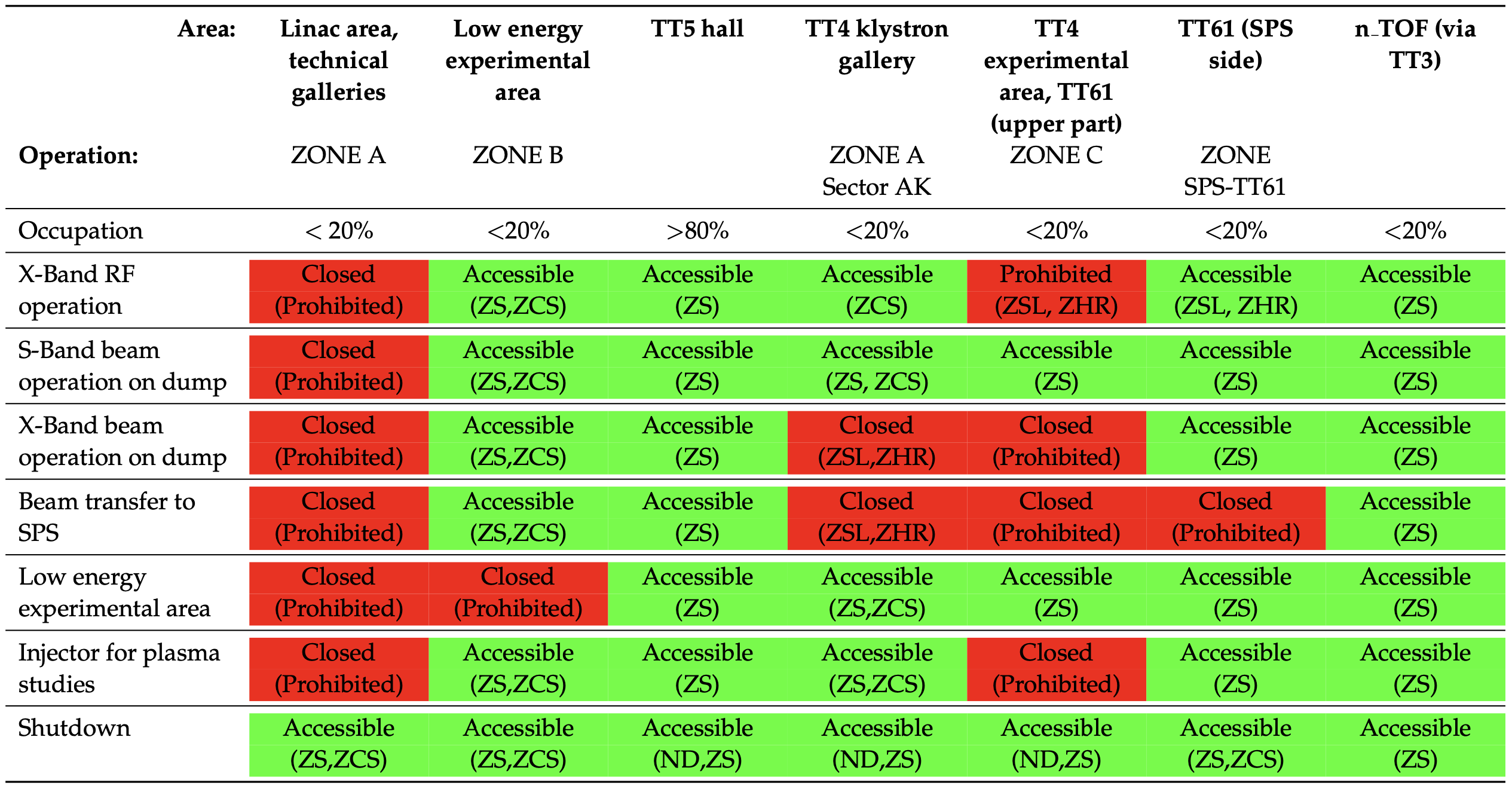}
\caption{\small Access conditions to different areas, depending on the operational scenarios. Radiation area classification: ND: Non-designated area, ZS: Supervised area, ZCS: Simple controlled area, ZSL: Limited stay area, ZHR: High radiation area.}
\label{fig:access}
\end{figure}

Access to classified radiation areas must be controlled through an access control system.
Figure~\ref{fig:access} summarises the access status to all areas depending to the operational conditions. The table includes as well the radiological area classification. All generally accessible areas will be classified as Supervised Radiation Area. An exception is the sector of the klystron gallery next to the X-Band linac (Sector AK). This part will be accessible during RF conditioning in the X-Band linac and will be classified as a 
Simple Controlled Radiation Area. During X-Band linac beam operation the area will be closed and access is prohibited. The high energy experimental area and the upper part of TT61 will remain closed during X-Band RF conditioning, injector operation for plasma studies (see Section~\ref{sec:LINAC_Plasma}) and X-Band linac beam operation. Access to the low energy experimental area will be possible independently of the operational modes of the linac.

The area in TT4 adjacent to TT3 and the n\_TOF Experimental Area 1 in TT2 is classified as Supervised Radiation Area and will remain accessible during all operation modes. This will allow personnel access to n\_TOF independent of eSPS operation. Material access through the access gallery 852 and TT4 will only possible without X-Band linac beam operation, injector operation for plasma studies or RF conditioning.

The exact location of the separation door between the SPS and the X-Band linac in TT61 has not yet been defined. It will be placed such as to allow access to each part of the TT61 tunnel while the adjacent zone is in operation. During X-Band linac operation without ejection to the SPS, the lower part of TT61 shall remain accessible, depending only on the operation mode of the SPS. Access to the upper part of TT61 shall be possible while the SPS is in operation with ejection to HiRadMat or the LHC.


\paragraph{Prompt radiation}
The shielding design in the linac part is challenging due to the increasing beam energy, high beam intensities, different beam destinations, operational conditions and space constraints. To maximise the available space a high density iron-ore based concrete (magnetite type with density of \SI{3.9}{\gram\per\cm^3}) is used as shielding material in most of the locations. This type of concrete is used throughout for the lateral and top shielding on both, the S-Band linac including its experimental area and the X-Band linac. A general wall and roof thickness of \SI{1}{m} is used on the S-Band linac and \SI{0.8}{m} around the X-Band linac. The shielding at the X-Band linac is thinner, despite the higher beam energy, because of the lower beam intensities, the foreseen access restrictions next to the linac and the distance to the next accessible areas.

The proposed layout limits access to areas in the forward direction of the beam. One exception 
is the S-Band linac beam, which is bent by \SI{180}{\degree} to deliver beam to the low energy experimental area. Shielding walls which could be hit directly by beam losses with angles close to \SI{0}{\degree} are made of \SI{1.5}{m} cast iron to provide sufficient protection in case of an incidental beam loss. In the next design iteration, once the beamline elements are known to more detail, the use of iron or of a composited shielding with heavy concrete can be further optimised to reduce the amount of heavy shielding, especially iron.

The beam dumps in the low energy experimental area and at the end of the X-Band linac are dimensioned to allow access behind these beam dumps. They also cover a large part of losses at small angles in the forward direction from the beamline itself.

Different beam loss and irradiation scenarios have been studied:
\begin{itemize}
\label{item:scenarios}
\item Distributed beam loss along the S-Band linac over a distance of \SI{15}{m} between \SI{5}{\MeV} and \SI{250}{\MeV} and \SI{2}{\percent} of the nominal beam intensity;
\item A point-like beam loss in the return chicane towards the low energy experimental area, pointing perpendicular to the linac beam direction at \SI{250}{\MeV} and \SI{1}{\percent} of the nominal beam intensity;
\item Distributed beam loss along the the low energy experimental area beamline over a distance of $\sim$\,\SI{10}{m} between \SI{250}{\MeV} and \SI{3.65}{\GeV} and \SI{2}{\percent} of the nominal beam intensity;
\item Distributed beam loss along the X-Band linac over a distance of $\sim$\,\SI{70}{m} between \SI{250}{\MeV} and \SI{3.65}{\GeV} and \SI{2}{\percent} of the nominal beam intensity;
\item Point-like beam losses in the vacuum pipe after the X-Band linac at \SI{3.65}{\GeV} and each with \SI{1}{\percent} of the nominal beam intensity;
\item Nominal beam operation on both beam dumps in the low energy experimental area at \SI{250}{\MeV} and on the linac main dump at \SI{3.65}{\GeV}.
\end{itemize}

Currently not included was a potential future upgrade for plasma related studies as described in Section~\ref{sec:LINAC_Plasma}. This additional injector may have an impact on the location of the inter-machine door in TT61 and the shielding wall towards the n\_TOF area.

\begin{figure}[!hbtp]
\centering
\includegraphics[width=\textwidth,trim={0cm 3.2cm 0cm 0cm},clip]{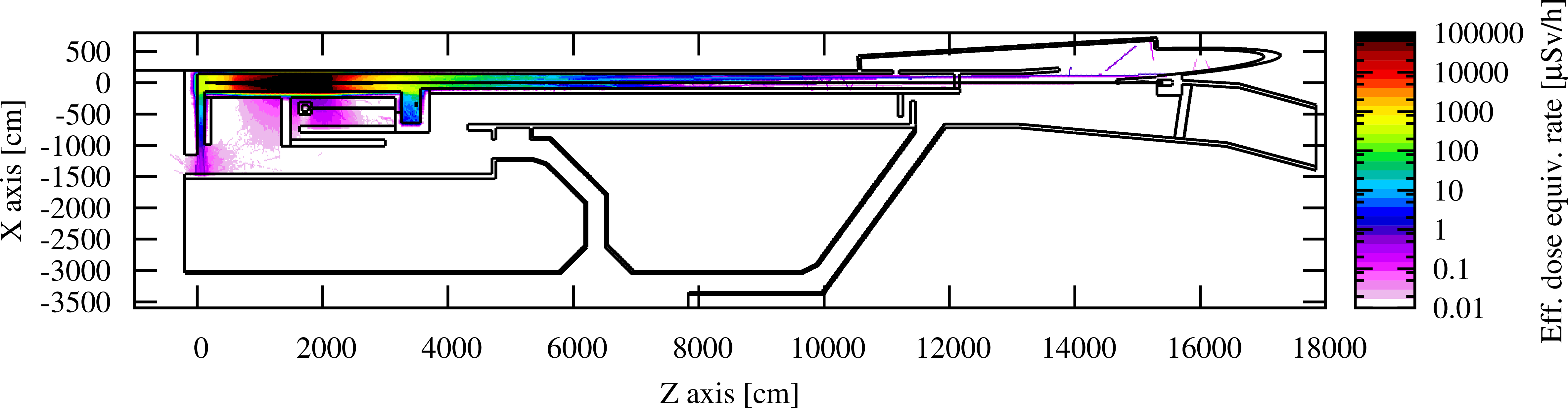}
\includegraphics[width=\textwidth,trim={0cm 3.2cm 0cm 0cm},clip]{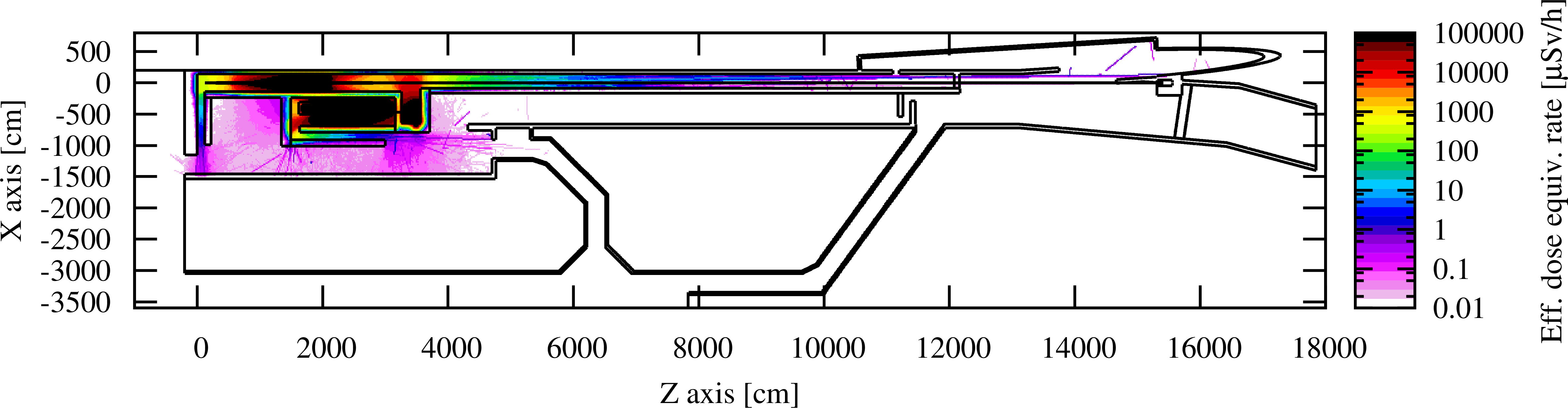}
\includegraphics[width=\textwidth,trim={0cm 3.2cm 0cm 0cm},clip]{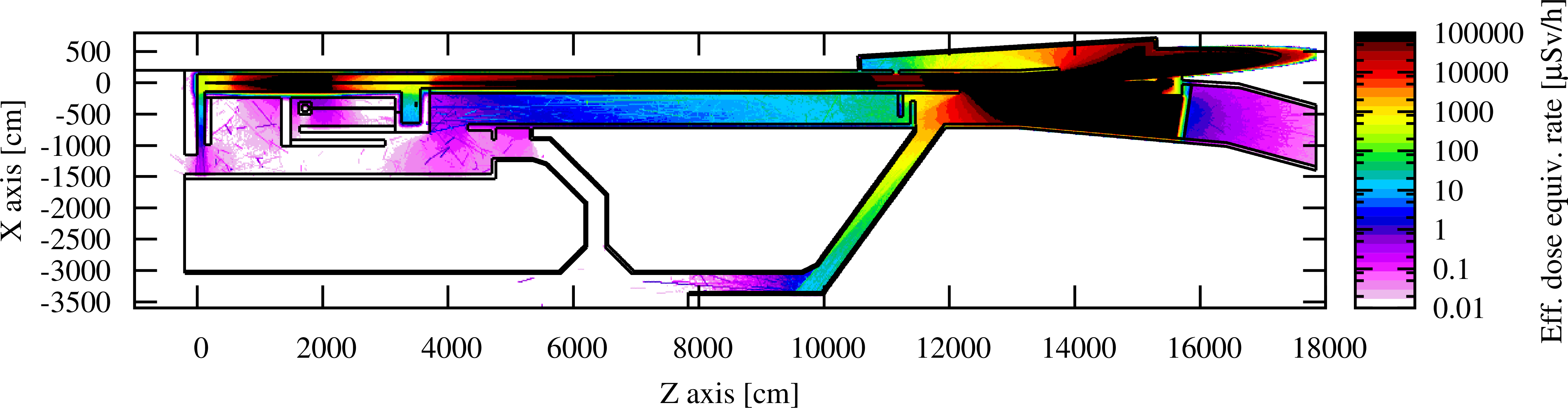}
\includegraphics[width=\textwidth,trim={0cm 3.2cm 0cm 0cm},clip]{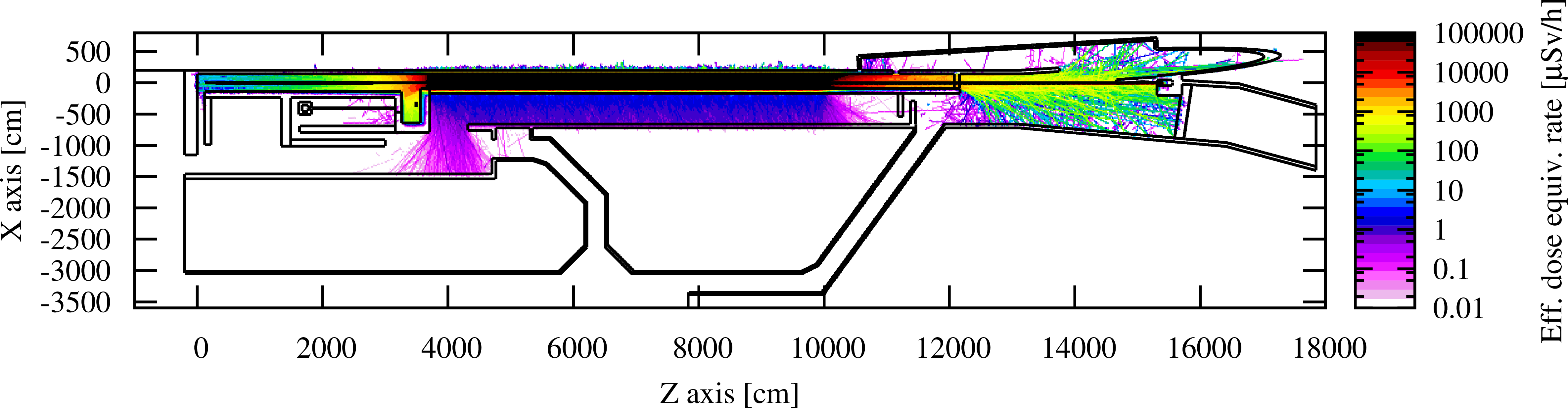}
\includegraphics[width=\textwidth,trim={0cm 0cm 0cm 0cm},clip]{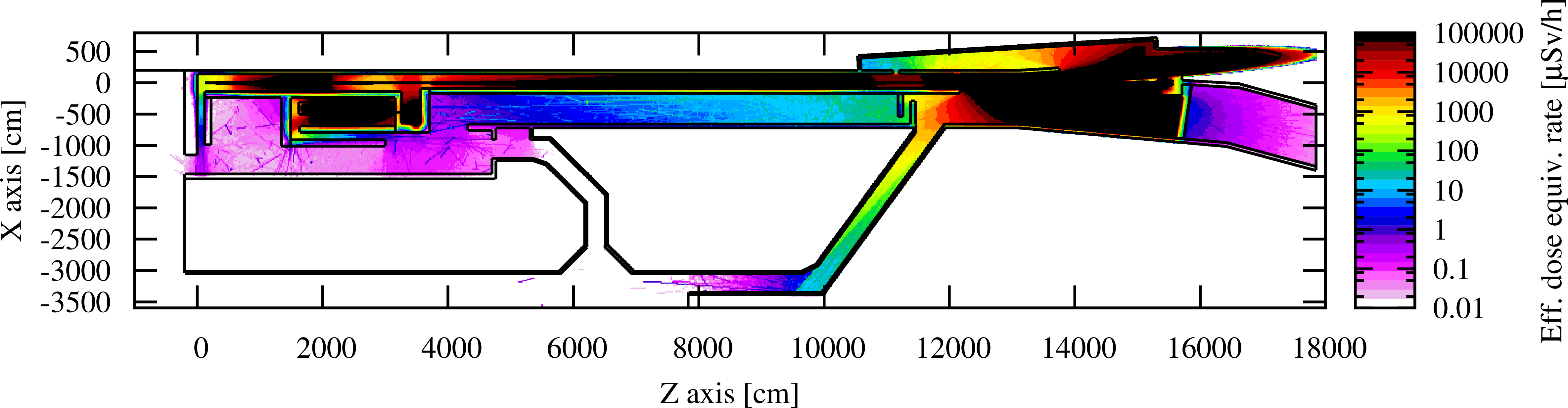}
\caption{Dose rate levels for different loss scenarios. From top to bottom: S-Band linac operation, low energy experimental area operation, X-Band linac operation to the high energy experimental area, X-Band linac RF conditioning, simultaneous beam operation to both experimental areas with maximum nominal beam losses.}
\label{fig:combined}
\end{figure}

Figure~\ref{fig:combined} shows the combined loss scenarios at nominal parameters with maximum admissible beam loss intensity. For such limiting operation scenarios at maximum nominal intensities to all destinations (except the injection towards the SPS), dose rate levels in accessible areas in TT5 and TT4 (n\_TOF side) remain within the defined design limits.

\paragraph{RF conditioning X-band linac}
High-energy X-ray radiation and neutrons may be generated from the dark current which is produced, captured and accelerated in the RF structures during conditioning. During beam operation the klystron gallery next to the X-Band linac will remain inaccessible, while access is needed during RF conditioning periods. Dose rates must hence comply with the defined area classification. The dark current source term, both in terms of absolute current and energy distribution, has a large uncertainty. An approximate source term input has been derived from measurements and simulations at the XBOX facility during conditioning for CLIC structures~\cite{Banon-Caballero2019}. It is assumed that the dark current can be accelerated over one module consisting of four RF structures to limit the current and the maximum energy gain. A dark current of \SI{76}{nA} per module and a continuous X-ray spectrum up to \SI{140}{MeV} was considered in this study. The dark current must be effectively stopped in between modules, e.g.\ through closed vacuum valves or dephasing of the RF between modules.

RF wave guides will connect from the klystrons to the RF structures through ducts roughly every \SI{2}{m}. The wave guide cross section is relatively small, so the effective duct diameter will be much smaller than illustrated in the integration drawings. Shielding inserts around the wave guides are required to reduce the radiation streaming through the ducts.

The high energy experimental area in TT4/TT61 must remain inaccessible because to the radiation propagating from the RF structures down the linac tunnel. However, with a modified access scheme at the end of the linac, e.g.\ a passage towards TT61 and an additional shielding wall, the area could be made accessible also during RF condition of the X-Band linac.

\begin{figure}[!hbt]
\centering
\includegraphics[width=\textwidth]{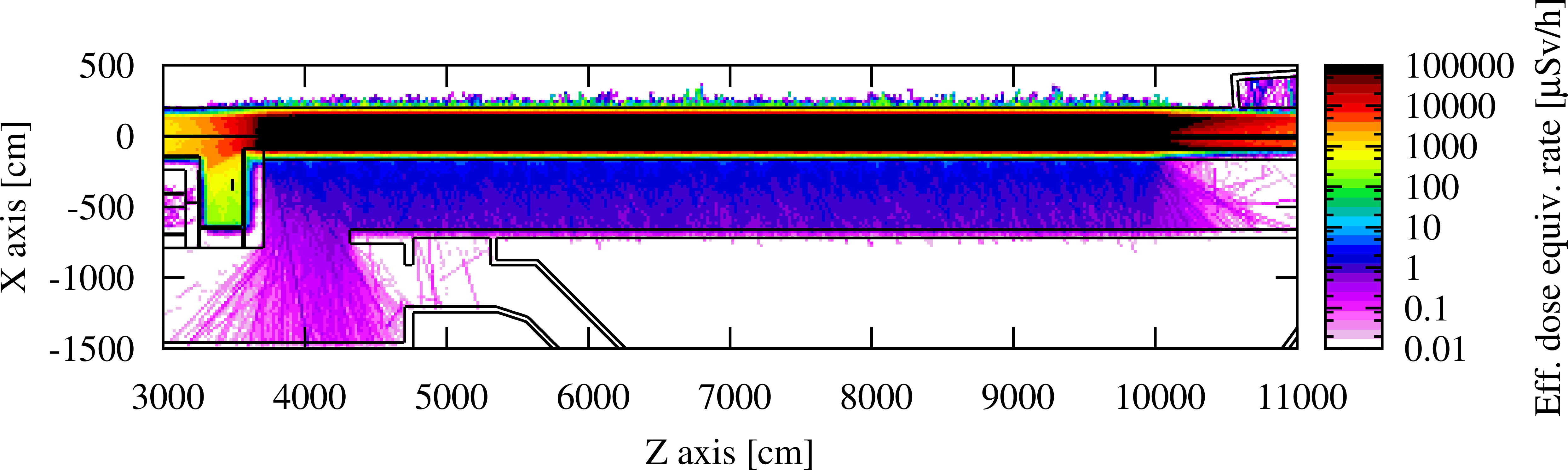}
\caption{Effective dose equivalent rate from RF conditioning, generating a dark current of \SI{76}{nA} and a continuous spectrum up to \SI{140}{\MeV} in each of the 24 modules.}
\label{fig:linRFcond_nom}
\end{figure}

Figure~\ref{fig:linRFcond_nom} shows the expected dose rate in the klystron gallery during RF conditioning in the X-Band linac with the assumed parameters.

\begin{figure}[!hbt]
\centering
\includegraphics[width=\textwidth]{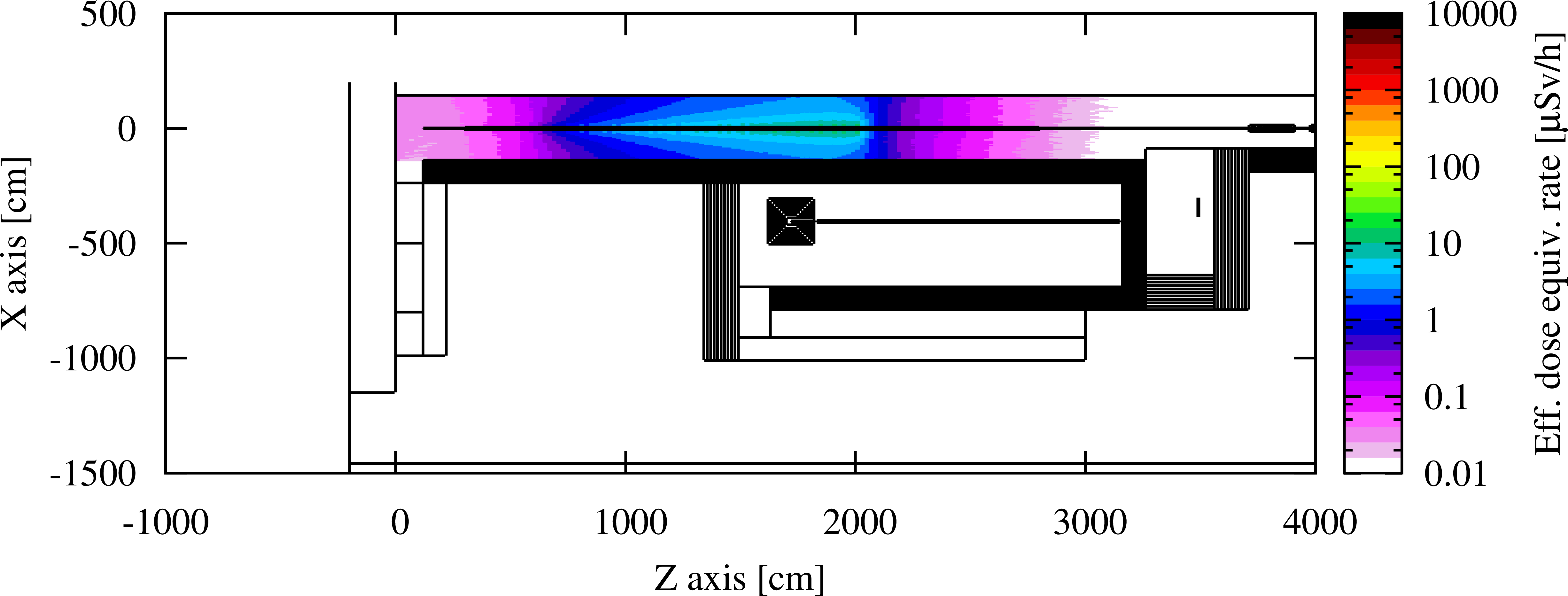}
\includegraphics[width=\textwidth]{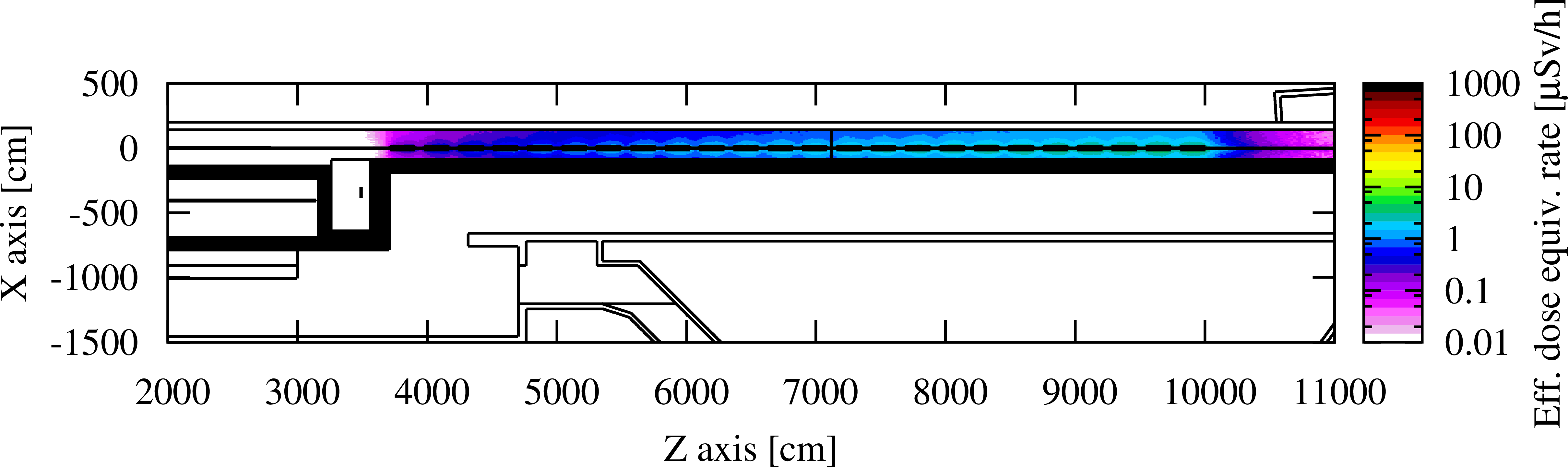}
\caption{Effective dose equivalent rate from induced activity after 1 day of cool-down following 180 days of beam operation at a \SI{2}{\percent} distributed beam loss rate at nominal intensity in the S-Band (\SI{1.85e13}{e^-\per\second}) and X-Band (\SI{1.25e12}{e^-\per\second}) linac.}
\label{fig:inducedactivity}
\end{figure}

\paragraph{Activation}
Induced activity and the resulting remnant radiation levels are usually less critical in electron accelerators compared to proton accelerators. The admissible beam losses in the linac are very much constrained by the available shielding against stray radiation. The permanent beam losses hence remain relatively low. The radiation levels after 180 days of operation with maximum admissible nominal beam losses along the linac remain in the order of \SI{10}{\micro\sievert\per\hour} after 1 day of cool-down, Fig.~\ref{fig:inducedactivity}.

The exposure of workers to activated air intervening directly in the accelerator after beam stop has been evaluated as well and is insignificant. Reference inhalation doses for a one hour stay after beam stop are negligible well below \SI{1}{\micro\sievert}. 

\paragraph{Environmental impact}
Accelerator installations may generate a direct radiological impact on the environment from exposure to stray radiation during operation and releases through air and water. The exposure path through water has not yet been studied due to lack of detailed implementation of accelerator components. It is however considered as negligible regarding the activation potential of the linac in its planned configuration. The stray radiation impact is negligible considering the existing shielding of TT5 and TT4. The main environmental impact from the linac would be induced by releases of radioactivity in the air. 

The production rates for radioisotopes in the accelerator air volume were calculated for a nominal operation scenario. The ventilation scheme is considered to operate mainly in a recycling mode with a permanent air extraction flow rate of \SI{200}{m^3\per\hour} during beam operation and \SI{400}{m^3\per\hour} during access mode.

The production rate, released activity, dose conversion coefficients and effective dose to a reference group for each relevant radioisotope are detailed in Ref.~\cite{Widorski2020}. The annual committed effective dose would arise to about \SI{13}{\nano\sievert}. In case of a non-recycling mode, the effective dose would increase by about a factor 10. These dose levels are acceptable from an environmental impact perspective and are insignificant in their contribution to the overall impact by the installations on the CERN Meyrin Site.

\paragraph{Radiation monitoring system}
A monitoring system measuring the ambient equivalent dose rates in accessible areas is needed to protect against high radiation levels in case of unrestricted beam losses in the various parts of the accelerator. A rather dense network of detectors is required to cover the different loss scenarios and operation modes of the accelerator.

The shielding was designed to limit doses from full beam losses, but dose rates may exceed admissible limits by up to a factor 100. The mixed field monitors and X-ray monitors need to be interlocked with the accelerator. A number of detectors is required to monitor the radiation levels from induced activity along the accelerator.

\begin{figure}[!hbt]
\centering
\includegraphics[width=\textwidth,trim={0cm 0cm 0cm 0cm},clip]{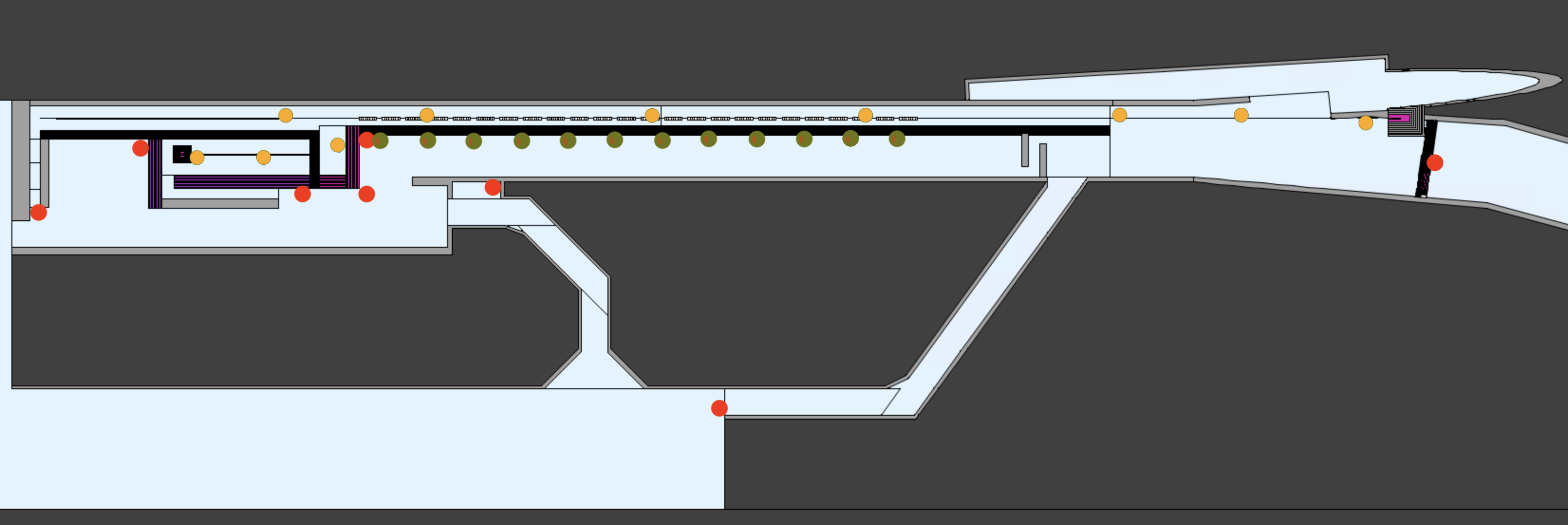}
\caption{Proposed locations of radiation monitors: Mixed field monitors (red), induced activity monitors (orange), X-ray monitors (green).}
\label{fig:rpmonitoring}
\end{figure}

Figure~\ref{fig:rpmonitoring} illustrates the proposed positions of different monitor types for X-ray, gamma and neutron radiation in TT5 and TT4.

\paragraph{Further studies}
In a more detailed technical design of the eSPS project, further optimisation of the shielding design can be achieved. As the design of the accelerator is more refined and more is known about beam loss terms, the shielding design can be adapted, both in terms of volume and shielding materials.

\begin{itemize}
    \item The magnet design in the return chicane to deliver beam to the low energy experimental area may have a positive impact to optimise the lateral shielding;

    \item As the beam dump designs will be developed, dump shielding and integration will potentially have an impact on the surrounding shielding wall. The current design is considered as sufficiently conservative such that the overall integration layout will not be jeopardised;

    \item A shielding separation to be introduced between the linac and the TT4/TT61 area will allow access to the latter area in case of RF conditioning operation. The feasibility in terms of safety, ventilation and operation modes has to be studied;

    \item More detailed knowledge about beamline equipment between the X-Band linac and the beam dump will allow to tailor the shielding wall towards n\_ToF. More detailed technical solutions have to be studied and implemented to allow the passage of cable trays and ventilation through this shielding wall;

    \item The location of the separation door between the SPS and the linac area in TT61 has to be defined. This must be based on more detailed studies on the radiation scattering down TT61 and upstream from possible stray radiation from protons extracted from the SPS towards the HiRadMat installation and TI2.
\end{itemize}

\subsubsection{Transport and handling}

\paragraph{TT4 handling infrastructure}

\begin{figure}[!hbt]
\centering
\includegraphics[width=\textwidth]{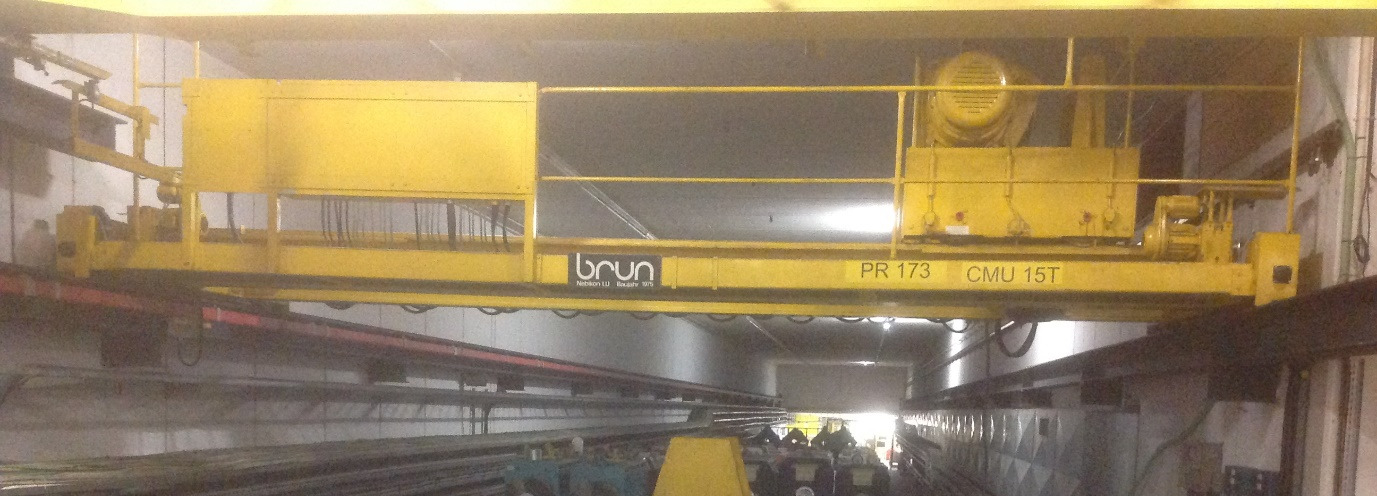}
\caption{Current EOT crane installed in TT4.}
\label{fig:currentEOTCraneTT4}
\end{figure}

Currently the TT4 tunnel is equipped with an electrical overhead travelling (EOT) crane CRPR-00173, shown in Fig.~\ref{fig:currentEOTCraneTT4}, that was installed in 1975 with a safe working load (SWL) of 15\,t.

\begin{figure}[!hbt]
\centering
\includegraphics[width=\textwidth]{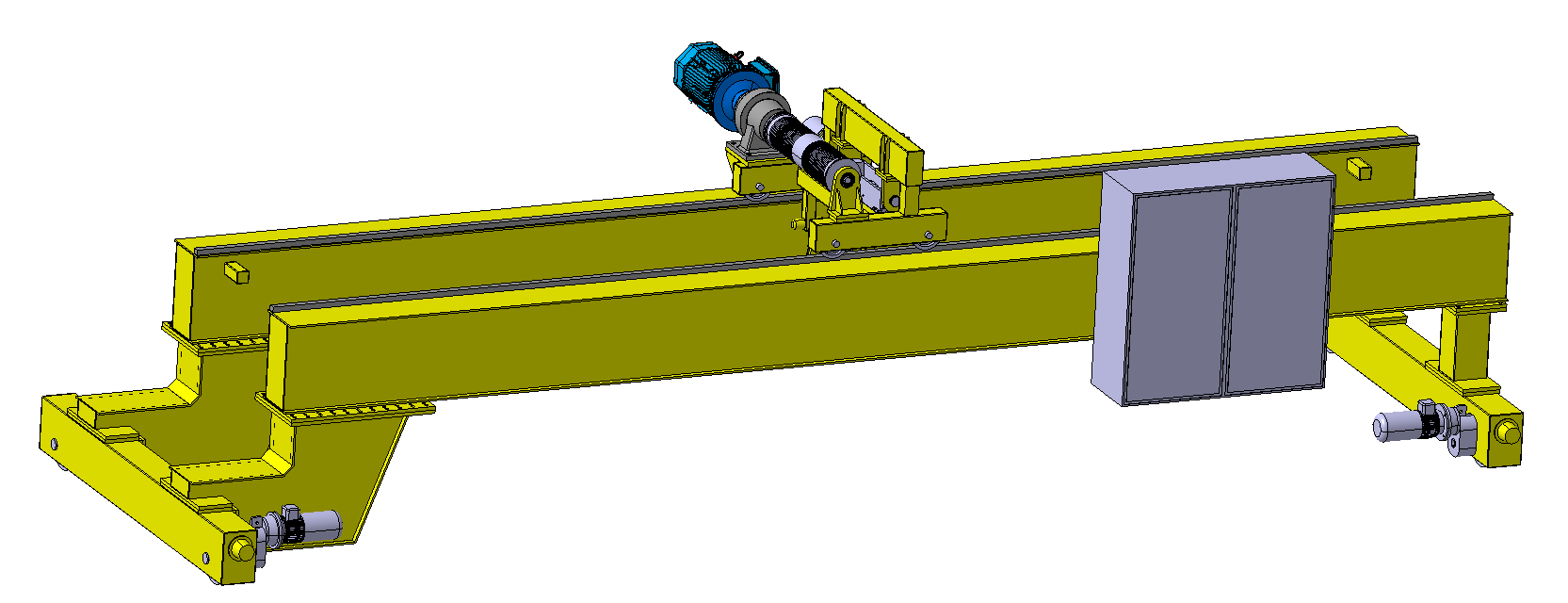}
\caption{A 3D view of the new EOT crane for TT4.}

\label{fig:newEOTCraneTT4}
\end{figure}

This EOT crane needs replacing with a new EOT crane with a reduced SWL of 10\,t but an increased coverage zone and lifting height in order to install the eSPS machine components in the TT4 tunnel. The design of the new EOT crane is shown in Fig.~\ref{fig:newEOTCraneTT4} and its characteristics are given in Table~\ref{tab:NewEOTCraneStatsTT4}.

\begin{table}[!hbt]
    \begin{center}
    \caption{New TT4 EOT crane characteristics.}
    \label{tab:NewEOTCraneStatsTT4}
        \begin{tabular}{cccc}
        \hline\hline
 \textbf{SWL} & \textbf{Span} & \textbf{Lifting height} & \textbf{Power} \\
        \hline 
 10\,t & 7500\,mm & 4500\,mm & 15\,kW \\  
        \hline\hline 
        \end{tabular}
    \end{center}
\end{table}

\paragraph{TT5 handling infrastructure}

\begin{figure}[!hbt]
\centering
\includegraphics[width=\textwidth]{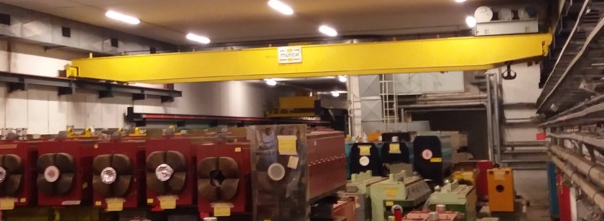}
\caption{Current EOT crane in TT5.}
\label{fig:TT5EOTCrane}
\end{figure}

Currently the TT5 tunnel is equipped with an EOT crane CRPR-00144 that was installed in 1971 with a SWL of 20\,t, as shown in Fig.~\ref{fig:TT5EOTCrane}. This obsolete EOT crane needs to be replaced by a new EOT crane with the characteristics shown in the Table~\ref{tab:TT5EOTCraneStats}.

\begin{table}[!hbt]
    \begin{center}
    \caption{Current TT5 EOT crane characteristics.}
    \label{tab:TT5EOTCraneStats}
        \begin{tabular}{cccc}
        \hline\hline
 \textbf{Lifting capacity} & \textbf{Span} & \textbf{Lifting height} & \textbf{Power} \\
 \hline
 20\,t & 15300\,mm & 5000\,mm & 35\,kW \\  
        \hline\hline 
        \end{tabular}
    \end{center}
\end{table}

\begin{figure}[!hbt]
\centering
\includegraphics[width=\textwidth,height=8cm,keepaspectratio]{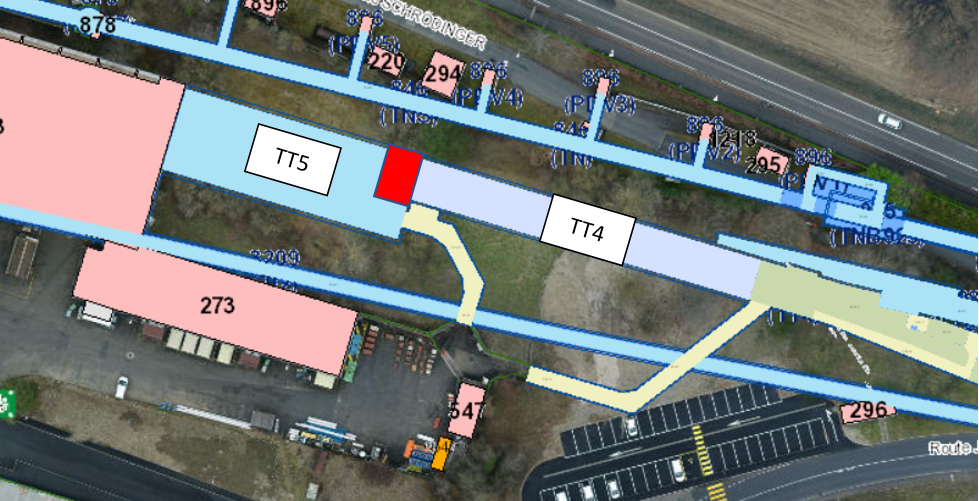}
\caption{Crane overlap region between the EOT cranes in TT4 and TT5.}
\label{fig:TT5/4EOTCraneOverlap}
\end{figure}

The overall dimensions of the new EOT crane will remain roughly the same as the existing one.  The two cranes in TT4 and TT5 overlap in order to transfer equipment’s at the junction of the two tunnels, as shown in red in Fig.~\ref{fig:TT5/4EOTCraneOverlap}.

\paragraph{TT4 installation}

\begin{figure}[!hbt]
\centering
\includegraphics[width=0.6\textwidth]{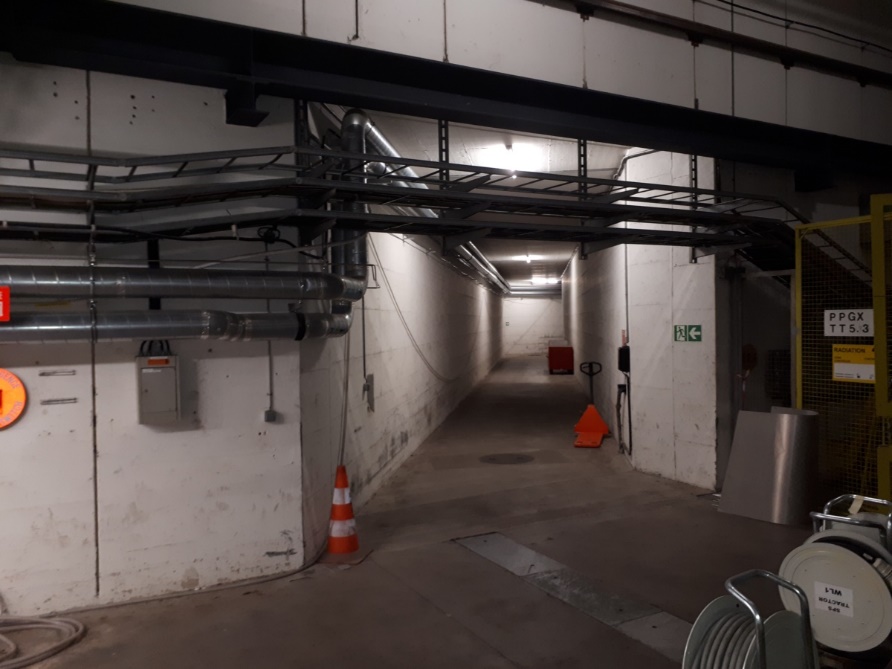}
\caption{Access gallery TTA4 entering TT4.}
\label{fig:AccessGalleryTTA4}
\end{figure}

The delivery of eSPS machine components for installation in TT4 takes place by accessing via the TTA4 access gallery shown in Fig.~\ref{fig:AccessGalleryTTA4}. The new EOT crane will pick up the eSPS machine components at the junction of the TTA4 access gallery and the TT4 tunnel and then will lift them to their final installation position.

\begin{figure}[!hbt]
\centering
\includegraphics[width=0.8\textwidth]{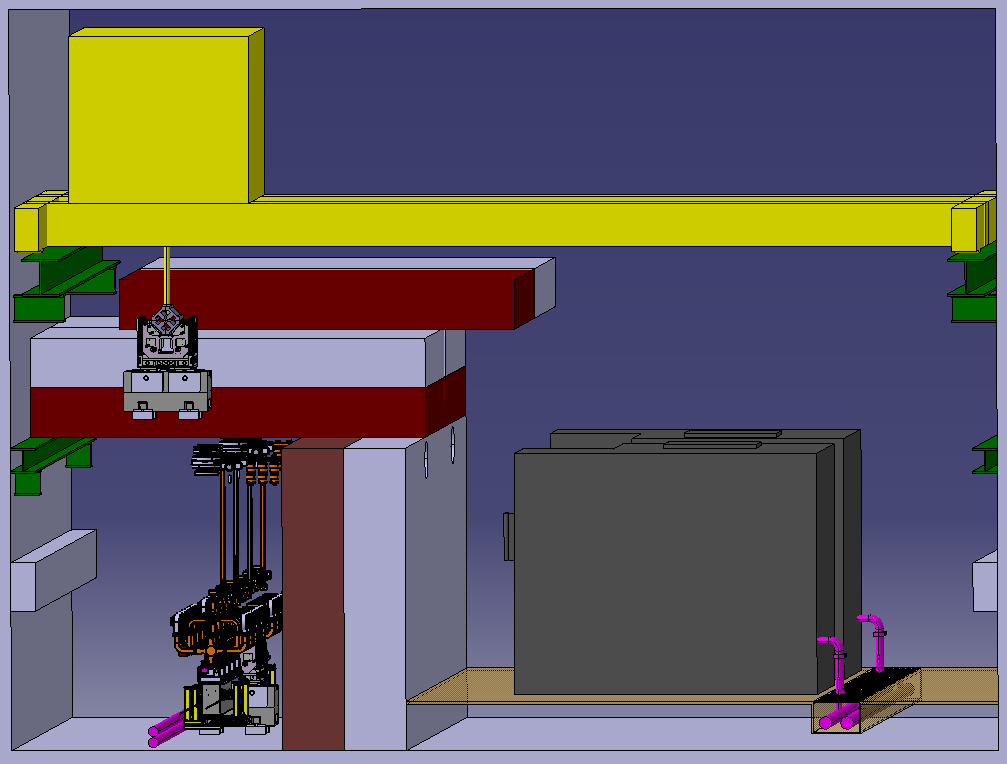}
\caption{Example of cavity handling in TT4.}
\label{fig:exampleCavityHandling}
\end{figure}

The imposed strategy for the installation of the modulators is to install them starting from the junction between TT4 and TT5, and working towards TT61. This is due to the limited headroom for the passage of modulators on top of each other, which is not possible. It is not foreseen to change the complete modulator unit for maintenance or repair purposes.
The SWL and the extended coverage zone of the new EOT crane will allow installing all other eSPS machine components in TT4.  Special lifting beams will be required to optimise the height clearance during lifting operations for the installation of the modulators and accelerator components as shown in Fig.~\ref{fig:exampleCavityHandling}.

\paragraph{TT5 installation}

\begin{figure}[!hbt]
\centering
\includegraphics[width=0.6\textwidth]{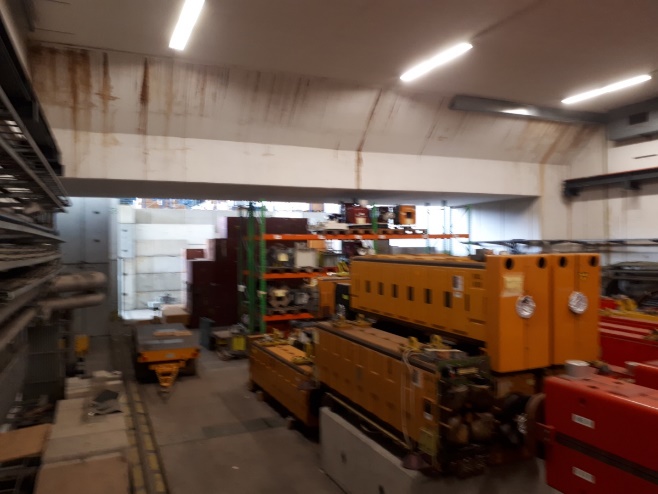}
\caption{Height restriction to enter TT5 via B183.}
\label{fig:HeightRestictTT5}
\end{figure}

The access to TT5 via B183 is limited due to the massive concrete beam that reduces the clearance to 4.5\,m, Fig.~\ref{fig:HeightRestictTT5}. There is no direct crane access to TT5 via B183. All the eSPS machine components needs to be transferred on a trailer via B183 to TT5. The SWL and coverage zone of the new EOT crane in TT5 will then allow installation of all the eSPS machine components. As in TT4, special lifting beams will also be required in TT5 to optimise the height clearance during lifting operations for the installation of the modulators and accelerators components.

\begin{figure}[!hbt]
\centering
\includegraphics[width=\textwidth]{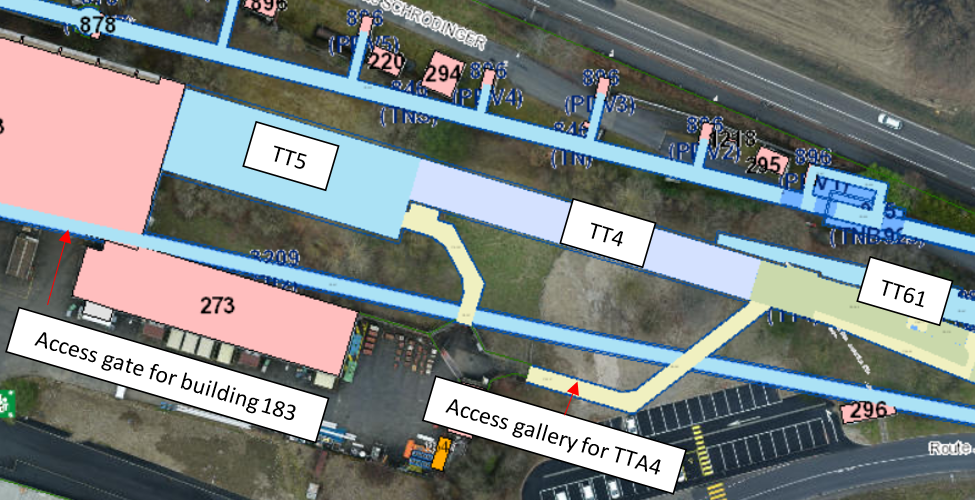}
\caption{Access points in TT4 and TT5 used to deliver equipment.}
\label{fig:TT4TT5AccessPoints}
\end{figure}

The current access routes, via B183 and via the TT4A access gallery, Fig.~\ref{fig:TT4TT5AccessPoints}, are suitable to deliver equipment into both TT4 and TT5.

\subsubsection{Safety engineering}

\paragraph{Fire safety}
The following requirements were outlined and implemented (as set out in Secs.~\ref{sec:ICE-Gen-safety} and \ref{sec:CV_TT4}) for the TT4 \& TT5 sectors:

\subparagraph{Access and egress}
A push-and-pull ventilation system is adopted the klystron and accelerator compartments. There is fire fighter access from two tunnels, upstream of the fire, into the fresh air. Due to the restricted cross-sectional geometry, the access on the side of the modulators close to the shielding wall is only 600\,mm. This is deemed acceptable as any casualties can be extracted into the transport access way for evacuation. For the TT4 and TT5 areas, it is additionally important that fire equipment can be freely moved through the "chicanes" without any problem, especially for a rescue operation. This has been discussed with CERN's Fire and Rescue Service.

\begin{figure}[!hbt]
\centering
\includegraphics[width=\textwidth,trim={0.3cm 0.3cm 0.3cm 0.3cm},clip]{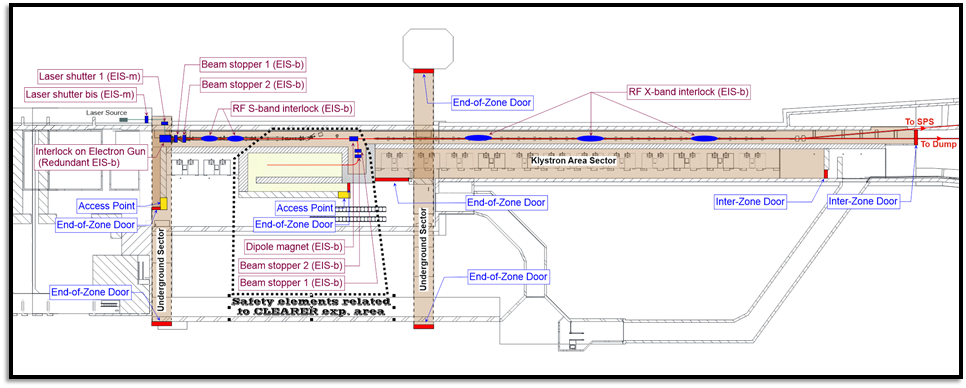}
\caption{Safety and access elements of the linac, adjacent areas and CLEARER experiment.}
\label{fig:Accesssafetyplan}
\end{figure}

\subparagraph{Compartmentalisation}
Two fire compartments are foreseen, one fully containing the linac, and one fully containing the klystrons (as shown in Fig.~\ref{fig:CV_Fire_Compartments}). As shown in Fig.~\ref{fig:Accesssafetyplan}, smoke-proof sector doors (rated to EI90 or greater) will be required at the following points (normally open for the air flow):

\begin{itemize}
    \item The entrance to the linac compartment;
    
    \item The entrance to the Klystron compartment.
\end{itemize}

A 2\,m thick dividing wall between B183 and TT5 will be constructed to provide compartmentalisation for fire safety and as additional radiation shielding to equipment located in B183. The entrance door to TT5 via B183 shall be rated to EI120. Foam shall be used to prevent smoke leakage from TT5 between the concrete blocks. This will need to be replaced when the blocks are removed.

\subparagraph{Fixed suppression means}
A fixed fire suppression system is not necessary for life safety; the project has opted not to install such a system for the protection of the equipment within the fire compartment through discussion of the costs and benefits. The cost estimation made for this has, therefore, been de-scoped from the CDR cost estimation. 
    
A dry riser/standpipe is foreseen for the full lengths of the klystron gallery, with outlets every 20\,m, as shown in Fig.~\ref{fig:CV_Fire_Compartments}. This is the area where the fuel load is the most significant. The flow rate for fire attack shall be 60~m$^3$ per hour. There shall be an outlet very close to the entry point of the linac for immediate use by the CERN Fire and Rescue Service. In case of fire, valves can be opened at the surface in order to provide the desired flow. The full requirements for this equipment are summarised in Ref.~\cite{2017CommonVentilation}.

\subparagraph{Smoke extraction}
Hot smoke extraction is foreseen, allowing the CERN Fire and Rescue Service to employ this tool in line with their strategy to tackle a fire.  During smoke extraction, fresh air is supplied to the fire compartments, displayed in Fig.~\ref{fig:CV_Ventilation_SmokeExtraction} without any treatment. Radiation Protection studies show that this smoke can be released to the atmosphere safely without control or filtration.

\paragraph{Electrical safety}
The HV oil tank will be in place to insulate the klystrons and modulators, with pulse quality monitoring to identify any breakdown in the oil. NF C 18-510 compliant covers, and restriction of access to those with the appropriate electrical training and certification shall also be in place. 

At the conceptual stage of design, it is known that the existing electrical infrastructure in TT4 and TT5 is aged, and will require rejuvenation work to bring it into an acceptable level of conformity suitable for the project. 

\paragraph{Laser safety}
B183 will house a pulsed UV laser, used to generate the required electron bunches. This will be located in a controlled access laser room, and the safety system will therefore need to meet the appropriate IEC required control measures:

\begin{itemize}
    \item IEC 60825-1: Safety of laser products - Part 1: Equipment classification and requirements;
    
    \item IEC/TR 60825-14: Safety of laser products – Part 14: A user’s guide. 
\end{itemize}

\paragraph{Noise nafety}
Noise generated at CERN shall comply with the safety requirements provided in the following Safety rules: 

\begin{itemize}
    \item GSI-SH-4: Protection of workers against noise; 
    
    \item SG-SH-4-0-1: Noise at the workplace.
\end{itemize}

Emissions of environmental noise related to neighbourhoods at CERN, on Swiss or French territory, shall comply with the requirements provided in the following safety rule: 

\begin{itemize}
    \item Noise footprint reduction policy and implementation strategy~\cite{CERN-HSE}.
\end{itemize}

As the cooling and ventilation systems will use a pre-existing cooling tower, the increase in noise levels at the CERN site as a result of the additional cooling load is expected to be very low. An additional chiller will be installed outside for cooling the air for the beam injection, which based on the current cooling and ventilation specification is expected to be approximately 150\,kW, with only one chiller in operation at any time. An initial simulation of the noise impact predicts a total increase in noise at the CERN border onto the Route de Meyrin of 1\,dBA, which is expected to be acceptable; this will need to be confirmed as the project moves into detailed design.

\paragraph{Chemical safety}
The most significant hazard identified in the chemical domain is the large quantity of oil in the HV tanks for the klystron and modulator assemblies. There will be 800--1000~litres of oil per modulator assembly. This is currently expected to be a highly refined mineral oil, capable of withstanding the high voltages. The Safety Data Sheet (SDS) must be obtained from the supplier, a copy of which must be available for consultation by users. The oil must be registered in the CERN Chemical database, CERES, with the quantity, location and any hazards recorded. A copy of the up-to-date SDS must be uploaded into the database and a Chemical risk assessment performed using CERES, if appropriate. The mitigation strategies will consist of a removable retention basin for each HV tank, with sufficient capacity for the entire quantity of oil stored in the tank. Coupled with this, an oil level detector shall be used to indicate any drop in level. As a further measure, consideration shall be given to the potential leak path should oil be spilled outside of the retention basins. Appropriate procedures should be put in place to prevent spillage during filling and draining of the tanks.

\paragraph{Environmental safety}
\subparagraph{Greenhouse gases}
Accelerating Klystron systems attached to waveguides can contain the fluorinated greenhouse gas SF6. The replacement with SF6 by dry air, vacuum or any other alternative shall be considered during the detailed design phase. If not possible, the installation of leak control systems and a recovery system for any maintenance operation shall be foreseen and personnel involved in the activities shall be trained and certified according to the regulation in force at the time of the construction/operation of the project.

\subsubsection{Personnel protection system and access control}

\paragraph{Electron linac area}

According to the RP assessment, the radiation levels of the electron beam in the linac tunnel and in its adjacent technical galleries, as well as the CLEARER experimental area, forbid the presence of people during beam operation. Therefore, new safety chains dedicated to the linac and to CLEARER will be integrated and managed by the SPS Personnel Protection System currently under renovation.

The access to the `electron linac' zone will be done through one Personnel Access Device. During beam operation, the personnel access will be forbidden in the two TT4 technical underground galleries and the Klystron area, which will be closed by new end-of-zone doors. As there is no access point to access to these sectors, a patrol procedure should be initiated before passing the zone into beam mode if one of the end-of-zone door delimiting the sector has been opened.

To ensure the protection of the users accessing the zone from radiation hazard, it is considered to act on the following Important Safety Elements (EIS-b): 

\begin{itemize}
    \item Two Beam Stoppers able to stop the electrons coming from the electron gun;
      \item The S-band and X-band RF preventing beam acceleration.
\end{itemize}

In case of intrusion during operation with beam or if the safe position of a safety element is lost during access, a safety interlock will act on the electron gun. As it is foreseen to use lasers inside the linac tunnel, the laser lines should be equipped with shutters, whose positions should be monitored by the PPS, thus protecting the personnel from the lasers hazards.

If needed, additional local safety functions could be considered such as an `RF Test' mode or a `Laser Alignment' mode.

\paragraph{CLEARER experimental area}

The CLEARER experimental area will be accessible through an access point made of one personnel access device. Aside the access point an "End-of-Zone" door will be installed to allow the exit of people in case of emergency.

To protect people accessing the zone from radiation hazard, one bending magnet and two beam stoppers will be used as EIS-b.

In case of intrusion during beam operation or if the safe position of a safety element is lost during access, the safety chain will act on the upstream zone: the electron linac.

\paragraph{TT5 hall}

\begin{figure}[!hbt]
\centering
\includegraphics[width=0.8\textwidth]{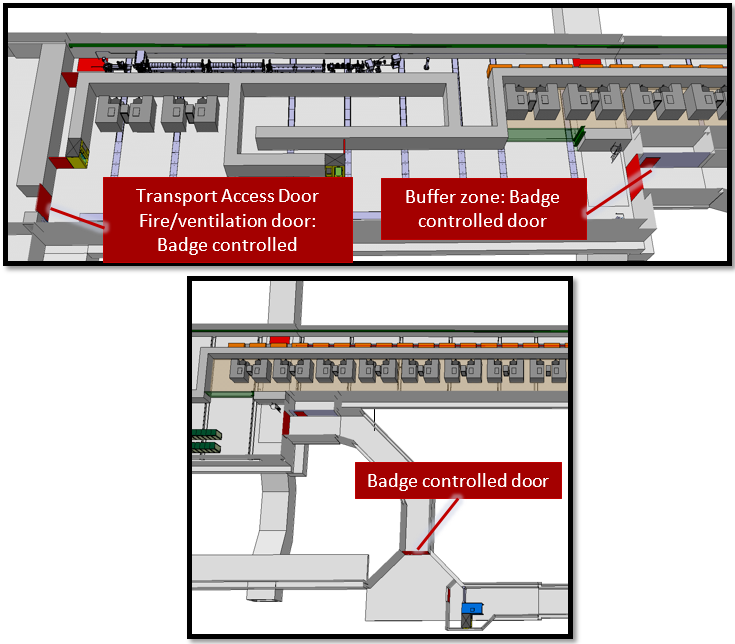}
\caption{Access controlled doors to the TT5 hall.}
\label{fig:Accesssafetyplan2}
\end{figure}

In order to control the access to the TT5 hall and to the buffer zone, the three doors in the following figure will be equipped with an electrical lock controlled by a dosimeter badge reader, as shown in Fig.~\ref{fig:Accesssafetyplan2}.

\subsubsection{Considerations related to existing infrastructure} \label{sec:Widercerncontext}

The installation of the linac in TT5 and TT4 has an impact on the current access path to the n\_TOF Experimental Area 1 (EAR1) situated at the extension of TT4 and upper part of TT2. Requiring the integration of the linac dump in TT4, space for accelerator R\&D experiments and allowing sufficient clearance for material access towards TT61, no solution was found for a local shielding which would have been compatible with an independent access to EAR1 at the same time.

A new potential access path to EAR1 could be provided via ISR T7 and TT3. This access requires additional fencing in the ISR, modifications of the access control system, securing the path with barriers and marking, and a new staircase towards EAR1. The access path would be only suitable for personnel and light equipment. Heavy material access to EAR1, which is required only a few times per year, could still be made through the access gallery 852 and TT4. For that purpose a part of the shielding wall in TT4 must be removable to allow material passage towards EAR1. Typically, such interventions would be scheduled during extended accelerator shutdown periods.

The area in B183 adjacent to TT5 is currently used for the storage of radioactive shielding blocks. To free this area for the required installation of the eSPS control and laser room, an alternative indoor storage location has to be found. It is assumed an area of 1500\,m\textsuperscript{2} is needed to accommodate displaced material from TT4,TT5 and B183. This is a condition of the project and if such space is not available in the flex-storage building or similar, then an additional cost will be incurred to re-provide this space.

%% file: include/05-ICE/TransferTT61.tex
\subsection{Transfer and SPS}
\label{sec:ICE_Transfer}

\subsubsection{Civil engineering}

\paragraph{Existing structure}

\begin{figure}[!hbt]
\begin{center}
    \includegraphics[width=.85\linewidth]{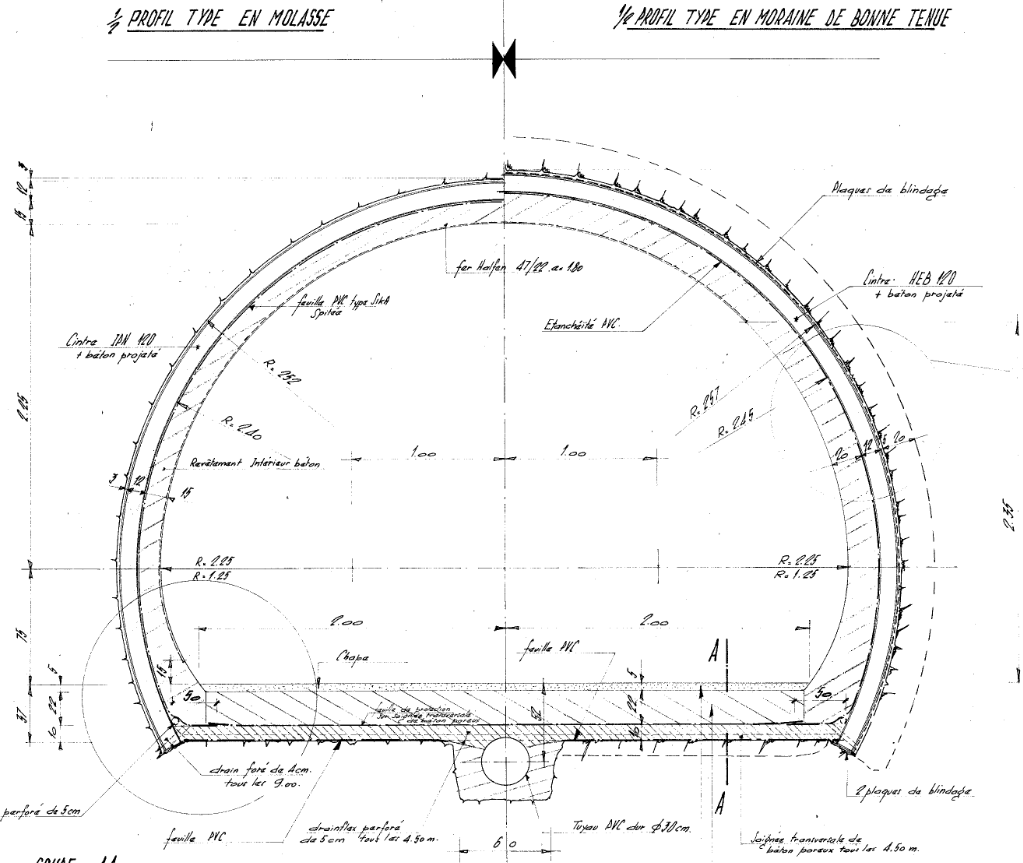}
    \caption{TT61 typical cross section.}
 \label{fig:TT61Typicalxsec}
\end{center}
\end{figure}

TT61 is a horseshoe shaped tunnel with horizontal invert over the majority of its length, measuring 4.5\,m wide (4\,m wide at floor slab level) and 3\,m tall at the crown. The tunnel was constructed predominately by mining into the molasse rock. The tunnel's primary lining consists of steel centring at regular spacings with sprayed concrete infill and a secondary lining of cast in situ mass concrete. A PVC waterproofing layer separates the primary and secondary lining. The tunnel invert is formed with a PVC waterproofing layer below a lean-mix concrete base 100\,mm thick, above that there is a 220\,mm thick cast in situ slab with a 50\,mm screed forming the invert. A 300\,mm diameter drain runs along the length of the tunnel, located centrally below the invert construction. The typical cross section in the molasse is shown in Fig.~\ref{fig:TT61Typicalxsec} with an additional stone and corrugated steel preliminary liner where non-cohesive ground was found.

Constructed in 1975, TT61 was originally part of TT60 until works to enable the LHC upgrade required construction of junction cavern TCC6 to link with the new TI2 injection tunnel, splitting TT60 into two parts (now named TT60 and TT61).
TT61 is now 488\,m long connecting TT4 and TCC6. The tunnel also has several alcoves dating from the original construction which housed beamline equipment. These may be put back into service to house off-axis experiments but this is to be confirmed. The tunnel houses some monitoring equipment for the HiRadMat experiment which is located in the adjacent TN tunnel. The monitoring equipment and shielding takes up a substantial portion of the tunnel cross-section. The need to relocate this is discussed later in the chapter.

\begin{figure}[!hbt]
\begin{center}
    \includegraphics[width=.65\linewidth]{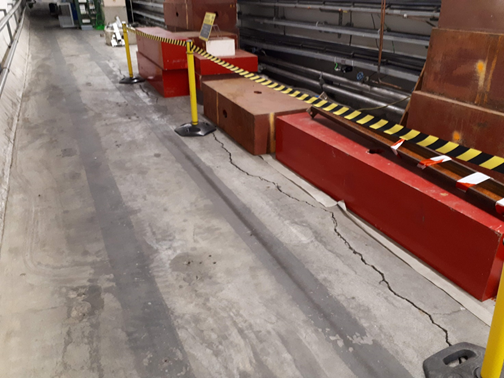}
    \caption{TT61 typical cracking in invert.}
 \label{fig:TT61_Cracking}
\end{center}
\end{figure}

\begin{figure}[!hbt]
\begin{center}
    \includegraphics[width=.65\linewidth]{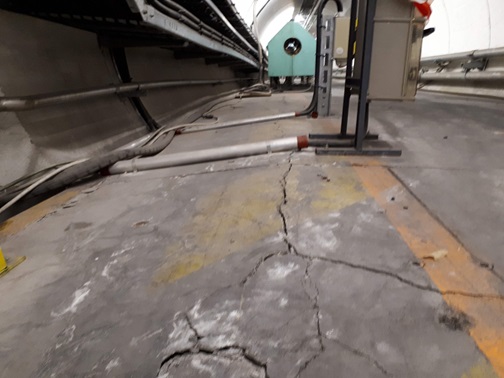}
    \caption{TT61 area of significant floor heave next to alcove.}
 \label{fig:TT61_Floor_heave}
\end{center}
\end{figure}

The tunnel was found to be in very poor condition at the latest inspection with significant cracks running along the centre of the invert up to 30\,mm wide and numerous drainage issues. The invert is also suffering from ground heave with some areas showing uplift of up to 100\,mm. This is likely due to the failure of tunnel drainage and the fact that the tunnel descends through the moraines into the molasse, forming a water pathway into the normally dry molasse. The molasse is known to swell on contact with water. This, in combination with the flat invert which provides little or no resistance to ground heave, has left the tunnel in quite poor condition as shown in Figs.~\ref{fig:TT61_Cracking} and \ref{fig:TT61_Floor_heave}. 

\paragraph{Enabling works}

In order to enable the eSPS project to go ahead, no works would be needed to change the geometry of TT61 which is suitable for replacement of infrastructure to re-implement a beam transfer line. 

Capital maintenance of the existing tunnel would still be important. Existing drainage systems should, as a minimum, be cleaned, surveyed with cameras to check their condition and repaired as required. 

Further work should also be carried out to investigate the causes of cracking and floor heave in the tunnel as well as a campaign of monitoring to determine to what extent the condition is deteriorating. If the rate of uplift is continuing this could pose problems for the overall stability of TT61 as well as for beam alignment during use as a transfer line to the SPS. 

At this stage, maintenance of TT61 has not been studied in detail. An assumption has been made at this stage that superficial repairs to the invert and minor drainage repairs will be required. This is effectively a provisional sum and must be re-valuated in full once monitoring and investigation works have been carried out.

\subsubsection{Integration}

Transfer tunnel TT61 is an underground tunnel that was used to extract the SPS beamline to B180. It is connected to the TT4 transfer tunnel at one end  and the TCC6 junction cavern at the other end.

\paragraph{Connection with transfer tunnel TT4}

\begin{figure}[!hbt]
\begin{center}
    \includegraphics[width=0.9\linewidth]{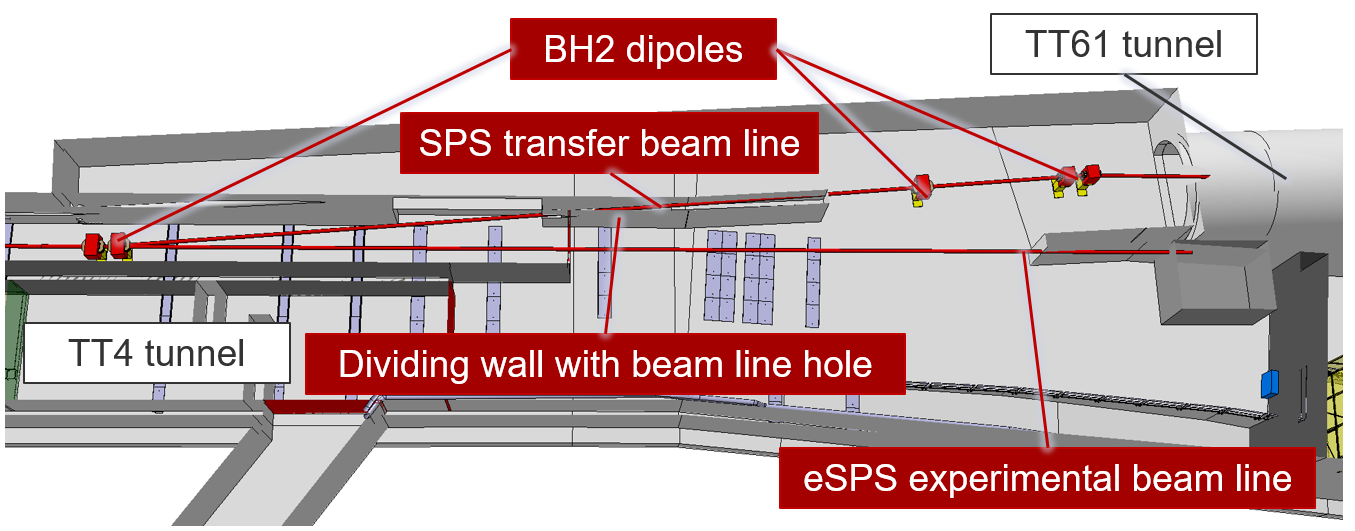}
    \caption{eSPS beamline transfer to the SPS from transfer tunnel TT4.}
 \label{fig:eSPS beam line transfer to the SPS from transfer tunnel TT4}
\end{center}
\end{figure}

The eSPS beamline is bent from transfer tunnel TT4 into transfer tunnel TT61 by horizontal and vertical BH2 dipole magnets as shown in Fig.~\ref{fig:eSPS beam line transfer to the SPS from transfer tunnel TT4}. 

The beamline goes through the wall dividing TT4 and TT61 via an existing hole. This hole will have to be enlarged as the height of the eSPS beamline is different than that of the hole. 

\paragraph{Transfer tunnel TT61}

Currently in TT61 there is some existing beamline equipment remaining from when the SPS beamline was extracted to B180. TT61 also houses some equipment for the HiRadMat facility of which includes removable shielding blocks at the TCC6 end of the tunnel. The tunnel is used by the transport group to access the LHC Injection Tunnel TI2.

\begin{figure}[!hbt]
\begin{center}
    \includegraphics[width=0.75\linewidth]{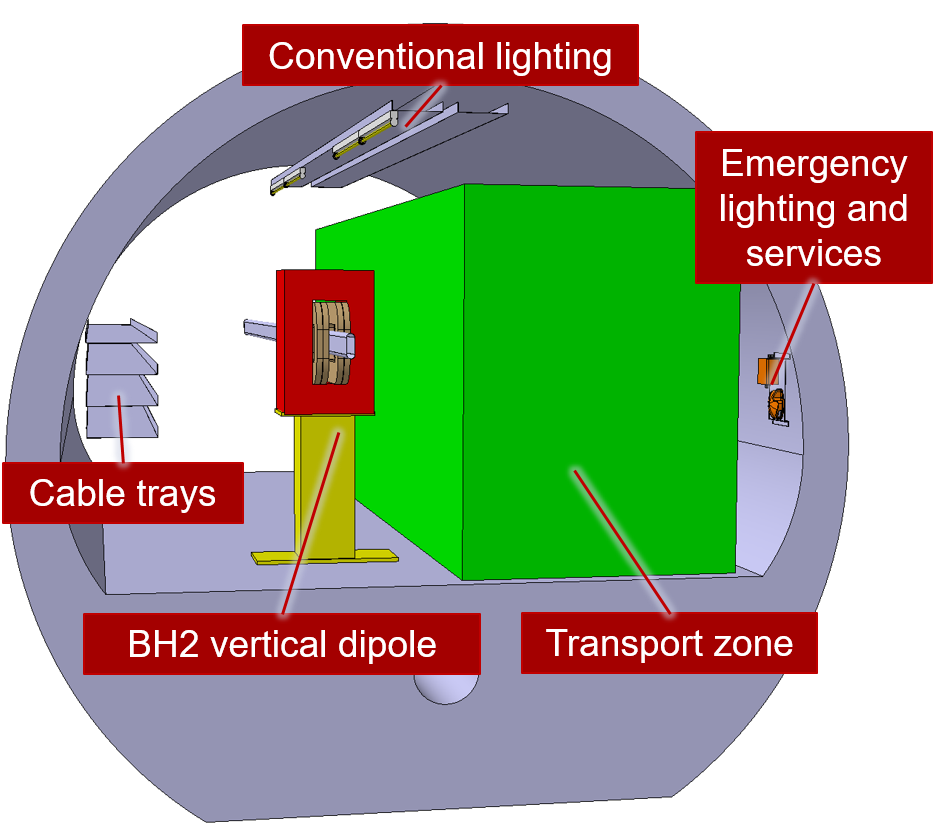}
    \caption{Cross-section of TT61.}
 \label{fig:Cross-section of TT61}
\end{center}
\end{figure}

TT61 will house the SPS transfer beamline from TT4 to TT61. The beamline is approximately 450~m long and consists of a few BH2 vertical dipoles and quadrupoles. As the tunnel’s cross-section remains uniform, a model of the typical cross-section was developed as shown in Fig.~\ref{fig:Cross-section of TT61}.

The cross-section includes:
\begin{itemize}

\item Allowance for safety lighting and services; 
\item Transport volume show in green as specified by the transport group;
\item Width of the dipole;
\item Allowance for personnel access for maintenance of the cable trays;
\item Width of the cable trays.

\end{itemize}

The positions of the dipoles and quadrupoles along the length of the tunnel were not studied at this stage if the project and does not pose any specific issues regarding the overall integration.

\paragraph{Junction cavern TCC6}

TCC6 is a junction cavern housing the TI2 LHC transfer beamline and the TT66 HiRadMat beamline. 

\begin{figure}[!hbt]
\begin{center}
    \includegraphics[width=0.75\linewidth]{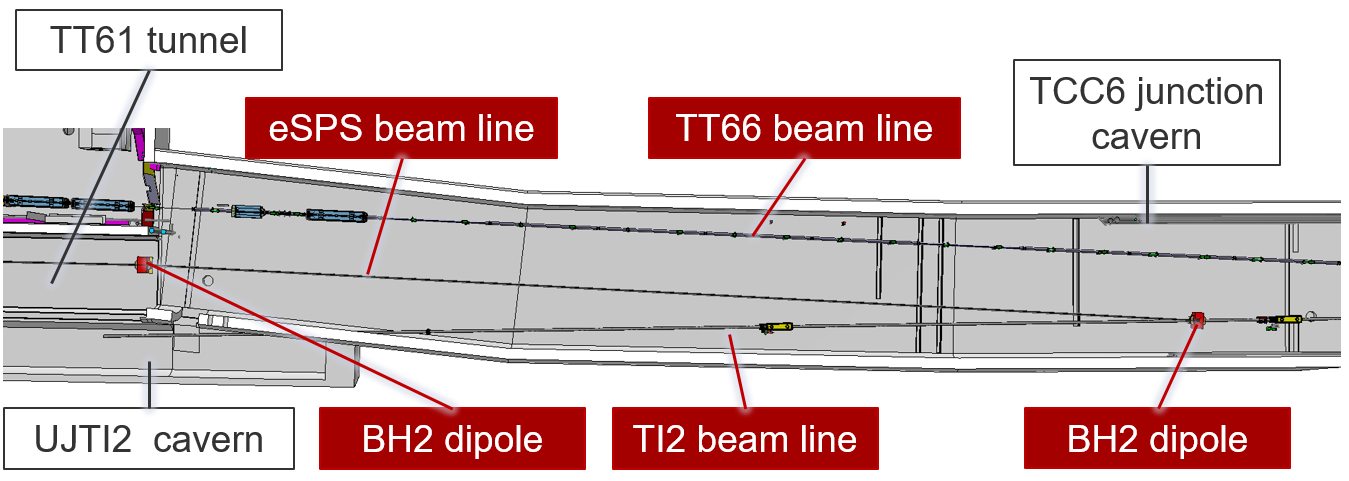}
    \caption{Plan view of the integration layout of TCC6.}
 \label{fig:Plan view of the integration layout of TCC6}
\end{center}
\end{figure}

\begin{figure}[!hbt]
\begin{center}
    \includegraphics[width=0.75\linewidth]{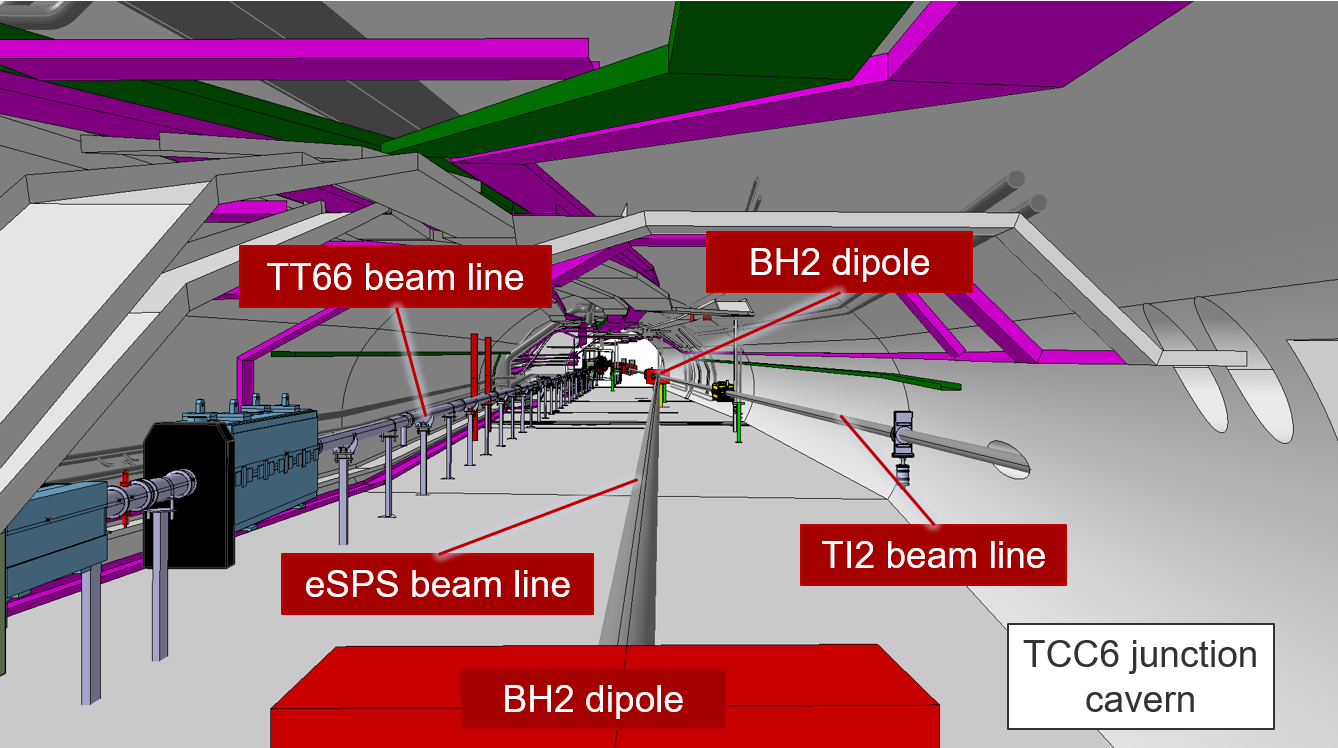}
    \caption{Integration layout of TCC6.}
 \label{fig:Integration layout of TCC6}
\end{center}
\end{figure}

TCC6 will house the eSPS SPS transfer beamline, the concept looks at merging this electron line with the TI2 LHC transfer in TCC6. Due to the low energy of the eSPS electron beam, a dipole is placed at the entrance of TCC6 and the beamline bent towards the TI2 beamline with a removable vacuum chamber allowing the transport to TI2 from TCC6 with minimal obstruction a shown in Figs.~\ref{fig:Plan view of the integration layout of TCC6} and~\ref{fig:Integration layout of TCC6}.

Within this section of eSPS beamline in TT66, quadrupoles and beam diagnostic equipment will be placed, its exact location is to be determined at the detailed design stage. Detailed integration studies of this area have not been undertaken as there has not been any integration studies of this area for several years. New scans of the area should be undertaken in the detailed stage of the project.

\subsubsection{Cooling and ventilation}
\begin{figure}[!hbt]
\begin{center}
    \includegraphics[width=12cm]{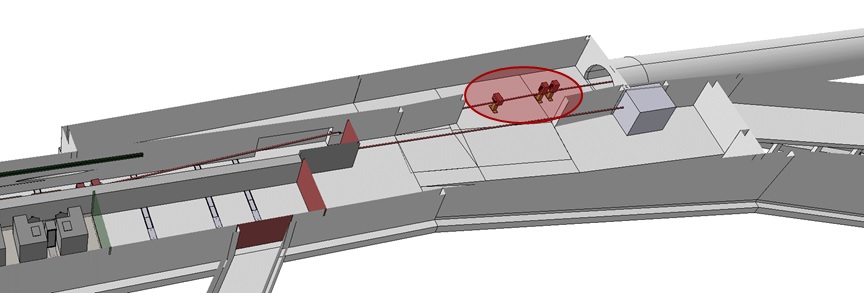}
    \caption{Magnets in TT4/TT61.}
 \label{fig:Magnets in TT4/TT61}
\end{center}
\end{figure}

The existing cooling and ventilation infrastructure for TT61 is not modified because there are no heat dissipating equipment added to this tunnel. Some magnets are installed in TT4/TT61, as displayed in Fig.~\ref{fig:Magnets in TT4/TT61}; however, they dissipate only 0.5\,kW of heat to air. Hence, the existing flow that ventilates TT61 is sufficient to deal with this load. These magnets are also water cooled, dissipating around 9.5\,kW to the demineralised water infrastructure dedicated to the klystron compartment, as displayed in Fig.~\ref{fig:CV_DEMI_TT4_TT5}.

\subsubsection{Radiation protection}

The interface between the linac and the SPS is situated in TT61. Access until this inter-machine door shall be possible from both sides while beam operation is ongoing on the respective other side. The exact location of the inter-machine separation door is depending on the potential beam loss scenarios in TCC6 by the LHC or HiRadMat beams and the radiation coming from electron beam losses in the Linac in TT4.

During electron beam injection from TT61 via TCC6 and TT60 into the SPS at TS65, the adjacent areas to TCC6, namely the access to the HiRadMat facility via BA7 through TJ7, TA7 and TNC must be prohibited. The same applies to the upper part of TI2 and the lower part of TT70. Radiation levels from the high energy electron beam in these areas are prohibitive for the presence of persons.

A more detailed radiological impact study of beam loss scenarios impacting the areas in vicinity of the beam transfer tunnel from TT4 to the SPS is required during the technical design phase.

The radiological impact of the high energy electron beam in the SPS has not been addressed in detail for this report. It is expected that the impact in terms of activation is negligible compared to the ongoing SPS proton operation with much more intense and radiologically relevant beams.

In case a dedicated internal electron beam dump needs to be implemented in the SPS, an appropriate radiological risk assessment shall be included in the technical design study. Similarly, all modification works in the SPS will have a radiological impact due to residual radiation levels and must be properly planned and optimised.

\subsubsection{Transport and handling}

\begin{figure}[!hbt]
\centering
\includegraphics[width=0.9\textwidth]{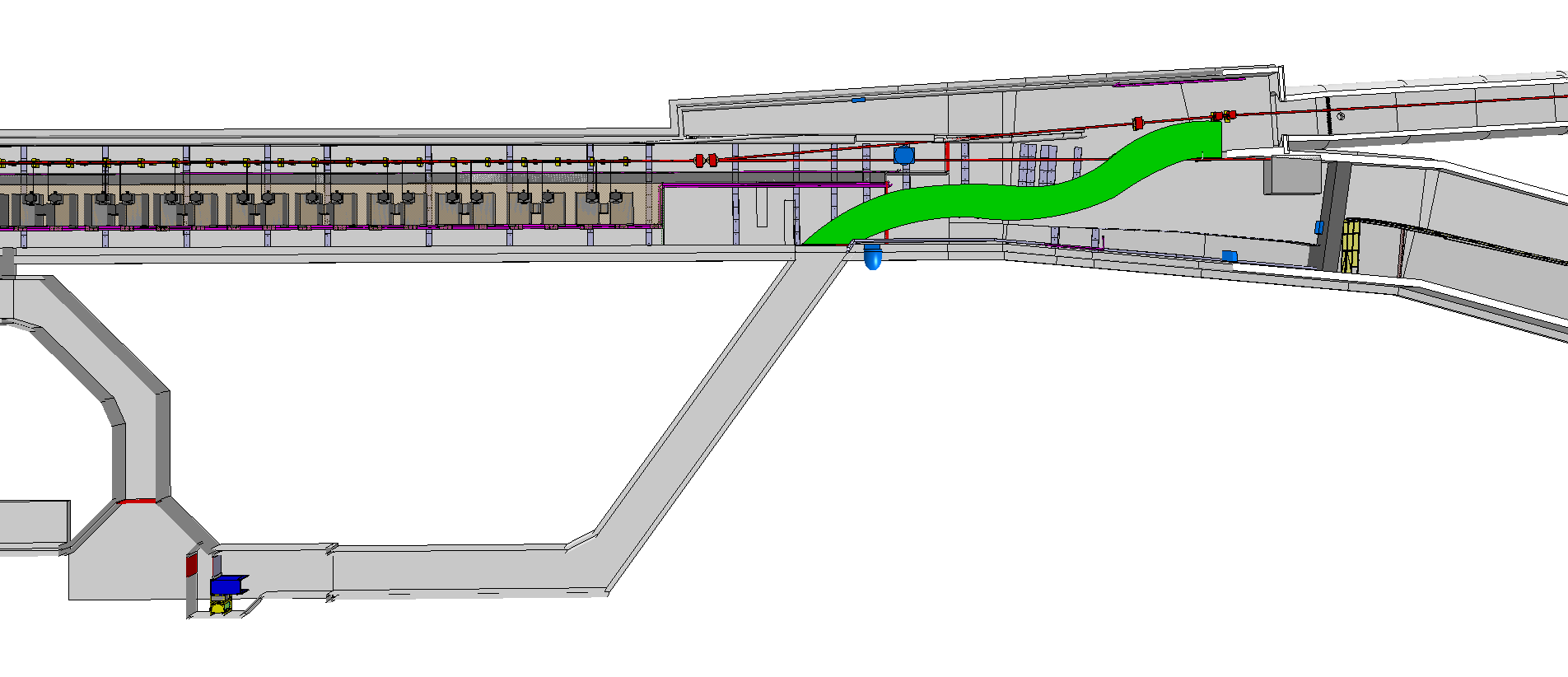}
\caption{3d view of the path to enter TT61 with handling equipment (green volume).}
\label{fig:PathToEnterTT61}
\end{figure}

To access TT61, equipment needs to be delivered through the access gallery TTA4 (804-U0-201, 804-U0-202). The transport volume in TT4 has been studied to allow the passage of the largest convoy for the installation of the magnets in TT61 and TCC6. This path, shown in Fig.~\ref{fig:PathToEnterTT61}, should remain free as it is the only path for the exchange of magnets for the transfer lines from SPS to TI2 (LHC injection) and from SPS to HiRadMat. For the installation of magnets in TT61 (heaviest magnet weight 1800~kg) a forklift equipped with a lifting jib with a capacity of 2\,t is required.
The transport volume design takes in consideration the largest required transport and handling equipment (special handling machine Dumont and tow tractors) for the maintenance of all transfer lines located in TCC6 (HiRadMat and TI2).

In the SPS accelerator, the following new components needs to be installed:
\begin{itemize}
    \item two kickers in the SPS LSS6;
    \item one electrostatic septum and one thin magnetic septum in the SPS LSS1.
\end{itemize}    
Those components will be installed with an existing custom-built forklift side loader (called PRATT) using the same procedure as for similar equipment installed in the SPS.

\subsubsection{Safety engineering}

\paragraph{Fire safety}
TT61, as a pre-existing tunnel without an active accelerator or experiment, has not yet had a new fire safety study carried out on it; a full fire safety study will, however, need to be to be made as the project moves into detailed design. The new installation in TT61 will also interface with ageing facilities: TCC6, T60 and BA7 and the associated infrastructure. The performance based fire safety design approach required for these areas shall have coherence with the SPS Fire Safety Study. It is not yet clear whether the implementation of the fire safety requirements for TT61 would fall under the scope of the eSPS project, or whether it would form part of a separate CERN consolidation project. The cost estimation made for this work (including the dismantling of the old fire safety installation) has, therefore, been de-scoped from the CDR cost estimation, but a study will need to be done, and an updated cost estimation made, at the detailed design stage.

\paragraph{Superconducting cavities}
The proposed installation of the superconducting cavities will use the pre-existing SPS crab cavities test stand in LSS6. This test stand was subject to checks by CERN's HSE Unit at its commissioning, and is, therefore, deemed to be compliant without any significant modifications. This shall be reviewed as the project moves into detailed design.

\subsubsection{Personnel protection system and access control}

According to the RP assessment, radiation levels from the high-energy electron beam in the TT4 experimental area and the upper part of the TT61 tunnel forbid the presence of personnel during beam operation. Thus, a new safety chain dedicated to the TT4 and the TT61 will be integrated and managed by the SPS access safety system.

\begin{figure}[!hbt]
\centering
\includegraphics[width=0.85\textwidth,trim={0.2cm 0.2cm 0.2cm 0.2cm},clip]{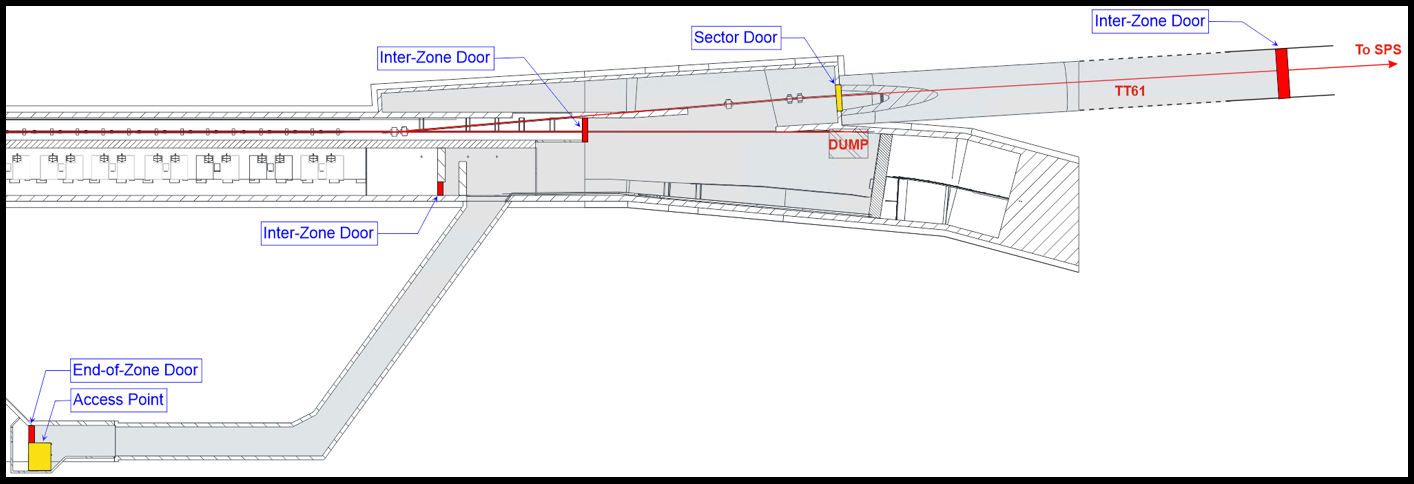}
\caption{Safety and access elements of the TT4 and TT61.}
\label{fig:SafetyandaccesselementsoftheTT4andTT61}
\end{figure}

To control the access to the TT4 and TT61 tunnels, the current TT61 access point, composed of a PAD and a MAD, will be moved at the entry of the gallery 852. A new “sector” door will be installed between the TT4 and TT61 tunnels. Moreover, three new “inter-zone” doors will be added to delimit these sectors as shown in Fig.~\ref{fig:SafetyandaccesselementsoftheTT4andTT61}.

\begin{figure}[!hbt]
\centering
\includegraphics[width=0.85\textwidth,trim={0.2cm 0.2cm 0.2cm 0.2cm},clip]{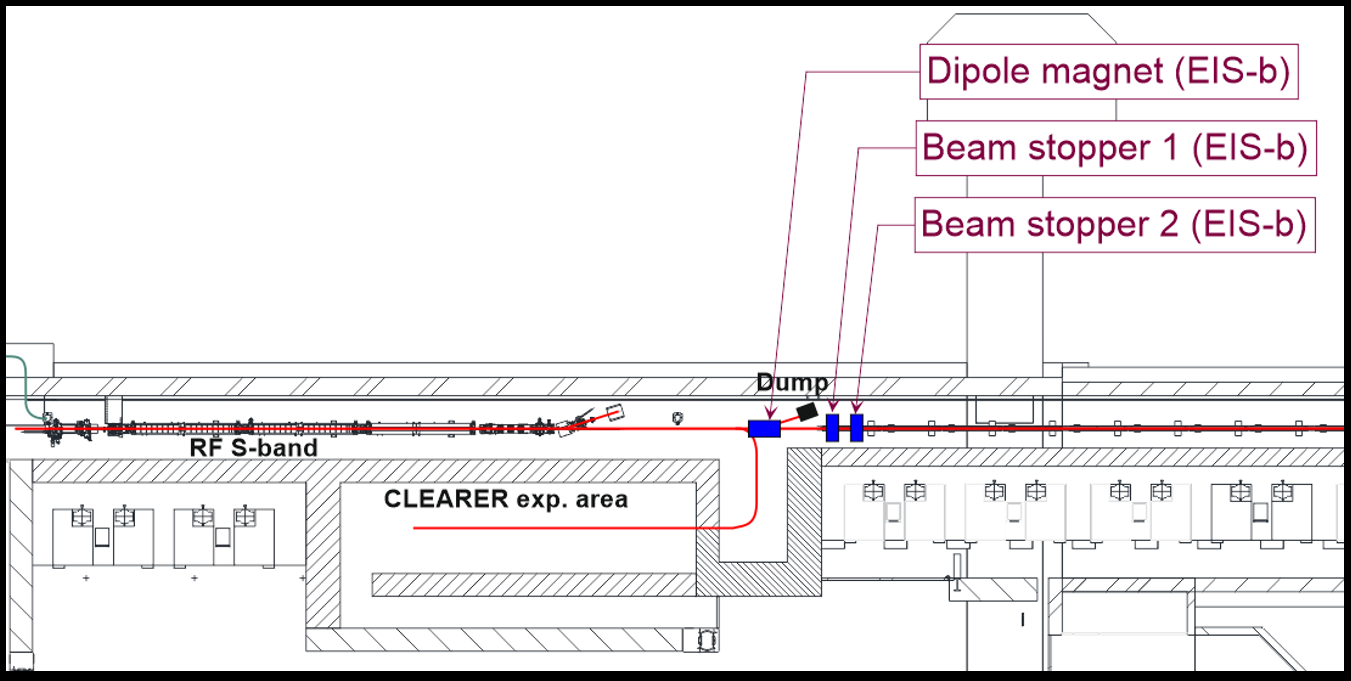}
\caption{Safety elements of TT4 and TT61.}
\label{fig:SafetyandaccesselementsoftheTT4andTT612}
\end{figure}

To protect people accessing the zone from radiation hazard, one bending magnet and two beam stoppers will be used as EIS-beam as shown in Fig.~\ref{fig:SafetyandaccesselementsoftheTT4andTT612}. The magnetic element being not fail-safe, it would be beneficial to review the EIS of this beamline in a future stage of the project.

Moreover, in case of intrusion during beam operation or if the safe position of a safety element is lost during access, the safety chain will act on the upstream zone: the electron linac.

The zoning of the current TT61 has to be adapted due to the removal of TT61 access point. The exact location of the “inter-zone” door inside TT61 will be defined according to RP calculations and depending on the potential beam loss scenarios in TCC6 from the LHC or HiRadMat beams and the radiation coming from electron beam losses in the TT4 linac.

\begin{figure}[!hbt]
\centering
\includegraphics[width=0.85\textwidth,trim={0.2cm 0.2cm 0.2cm 0.2cm},clip]{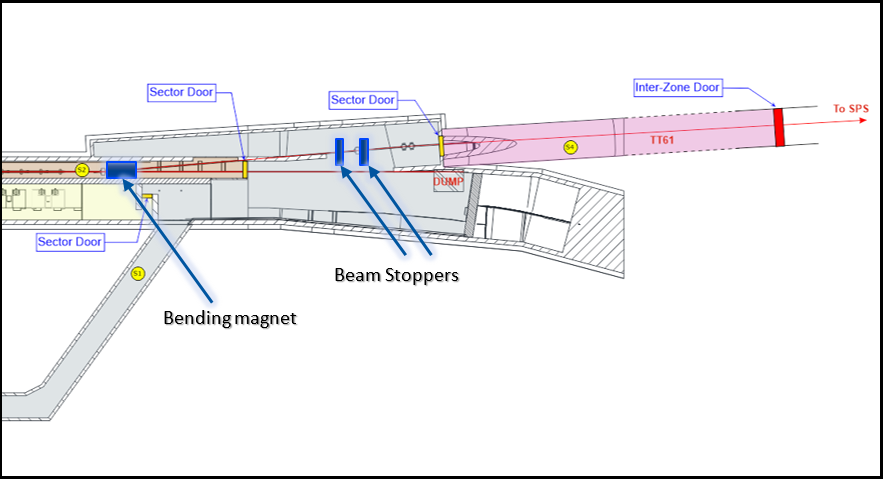}
\captionsetup{width=0.85\textwidth} 
\caption{Safety elements to avoid high energy electron beam injection into the SPS.}
\label{fig:SafetyelementstoavoidhighenergyelectronbeaminjectionintotheSPS}
\end{figure}

In order to protect people against radiation hazards coming from the injection of the electron beam from the eSPS injection into the SPS ring, new EIS-beam must added into the `West Extraction' and the `SPS Ring' interlock chains. To do that, it is foreseen to use one bending magnet and two beam stoppers, located just before the TT61 tunnel as shown in Fig.~\ref{fig:SafetyelementstoavoidhighenergyelectronbeaminjectionintotheSPS}.

%% file: include/05-ICE/AccelerationSPS.tex






%% file: include/05-ICE/ExtractionTT10_TT2.tex
\subsection{Extraction via TT10 and TT2}
\label{sec:ICE_Extraction}

\subsubsection{Civil engineering} \label{sec:ICE_Extraction_CE}

\begin{figure}[!hbt]
\begin{center}
    \includegraphics[width=12cm]{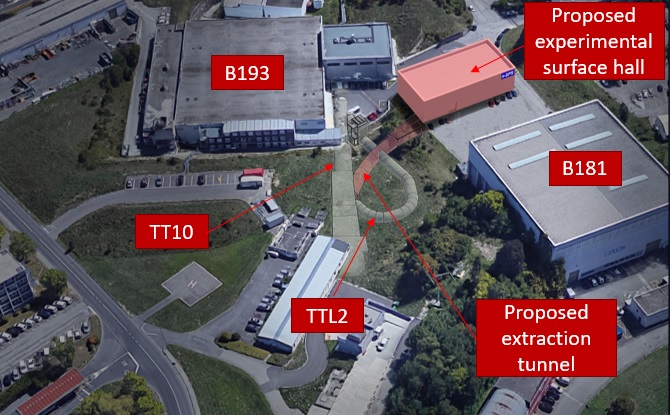}
    \caption{Birdseye visualisation of the proposed facilities in red showing surface and buried structures.}
\label{fig:3D}
 \end{center}
\end{figure}

A new extraction tunnel will be required to implement eSPS, located on the CERN Meyrin site between the AD building (B193) and the magnet assembly facility in B181 as shown in Fig.~\ref{fig:3D}. 

\paragraph{Existing site and infrastructure}

\begin{figure}[!hbt]
\begin{center}
\includegraphics[width=0.85\linewidth]{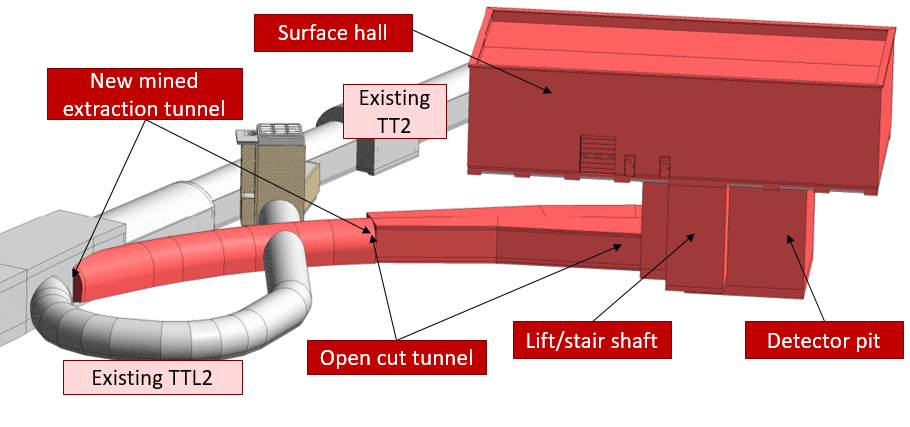}
    \caption{3D visualisation of the underground infrastructure showing the new infrastructure in red.}
 \label{fig:3DCEUndergroundlayout}
 \end{center}
\end{figure}

The site is currently characterised by a large earth mound around 7\,m above surrounding road levels with a covering of grass, shrubbery and trees. The proposed site is close to B181 and B193 (as well as its extensions and service Buildings: B93, B8854, B393) along with Tunnels TT2 and TT7. The site intersects the existing TTL2 tunnel, also known as 'The Ear' due to its distinctive shape as shown in Fig.~\ref{fig:3DCEUndergroundlayout}. 

The civil engineering works interface with the existing TT2 and TTL2 tunnels. To enable the concept design, a recent 3D scan of the area was used to update the models of the existing infrastructure to ensure it provides suitably accurate base for the design. 

The existing TT2 and TTL2 tunnels are both in good condition with no major defects. TT2 was constructed in 1969 as a transfer tunnel between the PS and the West area and has remained in use ever since. TTL2 was subsequently constructed in 1980 to allow the beam to be turned around in a loop and has been recently refurbished.

\begin{figure}[!hbt]
\begin{center}
\includegraphics[width=12cm]{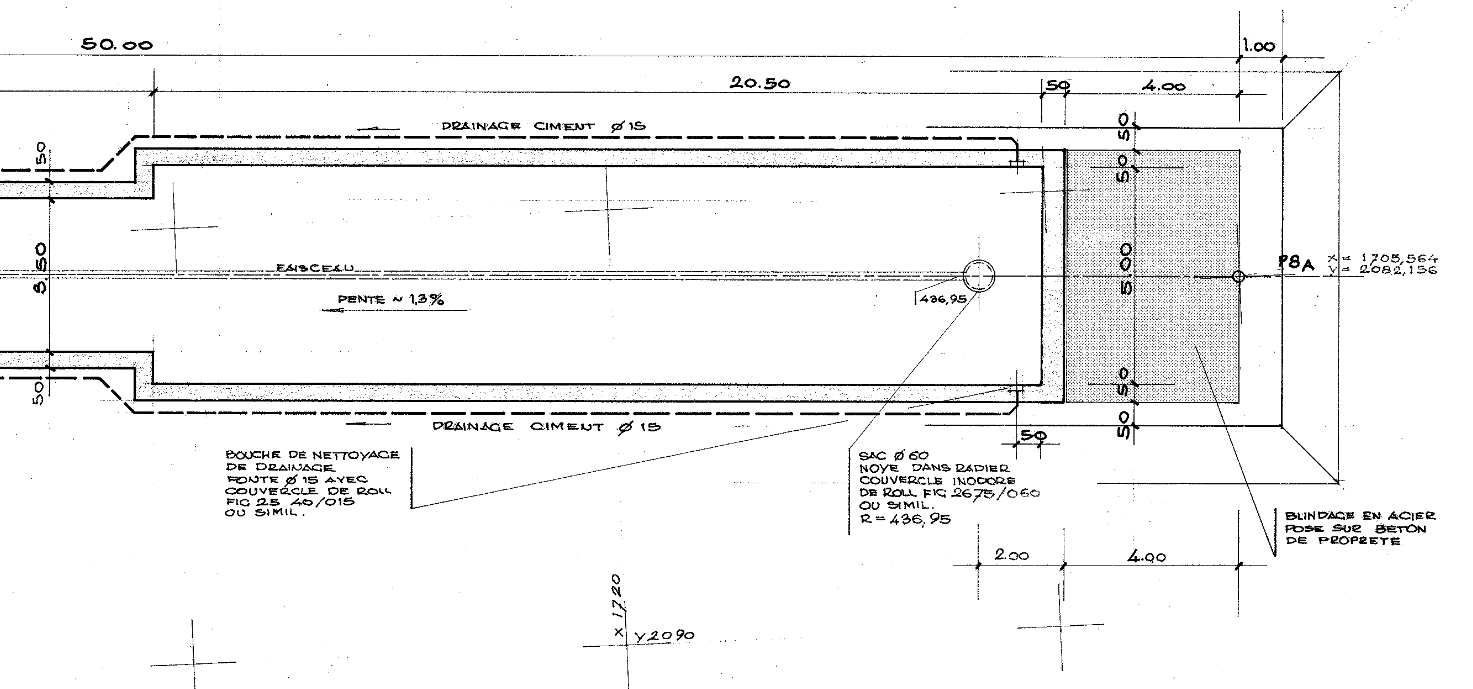}
    \caption{Plan view of TT7 and steel shielding block arrangement.}
\label{fig:TT7}
\end{center}
\end{figure}

The planned works are also very close to TT7 which formerly housed a beam dump with steel shielding immediately beyond the tunnel end as shown in Fig.~\ref{fig:TT7}. This has also been modelled to ensure it can be taken into account in the planning of works. 

\paragraph{Civil engineering layout}

This section covers the civil engineering required to enable the extraction from the SPS to a new experimental area. 

The new tunnel, approximately 55\,m long will form a link to deliver beam between the existing TT2 tunnel and the new detector underground hall as shown in Fig.~\ref{fig:3DCEUndergroundlayout}. 

A beamline core will be drilled through the wall of the existing TT2 tunnel, allowing the beam to pass within a vacuum tube. 

\begin{figure}[!hbt]
\begin{center}
\includegraphics[width=0.75\linewidth]{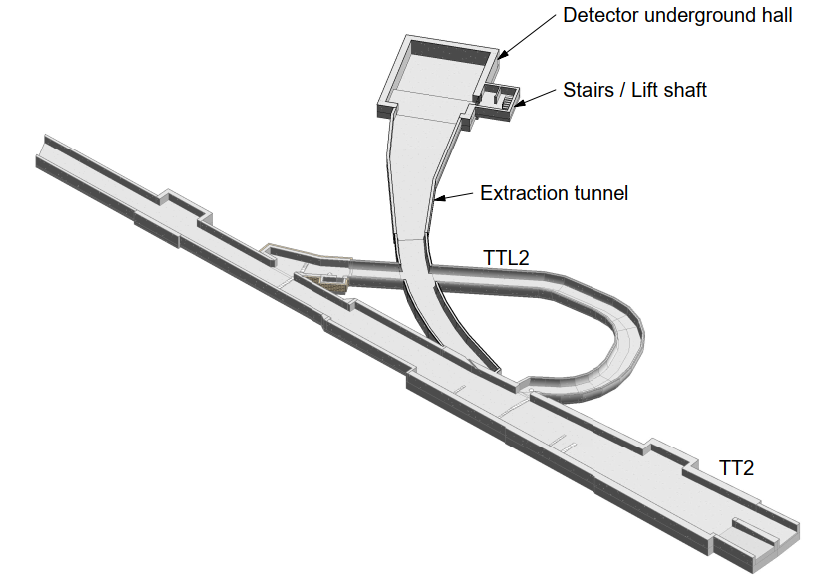}
    \caption{3D visualisation cut at tunnel invert level showing new extraction tunnel layout in relation to existing infrastructure.}
 \label{fig:Newtunnel}
 \end{center}
\end{figure}

Connections will be formed (as shown in Fig.~\ref{fig:Newtunnel}) with the existing TTL2 tunnel.

\begin{figure}[!hbt]
\begin{center}
\includegraphics[width=0.75\linewidth]{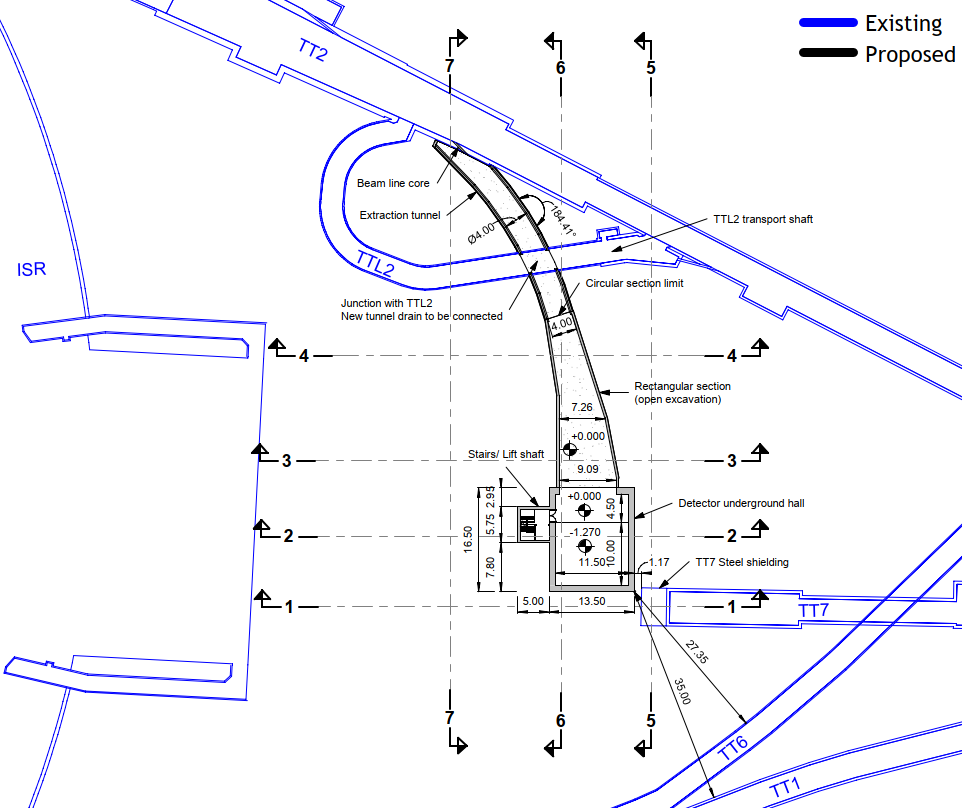}
    \caption{Plan view showing geometry of proposed CE layout and existing infrastructure.}
 \label{fig:Planview2}
 \end{center}
\end{figure}

The tunnel has been designed at a constant 3.57\,m wide internal cross section over the first 36\,m followed by a tapered section of 19\,m length widened to 9.09\,m at the connection with the building to allow beam delivery to two separate experiments as shown in Fig.~\ref{fig:Planview2}. The tunnel is effectively flat, with no slope over the full length to accommodate beamline requirements. 

\begin{figure}[!hbt]
\begin{center}
    \includegraphics[width=0.8\linewidth]{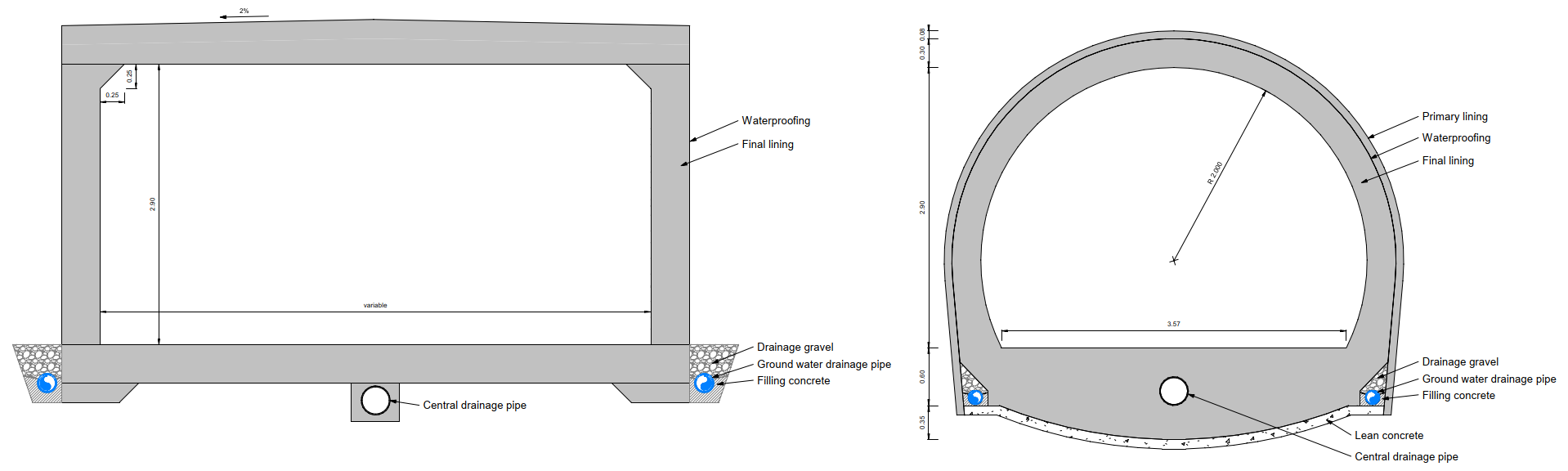}
    \caption{Typical Cross-sections of the tunnel showing the the cut and cover section (left) and the mined section (right).}
 \label{fig:Typsections}
 \end{center}
\end{figure}

One section will consist of a horseshoe cross-section  while the 'cut and cover' section will be rectangular, both 3.75\,m wide an the minimum as shown in Fig.~\ref{fig:Typsections}. 

The horseshoe shaped mined tunnel section will be constructed using a sprayed concrete primary lining around the arch incorporating rock bolts followed by a cast in situ mass concrete secondary lining. In this case, the curved invert slab is 600\,mm thick at the minimum and 950\,mm at the maximum cast on a 100\,mm blinding of lean-mix concrete. Drains would be included at the base of the arch with free draining material surrounding a perforated drainage pipe located on a concrete plinth. 

The cut and cover rectangular tunnel section will be reinforced cast in situ concrete comprising an invert slab 300\,mm thick, walls 400\,mm thick and a roof slab varying from 300\,mm at the edge to 350\,mm deep at the centre. The floor slab is haunch thickened to 450\,mm deep directly below walls and to accommodate shear forces efficiently, the joints between walls and roof slab have additional 250\,mm by 250\,mm haunches. 

Along the full length of the tunnel, a tunnel waterproofing system will be provided to prevent water ingress. 

\begin{figure}[!hbt]
\begin{center}
    \includegraphics[width=0.9\linewidth]{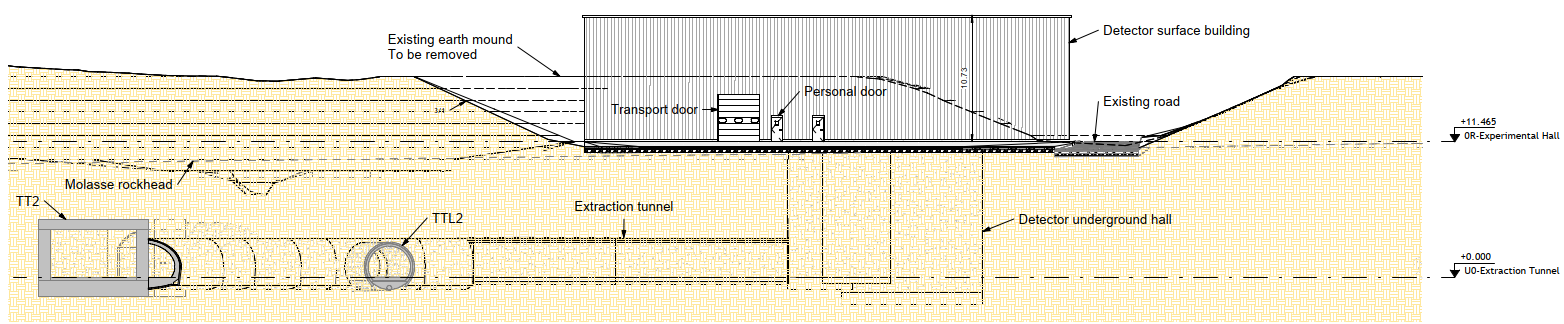}
    \caption{Section 7-7 showing extraction tunnel abutting TT2, junction with TTL2 and flat profile through to its connection with the detector underground hall.}
 \label{fig:Tunnellayout}
 \end{center}
\end{figure}

A drain will run along the centre of the tunnel, with a fall of at least 1\% to ensure water flows without creating blockages due to sediment build-up. The fall will be artificially created in the drain alone, with the tunnel invert remaining horizontal as shown in Fig.~\ref{fig:Tunnellayout}. The drain will connect into the existing tunnel drain in TTL2. Therefore, the section of extraction tunnel between TT2 and TTL2 will drain to the south while the remainder of the extraction tunnel will drain to the north.

The tunnel width will allow sufficient space to accommodate the machine, services and suitable space reservations for safe transport and maintenance. 

\paragraph{Civil engineering design considerations}

Tunnel cross-sections have been based on recent experiences and designs used for the High-luminosity LHC upgrade which is located relatively nearby at the LHC Point 1. The cross section design will need to be reviewed following detailed ground investigation and adjusted in the case ground conditions vary significantly from expectation.  

The new extraction tunnel will be constructed to abut TT2 without forming a junction between tunnels. The exact structural form here will need to be determined following assessment of the structural capacity of TT2. The most likely options are either a free standing tunnel with significant reinforcement in the tunnel lining to avoid imposing any additional loads on TT2; or some form of heavy duty structural connection between the crown of the new tunnel and TT2 to allow some transfer of loads, supporting the new tunnel. 

Connections between TTL2 and the new extraction tunnel are necessary for several reasons :

\begin{itemize}
 \item To facilitate connections with electrical services;
      \item To allow ventilation of the tunnels and avoid any 'dead leg' area without sufficient air movement;
  \item For transportation of some large magnets via TTL2 transport shaft, the operation of the ear needs to be preserved; 
\item Although not a requirement, in line with best practice, this also allows an alternative means of emergency access and egress via the TTL2 transport shaft. 
        
\end{itemize}

The tunnel will be constructed entirely within the molasse. To enable the CE tunnelling works, significant earthworks must first be carried out to clear the site down to a construction platform level. This has been set at a suitable level to allow the finished works to tie into the surrounding road levels to enable access to the site. 

\begin{figure}[!hbt]
\begin{center}
\includegraphics[width=0.55\linewidth]{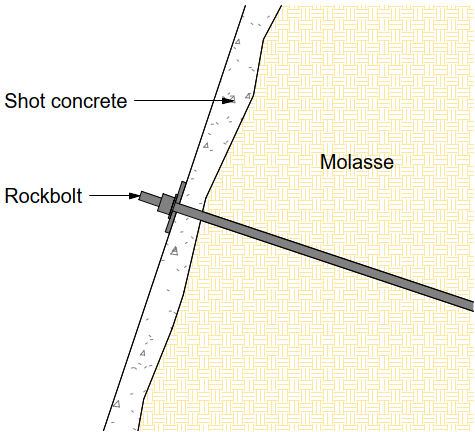}
    \caption{Indicative molasse face during excavation.}
 \label{fig:indicative detail}
 \end{center}
\end{figure}

\begin{figure}[!hbt]
\begin{center}
\includegraphics[width=0.95\linewidth]{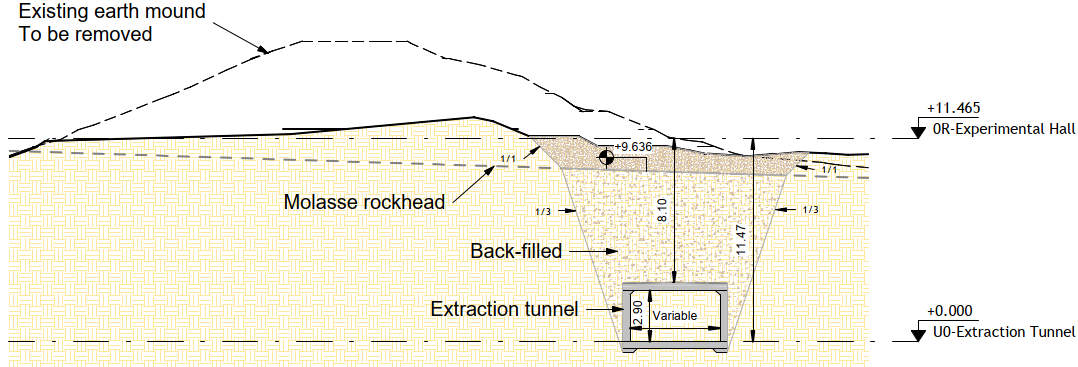}
    \caption{Section 4-4 through cut and cover tunnel showing likely excavation profile and backfill.}
 \label{fig:Cutandcover}
 \end{center}
\end{figure}

The section excavated as part of the cut and cover tunnel will be back-filled with excavated material to reduce the amount of spoil which must be taken off site as shown in Fig.~\ref{fig:Cutandcover}. During excavation in the molasse, both for the tunnel and experimental hall, rock bolts and shotcrete sprayed concrete will be used to provide temporary excavation face support to maximise the angle of excavation and minimise the volume of material. An indicative detail is shown in Fig.~\ref{fig:indicative detail}.

\subsubsection{Integration}
\label{sec:ICE:extract:integration}

\paragraph{Transfer tunnel TT10}

Transfer tunnel TT10 is an 800\,m long tunnel that houses the TT10 beamline that runs from transfer tunnel TT2 to the SPS. The eSPS beamline is extracted from the SPS via TT10. There are no integration changes in the tunnel for the eSPS project. 

\begin{figure}[!hbt]
\begin{center}
    \includegraphics[width=0.9\linewidth]{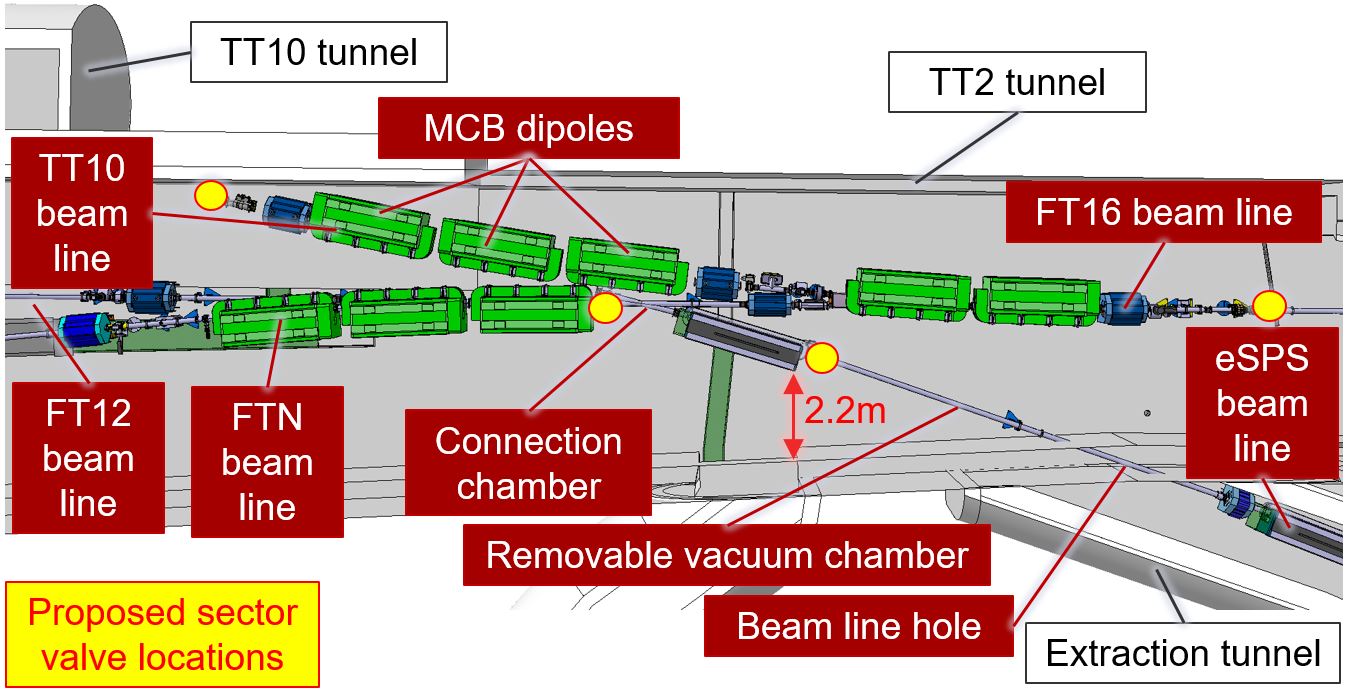}
    \caption{Integration layout of TT2.}
 \label{fig:Integration layout of TT2}
\end{center}
\end{figure}

\paragraph{Transfer tunnel TT2}
Transfer tunnel TT2 is a 400\,m long tunnel that houses the FT16, FTA, TT10, FTN and FT12 beamlines. The eSPS beamline crosses through the tunnel from TT10 into the new extraction tunnel at an angle of approximately \SI{20}{\degree} as shown in Fig.~\ref{fig:Integration layout of TT2}.

In TT2, approximately 10.5\,m of the TT10 beamline will be modified of which includes removing some beamline equipment and replacing it with three MCB dipoles. An additional MCW dipole will be installed in the TT2 transport zone for which the distance between the wall and the magnet is 2.2\,m, allowing enough space for transport vehicles to pass by. The removable vacuum chamber will pass from the MCW dipole into the extraction tunnel via a hole in the TT2 wall. The vacuum system will be modified such that a new four-way separation chamber will be installed as well as additional sector valves.

The MCB and MCW dipoles will be installed via the TT2 access shaft. This shaft is currently used to install the existing MCB magnets. As the MCW dipoles are longer, a transport integration study was undertaken to ensure that the magnets could be installed (see Fig.~\ref{fig:simOfMagnetLowering}).

\paragraph{Extraction tunnel}

\begin{figure}[!hbt]
\begin{center}
    \begin{subfigure}{0.49\linewidth}
        \begin{center}
           \includegraphics[width=\textwidth]{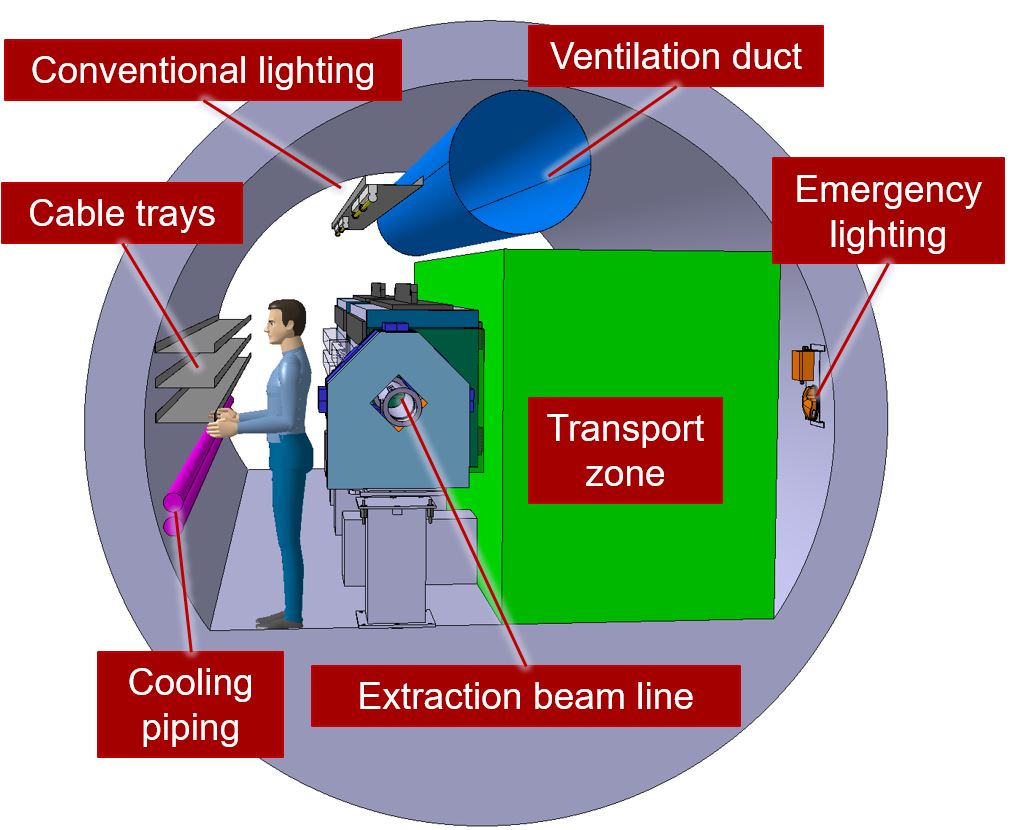}
           \caption{}
        \label{fig:Integration layout of the extraction tunnel}
        \end{center}
    \end{subfigure}
   \begin{subfigure}{0.49\linewidth}
        \begin{center}
            \includegraphics[width=\textwidth]{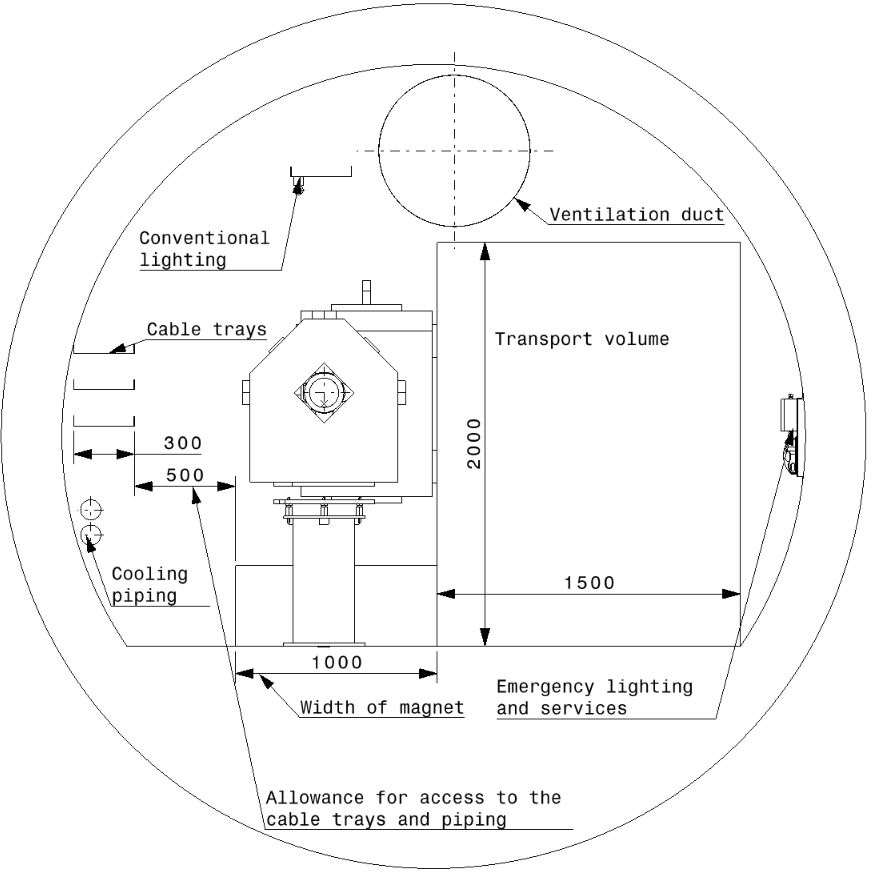}
            \caption{}
        \label{fig:Cross-section of the extraction tunnel}
        \end{center}
    \end{subfigure}
    \caption{3D layout \subref{fig:Integration layout of the extraction tunnel} and cross-section drawing \subref{fig:Cross-section of the extraction tunnel} of the extraction tunnel.}
\end{center}
\end{figure}

The extraction tunnel (shown in Fig.~\ref{fig:Integration layout of the extraction tunnel}) is a new tunnel that houses the eSPS beamline and its corresponding services. It allows the beamline to deviate away from the existing beamlines in TT2 to the experimental hall. The beamline services include cooling and ventilation, cable trays, and lighting, among other things. 

During technical stops and long shutdowns of the beamline, personnel and transport vehicles may access the extraction tunnel for maintenance purposes, using the transport access way located on the right-hand side of the tunnel (with respect to the beam direction). Personnel and transport access into the extraction tunnel is via the experimental hall connected to the downstream of the extraction tunnel. 

The tunnel has a length of approximately 60 m and is shaped with an internal height of about 3 m and a width of about 4\,m. The width and height of the tunnel are based on the dimensions required for equipment and personnel/transport access, as shown in Fig.~\ref{fig:Cross-section of the extraction tunnel}:

\begin{itemize}

\item \SI{1500}{mm} allowance for transport vehicles (height of 2000\,mm);
\item \SI{1000}{mm} width for the magnets;
\item \SI{500}{mm} allowance for personnel access for maintenance;
\item \SI{300}{mm} allowance for cable trays;
\item \SI{50}{mm} allowance for cabling and the cable tray support structure.

\end{itemize}



\subsubsection{Radiation protection}

Analogue to the injection of protons from TT2 via TT10 into the SPS, the extraction of electrons in the opposite direction requires the transfer tunnels to be in a safe state. Extraction must be allowed only if the receiving side, the experimental cavern, is closed and ready to receive beam. The safety elements allowing access to the experimental cavern must be located sufficiently far away in TT2 to limit an accidental exposure of persons in the experimental cavern in case of a faulty ejection from the SPS.

The failure modes of extracting the electron beam accidentally via TT2 towards the PS must be studied and, if required, mitigation measures must be taken to prevent such an event. The location of the inter-machine door in TT10 must be confirmed to be compatible with electrons circulating and dumped in the SPS, as access is possible from the TT2 side, while the SPS is operating with electrons.

The impact on the n\_TOF target area (TT2A) in case of exceptional beam losses in the upper part of TT10 or in TT2 must be studied. However, considering the configuration, beam directions and installed shielding, it is not expected that the n\_TOF target area will be impacted by the electron beam extraction through TT10.

\subsubsection{Transport and handling}
\label{sec:ICE:extract:transport}

\paragraph{Transfer line TT10}

\begin{figure}[!htb]
\centering
\includegraphics[width=0.6\textwidth]{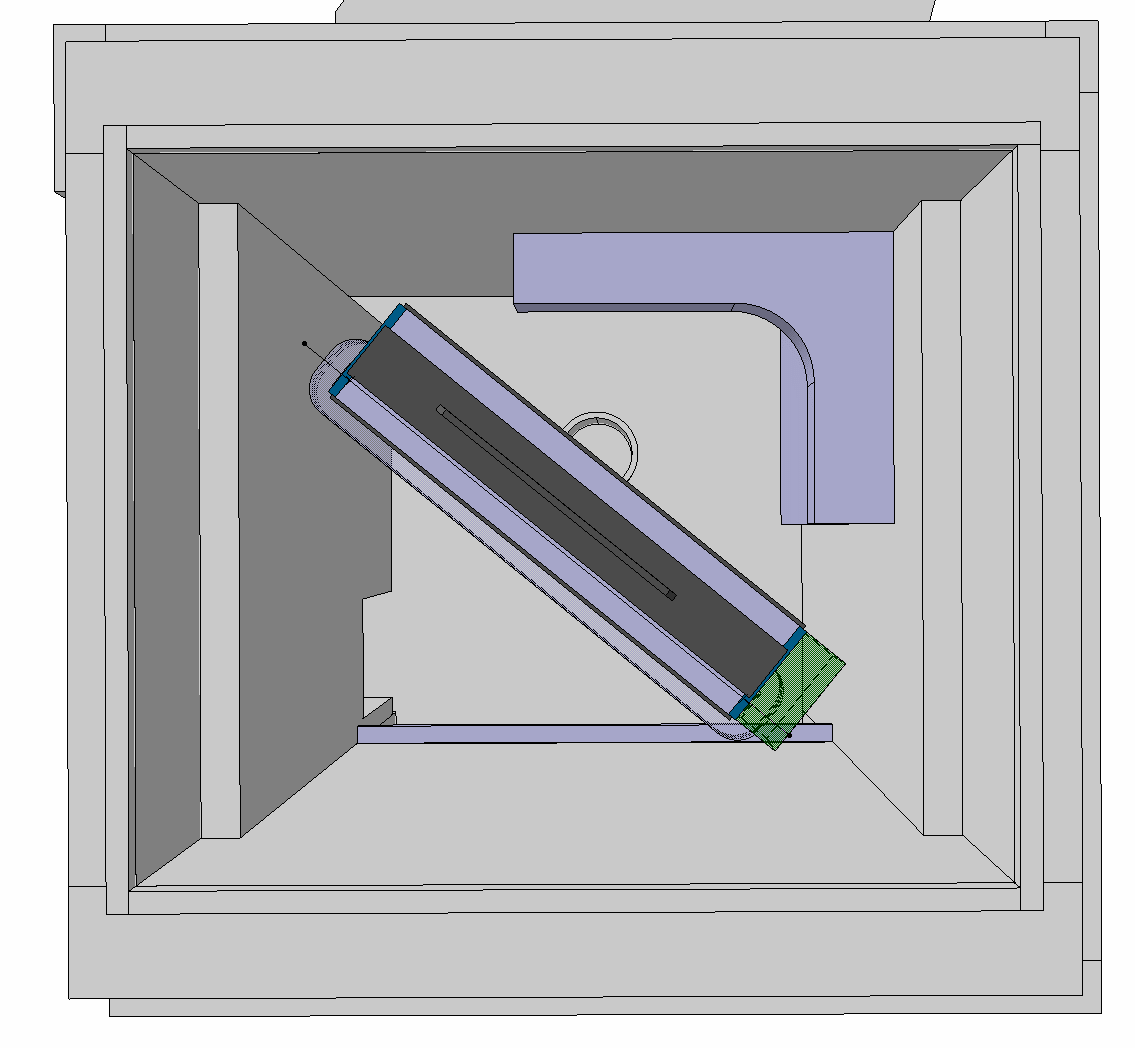}
\caption{3D simulation of one of the new magnets being lowered in the shaft.}
\label{fig:simOfMagnetLowering}
\end{figure}

At the junction region between TT10 and TT2 two magnets need to be removed and four new magnets installed.  All of these magnets enter the tunnel with the help of a mobile crane operating via the shaft located next to B193, a simulation of this procedure is shown in Fig.~\ref{fig:simOfMagnetLowering}.  The installation of the magnets on the beamline will be done with existing motorised bogies (Tortue).

\paragraph{Transfer line TT2 experimental area}
The magnets installed in this new tunnel will enter via the experimental hall by means of the existing EOT crane with 25\,t SWL of the building.
The installation of the magnets on the beamline will either be done using existing forklifts with lifting jibs or with existing motorised bogies (Kouba).

\subsubsection{Personnel protection system and access control}

\begin{figure}[!hbt]
\centering
\includegraphics[width=0.9\textwidth]{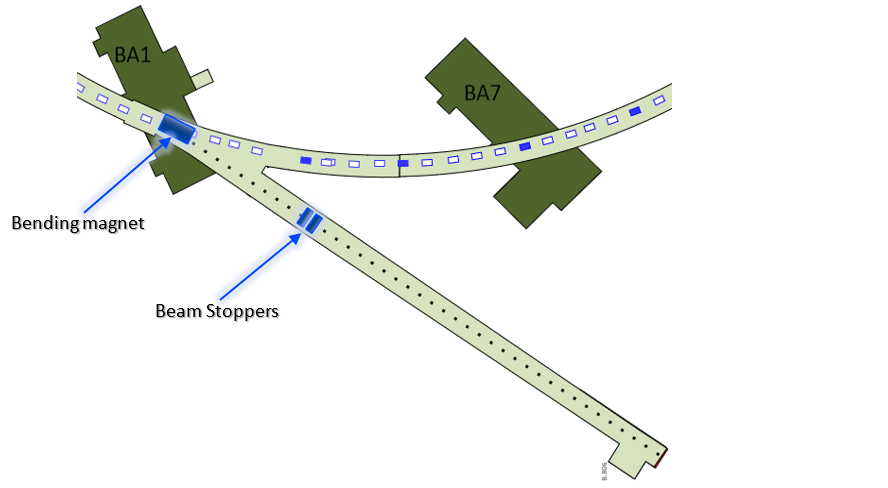}
\caption{Safety elements to protect TT2 from electron beam extraction via TT10.}
\label{fig:SafetyelementstoprotectTT2fromelectronbeamextractionviaTT10}
\end{figure}

To allow the extraction of electron beams from SPS ring via TT10 when the TT2 chain is ready for beam, one bending magnet and two beam stoppers must be included as safety elements (EIS-b) on the TT2 safety chain, as shown in Fig.~\ref{fig:SafetyelementstoprotectTT2fromelectronbeamextractionviaTT10}.

\begin{figure}[!hbt]
\centering
\includegraphics[width=0.9\textwidth,trim={0.2cm 0.2cm 0.2cm 0.2cm},clip]{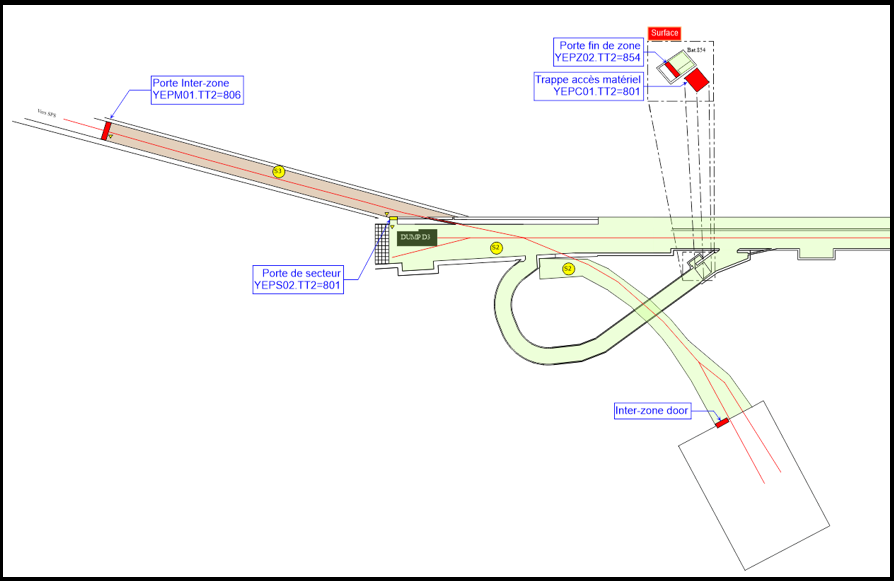}
\caption{Sector S2 with transfer tunnel to the eSPS exp. area included.}
\label{fig:Access4}
\end{figure}

In addition, the new transfer tunnel to the eSPS experimental area must be included in the sector S2 of the TT2 personnel protection system. An "inter-zone" door will be installed between this transfer tunnel and the eSPS experimental area to close this sector, as shown in Fig.~\ref{fig:Access4}.

%% file: include/05-ICE/ExperimentalArea.tex
\subsection{Experimental area}
\label{sec:ICE_ExperimentalArea}

\subsubsection{Civil engineering} \label{sec:ICE_CE}

The CE study has covered the creation of the space required for a detector underground hall, surface building and associated drainage, access and parking which are detailed here.

Following on from the optioneering exercise carried out at the feasibility stage of development, the chosen option has been further developed to create a full concept design. 

The study is based on that the experimental area should host two experiments. While one of them, the missing momentum experiment, is approximately known~\cite{whitepaper} with the physics potential described in Section~\ref{sec:PhysicsGoals}, not much information is available at this stage about the design of a second possible experiment. To model the most demanding scenario, this is assumed to be a beam dump experiment briefly discussed in Ref.~\cite{Torstenakesson2018} Section~III.B.

\paragraph{Existing site and infrastructure}

\begin{figure}[!hbt]
\begin{center}
    \includegraphics[width=\linewidth]{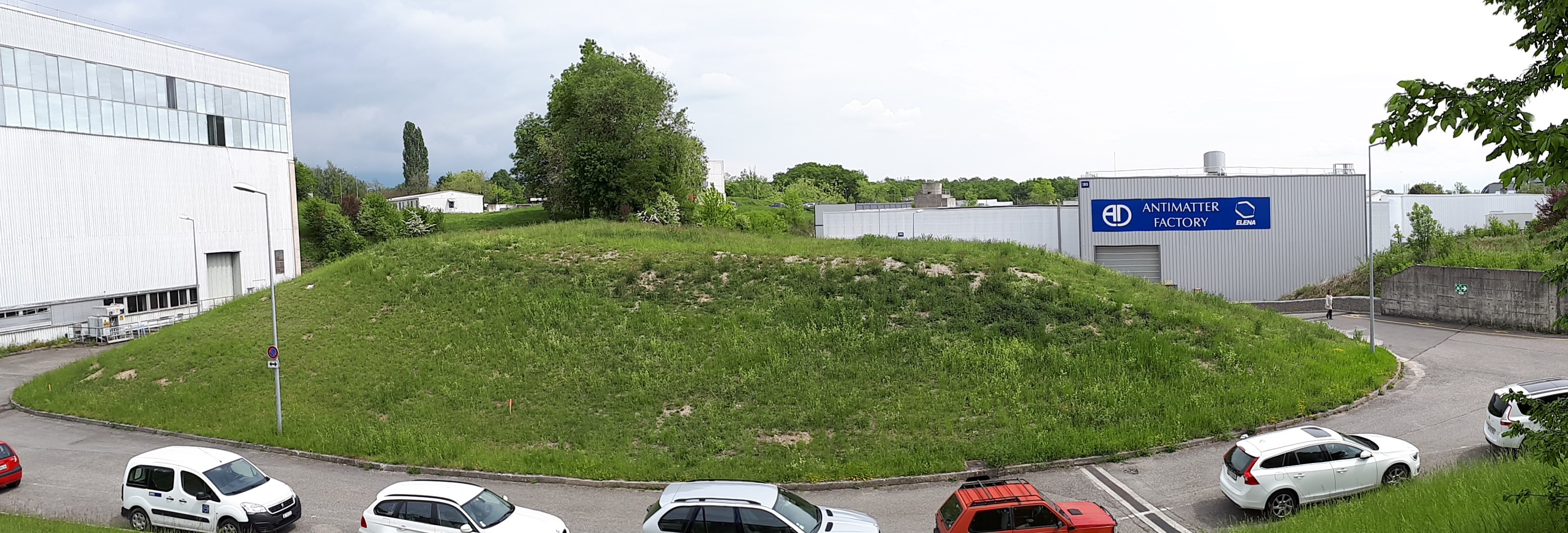}
    \caption{View of the proposed site looking north towards the existing earth mound.}
\label{fig:Mound}
 \end{center}
\end{figure}

The existing site is effectively the same as the extraction tunnel. The site is again close to the existing buildings described in Section~\ref{sec:ICE_Extraction_CE}. When discussing the experimental hall specifically, the site is dominated by the large mound shown in Fig.~\ref{fig:Mound} which is believed to be spoil from the construction of B181 and the ISR. The composition of the mound is not currently known and will need to be confirmed. This mound is also known to be home to a number of orchid species which are protected.

\paragraph{Civil engineering layout}

The proposed layout consists of a surface hall, underground detector hall and associated drainage, parking and access. 

All infrastructure has been designed and sized based on discussions with eSPS and the missing momentum experiment (LDMX) project leaders. A full 3D-model has been created in conjunction with CERN's integration team.  Space reservations have been  estimated and agreed between parties where details were not known. The layout of the building has been developed to accommodate sufficient space for installation, assembly, operation and maintenance of the two experiments and incoming beam line equipment.

\begin{figure}[!hbt]
\begin{center}
    \includegraphics[width=14cm]{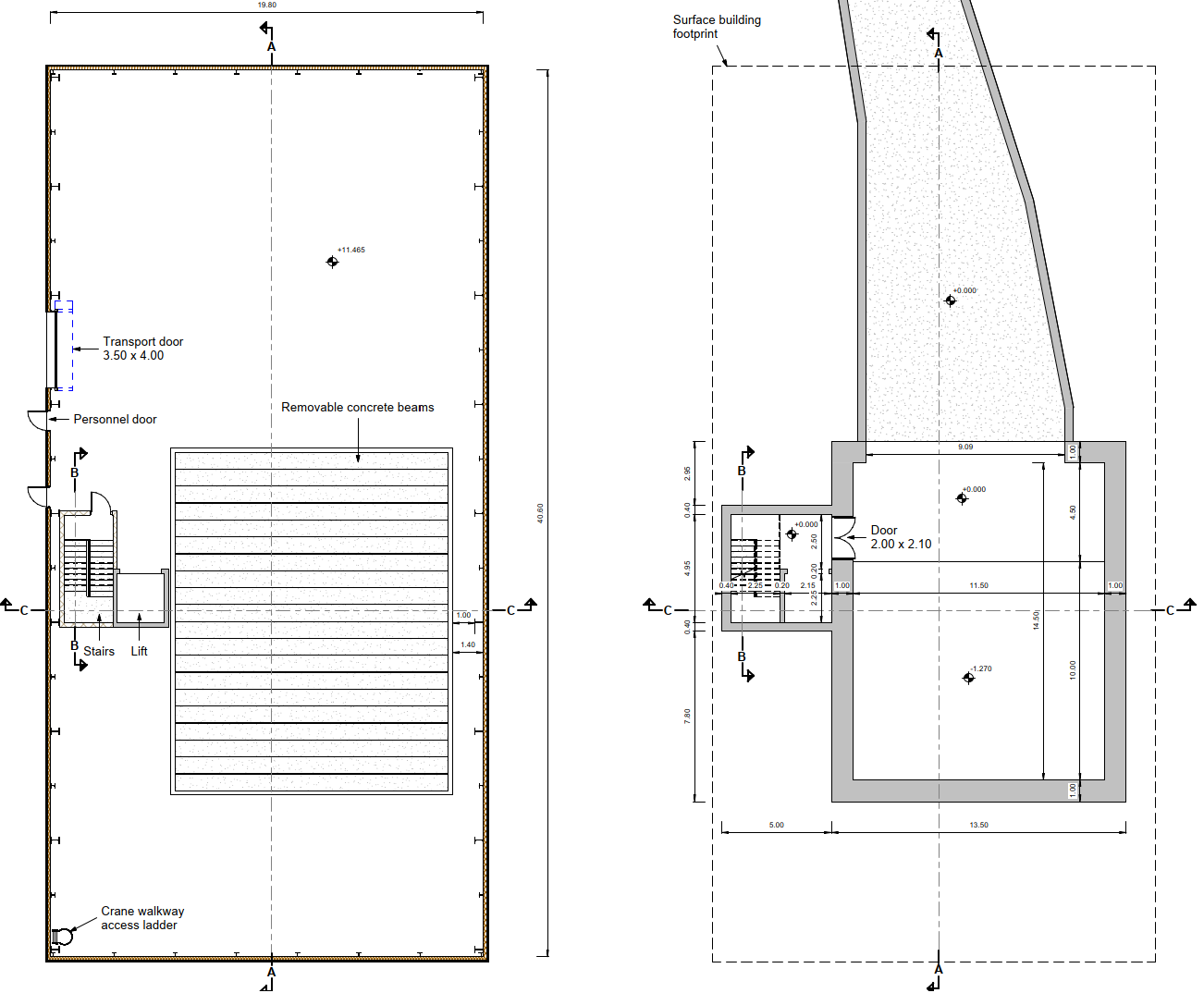}
    \caption{Layout plan of the experimental building at surface hall level (left) and detector underground hall level (right).}
\label{fig:Planlayout}
 \end{center}
\end{figure}

\begin{figure}[!hbt]
\begin{center}
    \includegraphics[width=\linewidth]{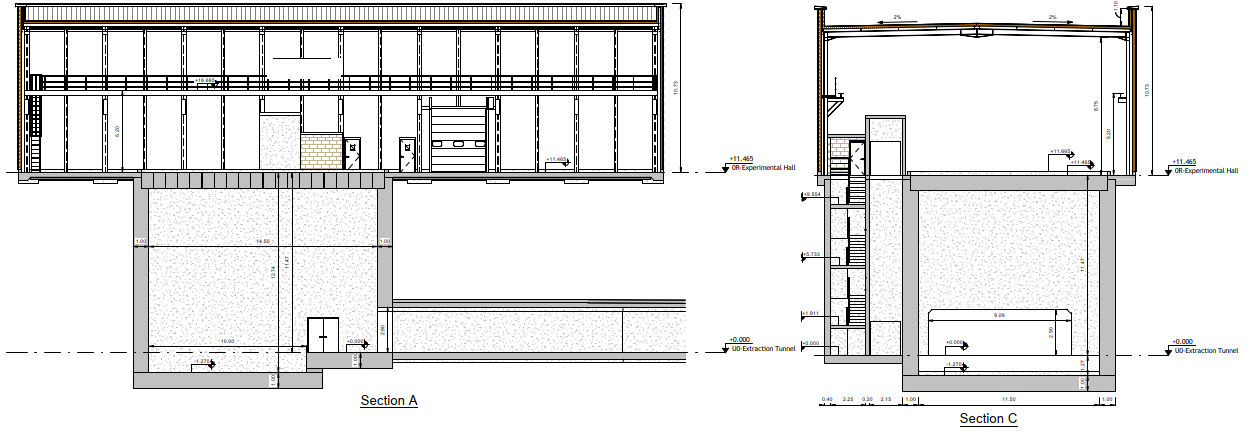}
    \caption{Sections through experimental surface building, detector pit and tunnel along beam axis (left) and transverse to it (right).}
\label{fig:EXPHallSections}
 \end{center}
\end{figure}

The underground detector hall  dimensions are 11.5\,m wide by 14.5\,m along the beam axis (internally), 12.74\,m tall at the maximum pit depth but 11.47\,m to the extraction beam tunnel invert level with the arrangement illustrated in Fig.~\ref{fig:EXPHallSections}. Adjacent to the hall is a shaft housing the stairs and lift to access the underground area. The shaft measures 4.6\,m by 4.95\,m (internally) and extends from the upper underground detector hall floor level to the surface hall as shown in Fig.~\ref{fig:Planlayout}. The functional layout of the surface hall is discussed in detail in Section~\ref{sec_ICE_ExperimentalArea_Integration} so is not covered here.  

The underground detector hall is formed from 1\,m thick reinforced concrete walls with a 1\,m floor slab cast onto a concrete blinding. A full width opening is provided to accommodate access to the extraction tunnel. The upper portion of the floor slab is at the same level as the extraction tunnel invert while there is a lowered section which is a further 1.27\,m deep to allow for the depth of the detectors, associated infrastructure and support structures. 

\begin{figure}[!hbt]
\begin{center}
    \includegraphics[width=6cm]{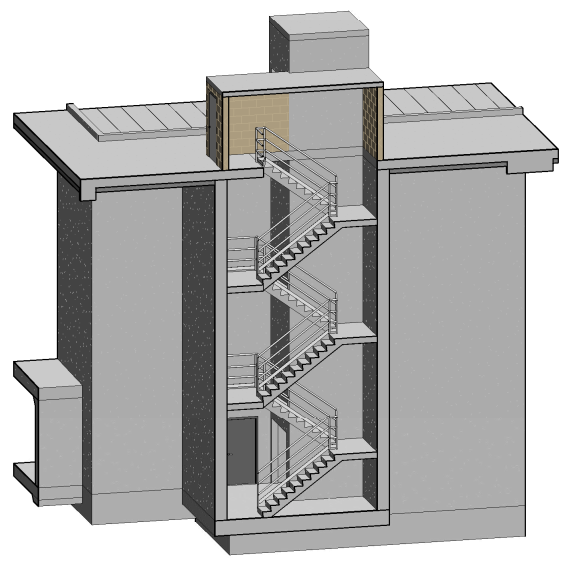}
    \caption{3D cut-away view of stair and lift shaft.}
\label{fig:3DviewStairsandliftshaft}
 \end{center}
\end{figure}

The lift shaft provisionally comprises 0.4\,m thick reinforced concrete walls with an internal dividing wall further separating the lift shaft as shown in Fig.~\ref{fig:3DviewStairsandliftshaft}. Standard dimensions are used for stair treads, risers and landings, conforming to safety requirements. The lift shaft is designed based on geometry to accommodate lifts used elsewhere around the CERN site for consistency. 

\begin{figure}[!hbt]
\begin{center}
    \includegraphics[width=\linewidth]{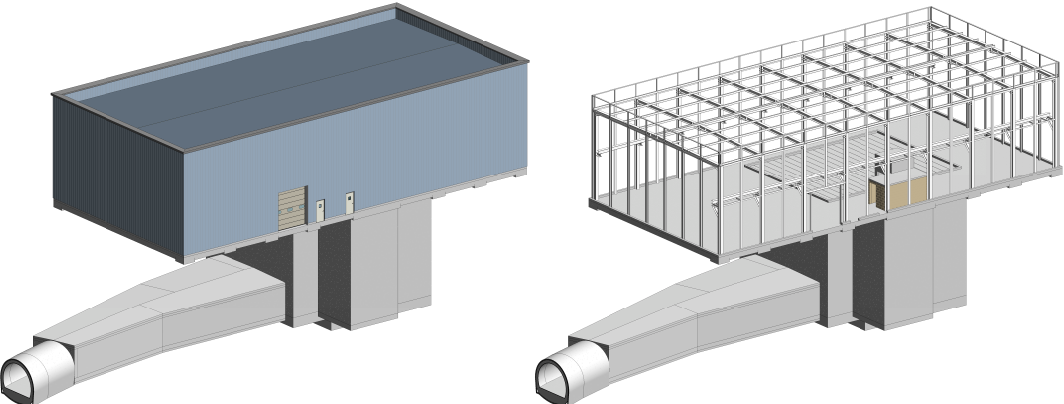}
    \caption{3D renderings of the experimental area surface building showing isometric view with cladding and access doors (left) and cut away view with underlying steelwork structure (right).}
\label{fig:3DviewEXPHALL}
 \end{center}
\end{figure}

The building above ground will be constructed as a steel portal frame structure with exterior cladding for insulation and weather-tightness. The Fig.~\ref{fig:3DviewEXPHALL} shows views of the building with and without cladding to reveal the preliminary  steel frame design. 

\begin{figure}[!hbt]
\begin{center}
    \includegraphics[width=0.8\linewidth]{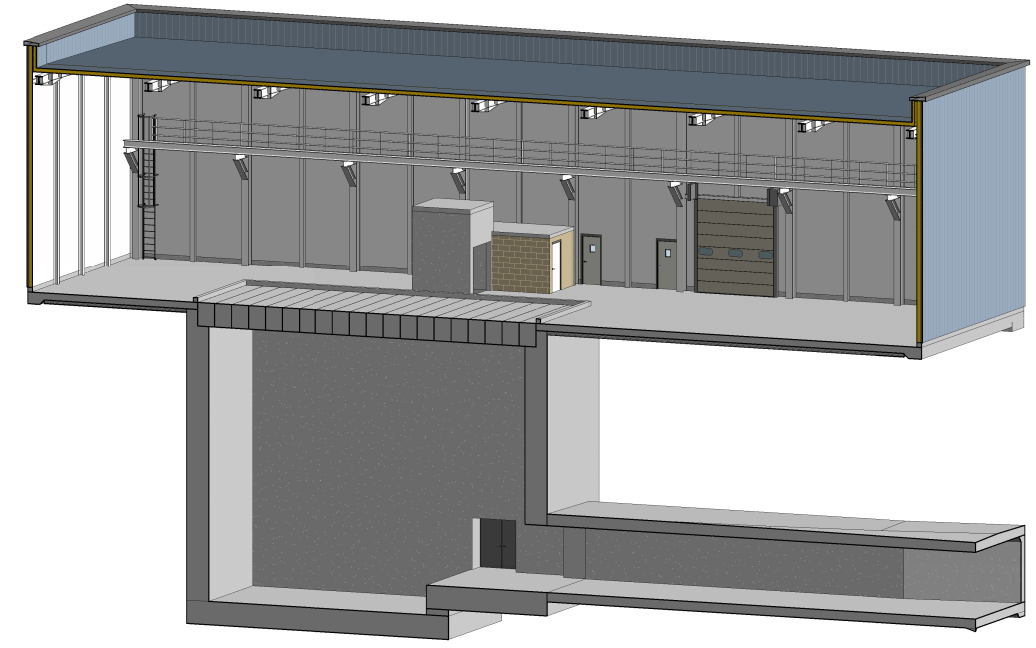}
    \caption{3D section view through experimental surface hall, detector pit and tunnel.}
\label{fig:3Dviewthroughmainbuildingandtunnel}
 \end{center}
\end{figure}

The equipment shaft used for transport access to the underground area is covered with 1\,m thick pre-stressed, pre-cast reinforced concrete beams, which span the opening to provide radiation shielding during beam operation. Beams are supported on a recessed plinth 0.5\,m wide on either side of the opening. Beams therefore sit at floor slab level when in position. When removed for access the building layout allows sufficient space to stack the beams to the side of the shaft opening. A removable socketed 1.1\,m tall guardrail will be installed around the opening for safety while the shaft is open. The beams are handled by the 25\,t crane within the building which is also used for transport of heavy equipment within the surface hall and underground areas. Figure~\ref{fig:3Dviewthroughmainbuildingandtunnel} shows the arrangement of shaft, beams and crane support structure.

The northern most portion of the hall floor will have an additional 400\,mm high false floor to allow for distribution of services. A technical gallery will be constructed just below slab level to carry services from the electrical and CV equipment on the false floor to the underground area via the stair and lift shaft. The technical gallery will be recessed in the floor slab: 1\,m deep, 2\,m wide with removable concrete  slabs, 400\,mm thick as covers during normal operation. 

An additional technical gallery to bring services to the hall may be required but at this stage, nothing has been confirmed. An allowance is made for costing purposes but this is a provisional sum and will need to be re-costed once confirmed. 

\begin{figure}[!hbt]
\begin{center}
    \includegraphics[width=0.9\linewidth]{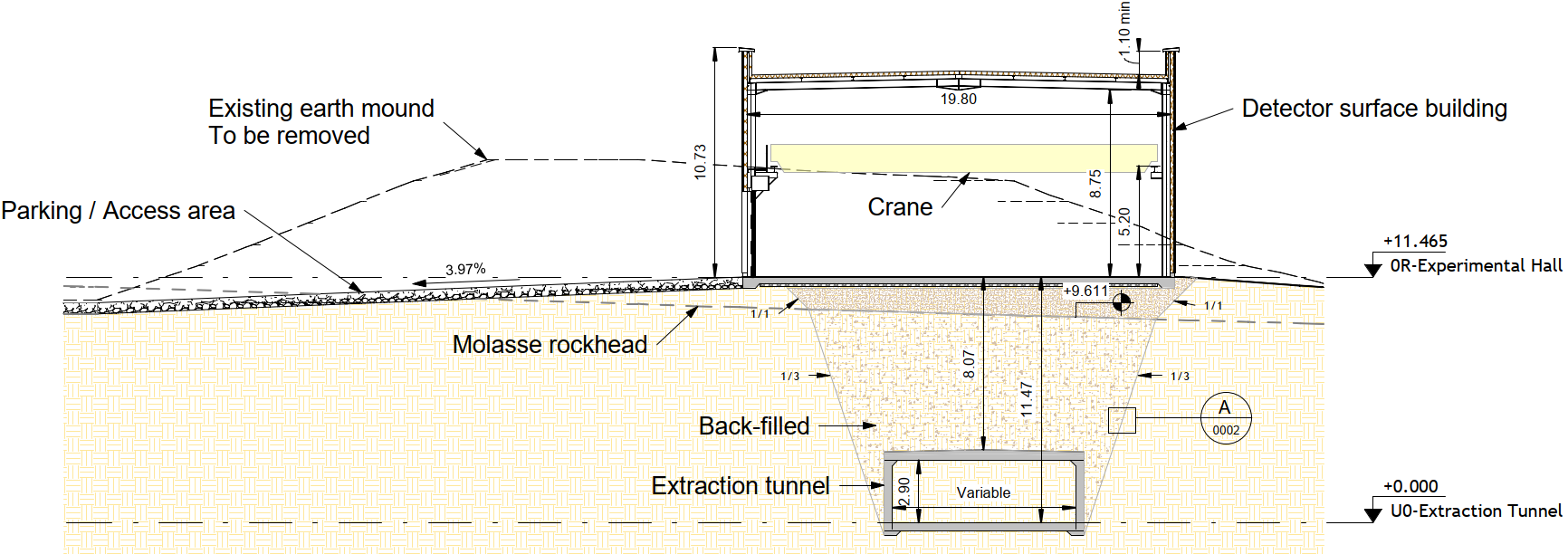}
    \caption{Section 2-2 through building showing parking, access and delivery area sloping down to existing road level. Dashed outline indicates the existing mound to be removed.}
\label{fig:EXPHallparking}
 \end{center}
\end{figure}

The existing earth mound shown in Fig.~\ref{fig:Mound} will need to be removed to enable the construction of the building and to allow the finished site levels to tie in with surrounding infrastructure.  It is assumed the levels of existing surrounding roads cannot be adjusted since they serve existing buildings and so these have effectively set the floor slab level. Outside the building, a parking and access area has been allowed for, although the space has not been further defined at this stage (see Fig.~\ref{fig:EXPHallparking}). The amount of earth to be removed has been minimised based on leaving slopes of 1 in 3 (as shown in Fig.~\ref{fig:Tunnellayout}) for the remaining mound. This could be further optimised following ground investigation by increasing the slope to the characteristic friction angle of the material when confirmed, or by introducing retaining structures to increase slope angles, for example soil nailed slopes or mechanically stabilised earth walls.

As part of the construction, a significant quantity of earthworks will be required in order to reduce the existing ground levels. A permanent storage site will be required for the spoil. Ideally this should be within CERN land and as close as possible to the existing site although a site has not been chosen at this stage. The composition of the earth mound will need be confirmed as soon as possible and any additional cost implications and constraints identified.

\paragraph{Construction methods}

An indicative construction sequence covering the extraction tunnel and experimental area has been envisaged as part of the cost estimation and in order to consider the timescales necessary to deliver the CE works. Initially, earthworks would need to be carried out with site facilities housed on a closed area of carriageway since there is no available accessible land in the vicinity. Prior to earthworks, and at a suitable time of year, the orchids would need to be relocated to an alternative site. 

Once earthworks were complete, the site compound, offices etc could be housed on the site itself. As the molasse rock is so close to the surface, it will be possible to use standard techniques such as rock bolting, shotcreting and traditional CE methods to provide temporary face support during excavation in the molasse. Once a space has been excavated and support put in place to work safely within, the reinforced concrete basement and shaft would be cast in situ using proprietary formwork systems or even slip-formed. Construction of the building thereafter would be straightforward using industry standard methods to erect and clad a steel portal frame building.

The widened section of tunnel will need to be constructed in reinforced concrete cast in situ via open cut while a rock breaker would be used to advance the final 36\,m of extraction tunnel using the New Austrian Tunnelling Method. The open cut section would be used as the construction access for  tunnelling due to schedule constraints requiring fit-out of the experimental hall to begin as soon as possible to reduce the overall duration. That requirement would mean the open cut section of tunnel would be constructed last then back-filled on completion. 

A core would be drilled using a wall mounted coring rig either from TT2 or the works side of the wall. In either case, a "sas" or airlock would be erected in TT2 to contain any dust produced during coring.

Tunnelling works close to existing infrastructure will have to be undertaken outside of beam operation periods both to avoid radiation close to existing tunnels carrying beam and to prevent vibration causing issues to the machine. In addition where works are required within existing tunnels, equipment will need to be removed to allow space for works and prevent contamination of equipment with dust. For this reason, the schedule will be planned around a long shutdown to give sufficient time to carry out:

\begin{itemize}
\item Coring into TT2;
\item Construction abutting TT2;
\item Demolition and construction of joints with TTL2;
\item Work within close proximity of the AD building. 
       
\end{itemize}

\begin{figure}[!hbt]
\begin{center}
    \includegraphics[width=12cm]{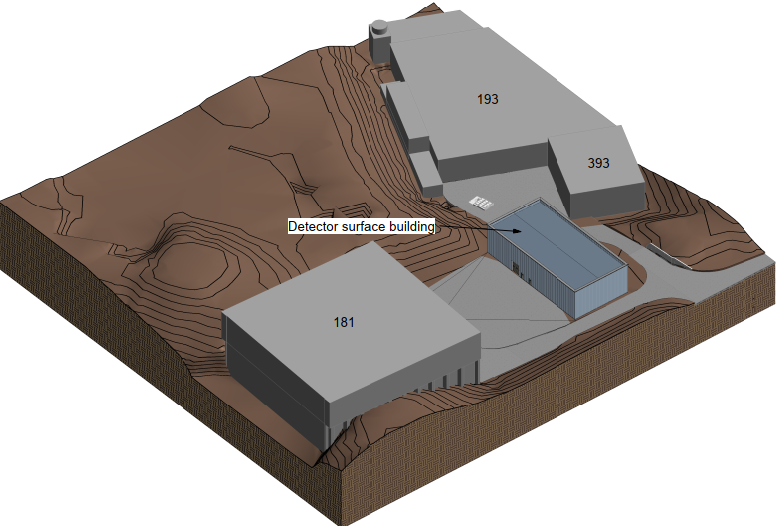}
    \caption{Overview showing siting of experimental surface building with existing adjacent infrastructure and contour plan of topography after proposed works.}
\label{fig:3Dlayoutwithexisting}
 \end{center}
\end{figure}

The site is very constrained in terms of space as can be seen from as shown in Fig.~\ref{fig:3Dlayoutwithexisting}, so further study will be required to ensure the impact on surrounding buildings, accesses and existing operations is minimised. The required off-sett to allow works during normal beam operations will need to be defined through study.

Radiation issues will also need to be carefully considered as part of the construction of the basement structure due to the minimal distance to TT7 and its shielding blocks. Further testing of soil activation and to confirm the shielding block arrangement will be needed in advance of works. The following hierarchy of considerations will be applied for works planning of this area specifically:

\begin{enumerate}
\item Avoid disturbing shielding blocks if possible;
\item Review possibility of casting concrete against shielding if required, giving consideration to future maintenance and demolition;
\item Remove if unavoidable, ideally reusing shielding in an area which will become activated in any case.
\end{enumerate}

All temporary facilities needed for the civil engineering work have been included in the cost estimate, but any temporary areas/buildings needed for machine or detector assembly/installation were not. It is assumed no existing roads will require diversion.

\subsubsection{Integration}\label{sec_ICE_ExperimentalArea_Integration}

The eSPS experimental area is composed of a surface hall located next to B393 and the underground experimental hall located below the surface hall next to transfer tunnel TT7 as shown in Fig.~\ref{fig:3DviewEXPHALL}. 

The purpose of these structures includes the following:

\begin{itemize}

\item Install and house the missing momentum and beam dump experiments;
\item Houses the control room, clean room and work area for the experiments; 
\item Service storage and distribution for the experiments and the extraction tunnel;
\item Beamline equipment access to the extraction tunnel;
\item Personnel access to the experiments and extraction tunnel.

\end{itemize}

The equipment includes magnets, beam instrumentation, vacuum equipment, and general electrical and safety equipment. The services include electrical services, cooling, and ventilation.

The surface building has a finished floor level of 444.5~m above sea level at ground level. The underground hall is 11.5~m below the surface building and is the same level as the extraction tunnel. There is second floor level 1.2~m below this to allow the large detectors of the missing momentum and beam dump experiments to be centred on the beam line. The surface hall’s east wall is approximately 10~m from B393 and experimental hall’s east wall is approximately 1.2~m from transfer tunnel TT7’s steel shielding wall.

\begin{figure}[!hbt]
\begin{center}
    \includegraphics[width=\linewidth]{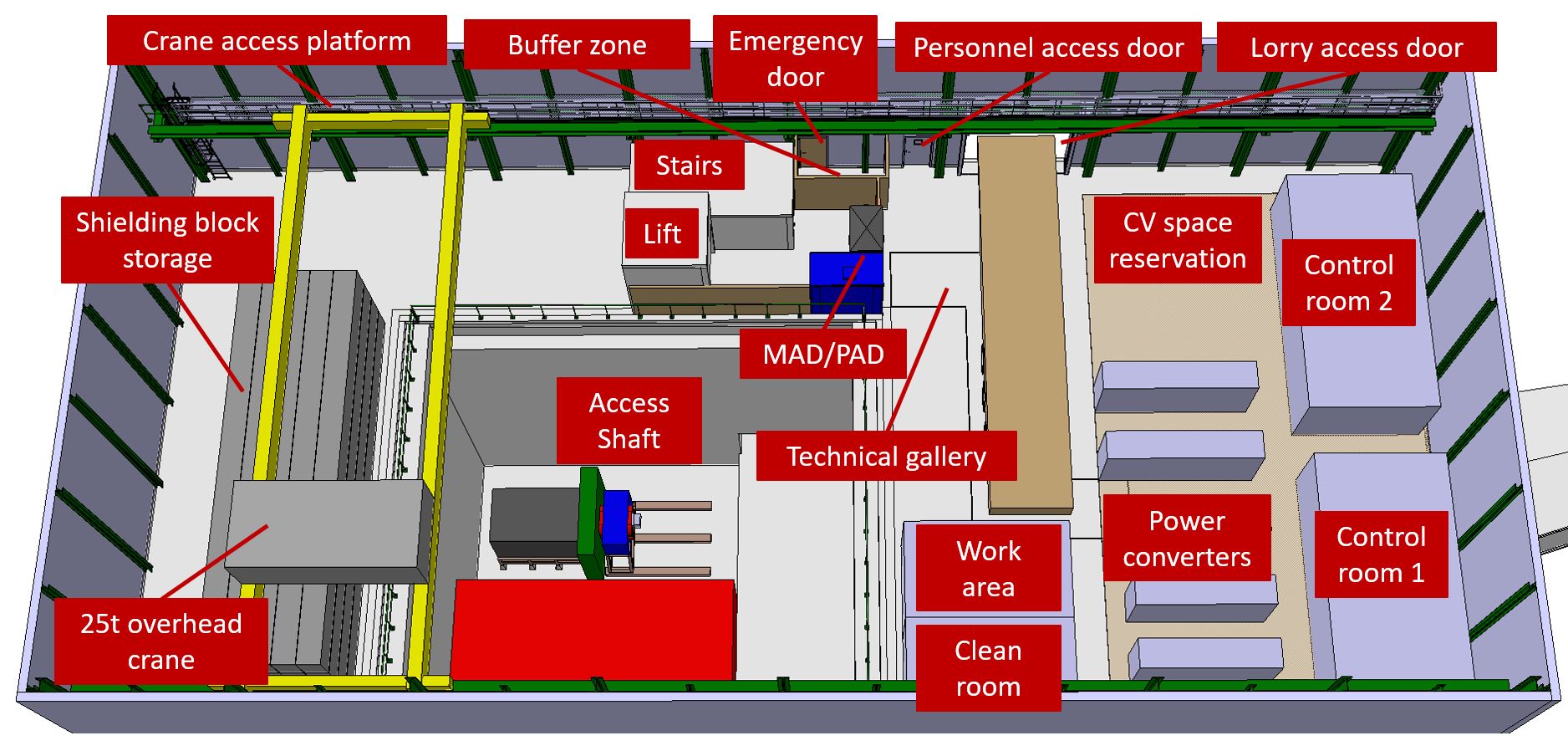}
    \caption{Integration layout of the surface building.}
 \label{fig:surfacebuilding}
\end{center}
\end{figure}

\begin{figure}[!hbt]
\begin{center}
    \includegraphics[width=0.8\linewidth]{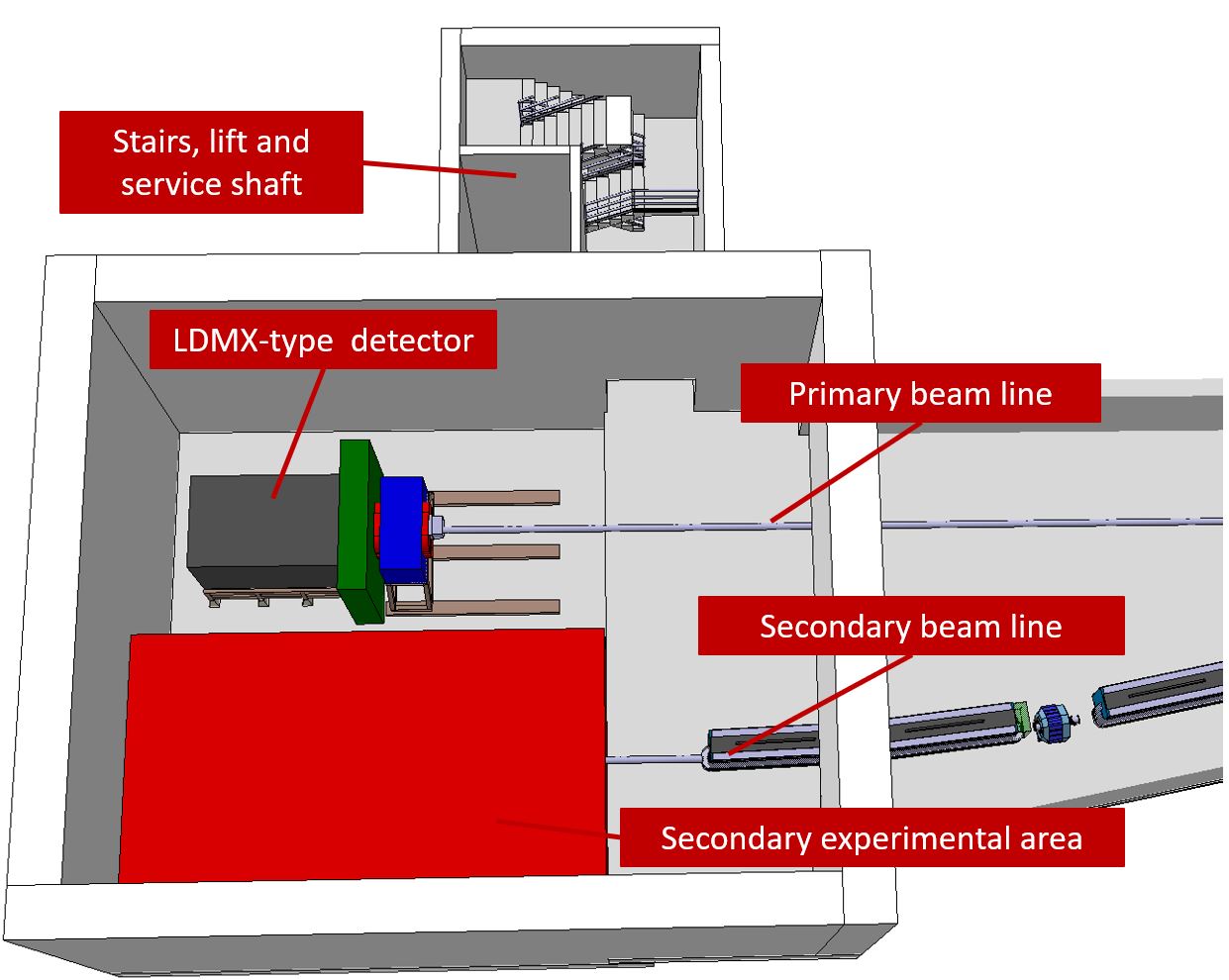}
    \caption{Integration layout of the experimental hall.}
 \label{fig:experimentalhall}
\end{center}
\end{figure}

The surface hall houses the power supply, control racks, and the cooling and ventilation equipment for the extraction tunnel and the experimental hall as well as the detector control room, clean room (for work on silicon detector) and work area as shown in Fig.~\ref{fig:surfacebuilding}. Within the building is the personnel access shaft and the equipment access opening. The experimental hall houses the missing momentum and beam dump experiments. The proposed integration layout is shown in Fig.~\ref{fig:experimentalhall}.

The cable trays, cooling piping, and ventilation ducts are distributed to the experimental hall and  extraction tunnel via the personnel shaft, from which services pass underneath the false floor in the surface building and into the technical gallery that runs into the personnel shaft. The technical gallery is 2.0\,m wide and 0.8\,m deep with a removable ceiling to allow personnel to access it from the floor level of the surface hall.

The access control for the experimental hall and the extraction tunnel is housed inside the surface hall and is composed of a material access door (MAD), a personnel access door (PAD), and an emergency door. Through the access control there is a personnel lift and stairs, which bring personnel and light equipment down to the experimental hall. A buffer zone for storing material brought up from the tunnel and experimental hall during maintenance work is located next to the access control. Located inside the surface hall is a large equipment access opening that allows equipment to be lowered directly onto the extraction tunnel floor and the experimental hall floor.

During the equipment installation, transport vehicles, including a 40\,t semi-trailer, a 19\,t truck, and a forklift, will move the equipment into the access building, from which a 25\,t overhead crane will lift and move equipment throughout the building as well as lower the equipment into the equipment access opening. A handling vehicle inside the extraction tunnel will collect the equipment for positioning along the beam line whilst the detector equipment will be manoeuvred into its final position using the overhead crane. Inside the building, there is a personnel platform at the height of the crane rail to allow personnel to access the crane. To access the platform, there is a ladder attached to the side wall. The equipment access opening will be filled with 20 concrete shielding beams during beam operation because of RP considerations, and surrounding the equipment shaft is a guard rail for personnel safety. During installation and de-installation of equipment in the tunnel, the shielding beams will be removed from the shaft and stored at the back of the building.

The personnel shaft is located inside the surface hall next to the access control. The shaft houses a personnel lift and a metallic personnel staircase, and the services (ventilation, cooling, and cable trays). The shaft brings personnel from the FFL of the surface hall to the FFL of the experimental hall. The services are brought into the personnel shaft from the technical gallery in the surface hall to a hole in the personnel shaft that allows the services to be distributed in the experimental hall.

The surface hall is a single-storey steel portal frame industrial building with a width of 19\,m between the inner edges of the columns. The building has a length of 40\,m between the inner edges of the gable columns with a longitudinal centre-to-centre column spacing of approximately 5\,m. The ceiling-to-floor height is approximately 8.8\,m, and there is a 0.4\,m deep false floor on the north side of the building with a technical gallery running from the false floor to the personnel shaft. 

The width of the structure (19\,m) is based on a personnel accessway (1.0\,m), upstand (0.2\,m), beam support (0.5\,m), the equipment access opening (11.5\,m), beam support (0.5\,m), upstand (0.2\,m), the lift shaft (2.65\,m) and the stairs shaft (2.45\,m). The length of the structure (40\,m) is based on the clearance of the overhead crane (3.365\,m), the shielding blocks (3.285\,m), a personnel accessway (1.0\,m), upstand (0.2\,m), the equipment access opening (14.5\,m), upstand (0.2\,m), an assembly area (3.75\,m), a~semi-trailer (2.5\,m), a personnel accessway (1.0\,m), length of power converter (4.8\,m), minimum power converter edge distance (1.4\,m), control room (4\,m).

The equipment access opening is 14.5\,m long and 11.5\,m wide, for transporting heavy equipment into and out of the experimental hall and extraction tunnel. The 14.5\,m is composed of the detector area is 10\,m long and the extraction tunnel area is 4.5\,m long (the largest magnets are approximately 3.5\,m long allowing for 0.5\,m clearance either side). The 11.5\,m is composed of the two detector widths (6\,m each with a 0.5\,m overlap).

The height of the structure (8.765\,m) is based on the height of the lorry access door (4.7\,m), the clearance between the top of the door and the underside of the crane (0.5\,m), the height of the crane (2.765\,m), the clearance between the top of the crane and the underside of the services (0.5\,m), and an allowance for lighting and services (0.3\,m). To access the overhead crane for maintenance, there is a personnel platform, for which the crane is offset from the wall/column by 0.8\,m. On the other side, the crane is offset by 0.35\,m to provide the minimum crane clearance. On the roof, there is a 1.1\,m tall parapet wall for personnel access safety.

The clean room, work area and control rooms sizes were specified by the users of the facility.

\subsubsection{Cooling and ventilation}

The cooling and ventilation conceptual design for the experimental area of eSPS is presented below. It has been designed in accordance with the heat loads estimated by each user group. It should be stated that since the design is conceptual and  estimates have been made conservatively, no safety factor was applied to the heat loads. This will be discussed at a later stage when more detailed studies can be made.
The scope of the cooling and ventilation needs of eSPS has excluded the SPS tunnel, TT10 and TT20 in line with the ICE scope. It is assumed that present cooling and ventilation systems in these areas which are sufficient for proton operation in SPS, can provide for the reduced needs of eSPS.

The experimental area premises are divided in three zones:

\begin{itemize}

\item Experimental hall, i.e. surface building;
\item Detector hall, i.e. underground area;
\item New tunnel, linking the detector hall to TT2 and crossing TTL2

\end{itemize}

\paragraph{Piped utilities}
The cooling system for the detector is composed of a primary circuit connected to the AD cooling towers and two secondary circuits:

\begin{itemize}

    \item One with demineralised water for the power converters and beam lines, 
        \item A second circuit with chilled water for detector cooling and air cooling.
    \item A third system composed by glycol water is foreseen for the cooling of part of the detector.

\end{itemize}

Table~\ref{tab:Water heat loads} summarises the heat loads for the different equipment. The cooling circuits are, therefore, sized with the operational parameters indicated in Table~\ref{tab:Working parameters cooling circuits}. 

\begin{table}[!hbt]
\centering
 \caption{Water heat loads for cooling circuits.}
  \label{tab:Water heat loads}
\begin{tabular}{ccccc}
 \hline\hline
 \textbf{ } & \textbf{ } & \textbf{Demin. water} & \textbf{Chilled water} & \textbf{Glycol water}\\
\hline
  Heat loads [kW] & Experimental hall & 150 & - & -\\

  & Detector hall  & - & 7 & 5.1 \\ 

   & New tunnel & 730 & - & - \\

   & TT10, TT20 & 200 & - & - \\ 

   & Air conditioning &  - &  60 & - \\
 \hline
 \textbf{Total heat load } & & \textbf{1080} & \textbf{67} & \textbf{5.1} \\  
 \hline\hline
\end{tabular}
\end{table}

\begin{table}[!hbt]
\centering
 \caption{Working parameters for cooling circuits in experimental area.}
  \label{tab:Working parameters cooling circuits}
\begin{tabular}{lcccc}
 \hline\hline
 \textbf{ } & \textbf{Cooling capacity} & \textbf{Flow rate } & \textbf{Temperature } & \textbf{Inlet pressure }\\
 \ & \textbf{[kW]} & \textbf{[m\textsuperscript{3}/h]} & \textbf{[$^\circ$C]} & \textbf{[bar]} \\
 \hline
  \ Primary circuit & \ 1200 & 130 & 24--32 & 5 \\ 

  \ Demin. water & 1100 & 120 & 25--33 & 12 \\

  \ Chilled water  & 80 & 11 & 12--18 & 10 \\ 

  \ Glycol water & 5.5 & 1 & 0--5 & 5 \\
  \hline\hline
\end{tabular}
\end{table}

\newpage 
The primary circuit shall be connected to the cooling towers of the AD experimental area, close to the future building. This connection is not studied yet and will be defined at a later stage.

The chilled water shall be produced by a dedicated chiller to be located outside the new building. In this case it is foreseen to have a redundant chiller to ensure continuity of operation for the cooling of equipment. This could be omitted at a later stage of project development if not considered essential in order to minimise costs. A second possibility would be to foresee a backup supply from the future chilled water station that will be installed during Long Shutdown 4 (currently planned for 2025-2026) for the ventilation needs in the AD complex.

All sections of the cooling plant would be installed in the surface building, close to control room 1 and the power converter area. An area measuring 13\,m in length and 6\,m in width is needed. A false floor below the station approximately 1\,m deep is needed to allow the routing of pipes.

Dry riser pipes made from stainless steel and ND100 for fire extinguishing purposes, will be installed in the underground areas, i.e. the detector hall and the tunnel up to TT2. Initially it is estimated that 20 connections will be made available to allow fire brigade interventions along the length of the pipeline. They will be connected to the Meyrin site hydrant distribution network. At the next stage of development, when the routing and length of the pipe will be defined, the available pressure will need to be checked and, if needed, an overpressure pumping system can be included. At the present level of the study such a system is not foreseen. Similarly, the installation of hydrants outside the building and fire hoses inside will be discussed at a future stage of the study.

Compressed air will be provided by the Meyrin site distribution network. The precise connection point and the distribution network in the experimental area will be defined in the future.

\paragraph{Heating, ventilation and air conditioning}
Ventilation in the experimental area is provided by several different systems in order to cover the functionalities required in the different areas for operation during running and maintenance periods. 
The different ventilation areas are: 

\begin{itemize}

\item Ventilation of the experimental hall;
\item Ventilation of the detector hall;
\item Ventilation of control room no.\ 1;
\item Ventilation of control room no.\ 2;
\item Ventilation of clean room;
\item Pressurisation of the lift cage;
\item Ventilation of the new tunnel and TTL2;
\item Smoke extraction from the experimental hall;
\item Gas extraction from the detector hall.

\end{itemize}

For the time being, it is not foreseen to allow for different flow rates for run and access modes since the requested values during run do not warrant it. In order to save energy, air re-circulation is foreseen and a maximum of 20\% fresh air can be provided in each installation with the exception of the pressurisation of the lift cage which will be fully supplied with fresh air.

All other considerations made for the ventilation systems of the injector and transfer areas are valid for the experimental area, in particular the redundancy level.

The ventilation of the experimental hall is performed via an air handling unit (AHU) located inside the building, with supply ducts and diffusers along the perimeter at ground floor level and an extraction line in the upper part of the building to recycle the air. Smoke extraction from the building will be allowed via 4 smoke extractors located on the ceiling of the building.

The control rooms and the clean rooms in the surface building will have a dedicated air conditioning system each, with fresh air and exhaust air ducts connected to the exterior of the building.

The lift cage and the stairs will have an overpressure of at least 40\,Pa above the underground detector hall with the pit closed. It is essential that an effective air-tight seal is created with the  shielding blocks closing the pit. For this, an AHU will be located on the roof of the lift cage and will distribute fresh air to the top and the bottom of the lift cage thus ensuring an overpressure  corresponding to the door locations.

The ventilation of the detector hall is effected using:

\begin{itemize}

\item AHUs located close to those dedicated to the ventilation of the building. 
\item A supply duct running through the pit in the underground area and supplying air via diffusors at floor level
\item An extraction duct in the upper part of the underground hall will allow the recycling of the air via dedicated fans.
    
\end{itemize}

The extraction duct and the fans can also be used for smoke extraction in case of fire with the pit closed by shielding beams. If a fire occurs with the pit open, smoke extraction will be via the smoke extractors in the building. Although the type of gas used has not been finalised, the detector will require the presence of some gas (Helium or carbon dioxide) and, therefore, a gas extraction system is foreseen from the detector hall venting to the roof of the building. The layout of this system is not defined and will be studied when more detailed information will be available.

Finally, ventilation of the new tunnel and TTL2 will be implemented using AHUs located in the surface building, with a supply duct linking AHUs, via the detector hall, to the far end of TTL2 and the new extraction tunnel. In order to reach one of the ends of TTL2, a hole will be drilled to allow the duct to be installed through the rock or tunnel lining between one end of TTL2 and the end of the new tunnel. Air will be blown at this end and at the beginning of the new tunnel, close to the detector hall. Then, air is recuperated via a duct extracting air at the level of the crossing between the new tunnel and TTL2 and circulated back to ventilation units on the surface.

Table~\ref{tab:ventilation loads} shows the heat loads to be removed by the ventilation systems. The operational parameters of the ventilation systems are in Table~\ref{tab:ventilation parameters}.

\begin{table}[!hbt]
\centering
 \caption{Technical loads on ventilation for the experimental area}
  \label{tab:ventilation loads}
\begin{tabular}{lccc}
 \hline\hline
 \textbf{ } & \textbf{Area [m\textsuperscript{2}]} & \textbf{Volume [m\textsuperscript{3}]} & \textbf{Technical load} \\
\hline
  \ Experimental hall & 808 & 7430 & 60 kW \\ 

  \ Underground Detector hall & 176 & 1922 & 25 kW \\

  \ Control room 1  & 32 & 96 & 10 PCs, 10 screens, 10 people \\ 

  \ Control room 2  & 32 & 96 & 10 PCs, 10 screens, 10 people \\ 

  \ Clean room  & 15 & 45 & 5 kW \\ 

  \ Lift / stair cage  & 23 & 251 & none \\ 

  \ New tunnel  & 316 & 778 & 37 kW \\ 

  \ Smoke extraction  & 808 & - & 3 m$^{3}$/s per 100 m$^{2}$ \\ 

  \ Gas extraction  &  - &  - & not available \\ 
  \hline\hline
\end{tabular}
\end{table}

\begin{table}[!hbt]
\centering
 \caption{Working parameters for ventilation systems in the experimental area}
  \label{tab:ventilation parameters}
\begin{tabular}{lccccc}
 \hline\hline
 \textbf{ } & \textbf{Technical load } & \textbf{Cooling} & \textbf{Heating} & \textbf{Flow rate} & \textbf{Number AHUs}\\
   \ & \textbf{[kW]} & \textbf{[kW]} & \textbf{[kW]} & \textbf{[m$^{3}$/h]} & \\
\hline
  \ Experimental hall & 60 & 20 & 47 & 30000 & 2 \\ 
  \ Detector hall & 25 & 10 & 23 & 15000 & 2 \\
  \ Control room 1  & 4.5 & 2 & 5 & 3000 & 1 \\ 
  \ Control room 2  & 4.5 & 2 & 5 & 3000 & 1 \\
  \ Clean room  & 5.6 & 1 & 2 & 10000 & 1 \\ 
  \ Lift / stair cage  & 0 & 10 &23 & 3000 & 2 \\ 
  \ New tunnel  & 37 & 15 &34 & 22000 & 2 \\ 
  \ Smoke extraction  & 0 & 0 & 0 & 100000 & 4 \\ 
  \ Gas extraction  & n.a.  & n.a. & n.a. & 5000  & 2 \\ 
  \hline\hline
\end{tabular}
\end{table}

\subsubsection{Radiation protection}

\paragraph{Access conditions}

Access to the experimental cavern, where the detectors for the dark matter and the beam dump experiment are installed, will be prohibited during beam operation. The experimental hall on top of the cavern is classified as non-designated area, with exception of the fenced area directly on top of the shaft.

Access to the experimental cavern itself shall be possible while proton beam operation is continuing in the TT2 transfer tunnel for beam delivery to n\_TOF and the SPS. The zoning, local shielding and distances are defined such to assure that this access is possible without impacting the proton beam operation.

\paragraph{Prompt radiation}

While the design of the missing momentum experiment is approximately known, and is discussed in Ref.~\cite{whitepaper}, not much information is available at this stage about the design of a second possible experiment. To model the most demanding scenario, this is assumed to be a beam dump experiment absorbing all electrons that eSPS can deliver. In order to plan with a minimum shielding on top of the cavern, a large amount of shielding is considered around the beam dump experiment target. This shielding is required as well to reduce the radiation background in the missing momentum experiment. Note, however, that a dumped beam into a second experiment, will not arrive at the same time as an electron beam is delivered into the missing momentum experiment. A large part of the prompt radiation from the interacting electrons is absorbed in the detectors of the experiment themselves and the bulk shielding. The protection of the experimental hall is achieved by a shielding on top of the experimental cavern made of \SI{1}{m} thick removable concrete slabs. The shielding design must be confirmed once more details are known about the design of this second experiment.

\begin{figure}[!hbt]
\centering
\includegraphics[width=0.85\textwidth,trim={0cm 0cm 0cm 0cm},clip]{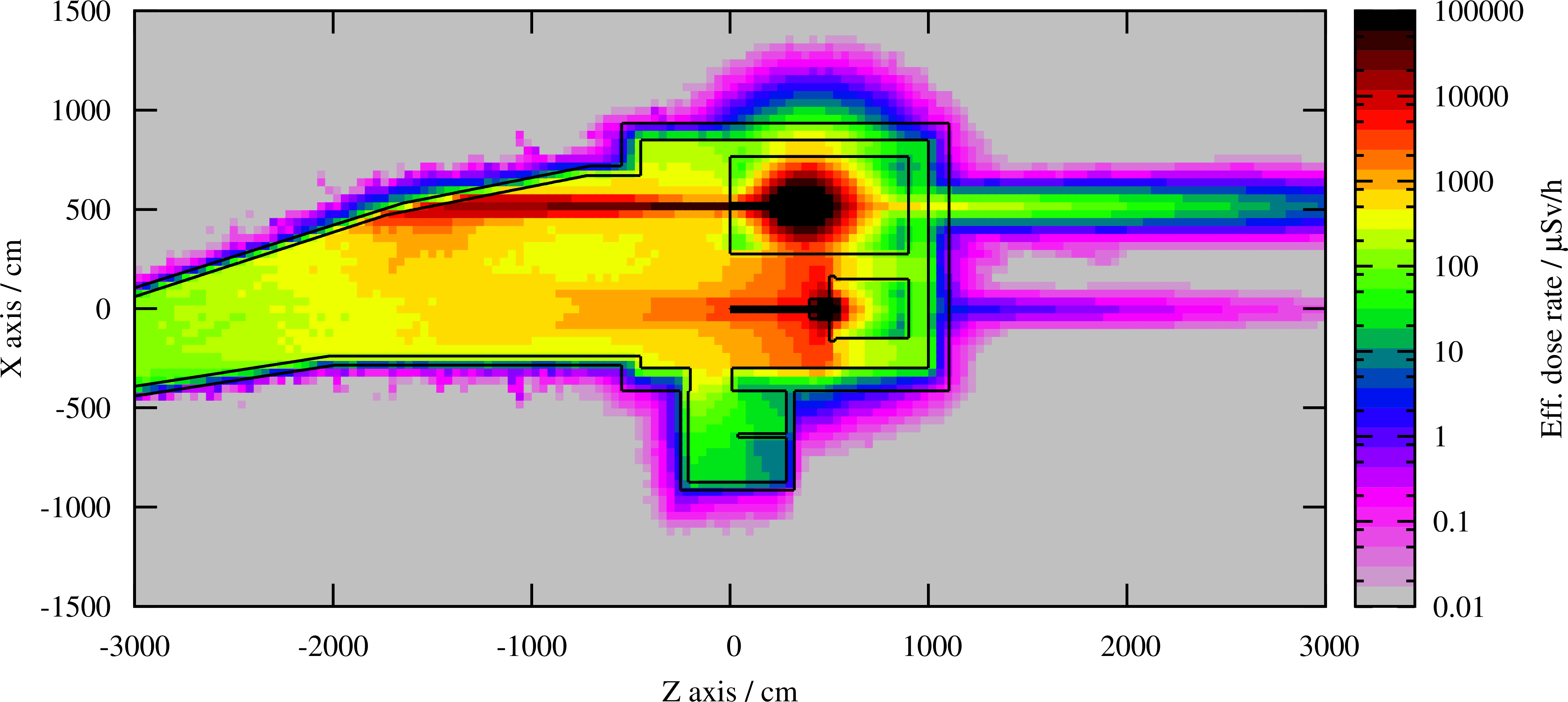}
\caption{Effective dose rate inside the experimental cavern from beam on the missing momentum experiment (\SI{3.25e9}{e^-\per\second}) and the beam dump experiment (\SI{5e11}{e^-\per\second}).}
\label{fig:expcav}
\end{figure}

\begin{figure}[!hbt]
\begin{center}
   \begin{subfigure}{0.49\linewidth}
        \begin{center}
            \includegraphics[width=\textwidth]{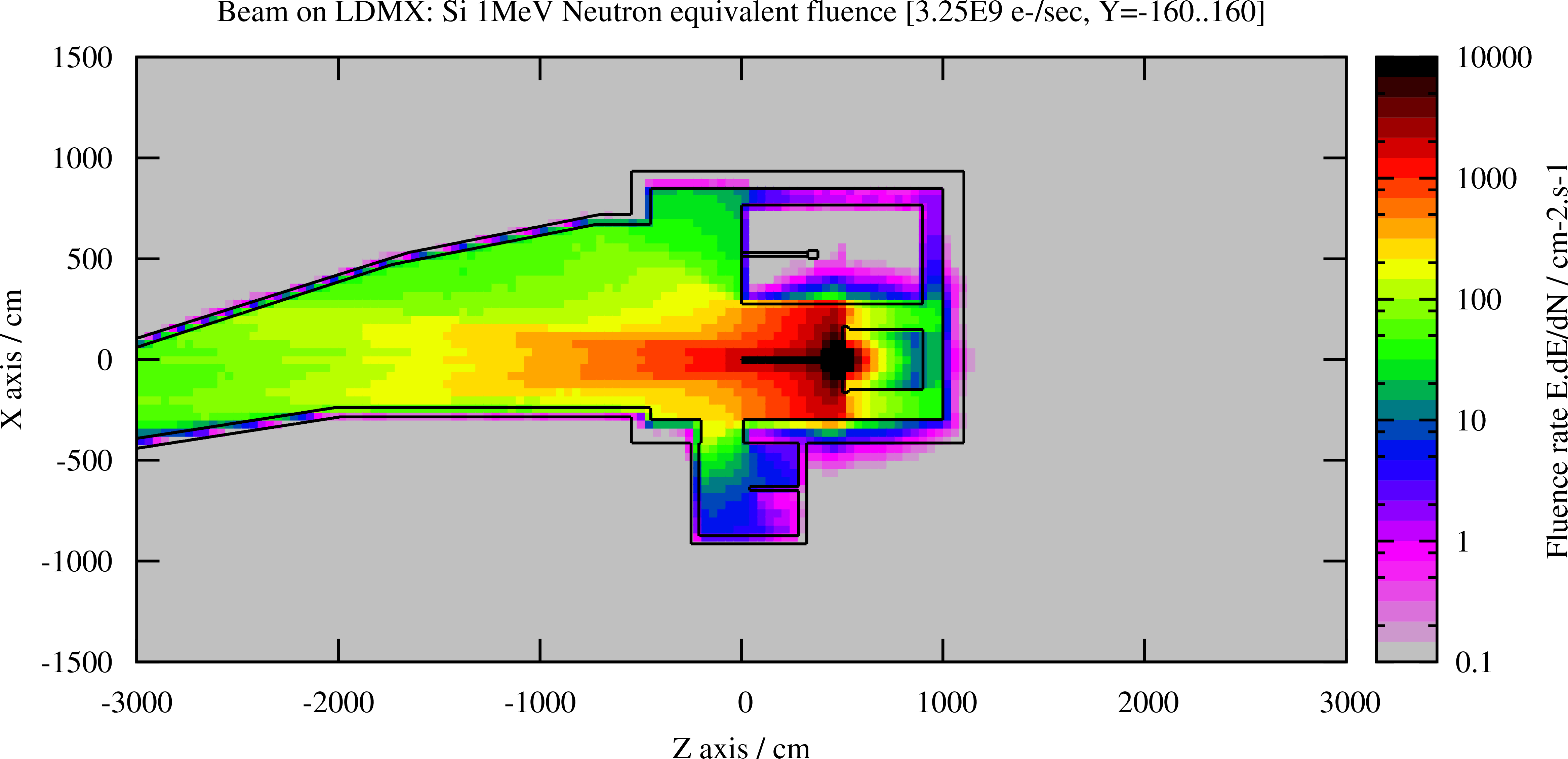}
            \caption{}
        \label{fig:RP_esps_exp_ldmx_cav_Y_SI1MeV}
        \end{center}
    \end{subfigure}
    \begin{subfigure}{0.49\linewidth}
        \begin{center}
           \includegraphics[width=\textwidth]{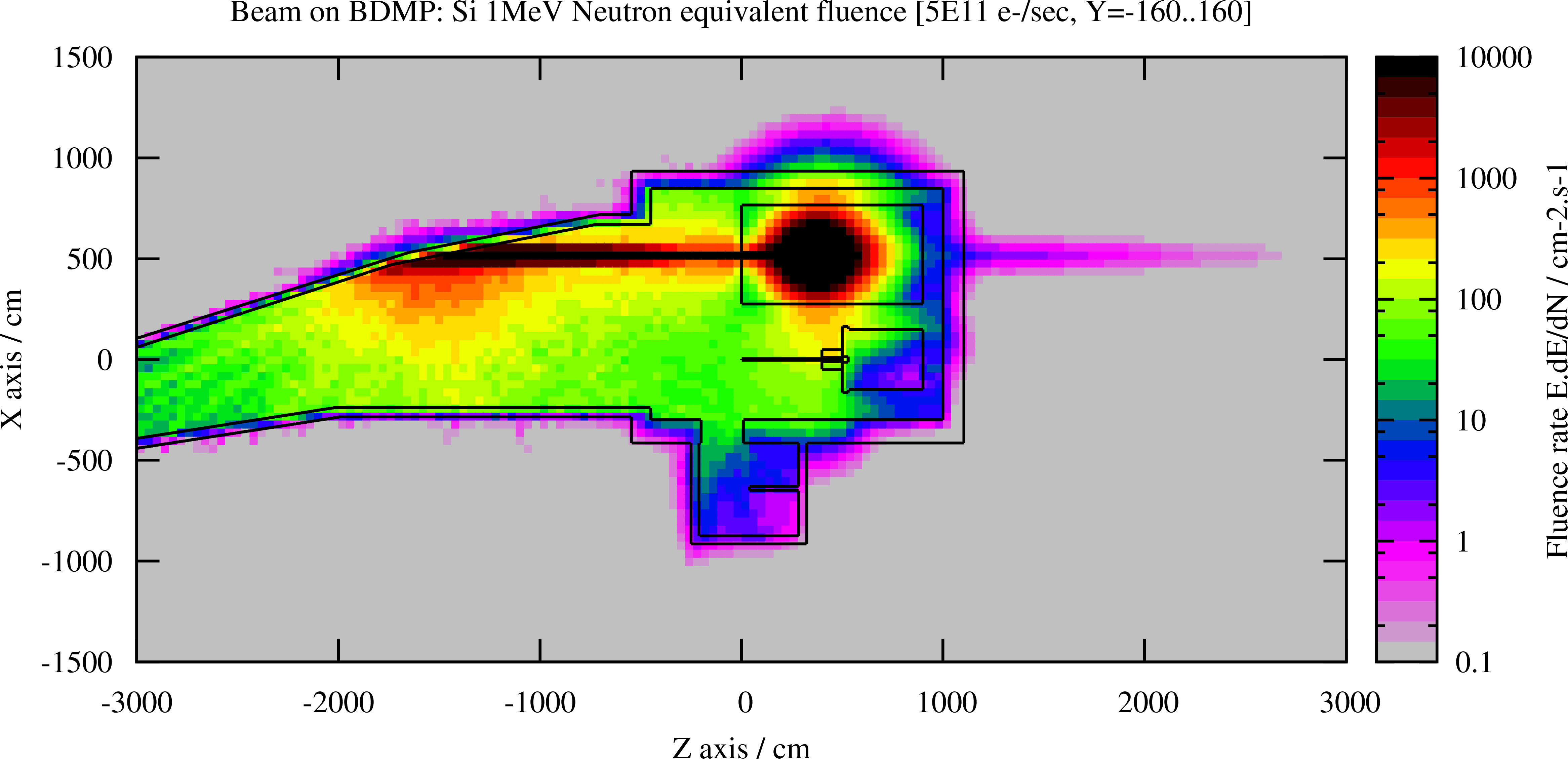}
           \caption{}
        \label{fig:RP_esps_exp_bdmp_cav_Y_SI1MeV}
        \end{center}
    \end{subfigure}
    \caption{\SI{1}{MeV} neutron equivalent fluence in Silicon inside the experimental cavern from \subref{fig:RP_esps_exp_ldmx_cav_Y_SI1MeV} beam on the missing momentum experiment (\SI{3.25e9}{e^-\per\second}) and \subref{fig:RP_esps_exp_bdmp_cav_Y_SI1MeV} the beam dump experiment (\SI{5e11}{e^-\per\second}).}
\label{fig:expcav_si1mev} 
\end{center}
\end{figure}

\begin{figure}[!hbt]
\centering
\includegraphics[width=0.85\textwidth,trim={0cm 0cm 0cm 0cm},clip]{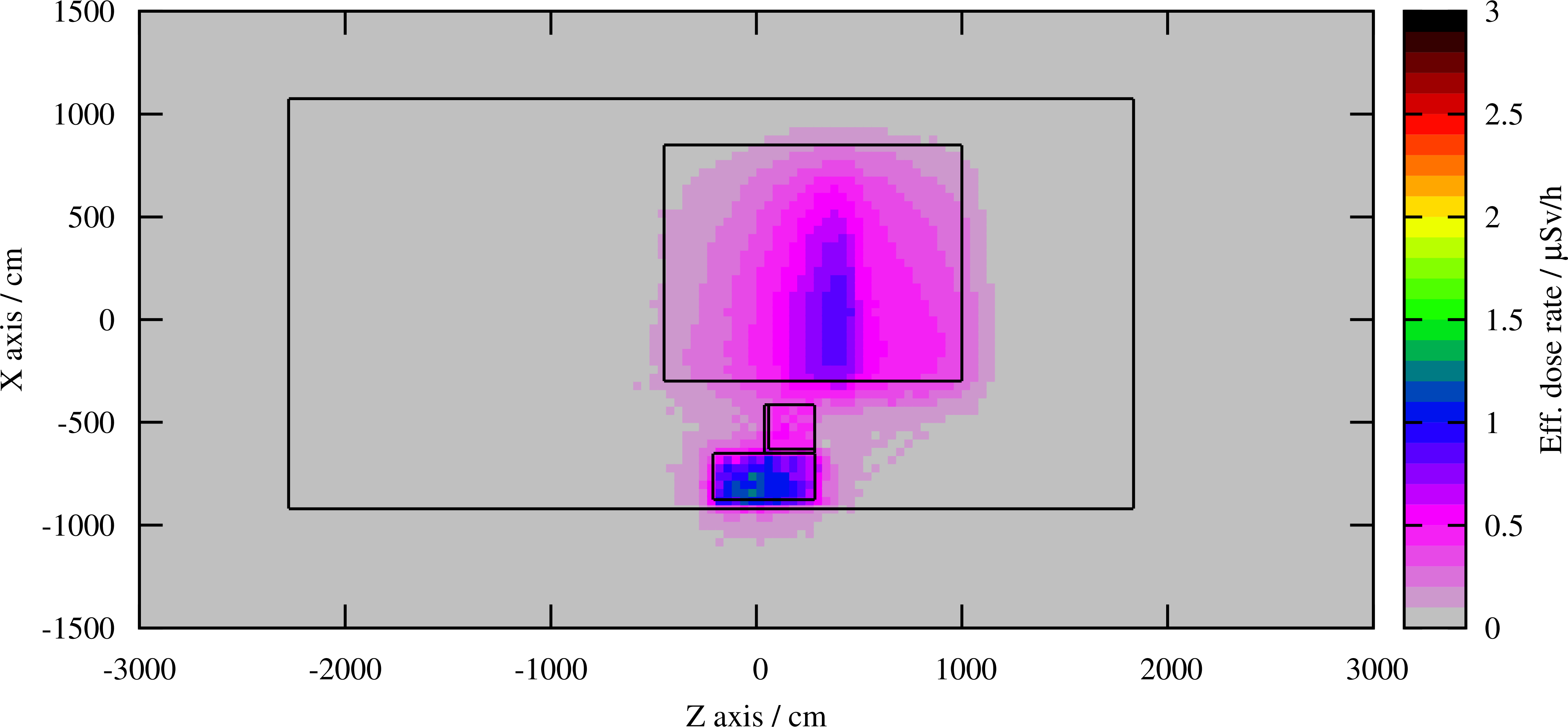}
\caption{Effective dose rate in the experimental hall (on top of the cavern) from beam on the missing momentum experiment (\SI{3.25e9}{e^-\per\second})and the beam dump experiment (\SI{5e11}{e^-\per\second}).}
\label{fig:exphall}
\end{figure}

Figures~\ref{fig:expcav} and \ref{fig:exphall} show the expected dose rates inside the cavern and on top of the cavern at the ground level of the experimental hall. The dose rates in the accessible areas in the hall remain below the design target of \SI{0.15}{\micro\sievert\per\hour} for a non-designated area. Figure~\ref{fig:expcav_si1mev} shows the \SI{1}{MeV} neutron equivalent fluence in Silicon inside the experimental cavern from beam on both experiments.

\begin{figure}[!hbt]
\centering
\includegraphics[width=0.8\textwidth,trim={0cm 0cm 0cm 0cm},clip]{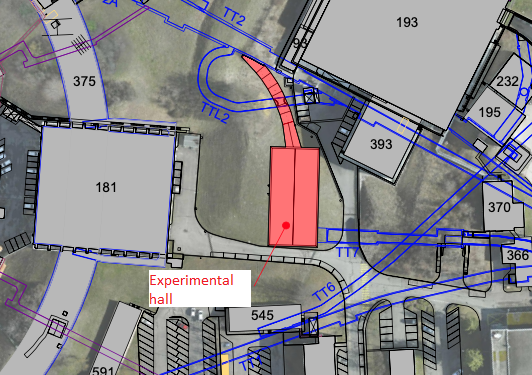}
\caption{Location of the experimental hall (red) on the Meyrin Site with respect to other underground infrastructure.}
\label{fig:exp_topview}
\end{figure}

\begin{figure}[!hbt]
\centering
\includegraphics[width=\textwidth,trim={0cm 0cm 0cm 0cm},clip]{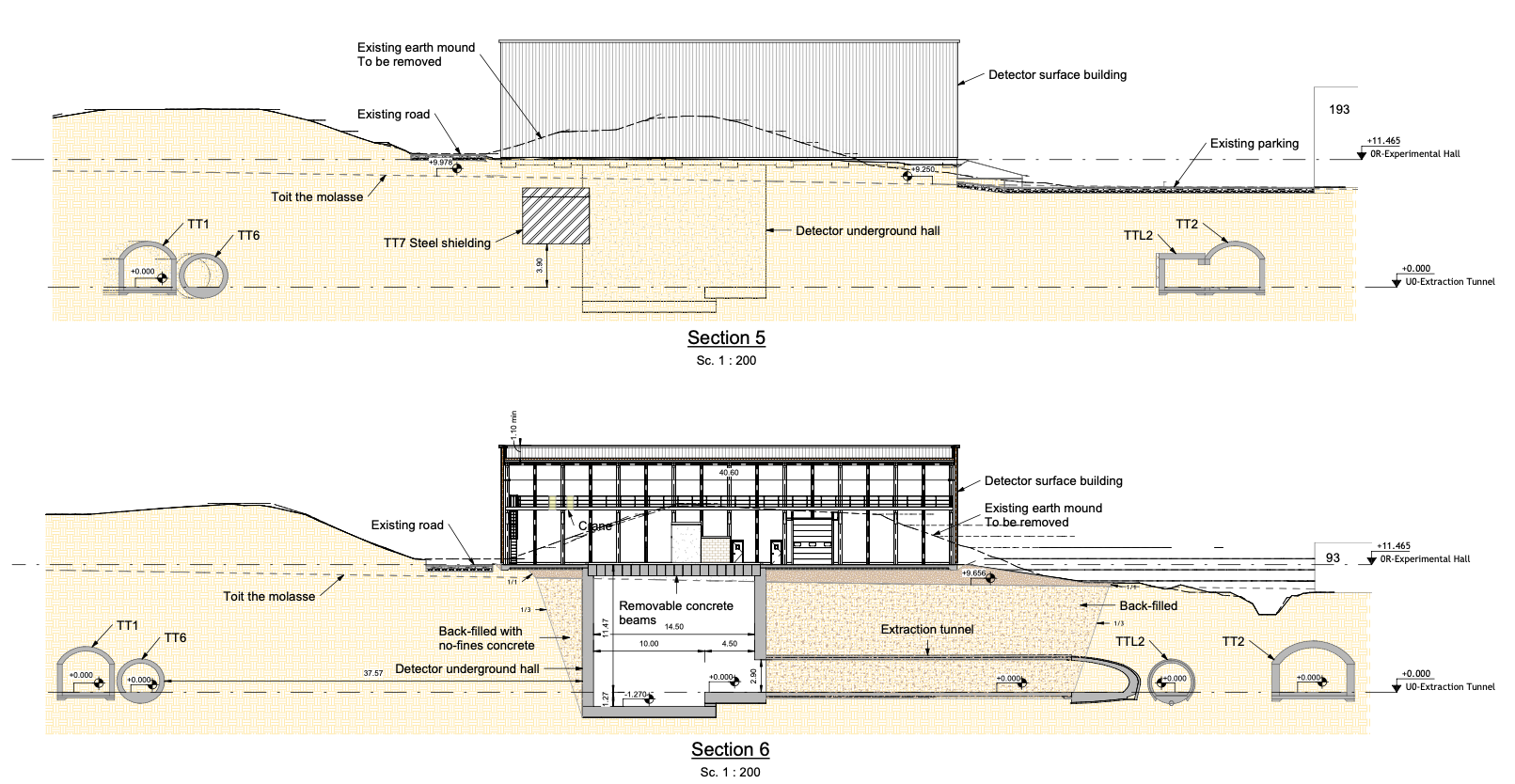}
\caption{Location of the TT1, TT6, TT7 and TT2 tunnels with respect to the experimental cavern. Vertical cuts parallel to the beamline directed on the experiment, through the experimental cavern (bottom) and with some offset towards TT7 (top).}
\label{fig:exp_xsection}
\end{figure}

Three underground infrastructures are located in the vicinity of the experimental cavern. TT7 is located laterally and TT1/TT6 in the forward direction of the beam, as shown in Fig.~\ref{fig:exp_topview}. Figure~\ref{fig:exp_xsection} shows vertical cuts aligned with the beam line directed on the experiments and illustrate the distances and levels with respect to TT1/TT6, TT7 and TT2.

The unused tunnel TT7 is located at only about \SI{1}{m} lateral distance from the cavern, but after the cavern itself. The massive steel shielding at the end of TT7 protects against radiation inside TT7 such that it is of no concern during beam operation (see as well Fig.~\ref{fig:expcav}).

\begin{figure}[!hbt]
\centering
\includegraphics[width=0.7\textwidth,trim={0cm 0cm 0cm 0cm},clip]{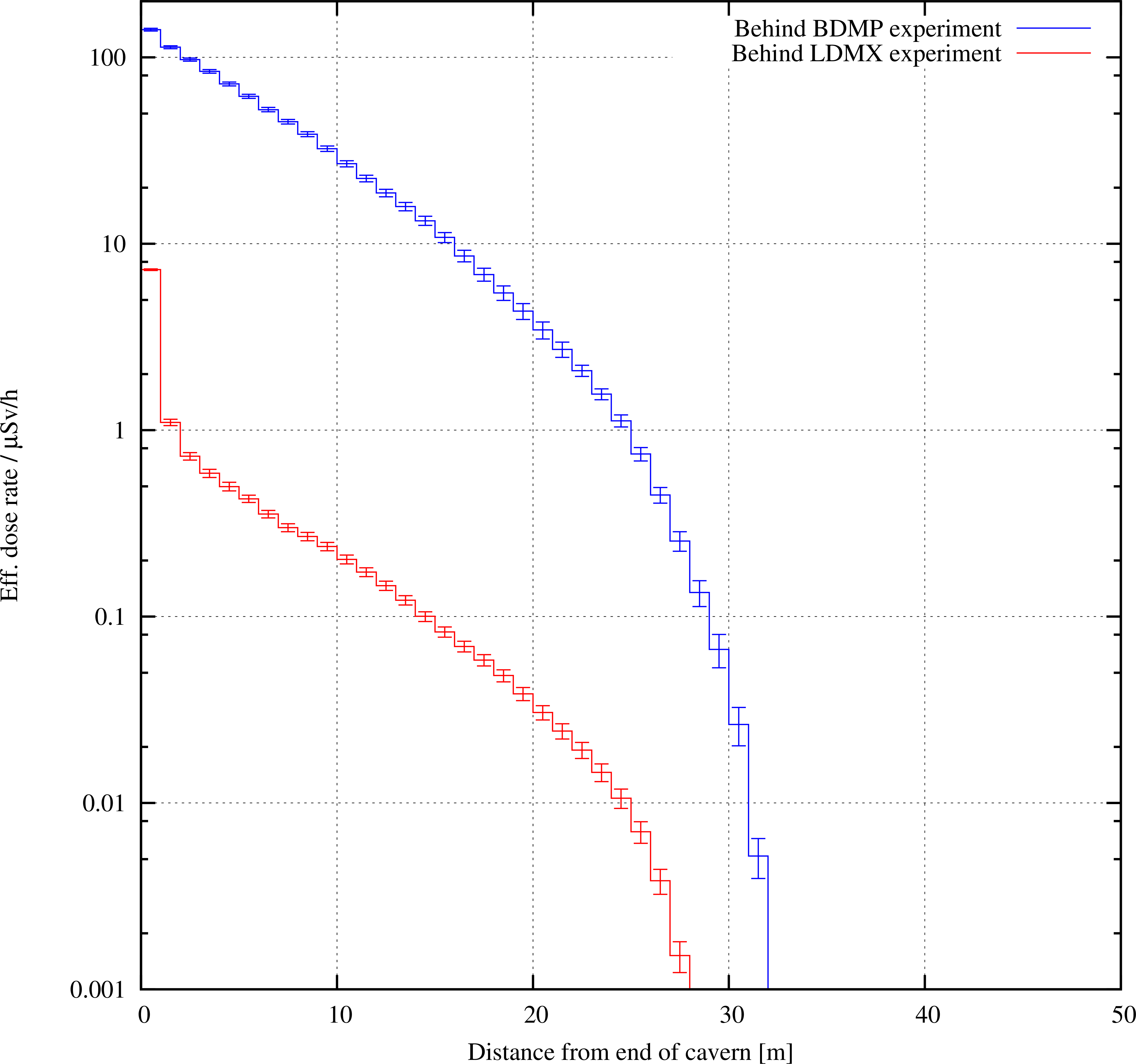}
\caption{Dose rates in beam forward direction behind the missing momentum experiment and beam dump experiment for nominal beam intensity of \SI{5E11}{e^-\per\second}, mainly coming from penetrating muons.}
\label{fig:exp_muons}
\end{figure}

\newpage  
TT1 and the junction with TT6 are located in the forward direction of the beams in the experimental caverns and at the same level. The distance is about \SI{38}{m}. Muons produced by the high-energy electron beam will penetrate and potentially reach the TT1/TT6 tunnel, which is visible in Fig.~\ref{fig:expcav}. The radiation transport calculations show that the radiation levels drop at about \SI{30}{m} distance from the end of the experimental cavern (Fig.~\ref{fig:exp_muons}). Access to TT1/TT6 during beam operation should hence be safe.

\paragraph{Impact from proton operation in TT2}

The experimental cavern shall remain accessible during proton operation in TT2. The dose rate contribution inside the cavern from proton beam operation shall at least remain to Supervised Radiation Area design target values. This is to avoid unjustified exposure of persons inside the experimental cavern whose work is not linked to the proton operation of the accelerator complex.

\begin{figure}[!hbt]
\centering
\includegraphics[width=\textwidth,trim={0cm 0cm 0cm 0cm},clip]{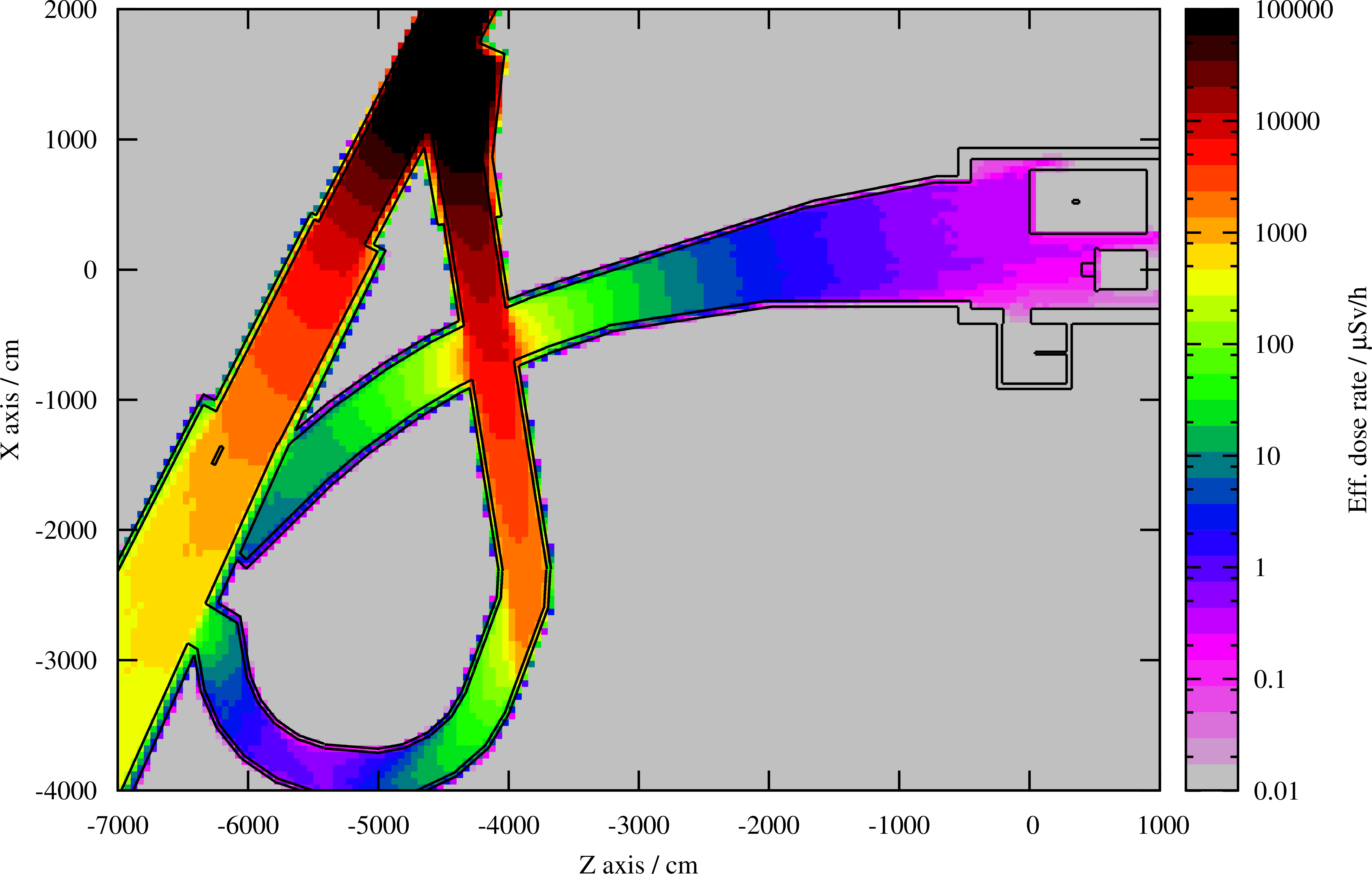}
\includegraphics[width=\textwidth,trim={0cm 0cm 0cm 0cm},clip]{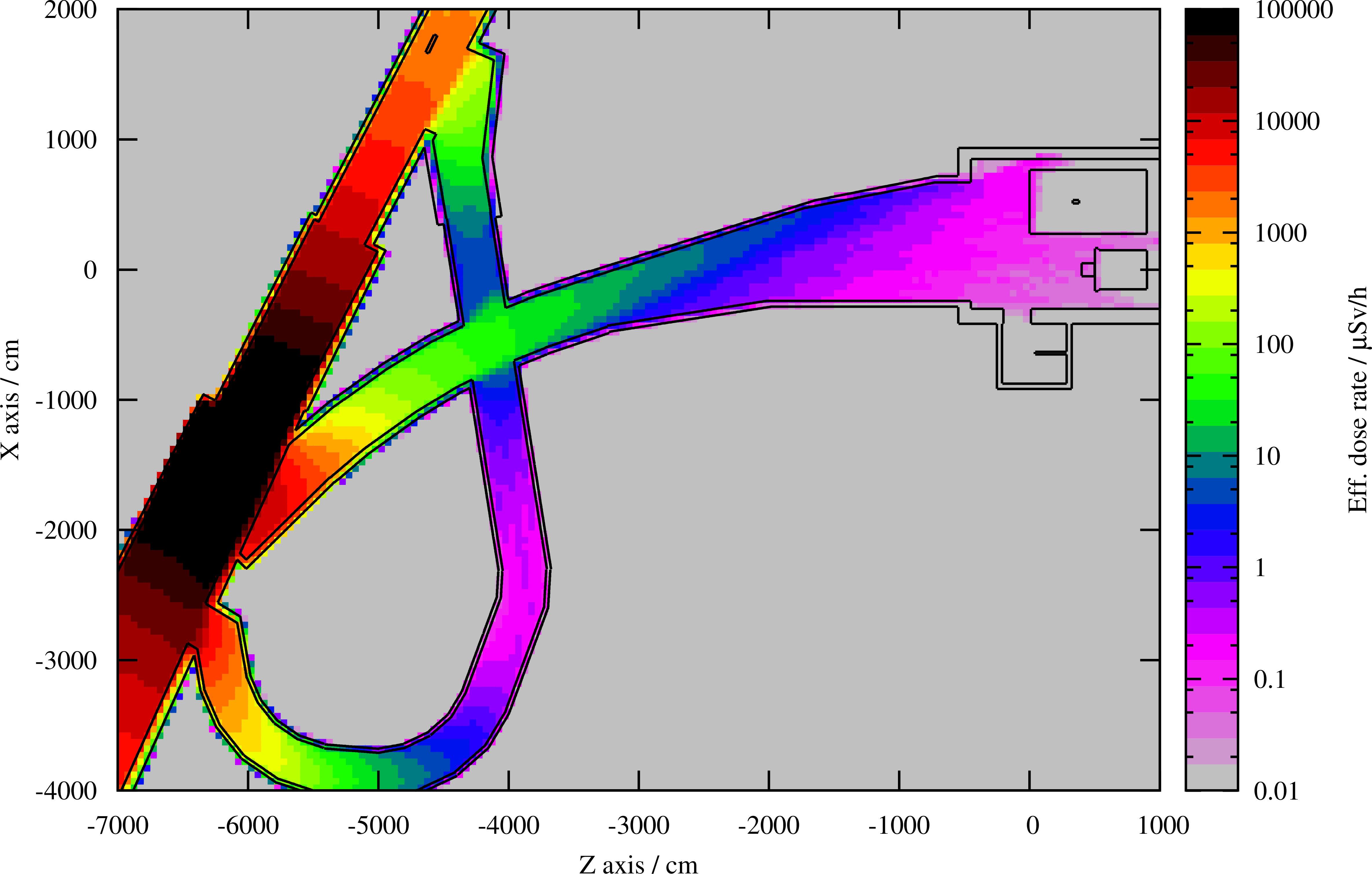}
\caption{Effective dose rates from \SI{14}{GeV} proton operation in TT2 with 2~$\times$~10~W localised beam loss points, corresponding to \SI{4.46E9}{protons\per\second}.}
\label{fig:TT2protons}
\end{figure}

Figure~\ref{fig:TT2protons} shows the effective dose rate impact on the experimental cavern from proton operation in TT2. Two scenarios were considered: Permanent beam operation with \SI{4.46E9}{protons\per\second} (=\SI{10}{W}) point-like losses in TT2 and a full beam loss localised at the most penalising location with \SI{2E13}{protons}, which corresponds to two high-intensity PS cycles destined to the SPS for fixed target operation.

In both scenarios the dose rates and doses are compliant with the design target inside the experimental cavern leading to less than \SI{1}{\micro\sievert\per\hour} and \SI{1}{\micro\sievert} per \SI{2E13}{protons} lost respectively. From actual experience, transmission in TT2 is much cleaner, hence the dose rates should be well within these values. A radiation monitor will be required at the zone delimitation to detect potential proton beam losses while access is possible to the experimental cavern.

\paragraph{Residual radiation} \label{sec:ICE_ExperimentalArea_ResidualRadiation}

The beam intensity on the missing momentum experiment represents only a fraction of the total beam power sent to the experimental area. The majority of electrons and the resulting secondary radiation from the interaction with the target will be absorbed in the massive shielding around the beam dump experiment. 

The residual dose rates in the experimental cavern from beam operation on the missing momentum experiment and the beam dump experiment remain hence at a very low level, well below \SI{1}{\micro\sievert\per\hour}. The expected dose rates after \SI{180}{d} of operation and \SI{1}{d} cool-down are shown in Ref.~\cite{Widorski2020}.

\paragraph{Environmental impact}

As noted in Section~\ref{sec:ICE_ExperimentalArea_ResidualRadiation}, on the low level of activation of the experimental cavern, the activation of air, the principal path of production of radioactivity which could impact the environment, will remain low. Compared to the activation in the linac in TT5/TT4, the levels will be considerably lower in the sub-nanosievert per year range for the committed dose to the reference group.

\paragraph{Radioactive Waste}

Under nominal operation conditions the activation levels in the experimental cavern will be low as shown above. A more detailed study of the activation of the experiments can be conducted once the detailed design and material compositions are known. A full life-time analysis of the experiments is required to estimate volume, mass and production rate of radioactive waste. Such details are expected to be known and addressed during the technical design phase of the project.

\paragraph{Radiation monitoring system}

A small number of detectors will be required to monitor the surroundings of the experimental area. One mixed-field detector will be required at the interface between the experimental cavern and the TT2/TT2L tunnel connection. This detector will be conditioned by the access mode to the experimental cavern. Another mixed-field monitor will be required in the experimental hall on top of the shaft to detect potential beam losses potentially occurring just before the experiment targets.

\subsubsection{Transport and handling}

\begin{figure}[!hbt]
\centering
\includegraphics[width=\textwidth]{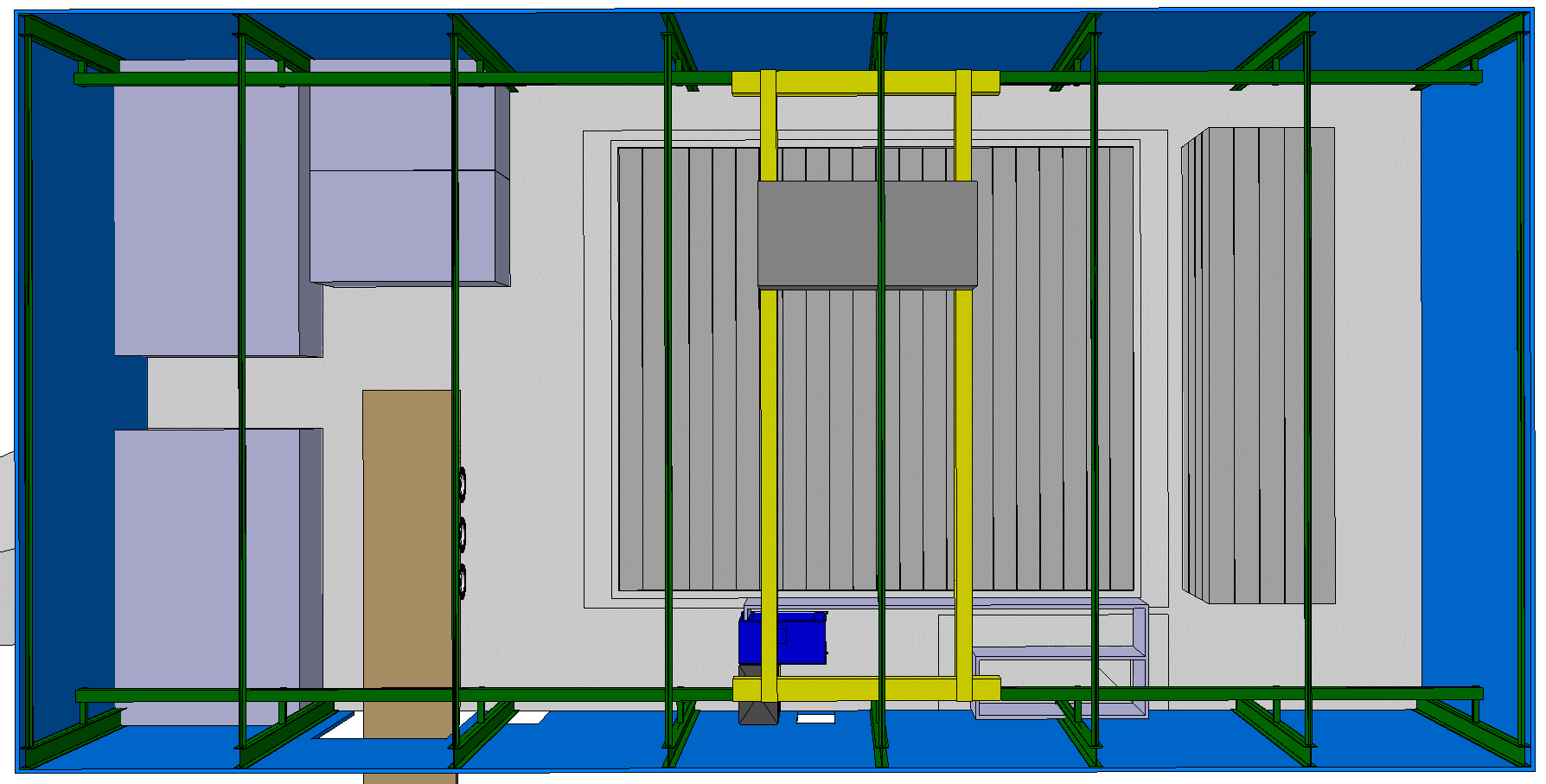}
\captionsetup{width=0.85\textwidth} 
\caption{3D view of the new EOT crane inside the experimental hall.}
\label{fig:EOTCraneExperimntalHall}
\end{figure}

The new building will be equipped with an EOT crane, shown in Fig.~\ref{fig:EOTCraneExperimntalHall}, with the characteristics described in the Table~\ref{tab:EHEOTCraneStats}. The EOT crane will allow the construction, the installation of the detectors in the pit and the lowering of all magnets for the new beam line.

\begin{table}[!hbt]
\centering
 \caption{Experimental hall EOT crane characteristics.}
  \label{tab:EHEOTCraneStats}
\begin{tabular}{cccc}
 \hline\hline
 \textbf{SWL} & \textbf{Span} & \textbf{Lifting height} & \textbf{Power} \\
 \hline
 25\,t & 1500\,mm & 19000\,mm & 50\,kW \\  
 \hline\hline
\end{tabular}
\end{table}

In addition, the building will be equipped with a lift with the characteristics described in Table~\ref{tab:EHEOTLiftStats} below being capable to lower the size of a Euro-palett including the pallet truck.

\begin{table}[!hbt]
\centering
 \caption{Experimental hall lift characteristics.}
  \label{tab:EHEOTLiftStats}
\begin{tabular}{ c  c  c }
 \hline\hline
 \textbf{Capacity} & \textbf{Cabin size} & \textbf{Door width} \\
 \hline
 25\,t & 1500\,mm & 19000\,mm  \\  
 \hline\hline
\end{tabular}
\end{table}

\newpage  
\subsubsection{Safety engineering}

\paragraph{Fire safety}
\subparagraph{Access and egress}
\begin{itemize}
    \item The lift and stairs shall be protected against fire, not connected to the general electrical circuit (i.e., can be used at any time);
        
    \item A safe area, with an over-pressure relative to the surroundings shall be available at the base of the lift and stairs. The size of this area shall be commensurate with the number of occupants, in addition to the time taken for evacuation, and shall be determined as the project moves to the technical design report stage.
\end{itemize}

\subparagraph{Compartmentalisation}
A fire compartment must be established between the new transfer tunnel to the eSPS and the eSPS experimental area. This will be subject to further analysis as the project moves into detailed design;

\subparagraph{Fire suppression means}
A dry riser is foreseen for the experimental area. The full requirements for this equipment are summarised in Ref.~\cite{2017CommonVentilation}.
    
\subparagraph{Smoke extraction}
Hot smoke extraction is foreseen, allowing the CERN Fire and Rescue Service to employ this tool in line with their strategy to tackle a fire.  

\paragraph{Chemical safety}
No chemical agents or gases are currently foreseen for use within the missing momentum experiment; should any additional chemicals be proposed for use in the facility, the chemical specialists within the HSE group must be consulted. 

Lead is not currently part of the shielding design for the experimental area. However, as the design is still at a preliminary stage, it is important to note that lead can present significant hazards. Care must be taken that the necessary procedures are followed for purchasing, shipping, storing and handling of the blocks to limit the dangers of lead poisoning or exposure to activated materials. In particular, blocks should arrive at CERN pre-painted or adequately protected by equivalent means, to ensure that risks from dust are contained. Should lead be required, it must be registered in the CERN Chemical database, CERES, with the quantity, location and hazards recorded. A copy of the up to date SDS must be uploaded into the database and a Chemical risk assessment performed using CERES, if required. The following safety form provides guidance for the safe handling, storage and use of lead:

\begin{itemize}
    \item Safety Guideline C-0-0-3 – Lead.
\end{itemize}

\paragraph{Electromagnetic safety}
The analysing magnet for the missing momentum experiment is intended to be operated with a 1.5 T central field. The restrictions around this magnet are expected to be enforced as a natural result of the radiation levels around the magnet when it is in operation. The expected risk is, therefore, only foreseen for commissioning, testing, and maintenance of the magnet, which shall be mitigated by ensuring that these operations are only carried out by appropriately qualified personnel. This shall be reviewed as the project moves into detailed design.

\paragraph{Preservation of the natural environment}
The mound in which the experimental area is to be located is known to be home to a number of orchid species which are protected under the following regulation; they shall be protected, restored or adequately replaced:

\begin{itemize}
    \item \textit{Code de l’environnement, Art. L411-1}.
\end{itemize}

\subsubsection{Personnel protection system and access control}

A new safety chain dedicated to the eSPS experimental area will be integrated in the personnel protection system of the PS complex.

\begin{figure}[!hbt]
\centering
\includegraphics[width=\textwidth]{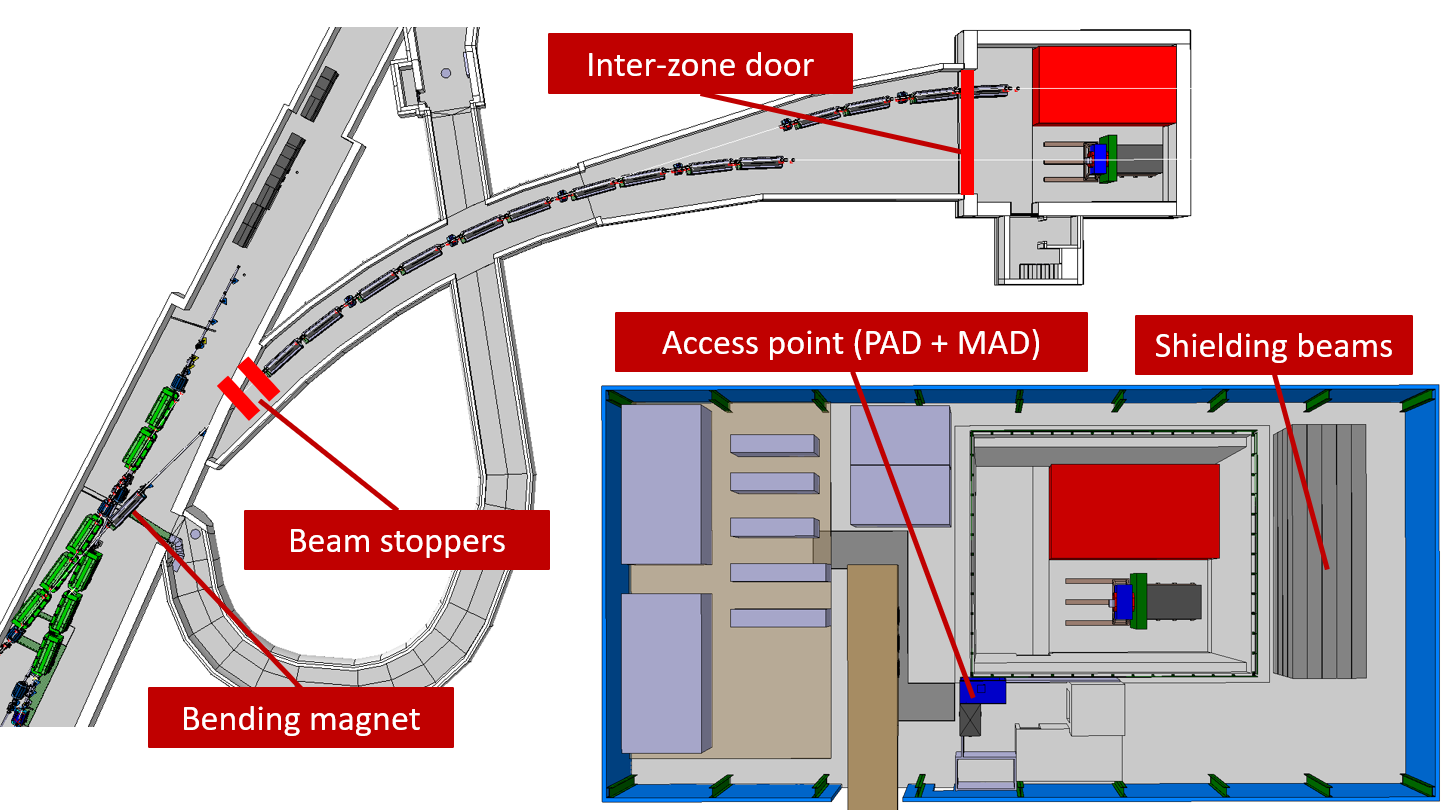}
\caption{Safety elements and access equipment of the eSPS Experimental area.}
\label{fig:Access9}
\end{figure}

The access to the eSPS Experimental area will be made through an access point composed of one personnel access device and one material access device. Aside the access point an `end-of-door' zone is installed to allow the exit of people in case of emergency. The protection of the experimental hall is ensured through a specific shielding on top of the experimental cavern. An operational procedure or a specific interlock shall check the presence of this shielding before authorising beam operation. A new `inter-zone' door will be installed to close the end of the TT2 tunnel before entering the eSPS experimental area. The access control layout is shown in Fig.~\ref{fig:Access9}.

To protect people accessing the eSPS experimental area from radiation hazard, it is proposed to use one bending magnet and two beam stoppers in the TT2 tunnel as EIS-beam. In case of intrusion during beam operation or if the safe position of a safety element is lost during access, the safety chain will also secure the upstream zone: the TT2.

The entrance to the eSPS experimental hall will be equipped with an electrical lock controlled by a badge reader.

%% file: include/06-Acc_RnD/OverviewAcc_RnD.tex
\section{Accelerator facility research and development}
\label{sec:ACCRnD}

\subsection{Overview} 
The proposal for the eSPS accelerator described in Sections~\ref{sec:LINAC} and~\ref{sec:SPS_Transferlines} was initiated as an electron beam facility optimised for a missing momentum experiment to reach the physics goals described in Section~\ref{sec:PhysicsGoals}. However, in addition the \SI{3.5}{GeV} electron linac and the re-introduction of electrons in the SPS provide CERN with completely new possibilities for accelerator project development for its future programme and for general accelerator R\&D. These possibilities are discussed in this chapter. 

During the eSPS operation the linac beam at \SI{3.5}{GeV} is used less than 5\% of the time for injection into the SPS and is available for other uses the rest of the time. Two experimental areas will be available for a broad range of accelerator R\&D using the injector (see Section~\ref{sec:LINAC_GunInjector}) and/or the full linac beam (as for example envisaged for the plasma accelerator R\&D discussed in  Section~\ref{sec:LINAC_Plasma}). A brief introductory guide to the possibilities is given in this section. More details about the different aspects of accelerator and future project development based on the eSPS facility are then described in Section~\ref{sec:ACCRnD_LINAC} to Section~\ref{sec:ACCRnD_Positrons}. 

\paragraph{Studies with relevance for future facilities} \label{sec:acc:int:Future}
The X-band linac and operation of the SPS with electrons open key strategic possibilities for future e$^+$e$^-$ facilities at CERN. First of all, the construction and operation of the linac is in itself a natural next step for the X-band high-gradient technology towards a Compact LInear Collider (CLIC)~\cite{CLIC}. The resources needed are similar to those invested annually during the last decade in the CLIC study. In addition the FCC-ee~\cite{FCC-ee}, LEP3~\cite{LEP3} and LHeC~\cite{LHeC} concepts rely on electron and positron injectors with important challenges that could be addressed with the eSPS facility. In particular the SPS RF system proposed would serve as a prototype for FCC-ee, see Section~\ref{sec:SPS:SRF}. 

Studies of positron production at the end of the X-band linac are an interesting future addition which would benefit any future e$^+$e$^-$ collider, be it linear or circular, inside or outside CERN. For any future e$^+$e$^-$ facility, positron production is one of the most important parts. New ideas being discussed in this context include an implementation of the LEMMA~\cite{LEMMA}) (Low EMittance Muon Accelerator) concept for a muon collider to be implemented at CERN by re-introducing positrons at 45 GeV in the LEP/LHC tunnel. The muon production scheme could in this case be studied in parallel for example with LEP3 operation. While such a phased approach to a muon collider is well beyond the scope of this CDR and also current planning, positron production and target studies will be both relevant and crucial for this possible avenue. The relevance of eSPS to the future facilities mentioned is described in Sections~\ref{sec:ACCRnD_Ring} and~\ref{sec:ACCRnD_Positrons}, while the additional studies made possible by positron production are described in Section~\ref{sec:ACCRnD_Positrons}. 

\paragraph{Plasma acceleration} 
The proposed facility opens the possibility of a significantly broader plasma acceleration programme at CERN, in line with European priorities for this field, primarily addressing key challenges as needed for the potential use of the technology in colliders. Such a programme would naturally build on or be an extension of the existing AWAKE collaboration.
Use of the linac electron beam is considered for plasma acceleration as both driver and probe. As mentioned, positron production studies are of vital interest for any e$^+$e$^-$ machine, as well as new muon collider ideas, but such a positron beam would also provide unique opportunities for studies of plasma acceleration of positrons.  
The relevant possibilities are described in Sections~\ref{sec:ACCRnD_Plasma} and~\ref{sec:ACCRnD_Positrons}. 

\paragraph{General accelerator R\&D} 
The facility will provide a test bed for accelerator R\&D covering a wide range of topics potentially serving an important user community in its own right. Many of the general accelerator studies that can be envisaged are natural continuations of the studies currently carried out in the CLEAR facility at CERN, see Section~\ref{sec:ACCRnD_CLEARER}. Examples of potential R\&D studies include, high gradient and plasma lens studies, instrumentation and impedance studies, medical accelerator developments for example for VHEE~\cite{VHEE} irradiation, component irradiation, THz acceleration, and educational activities. With the eSPS linac parameters, the capabilities are significantly increased with respect to what can be done today in the CLEAR facility. The possibilities are described in Section~\ref{sec:ACCRnD_CLEARER}. 
 
\paragraph{The SPS electron beam} \label{sec:acc:int:ebeam}   
The use of the SPS electron beam is also possible but will be in competition with other users of the SPS. For the SPS beam one can take advantage of small equilibrium emittance optics using the existing SPS lattice. Low emittance ring studies and the use of SPS as a damping ring can be pursued, as have been considered for both linear and circular future e$^+$e$^-$ machines at CERN. One could also consider pursuing final focus studies beyond the ATF2~\cite{ATF2} if this becomes a priority. A realistic implementation of the latter is not studied at this point. The possibilities offered by the SPS electron beam are briefly outlined in Section~\ref{sec:ACCRnD_Ring}.

\vspace{\baselineskip}

Overall, the facility can significantly extend the strategic possibilities for future machines at CERN and also provide a platform to carry out important studies for future facilities outside CERN as part of a European contribution to such facilities. 
The possibility of achieving this at a resource level which is compatible with that already existing, profiting of the investment made in the last decades for CLIC, and electrons in the SPS during the LEP era, while performing physics studies that are essential to pave the way towards future larger machines, is very attractive and a unique opportunity for the organisation at this time.  
 
\paragraph{The accelerator community involved as developers or users}

The scientific and technical community capable of contributing to and interested in the X-band machine development and construction, as well as the accelerator user community, are both large. 
The potential links between the eSPS facility and X-band accelerator developments outside CERN are many, in particular the on-going design study for an X-band based FEL, CompactLight~\cite{Latina2018a}, and the collaboration with INFN-LNF for building a 1 GeV X-band linac~\cite{Ferrario2018} as part of the LNF EuPRAXIA~\cite{Diomede:IPAC18} efforts. 
These ongoing developments already provide a network of 25 collaboration partners, many of which are developers and users of X-band technology in their local facilities. 
One can also expect many additional CLIC collaboration partners to actively direct their collaboration efforts towards studies and technology developments directly applicable to the eSPS linac. 
As shown above, it is not only the CLIC collaboration partners that will engage in this facility, AWAKE~\cite{AWAKEacc} collaboration partners and CLEAR users would be ready to pursue the facility build-up and its scientific programme in the area of accelerator R\&D. Both these communities are large, in particular the very large novel accelerator technology community consider test-facilities at CERN crucial for making these technologies applicable for high energy colliders. The relevance of the facility for FCC-ee RF and more general circular e$^+$e$^-$ accelerator studies, as well as possible studies for novel muon colliders, bring in additional groups of potential users.  

%% file: include/06-Acc_RnD/Linac_RnD.tex
\subsection{Linac related research and development}
\label{sec:ACCRnD_LINAC}

\subsubsection{The Compact Linear Collider}

The project implementation plan for CLIC foresees an initial 5 year preparation phase prior to a potential construction start. The preparation phase will focus on further technical and industrial developments as well as production and preparation of key components, system verifications (not necessarily at CERN), strategic developments focused on risk, cost and power reduction, as well as developments towards the technical proposal for the detector. The governance structure and the international collaboration agreements for the construction and operation will be set up during this time. Site authorisations will also be established during this period and site/civil engineering and infrastructure preparation will become increasingly detailed. All of these considerations will be folded into the final design and parameters for the first CLIC stage. 

The technical developments needed for the preparation phase of CLIC have a large overlap with what is needed for the X-band linac of the eSPS facility described in Sections~\ref{sec:LINAC}--\ref{sec:ICE}. In Table~\ref{tab:CLICprototype} the key CLIC accelerator programme for the next phase are considered in view of the technical overlap with the construction of an 3.5 GeV X-band linac for eSPS.

\begin{table}[!hbt]
    \centering
    \caption{Main CLIC related activities and their relation to the 3.5 GeV linac for eSPS.}
    \label{tab:CLICprototype}
    
    \begin{tabularx}{\linewidth}{L L L L}
    \hline
    \hline
    \textbf{Details} & \textbf{Purpose} & \textbf{eSPS Equivalent} & \textbf{Comment} \\
    \hline
    \multicolumn{4}{l}{\textbf{Main linac modules}}\tabularnewline
    \hline
    Build ten prototype modules in qualified industries, two beam and klystron versions & Final technical design, qualify industry partners, verify performance & 12 X-band klystron modules & Covered by eSPS but adaptations to two beam modules need to be considered \\
    \hline
    \multicolumn{4}{l}{\textbf{Accelerating structures}}\tabularnewline
    \hline
    Around 50 structures incl. for modules above & Industrialisation, manufacturing and cost optimisation & Same number needed & Programmes overlapping \\
    \hline
    \multicolumn{4}{l}{\textbf{Operating X-band test-stands, high efficiency RF}}\tabularnewline
    \hline  
    X-band test-stands at CERN and collaborating institutes, cost optimised X-band RF & X-band component test, validation and optimisation, cost reduction and industrially available RF units & Similar test capacity needed for eSPS, 24 X-band RF units needed for eSPS & Programmes overlapping \\
    \hline
    \multicolumn{4}{l}{\textbf{Technical components}}\tabularnewline
    \hline  
    Magnets, instrumentation, alignment, stability, vacuum & Luminosity performance, costs and power, industrialisation & These components are also needed for eSPS & eSPS specifications less stringent, however significant advantage to implement in smaller complete system \\
    \hline
    \multicolumn{4}{l}{\textbf{Design \& Parameters}}\tabularnewline
    \hline  
    Beam dynamics studies, parameter optimisation, costs, power & Luminosity performance, risk, costs and power reduction & Needed for eSPS linac & Specific studies for CLIC needed but good reality check \\
    \hline
    \hline
    \end{tabularx}
\end{table}

Additional studies are needed for CLIC, the most prominent is the drivebeam front end optimisation and system tests to around 20 MeV. The purpose is a careful verification of the most critical parts of drivebeam concept and to develop the industrial capabilities for L-band RF systems as needed for the CLIC drivebeam. Another area where the CLIC studies do not overlap is in the area site and civil engineering studies. These activities are, however, not expected to be resource consuming until closer to the beginning of construction when increased effort will be needed. Furthermore, system tests in low emittance rings, FELs, etc. should be pursued and are relevant for eSPS and CLIC. These tests are not resource intensive and are usually carried out in collaboration with outside institutes.   

The programme outlined for the preparation phase of CLIC above, 
has, for the most resource demanding parts, a large overlap with constructing a 3.5~GeV linac for eSPS. Hence, from the CLIC development point of view, constructing a 3.5~GeV X-band linac for the eSPS facility as a next stage is a very attractive possibility. It enables important physics and accelerator studies of vital interest for CERN, and with limited additional CLIC specific studies, also moves the CLIC accelerator preparation efficiently and in a timely manner towards a more technically and industrially mature state. 

In addition, collaboration with external groups, for example INFN for the SPARC 1~GeV X-band linac, a potential energy upgrade of the Clara facility at Daresbury and future CompactLight FEL implementations, provide a network of important external collaborators that need a linac module and components on a similar timescale as that of the eSPS. More generally, the CLIC collaboration partners would feel that these concrete construction projects, as well as R\&D opportunities offered by the linac, would provide them with completely new possibilities for contributions and participation in the coming phase(s).  

\subsubsection{The International Linear Collider}
The ILC in Japan is being considered for implementation in the coming decade~\cite{ILC}. The ILC is a 250~GeV e$^+$e$^-$ linear collider based on super-conducting RF technology. In the scenario of the ILC being constructed in the near future it is expected that Europe will make a contribution to the project. Possible contributions from CERN in relation to the eSPS include a CERN platform for injector, positron production, stability, alignment and high efficiency RF studies. Longer term there will be a push to increase the ILC energy and luminosities, and CERN along with its collaborators would be in a good position to provide testing grounds for such developments using the eSPS installations. In addition, the damping ring studies mentioned above, possibly extracting a low emittance beam for final focus studies, are relevant in this scenario. It it scientifically very attractive to combine an in-kind activity at CERN towards the development of the ILC with a strong R\&D effort for higher energies and higher luminosities for the ILC, as well as any future linear collider. 

%% file: include/06-Acc_RnD/Ring_RnD.tex
\subsection{Ring-related R\&D}
\label{sec:ACCRnD_Ring}

Circular Machine studies at CERN such as FCC-ee~\cite{FCC-ee} and LEP3~\cite{LEP3} would rely on cost-effective electron and positron production and injection. FCC-ee studies are currently an important R\&D activity, providing an initial option for a future ring based accelerator facility at CERN. The RF system proposed for eSPS is a prototype of an FCC-ee RF system as discussed in Section~\ref{sec:SPS:SRF}. Design, construction and operation of such a system can provide important experience in all the aspects mentioned above. 
Beyond these two straightforward considerations, more elaborate possibilities exist. The possible use of the SPS as a damping ring, building on the LEP experience and expanding on the studies made in the context of CLIC and modern low emittance rings, for these machines have been described in Ref.~\cite{Papaphilippou:1588122}. Studies considered in this context are tests of super-conducting wigglers, kickers, vacuum/coating, instrumentation, beam profile monitors (synchrotron light), halo monitoring, BPMs,  bunch-by-bunch and turn-by-turn feedback (LARP), RF systems (800 MHz cavities in particular), beam dynamics issues related to optics, non-linear dynamics, intra-beam scattering, instabilities, e$^-$ cloud (for e$^+$) and ions (for e$^-$).

%% file: include/06-Acc_RnD/Plasma_RnD.tex
\subsection{Plasma-based acceleration research and development}
\label{sec:ACCRnD_Plasma}

\subsubsection{Introduction}
\begin{figure}[!hbt]
\begin{center}
\includegraphics[width=12cm]{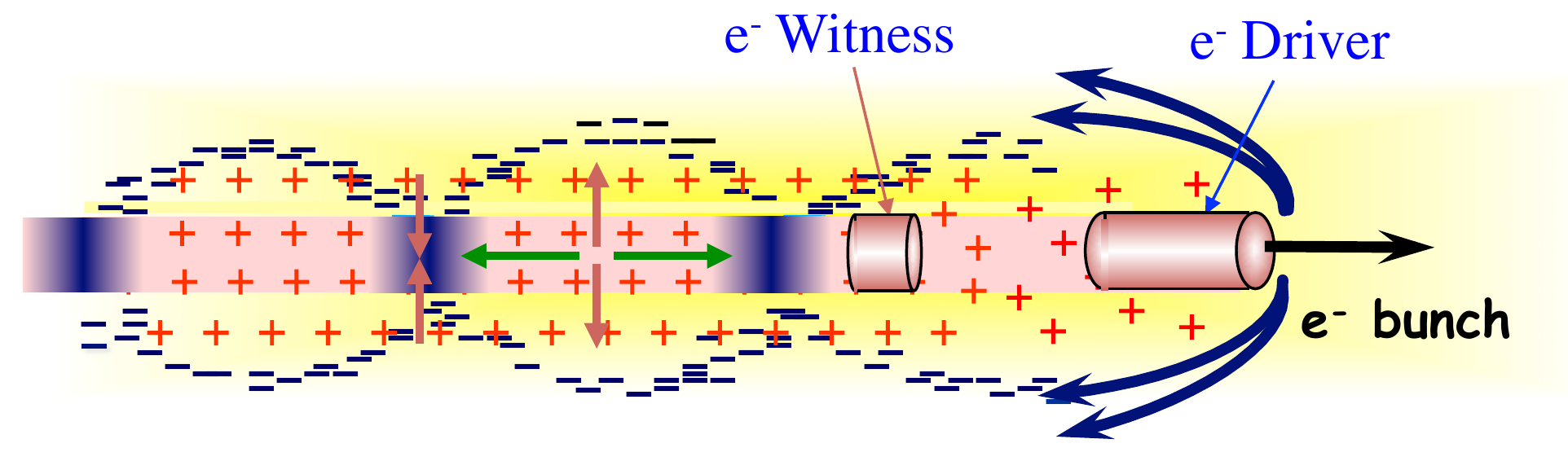}
\caption{Schematic of the PWFA: the drive electron bunch expels the plasma electrons that are attracted back towards the axis by the positive charge of the ion column and overshoot (blue arrows in the bunch frame of reference). %
The plasma electron density perturbation sustains the longitudinal wakefields that are decelerating inside the drive bunch (green arrow in the second bubble) and accelerating within the witness bunch. %
It also sustains transverse focusing fields inside the bubble (red arrows).
}
\label{fig:PWFAfig}
\end{center}
\end{figure}

Electron beam driven plasma-based accelerators, known as plasma wakefield accelerators (PWFA) (see Fig.~\ref{fig:PWFAfig}), have shown accelerating gradients in excess of 50\,GeV/m sustained over a metre-long plasma leading to a 42\,GeV energy gain~\cite{Blumenfeld2007} and the acceleration of a witness electron bunch with narrow energy spread (few \%-level) and good energy transfer efficiency ($\sim$\,30\%) with beam loading~\cite{Litos2014}. %
However, there are issues that need to be addressed for a successful PWFA based collider. These include:
\begin{itemize}
\item Preservation of the incoming emittance of the accelerated bunch; %
\item Acceleration with a narrow final relative energy spread (\%-level) using beam loading. %
This also contributes to the preservation of the beam emittance;%
\item Matching of the beam to the plasma strong focusing, including plasma density ramps at the entrance and exit of the plasma; %
\item Possibility of shaping the drive bunch to reach a transformer ratio larger than two, the current maximum reachable, with a symmetric current profile bunch; %
\item Operating with a plasma source with characteristics suitable to preserve emittance and energy spread (density uniformity). %
\end{itemize}

The use of the eSPS facility for PWFA studies could directly address some of the challenges related to the first electron acceleration PWFA stage of a possible linear collider. %
They are related to the acceleration and the quality of the electron witness bunch. %
Though the energy of the drive bunch (3.5\,GeV) is lower than that envisaged for such a collider stage ($\sim$\,25\,GeV), the challenges are not (or weakly) energy dependent. %

An experimental facility with a multi-GeV electron beam capable of driving wakefields in the non-linear regime and with an independent electron witness bunch is necessary to demonstrate the applicability of the concept to HEP. %
Challenges related to the quality of acceleration of a positron witness bunch will be addressed if/when positron bunches become available. %

Such a facility for PWFA experiments would be unique in the world to study collider related challenges. %
It may be the only facility in the world with a multi-GeV drive bunch and truly independent electron witness bunch, absolutely necessary features to tackle these challenges. %
The addition of a positron witness bunch would make the eSPS the facility for PWFA collider studies.

\subsubsection{General beam and plasma parameters requirements}
\label{sec:beam_and_plasma_parameters_requirements}

Considering the potential eSPS linac drive bunch energy of 3.5\,GeV, the beam and plasma parameters necessary to conduct a PWFA experiment in the non-linear regime can be determined from the following "recipe" for the plasma and drive bunch parameters. %
Reaching the blow-out regime is necessary to be able to preserve emittance and narrow relative energy spread over long plasma lengths and in the presence of large energy gains. %

One can target full energy depletion over a metre scale plasma, i.e.\, an accelerating field of $\sim$\,3\,GV/m. %
The maximum accelerating field is of the order of the wave breaking field,
\begin{equation}
    E_{WB}=m_ec\omega_{pe}/e\approx96\sqrt{n_{e0}[\textrm{cm}^{-3}]}  \textrm{[V/m]}.
\end{equation}
\newline 
Here $\omega_{pe}=\left(\frac{n_{e0}e^2}{\epsilon_0m_e}\right)^{1/2}$ is the plasma electron angular frequency in a plasma of electron density n$_{e0}$. %
This determines n$_{e0}$. %
To effectively drive wakefields, the rms length of the drive bunch $\sigma_z$ must be of the order of the cold plasma skin depth $c/\omega_{pe}$, which is expressed as: $\sigma_z/\left(c/\omega_{pe}\right)\approx\sqrt{2}$. %
The transverse drive bunch rms size $\sigma_r$ must also be smaller than  $c/\omega_{pe}$ to avoid filamentation of the bunch: \mbox{$\sigma_r/\left(c/\omega_{pe}\right)<1$}. %
In order to reach $E_{WB}$ and the nonlinear regime of the PWFA, the bunch density $n_b=\frac{1}{(2\pi)^{3/2}}\frac{N}{\sigma_z\sigma_r^2}$ must exceed n$_e$: $N>(2\pi)^{3/2}\sigma_z\sigma_r^2n_{e0}$. %
This "recipe" yields the parameters listed in Table~\ref{tab:recipe}. %
\begin{table}[!hbt]
\begin{center}
\caption{Parameters obtained from PWFA "recipe" for the plasma and drive bunch.}
\label{tab:recipe}
\begin{tabular}{p{6cm}cc}
\hline\hline
\textbf{Parameter}             & \textbf{Symbol} & \textbf{Value}\\
\hline
Bunch electrons energy   & $W_0$             & 3.5\,GeV \\
Wave breaking field amplitude     & E$_{WB}$             & 3\,GV/m \\
Plasma electron density     & n$_{e0}$             & $9.7 \times 10^{14}$\,cm$^{-3}$ \\
Plasma length     & L$_p$             & 1\,m \\
Bunch rms length     & $\sigma_{z}$             & 241\,$\mu$m \\
Bunch rms radius     & $\sigma_{r}$             & 174\,$\mu$m \\
Bunch population     & N             & $> 1.1 \times 10^{10}$ \\
Bunch charge     & Q             & $> 1.7$\,nC \\
\hline\hline
\end{tabular}
\end{center}
\end{table}
The bunch emittance and energy spread must be sufficiently low for the beam to be focused to the specified transverse size at the plasma entrance. %
This mostly depends on the final focus system characteristics. %
These last two parameters play no direct role in driving wakefields. %
However, lower emittance limits the erosion of the drive bunch head. %

We note here that the bunch parameters of Table~\ref{tab:recipe} are consistent with what can be produced by the eSPS linac (Section~\ref{sec:LINAC}). %

\subsubsection{Witness bunch}

Because the plasma density envisaged here (n$_{e0} = 10^{15}$\,cm$^{-3}$) is similar to that used in the AWAKE experiment, witness bunch parameters can be similar. %
Therefore, the design of the linac that produces this bunch can be similar to that for AWAKE. %
The witness bunch can be of lower energy than the drive bunch. %
An estimate for its minimum energy is obtained by requiring that with its incoming energy it does not de-phase with respect to the wakefields by more than a quarter plasma wavelength (i.e.\, it remains in the accelerating and focusing phase of the wakefields) over the plasma length. %
With a 1\,m-long plasma with density 10$^{15}$\,cm$^{-3}$ (plasma wavelength of 530\,$\mu$m), the incoming witness bunch relativistic factor must exceed 31. %
This corresponds to an energy of 16\,MeV. %

\begin{figure}[!hbt]
\begin{center}
\includegraphics[width=6cm]{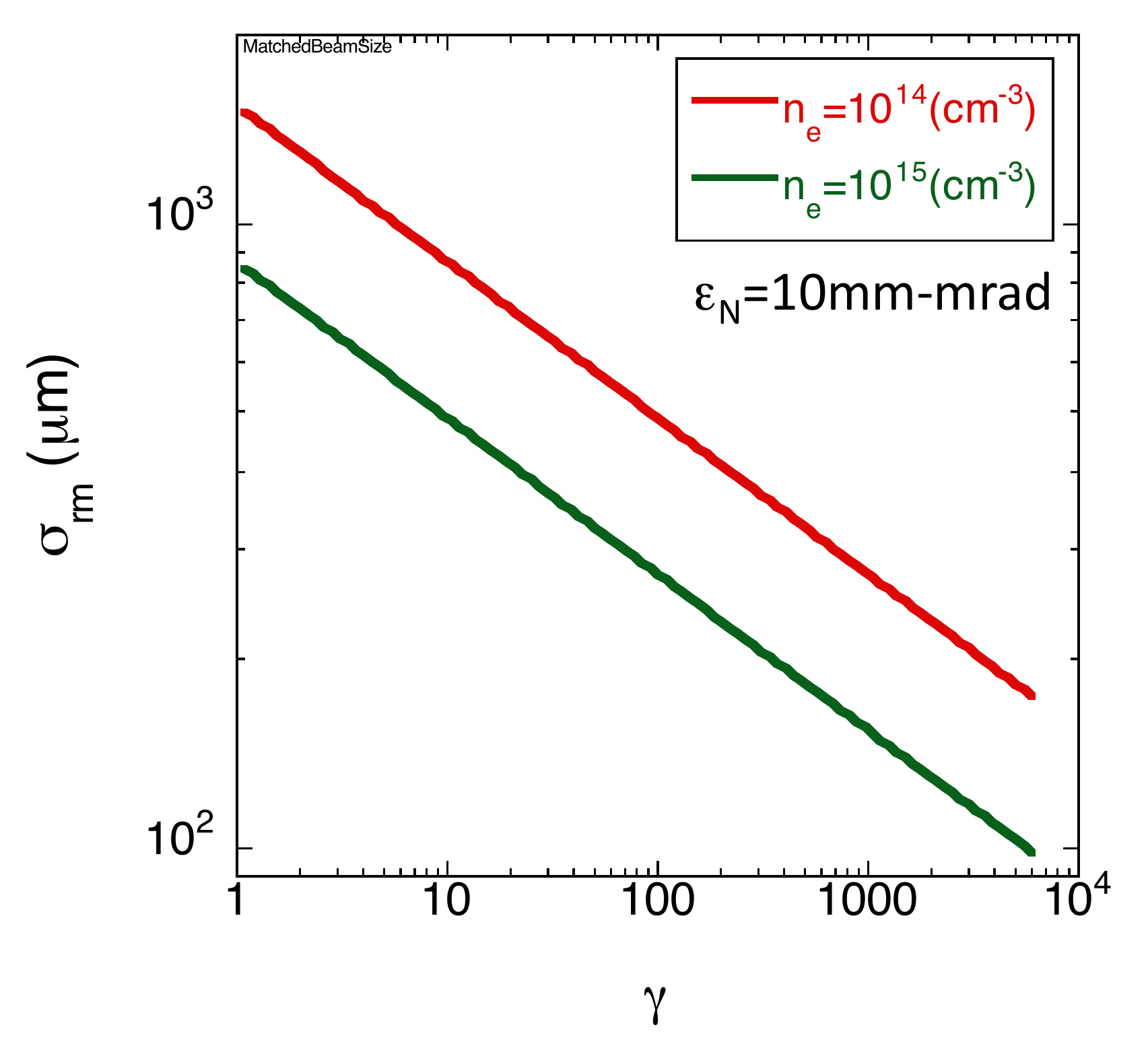}
\caption{Transverse bunch size $\sigma_m$ matched to the pure plasma ion column as a function of the beam relativistic factor $\gamma$ for two plasma densities n$_{e0} =10 ^{14}$ and 10$^{15}$\,cm$^{-3}$ and a normalised emittance $\epsilon_N = 10$\,mm.mrad.
}
\label{fig:PWFAmatching}
\end{center}
\end{figure}

Its normalised emittance $\epsilon_N$ may be small enough to match the witness bunch to the ion column focusing force. %
In this case, the bunch and plasma parameters must satisfy: $\sigma_m=\left(\frac{1}{2\pi r_e}\frac{\epsilon_N}{\gamma n_{e0}}\right)^{1/4}$, $\sigma_m$ being the matched transverse size. %
Here r$_e$ is the classical electron radius. %
Figure~\ref{fig:PWFAmatching} displays the beam matched radius as a function of the beam relativistic factor $\gamma$ for two plasma densities. %
It shows that for a low energy witness bunch ($\gamma \approx$\,100), the matched radius at the plasma entrance is between $\approx$\,250 and 400\,$\mu$m. %
Upon (adiabatic) acceleration, the transverse size reduces to less than 200\,$\mu$m at the final energy ($\gamma \approx$\,7000 plasma exit). %
The witness bunch energy is, therefore, determined by the ability to focus it to its matched size at the plasma entrance, rather than by the trapping energy, which is very low. %

The witness bunch charge must be a fraction of the drive bunch charge. %
Together with the charge, the bunch length must be optimised for beam loading and minimum final energy spread. %
Precise delaying and alignment between the drive and the witness bunch must be possible. %

\subsubsection{Witness bunch source}

Plasma and wakefield parameters are similar to those used in the AWAKE experiments. %
Witness bunch parameters will, therefore, be similar. %
AWAKE studies showed that a witness bunch length of the order of $\sim$\,5\% of the wakefield's period is suitable, 42\,$\mu$m (or 140\,fs) for parameters of Table~\ref{tab:recipe}. %
For matching of the bunch to transverse wakefields, its beta function must be of the order of 5\,mm. %
This then leads to a transverse beam size at the waist, located at the plasma entrance, in the order of a few to a few tens of microns. %
Producing such small and short bunches carrying the hundreds of picocoulombs necessary to load wakefields and limit the final relative energy spread requires relativistic electrons. %
Experience shows particle energy must be in the region of 150\,MeV. %
The injector could thus be similar to that which will used for AWAKE, CLEAR and EuPRAXIA@SPARC\_LAB (see Section~\ref{sec:LINAC_Plasma}). %
It consists of an S-band gun, followed by an X-band structure for velocity bunching and one or two X-band accelerating structures. %
With two accelerating structures the energy can reach 160\,MeV. %
The witness beam produced in this accelerator needs to be merged with the 3.5\,GeV drive beam. %
An achromatic and synchronous dog-leg was developed for AWAKE. %
A similar design could be adopted to bring the bunch from the low energy witness bunch linac to the plasma entrance. %
Integration of the witness injector is described in Section~\ref{sec:LINAC_Plasma} and its integration in Fig.~\ref{fig:Second_injector_layout}. %
Typical parameters are given in Table~\ref{tab:plasmainjectorpar}. %

\subsubsection{Simulation results}

At this point, numerical simulations must be used to determine the parameters more precisely. %
This includes drive and witness bunch parameters. %
In particular, the drive bunch needs to be shorter to accommodate the witness bunch in the accelerating cavity and the witness bunch parameters need to be optimised to reach optimum beam loading. %

\begin{figure}[!hbt]
\begin{center}
\includegraphics[width=12cm]{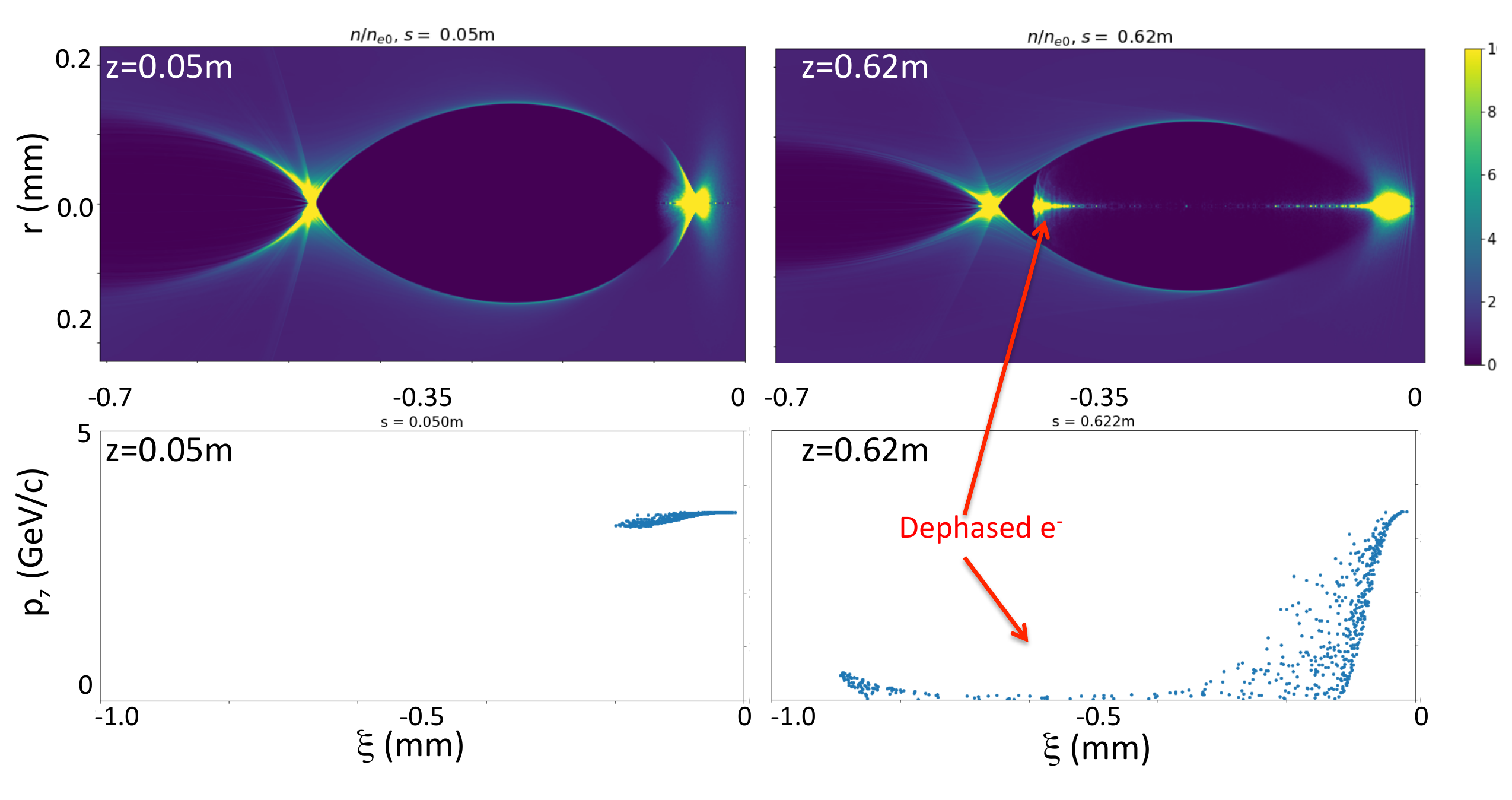}
\caption{
Top panels: snapshots of the drive bunch ($-0.1 \le \xi \le \,0.0$\,mm) and plasma densities (no witness bunch, colour scale for n$_e$/n$_{e0}$) at two locations along the plasma (z~$=5$ and 62\,cm). %
Bottom panels: corresponding snapshots of the drive bunch electrons longitudinal momentum. %
The simulation parameters used were: n$_{e0} = 5.6 \times 10 ^{15}$\,cm$^{-3}$, $\sigma_z = 100\, \mu$m, $\sigma_r =70\, \mu$m, N~$=4.3 \times 10^{10}$ (n$_b$/n$_e =1$), thus E$_{WB} =7.24$\,GeV/m. %
}
\label{fig:WakeFieldsNoWB}
\end{center}
\end{figure}

Figure~\ref{fig:WakeFieldsNoWB} shows snapshots of the drive bunch ($-0.1 \le \xi \le 0.0$\,mm, top panels) and plasma electron density at two locations along the plasma. %
In this case the beam and plasma parameters are chosen to reach the blow-out or bubble regime of the PWFA and for the drive bunch to be shorter than the bubble size. %
Near the entrance (z~$=0.05$\,m, top-left panel) the drive bunch (moving to the right, as in Fig.~\ref{fig:PWFAfig}) has been focused by its own wakefields and drives a clear bubble structure in the plasma electron density. %
The bottom-left panel shows that bunch electrons start losing energy driving wakefields. %
At about z~$=0.62$\,m into the plasma, the electrons at the back of the drive bunch have lost almost all their energy  (see bottom-right panel) and have de-phased towards the back of the bubble ($\xi \approx$\,$-0.40$\,mm, top-right panel). %
This is evidence of energy depletion of the drive bunch. %

The detailed parameters of the drive and witness bunch will be determined from similar simulations with bunch parameters consistent with the capabilities of the injector(s), as described for example in Section~\ref{sec:LINAC}. %
We note here that the PWFA requires short bunches, which seem to be less sensitive than longer bunches for example to misalignment (see 
Section~\ref{sec:ACCRnD_Positrons}). %
The drive electron bunch parameters can also be adjusted to drive wakefields in other regimes (e.g.\ quasi-linear, hollow channel) for example for acceleration of a positron bunch on the wake driven by the electron bunch (see Section~\ref{sec:ACCRnD_Positrons}). %

\subsubsection{Plasma source}

The plasma parameters of Table~\ref{tab:recipe} call for either a laser-ionised alkali metal vapour source~\cite{774685}, or for a discharge source. 
Each source has its pros and cons. %
Both require isolating vacuum windows or differential pumping. %
The vapour source is attractive because it is a passive and very reproducible system that can be finely tuned, but it requires an ionising laser system (excimer for lithium or Ti:sapphire for rubidium). %
The discharge source operates with a noble gas (argon), which simplifies operation, but it requires a (rather simple) discharge system. %

\subsubsection{Plasma physics experimental area}

Reaching drive bunch parameters suitable for the PWFA experiments requires further compression of the linac bunch by a chicane compressor. %
The main components of the experiment itself are a final focus system, the plasma source (metre-long), a magnetic imaging spectrometer capable of measuring up to $\sim$\,7\,GeV electron (and positron) energies, as well as a number of specific bunch diagnostics: optical transition radiation (OTR) imaging stations (before and after the plasma), single- and multi-shot emittance measurements, X-ray betatron radiation measurements, as well as standard beam diagnostics (BPMs, ICT, etc.). %
The experiment can accommodate both electron and positron witness bunches. %
The facility could be located at the end of the X-band linac, see Fig.~\ref{fig:Second_injector_layout}.

\subsubsection{Conclusion}

A PWFA experimental facility driven by the 3.5\,GeV electron bunch would play an essential role towards the development of a beam-driven, plasma-based, more compact and affordable linear collider, by focusing on collider-specific issues. %
Such a facility would most likely be the only one dedicated to this important topic. %
The availability of a positron witness bunch (see Section~\ref{sec:ACCRnD_Positrons}) would make it a true and complete plasma-based collider research facility. %
This facility would also be complementary to the high energy CLEARER facility. %
Possible extensions could include multiple two or more drive bunches and plasma sections to address the staging challenge. %

%% file: include/06-Acc_RnD/CLEARER.tex
\subsection{Proposed facility: CLEARER}
\label{sec:ACCRnD_CLEARER}
\subsubsection{Overview of CLEAR facility}
CLEAR~\cite{CLEAR,Sjobak2019} is a user facility at CERN, running in parallel with the main CERN accelerator complex, with the primary goal of enhancing and complementing the existing accelerator R\&D and testing capabilities at CERN. A workshop on the conversion of the probe beam line of the former CLIC Test Facility (CTF3) into a new test-bed was held in October 2016 with participation from 80 people covering a broad science community. The scientific and strategic goals set out were the following: 
\begin{itemize}
\item Providing a test facility at CERN with high availability, easy access and high quality bunched electron beams;
\item Performing R\&D on accelerator components, including beam-based impedance measurements, innovative beam instrumentation prototyping and high gradient RF technology advancement with realistic beam tests;
\item Providing a radiation facility with high-energy electrons, e.g.\ for testing electronic components in collaboration with ESA or for medical purposes, possibly also for particle physics detectors;
\item Performing R\&D on novel accelerating techniques - electron driven plasma acceleration and THz acceleration. In particular developing technology and solutions needed for future particle physics applications, e.g.\ beam emittance preservation for reaching high luminosities;
\item Maintaining CERN and European expertise for electron linacs linked to future collider studies (e.g.\  CLIC and ILC, but also AWAKE), and providing a focus for strengthening collaboration in this area;
\item Using CLEAR as a training infrastructure for the next generation of accelerator scientists and engineers.
\end{itemize}
The CLEAR facility was approved in December 2016 with the first beam being set up in August 2017.  After only a few weeks of commissioning, stable and reliable electron beams with energies between 60 and 220 MeV in single or multi bunch configuration at 1.5 GHz could be provided to users.

\paragraph{Irradiation studies}
Irradiation tests are being performed mainly in a dedicated spectrometer beam line (Very energetic Electron facility for Space Planetary Exploration missions in harsh Radiative environments - VESPER). The initial aim of VESPER was to characterise electronic components for operation in a Jovian environment - as foreseen in the JUpiter Icy Moon Explorer mission (JUICE) of the European Space Agency (ESA), in which trapped electrons of energies up to several hundred MeV are present with very large fluxes. Measurements in VESPER showed the first experimental evidence of electron-induced single event upsets (SEU) on electronic components in this kind of environment, thus justifying further studies. A dependency of SEU cross-section with energy and no dependency on radiation flux was observed, excluding prompt dose effects~\cite{VESPER_1}. Since then, a wider range of devices have been tested, showing a strong dependency on the device process technology. 

Interest in further measurements, both at low and high beam intensities and over a wide energy range, is quite strong in the space community. A contact with NASA has recently been established and the possibility of a dedicated measurement is being explored. In parallel, the local CERN group studying radiation damage of electronics components in an accelerator environment is interested in the continued use of the facility.  A first request for irradiation studies on electronics components for detector applications was also received from Uppsala University, and the first measurements have been completed.  Other internal CERN groups have investigated the effect of radiation on components for various detector materials and components. 

\paragraph{Medical accelerator studies}
The scope of VESPER has been further extended to medical applications.  The recent advances in compact high-gradient accelerator technology, largely prompted by the CLIC study, renewed the interest in using very-high energy electrons (VHEE) in the 50--250 MeV energy range for radiotherapy of deep-seated tumours. In order to understand the dosimetry of such beams and assess their viability for treatment, a group from the University of Manchester carried out energy deposition studies in VESPER using a set of EBT3 Gafchromic films submerged in water. The measured dose deposition agreed with simulations, and the longitudinal dose profiles with and without inserts of different materials showed that the electron beam is relatively unaffected by both high-density and low-density media, thus indicating the potential of VHEE to be a reliable mode of radiotherapy for treating tumours also in highly inhomogeneous and mobile regions such as lungs~\cite{VHEE}. Other groups, from the UK National Physical Laboratory and from Strathclyde University, also showed interest in performing VHEE studies in CLEAR and completed preliminary tests there. Further studies on the dose distribution of a converging beam as opposed to a parallel wide beam are in preparation. 

\paragraph{Novel accelerator research}
CLEAR's first exploration of new accelerator technologies has been on active plasma lenses, promising as strong focusing devices in novel accelerators, due to their compact size. Transverse field uniformity and beam excitation of plasma wakefields appear to be the more significant potential limitations. A collaboration lead by the University of Oslo and including CERN, DESY and the University of Oxford was set up to develop a novel low-cost, scalable plasma lens and assess such limitations. The setup, described in Ref.~\cite{Lindstrm2018OverviewExperiment}, consists of a 1~mm diameter, 15~mm long sapphire capillary. The capillary can be filled with He, Ar or other gases at a controllable pressure. The gas is ionised by a 500~A peak current discharge with a duration of up to a few hundred ns, provided by a 20~kV spark-gap compact Marx bank generator. The longitudinal discharge current creates the transverse focusing force in both transverse planes. The experimental set-up was installed in CLEAR in September 2017 and a clear focusing effect was rapidly observed. Extensive measurements, including transverse position scans of a pencil beam, revealed gradients as high as 350~T/m and beyond, compatible with use in a staged plasma accelerator. More studies measured the uniformity of the field and beam emittance preservation employing different gas species and gave clear indications on the origin of transverse non-linear effects and emittance growth, and on conditions for linear and non-linear regimes linked to emittance growth~\cite{Lindstrm2018EmittanceLens}.  Evidence of non-linear self-focusing at relatively high bunch charge (about 50~pC per bunch) was also observed on the beam after the plasma-forming discharge. This opens up another branch of studies on passive plasma lenses that will be further developed. 

\paragraph{Novel production of terahertz radiation}
Another novel technology being explored at CLEAR is the production of terahertz radiation. Apart from accelerator applications, this technology has a strong impact in many other areas of research, spanning the quantum control of materials, plasmonics, and tunable optical devices based on Dirac-electron systems to technological applications such as medical imaging and security. The aim at CLEAR is to characterise a linac-based THz source, exploiting relativistic electron bunches that emit coherent radiation in the THz domain. For such a source, sub-picosecond electron bunches are needed. This triggered a study and optimisation of the CLEAR injector in collaboration with LAL-Orsay, thanks to which sub-ps bunches down to 100~fs rms have been demonstrated in the machine, paving the way to the THz radiation generation. Current studies, carried out in collaboration with Rome University La Sapienza, CEA/CESTA and Royal Holloway University of London, are focused on the production of (sub-)THz radiation by Coherent Transition Radiation (CTR) in thin metal foils, by Coherent Cherenkov Radiation (CCR) mechanism in targets of different materials and shapes, and by Coherent Smith-Purcell Radiation (CSPR) in open periodic structures. A peak power of about 0.1 MW at 0.3 THz has been measured so far, in agreement with theoretical expectations, and test set-ups to reach the 1-10 MW level are in preparation. Moreover the coherent radiation has been used for longitudinal bunch diagnostics, providing reliable bunch length values consistent with other diagnostics. Transverse shaping of the radiation source has been also demonstrated via control of the size and divergence of the electron bunch at the source plane. The problem of electromagnetic shadowing has been investigated in view of experiments of (sub-)THz radiation production/utilisation involving multiple sources and/or compact setups. 

\paragraph{General accelerator R\&D}
Ongoing CLEAR activities allow the continuation of R\&D and beam test verifications of CLIC critical technologies, for example by measuring the resolution of CLIC cavity BPM prototypes and by checking the behaviour of the wakefield monitors installed in the CLIC accelerating structures.

An intense R\&D activity on beam instrumentation is also ongoing in the CLEAR facility, led by the BI group at CERN in conjunction with many external collaborators.  For example, in the last few years, several studies were performed on high-precision BPM systems based on RF cavities~\cite{CLIC_BPM} or on studying the limitations in Optical Transition Radiation (OTR) beam imaging systems due to shadowing effects~\cite{Kieffer2018a} that occur when using two consecutive OTR screens. CLEAR has also been used to validate beam instruments prior to their installation on other machines, like the testing of AWAKE BPMs and the AWAKE spectrometer screens.

\subsubsection{Proposals for the CLEARER facility}
An obvious synergy exists between the CLEAR facility and the electron beam facility proposed in this document. As already mentioned, the CLEAR injector, or some of its components, could be transported to and used as an injector for the 3.5~GeV linac. It has also been noted that the linac itself would have a strong impact on further developments of the X-band high-gradient technology and help to build-up an industrial basis for its extensive use in future projects, including CLIC and the ILC. The linac would also be a test bed for other technical components, for example diagnostics, or operational tools and procedures, developed for CLIC, and thus constitute an ideal extension of part of the CLEAR experimental programme.

Other experimental activities ongoing in CLEAR could also be made part of the programme of the 3.5~GeV linac, by sharing the beam time available when not used for SPS injection. Most of these activities would profit hugely from the enhanced beam characteristics provided by such facility with respect to the present CLEAR installation. In particular, a 3.5~GeV and better quality beam will extend the scope of beam diagnostics R\&D dramatically (e.g., for diffraction radiation studies for beam profile monitors and high-resolution BPMs). Having access to electron beam energies as high as 3 GeV at the end of the linac and 16 GeV in the SPS and its extraction line will open new opportunities in testing beam instrumentation methods for ultra-relativistic beams. For example, synchrotron radiation (SR) in the ring will allow development of further innovative techniques using randomly distributed interferometric targets in order to improve the performance of SR based monitors. At the end of the linac or in the extraction line, R\&D on non-invasive beam diagnostics could easily be pursued, for example instruments based on the detection of Cherenkov diffraction radiation (ChDR), an innovative scheme that was recently demonstrated. ChDR refers to the emission of Cherenkov radiation by charged particles travelling parallel to the surface of a dielectric material. This combines the already well-known advantages of Cherenkov radiation with a non-invasive technique, and would find applications as beam imaging systems or high directivity BPMs for a wide range of accelerators, i.e.\ from high energy lepton or hadron colliders to light sources (3rd generation synchrotron light sources or free electron laser facilities). This would also be a very attractive solution for novel accelerating technologies, like plasma or dielectric wakefield accelerators, into which standard beam instruments cannot be easily integrated.

The same is true for the studies on novel accelerator technologies. THz radiation sources at much higher power than in CLEAR would be possible, with direct applications to the acceleration and manipulation of charged particle beams. Higher beam energy will make easily accessible extended interaction regions and staging for both THz and plasma-based acceleration schemes. Potentially important issues to be explored in order to make even more attractive novel accelerations studies are the single bunch capability of the linac, positron production and acceleration, and the use of two sources with independent timing for drive/probe type experiments. 

%% file: include/06-Acc_RnD/Positron_RnD.tex
\subsection{Potential upgrade: positron related research and development}
\label{sec:ACCRnD_Positrons}

\subsubsection{Studies for future lepton colliders}
Future lepton colliders such as CLIC, ILC, FCC-ee, or a Muon collider based on high-energy positrons, rely on challenging positron source designs which are still at a conceptual level.
The CLIC positron source design can be seen as a reference for electron-positron colliders \cite{Aicheler2019,Aicheler2012a}, it was adopted for FCC-ee and serves as an alternative for ILC. 
A Muon collider based on the LEMMA scheme is of course much more demanding with respect to positron production but the fundamental challenges are similar.
Positron source R\&D is urgently needed in the areas of target construction,cooling and survival, magnetic field concentrators, and photon production enhancement through channelling radiation.
Such challenges could be addressed with a simple positron production area as described in Section~\ref{sec:LINAC_Positrons}.

\subsubsection{Upgrade with positron beam production}
\label{sec:LINAC_Positron}

The availability of intense positron beams with an energy of a few GeV would be vital to a large number of projects and experiments. For example, future lepton colliders will require both intense electron and positron beams, PWFA based colliders will need to overcome substantial challenges that exist in preserving the quality of the accelerated positron witness bunch and the interaction of positron beams with crystals to generate photon beams and their performance against electron beams should be studied. Yet, during the last decades, almost no work in these areas has been done due to a lack of available positron beams, since most of the limited number of positron beams currently generated around the world are serving high-energy physics experiments and are rarely available, if at all, for other experiments or accelerator R\&D. 
By providing electron and positron beams, the eSPS facility would therefore enable absolutely unique and crucial research in at least three areas:
\begin{itemize}
\item {positron production R\&D for future lepton colliders;}
\item {plasma wakefield acceleration;}
\item {photon production in crystals using positrons.}
\end{itemize}

The proposed 3.5~GeV electron linac could be used to produce positrons using a conventional fixed tungsten-target or a CLIC-like hybrid-target scheme followed by a capture linac. The CLIC-like hybrid-target is a highly innovative scheme, as the primary electron beam impinges on a single crystal target which is used as a radiator of intense gamma-rays, followed by an amorphous converter target placed downstream of the crystal. Such a scheme, baseline for the CLIC positron source as well as FCC-ee, promises high-intensity positron bunches. Other challenges related to positron production and capture would require experimental validation, e.g.\ the adiabatic damping device and the capture linacs, which ultimately define the initial positron beam parameters.%

In order to provide high-quality positron bunches with small transverse emittances, the low energy positron beam should be transported to a damping ring to reduce the transverse emittance and to make them suitable for re-acceleration. 
The design of such a positron damping ring could be based on the FACET II facility design~\cite{facet2}.
Possible parameters of the positron production and beam are collected in Table~\ref{tab:recipe_positrons}. These parameters have been checked with beam dynamics simulations and are within the capabilities of the linac described in Section~\ref{sec:LINAC}.

Summarising, to operate for production of high-quality high-current positron beams, the facility would need: a positron production target, a capture linac, a positron return line, and a small damping ring. Such a scheme is thought to be technically feasible but would represent serious integration challenges or require some civil engineering to provide the appropriate space for the different components.

\begin{table}[!hbt]
\begin{center}
\caption{Possible parameters for positron production.}
\label{tab:recipe_positrons}
\begin{tabular}{p{6cm}cc}
\hline\hline
\textbf{Parameter}             & \textbf{Symbol} & \textbf{Value}\\
\hline
Electron drive bunch\\
\hline
Energy   & $W_0$             & 3.5\,GeV \\
Charge     & Q             & 1.7\,nC \\
Bunch rms length     & $\sigma_{z}$             &200\,$\mu$m \\
\hline
Positron bunch\\
\hline
Energy   & $W_0$             & 3.5\,GeV \\
Charge     & Q             & $>$ 1\,nC \\
Bunch rms length     & $\sigma_{z}$             &200\,$\mu$m \\
Capture energy     &$W_c$              & 335\,MeV \\
Final emittance     & $\epsilon$             & $<$ 20\,mm mrad\\
\hline\hline
\end{tabular}
\end{center}
\end{table}%

\subsubsection{The LEMMA muon collider}

The LEMMA~\cite{LEMMA} muon collider concept has many attractive features. One possible implementation route of such a facility at CERN is a phased approach in three stages: 
\begin{itemize}
\item Phase 1: eSPS tests of positron production and targets, and injector studies for LEMMA;
\item Phase 2: LEP3 where 45 GeV positrons would be available in the LEP/LHC tunnel providing a test ground 
for muon production with positrons; 
\item Phase 3: A final booster and storage ring for muons which could be the SPS or a larger tunnel adapted to the physics requirements pertaining.  
\end{itemize}
Alternatives to phase 3 would be the higher energy stages of CLIC or using novel accelerator schemes to reach higher energies in e$^+$e$^-$ collisions.
The attractiveness of this is that at all three stages high-priority physics studies can be performed, and decisions about the future phases will be based on physics results and accelerator studies in the previous stages.
The initial stage physics at the eSPS is the main subject of this section, while the necessity of exploring the Standard Model in detail in e$^+$e$^-$ collisions is widely recognised.
More information about the studies for LEMMA that could be made at the eSPS and about the scenario above can be found in Ref.~\cite{LEMMACERN}. 

LEMMA is a novel scheme	to produce the muons for a muon	collider.
It uses	a positron beam	that is	sent through a target to produce muon pairs.
The produced beams have	much smaller emittances	than muon beams	produced
via pion decay.	This could avoid the complexity of cooling the	muon
beams and could	lead to	high luminosities with small muon beam currents.
Such a technology could	have the potential to reach very high lepton collision energies.

One of the challenges that this	approach has to	face is	the production of very high
average positron currents, well in excess of what is needed for linear colliders. Positron	source R\&D
is thus	instrumental for this approach.

A key issue of the LEMMA scheme	is the stress in the muon production target due	to the
impinging positron beam. Different technologies	could be considered to overcome	this.
They range from	conventional targets from robust materials to crystals or liquid targets.
Experimental studies of	these targets are of great importance in order to establish whether they are practical
for a muon collider.

Another challenge that has to be faced is the control of the positron and muon beam
emittance. The beams pass through the muon production target repeatedly, each time increasing
emittance by multiple scattering. Experimental	studies	of the scattering of positrons (and
also electrons)	will improve the reliability of	the predictions	of the emittances of the
produced muon beams. It	will also allow	the testing of specific target	shapes and materials that promise improved performances.

\subsubsection{Plasma wakefield experiments with a positron beam}
As introduced in Section~\ref{sec:ACCRnD_Plasma}, PWFA is under consideration as a possible high-gradient accelerator technology for an Advanced Linear Collider~\cite{anar,osti_1358081}. The electron beam-driven experiments described in Section~\ref{sec:ACCRnD_Plasma} are needed to show that the quality of the electron witness bunch can be preserved throughout the acceleration process, thus demonstrating a suitable first stage of a plasma-based linear collider (or advanced linear collider, ALIC). 

However, a linear collider requires both electron and positron beams, and there are substantial challenges in preserving the quality of the accelerated positron witness bunch in a plasma. 

\subsubsection{Physics of positron acceleration in plasma}

Plasma wakefield acceleration is the only charge-asymmetric acceleration mechanism currently under consideration as a future accelerator technology. The asymmetry arises from the fact that plasma is composed of light electrons and heavy ions. The ion-electron mass ratio ranges from 1836 for hydrogen (commonly used in laser-driven plasma acceleration~\cite{Leemans2014}) to 155800 for rubidium used in the AWAKE experiment~\cite{AWAKEacc}. When an intense drive bunch propagates into neutral plasma, it induces an oscillation in the mobile plasma electrons, while the plasma ions remain stationary on the time scale of an electron oscillation period (typically measured in femtoseconds or picoseconds depending on the plasma density). For electron beam-driven PWFA in the non-linear regime, this results in complete expulsion of the plasma electrons from the region surrounding the drive beam (referred to as blow-out). This region, filled with a uniform background of plasma ions, provided a strong focusing force on the transiting electron beam. The blow-out wake has attractive properties for accelerating a trailing bunch of electrons, namely the large gradients and strong focusing fields, which when combined allow for substantial energy gain in the plasma over distances much larger than the beam vacuum beta-function. Finally, since the ions are uniformly distributed in the blow-out region, it is possible to exactly calculate the parameters of a matched witness beam, as shown in Section~\ref{sec:ACCRnD_Plasma} and in principle to also maintain the emittance of the accelerated electron bunch.

Such a a pure ion column situation, ideal for an electron bunch, does not exist for a positron bunch. %
Despite this complication, significant experimental progress has been made over the years that demonstrates the possibility of accelerating positron beams in plasma~\cite{BluePosi,Corde2015,Gessner2016}, while also characterising the effects of nonlinear focusing and transverse wakefields on the quality of the accelerated beam~\cite{Muggli2008,Lindstrom2018}. However, there have been no experimental investigations of electron beam-driven/positron witness PWFA. We note that this is the proposed method for accelerating positron beams in current ALIC concepts~\cite{Adli2013}. The eSPS facility enables research on electron beam-driven wakes in the quasilinear, non-linear, and hollow channel regimes, which is a critical step in determining viable paths towards ALIC.

\subsubsection{Crystal undulators and photon production}
Innovative ideas to use crystals as undulators to produce gamma-rays would need positron beams to validate the concept. Positron beams with energy of a few GeV offer unique possibilities in investigating various effects related to interactions between high-energy particle beams and oriented crystals. Besides channelling radiation and coherent bremsstrahlung, radiation in crystalline undulators reaches its optimal characteristics in this energy range~\cite{tikho1}. It is expected that a sub-millimetre crystal undulator will allow one to generate a few MeV photon energy radiation with few \% spectrum width and the efficiency of one photon per thousand positrons. Planar channelling and coherent bremsstrahlung in crystals shows a high degree of linear polarisation, which can be exploited for application related to nuclear physics~\cite{lohm, pate}. In case of axial orientation, the emitted radiation becomes harder (10\,MeV and more), easily tunable, with narrow spectrum and can be made circularly polarised~\cite{guid}. All these innovative sources of radiation could open new possibilities for studies of both nuclei excitation and disintegration competing with the large infrastructures, in 
particular with inverse compton sources, such as ELI-NP, in both size and spectrum width and hardness. Availability of a linac providing a low emittance beam with energy of a few GeV combined with properly designed crystals looks a very promising source of intense high-energy radiation.%

Furthermore, it could be interesting to exploit bent crystals for GeV positron beam manipulation, for beam extraction, focusing etc. Indeed, in this energy range, studies with bent crystals at INFN-Ferrara have already demonstrated their capability to steer electron beams. In the case of electrons,the maximum deflection efficiency obtained up to now is above 35\% at 0.855\,GeV~\cite{syto}. One may expect a much higher deflection efficiency in case of positrons that are less affected by de-channelling than electrons. Indeed, being negatively charged, channelled electrons repeatedly oscillate across the nuclei of the crystal, leading to an increase of particle de-channelling compared to positive charges. 

%% file: include/07-Implementation/Implementation.tex
\section{Implementation}
\label{sec:IMPL}

The work for this CDR has been done during 2019 and the first months of 2020. Compared to the Expression of Interest~\cite{Torstenakesson2018} in 2018, major work has been done in three main areas: infrastructure and civil engineering for the linac, final beamline and experimental hall, as well as a revised SPS RF system solution. Other changes, to the linac RF layout and SPS specific parts, are also important but the main solutions were already presented in the Expression of Interest. The CDR is now a complete description of an eSPS facility and the physics and accelerator R\&D potential of it. 

\subsection{Schedule}
\begin{figure}[!hbt]
\begin{center}
\includegraphics[width=\textwidth]{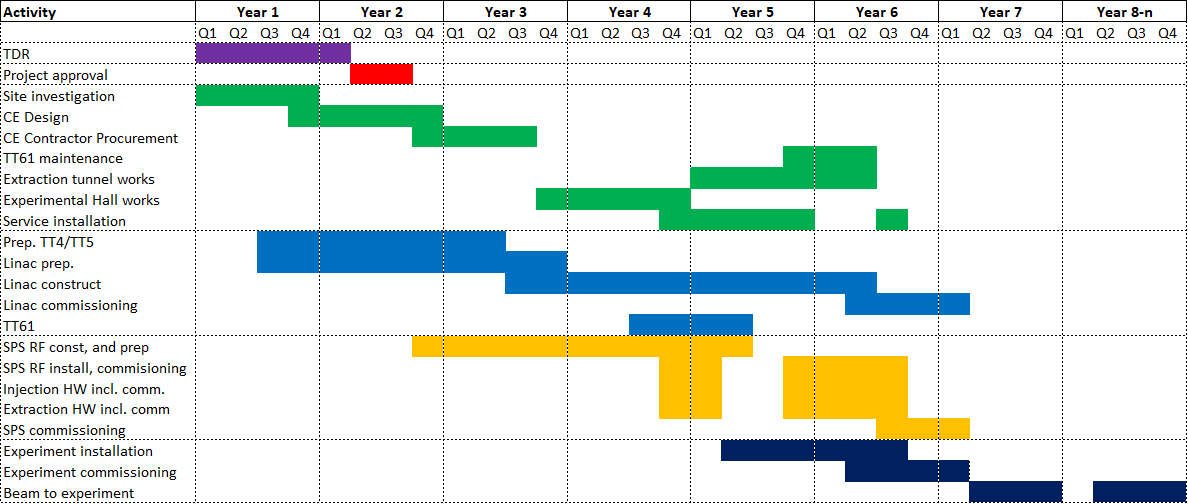}
\caption{Possible eSPS implementation schedule.}
\label{fig:esps-schedule}
\end{center}
\end{figure}

A first schedule for the project has also been established and is shown in Fig.~\ref{fig:esps-schedule}. The schedule drivers are: the project approval time; linac construction and commissioning; SPS SRF system design and production; construction of the experimental hall/area; and connection to it from the extraction tunnel. 

The schedule is technically based, but not linked to a specific starting date. This makes the proposed schedule flexible even though some aspects of the work, such as the new transfer tunnel to the experimental area, will need to be organised around shutdown periods. Most of the other aspects of the project, in particular in the SPS ring, can be done during short or extended end of year shutdowns. 

This program assumes that important investments for the project implementation can be made from year 1 with significant deliveries and payments from year 2.


\subsection{Cost}
The eSPS facility, i.e.\ the linac, transfer lines, SPS adaptations and experimental area, has been preliminarily costed. These cost estimates were produced by CERN's groups based on their expertise and experience from previous projects. The methodology, the level of detail and the resulting uncertainty on the estimated cost is not completely homogeneous across all aspects included. In particular the risks on the cost of the elements discussed Section~\ref{sec:towards_tdr} are not included. However, it has been the goal of this exercise to achieve an uncertainty in the order of \textcolor{black}{$\pm$ 30\%} on the costs presented here.



\begin{figure}[!hbt]
\begin{center}
    \begin{subfigure}{0.55\linewidth}
        \begin{center}
           \includegraphics[width=\textwidth,trim={0.15cm 0.15cm 0.15cm 0.15cm},clip]{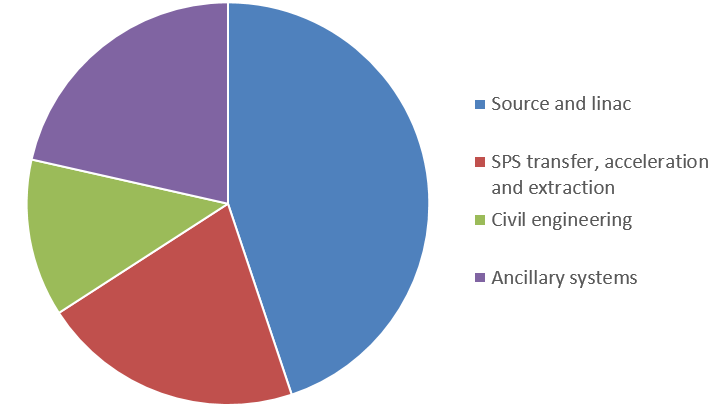}
        \end{center}
    \end{subfigure}
   \begin{subfigure}{0.44\linewidth}
\begin{tabular}{lr}
\hline\hline
\textbf{Item} &  \textbf{cost }    \\ 
\textbf{} &  \textbf{[MCHF]}    \\ 
\hline
Source and linac &  49.8 \\
SPS transfer,  & \\
acceleration and extraction & 23.4 \\
Civil engineering & 14.0 \\
Ancillary systems & 23.8 \\
\hline
\textbf{Sum} & 111.0 \\
\hline\hline 
\end{tabular}
    \end{subfigure}
    \caption{Summary of the cost estimate for the project.}
\label{fig:cost_global} 
\end{center}
\end{figure}

The top level cost summary is shown in Fig.~\ref{fig:cost_global}. The facility cost is dominated by the linac and source. Ancillary systems account for around a fifth of the cost of the project, with the majority accounting for heating, cooling and ventilation costs. The next largest item of this summary includes the beam transport to the SPS, its acceleration in the injector and SPS and its extraction towards the experimental area. Civil engineering costs account for around an eighth of the total.

Some of the more specific aspects relating to each item included in Fig.~\ref{fig:cost_global} are discussed in the following sections.

\subsubsection{Linac}
\begin{figure}[!hbt]
\begin{center}
    \begin{subfigure}{0.55\linewidth}
        \begin{center}
           \includegraphics[width=\textwidth,trim={0.15cm 0.15cm 0.15cm 0.15cm},clip]{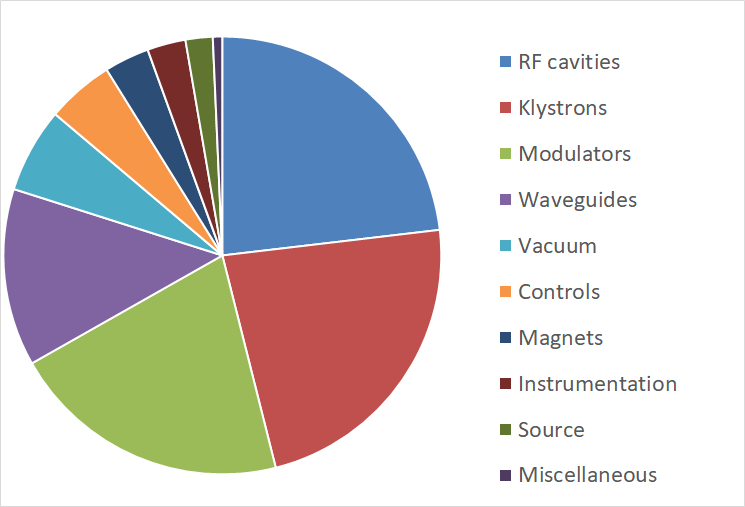}
        \end{center}
    \end{subfigure}
   \begin{subfigure}{0.44\linewidth}
\begin{tabular}{lr}
\hline\hline
\textbf{Item} &  \textbf{cost }    \\ 
\textbf{} &  \textbf{[MCHF]}    \\ 
\hline
RF cavities &	11.52 \\
Klystrons  &	11.42 \\
Modulators &	10.32 \\
Waveguides &	6.53 \\
Vacuum	& 3.12 \\
Controls &	2.47 \\
Magnet &	1.66 \\
Instrumentation &	1.41 \\
Source	&  1.00 \\
Miscellaneous &	0.35 \\
\hline
\textbf{Sum} & 49.80 \\
\hline\hline 
\end{tabular}
    \end{subfigure}
    \caption{Summary of the linac and source cost estimate.}
\label{fig:cost_linac} 
\end{center}
\end{figure}

The linac costs, shown in Fig.~\ref{fig:cost_linac}, are based on extensive prototype experience within the CLIC project and additional industrial quotes for the main components of the volume needed for eSPS, for example klystrons, modulators and accelerator structures. With further prototyping of high efficiency klystrons the linac costs can be reduced by around 10\%. If a new source optimised for 800 MHz is constructed the linac costs will increase by around 15\%. 

\subsubsection{SPS and transfers}
\begin{figure}[!hbt]
\begin{center}
    \begin{subfigure}{0.55\linewidth}
        \begin{center}
           \includegraphics[width=\textwidth,trim={0.15cm 0.15cm 0.15cm 0.15cm},clip]{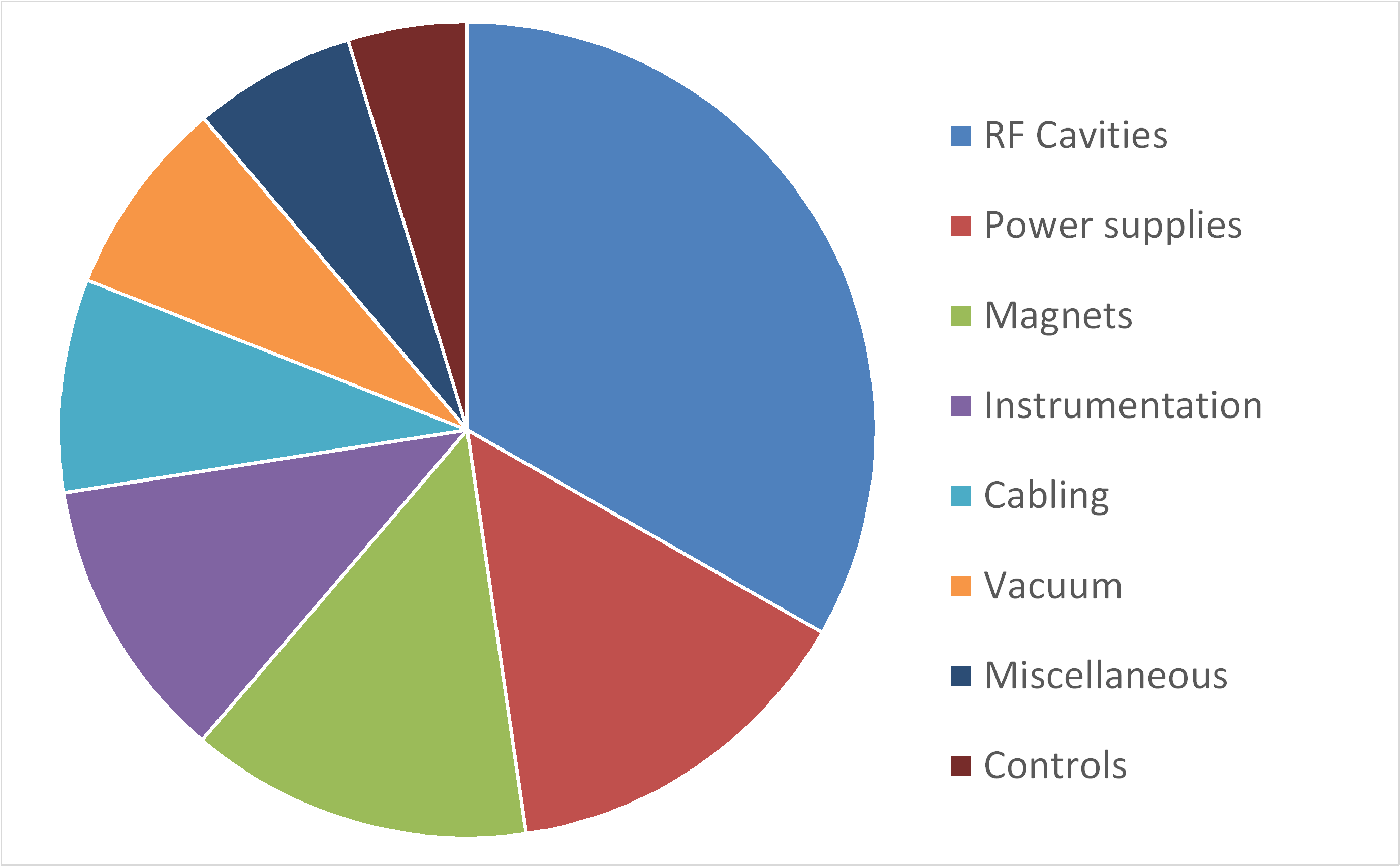}
        \end{center}
    \end{subfigure}
   \begin{subfigure}{0.44\linewidth}
\begin{tabular}{lr}
\hline\hline
\textbf{Item} &  \textbf{cost }    \\ 
\textbf{} &  \textbf{[MCHF]}    \\ 
\hline
RF Cavities &	7.75 \\
Power supplies	& 3.37 \\
Magnets	& 3.16 \\
Instrumentation	& 2.63 \\
Cabling	& 1.97 \\
Vacuum	& 1.83 \\
Miscellaneous &	1.50 \\
Controls &	1.10 \\
\hline
\textbf{Sum} & 23.30 \\
\hline\hline 
\end{tabular}
    \end{subfigure}
    \caption{Summary of the SPS and beam transfers cost estimate.}
\label{fig:cost_SPS} 
\end{center}
\end{figure}

The cost estimates for the electron beam transport and acceleration from \SI{3.5}{GeV} to the the experimental area are showed in Fig.~\ref{fig:cost_SPS}. The major driver of the cost is the RF system required to accelerate the electron beam in the SPS. The estimated cost of the RF cavities includes the design and production of the \SI{800}{MHz} superconducting cavities as well as all the supporting systems such as cryostats or power amplifiers.

The power supplies category includes an estimate of the magnet power converters. The powering of the new line to transport the beam to the experimental area from the TT10 line is the main contributor due to the large power required. Next are the power converter costs for the SPS injection and extraction systems, in particular the solid-state generators of the new fast injection kickers. Powering of the linac to SPS transport line is relatively economical due to the low energy and required magnet currents of the line magnets.

The main driver of the magnet costs is the design and production of injection and extraction devices. All  other transport beamline magnets used in the project make use of existing and available magnets, needing at most a refurbishment before their installation.

Instrumentation costs mainly come from development and production of devices for the SPS ring and high energy electron transport line to the experimental area. The low energy section, \SI{3.5}{GeV} electron beam transport and injection, can be implemented for a comparatively smaller cost due to the use of more common systems.

{Cabling}, {Vacuum} and {Controls} costs were estimated directly from the other items discussed above and based on experience from other similar projects conducted at CERN. The {miscellaneous} category in Fig.~\ref{fig:cost_SPS} is an estimation based on further items and studies required and includes beam dynamics studies, coordination and integration studies.

\subsubsection{Civil engineering}



The cost estimate has been based on the layout presented in this CDR and includes costs for detailed design, construction and construction management, excluding personnel costs for CERN resources. The estimate does not include development costs, permits, materials or personnel costs in advance of detailed design and construction. 

A detailed bottom up cost estimate has been carried out for civil engineering work required to implement eSPS. The cost estimation process has involved building up costs from labour, equipment and materials for each item. The cost for each item is multiplied by the quantity of that item to obtain a total cost. The total cost has then been compared with other similar projects carried out by CERN in recent years to ensure the costs are adjusted as accurately as possible to current market rates. The costs are all stated in terms of Q2 2020 prices.

In order to produce a cost estimate at this stage of the project, a certain number of assumptions had to be made:

\begin{itemize}

\item Ground conditions have been based on 33\% poor rock and 67\% good rock;
\item Costs have been based on spoil disposal on CERN land within 20\,km of the site with no tipping and disposal costs. If this were to change, the cost increase could be significant;
\item The proposed drainage can be connected into existing tunnel drainage without significant capacity enhancement or repairs;
\item Maintenance of TT61 has not been studied in detail. An assumption has been made at this stage that superficial invert and drainage repairs will be required. This must be re-valuated once monitoring and investigation works have been carried out;
\item The cost of bringing electrical or other service supplies to the experimental hall will be evaluated from scratch with the estimate replacing the provisional sum included to date;  
\item The existing earth mound has been assumed to be composed of unactivated, uncontaminated 'acceptable' material;
\item A space requirement of 1500\,m\textsuperscript{2} is needed to accommodate displaced material from TT4, TT5 and B183. This is a condition of the project and if such space is not available in the flex-storage building or similar, then an additional re-provision cost will be incurred;
\item Costs of moving or disposal of the material stored in  TT4, TT5 and B183 is assumed to be covered by the owners of the material;
\item No contingency, risk allowance or optimism bias have been included;
\item Prices are stated correct for Q2 2020 based on the stated schedule, if there are any delays, then cost inflation must be applied. 
\end{itemize}

The accuracy of the estimate is \textit{Class 4---study or feasibility}, which could be 15--30\% lower or \mbox{20--50\%} higher (in line with Ref.~\cite{christensen2005cost} as has been used for previous CERN projects).  Given the level of uncertainty and project development, the error bars for this estimate have been set between $-20$\% and $+30$\%). A full list of assumptions for the costing is noted in the detailed cost estimate. 

\begin{figure}[!hbt]
\begin{center}
    \begin{subfigure}{0.55\linewidth}
        \begin{center}
           \includegraphics[width=\textwidth,trim={0.15cm 0.15cm 0.15cm 0.15cm},clip]{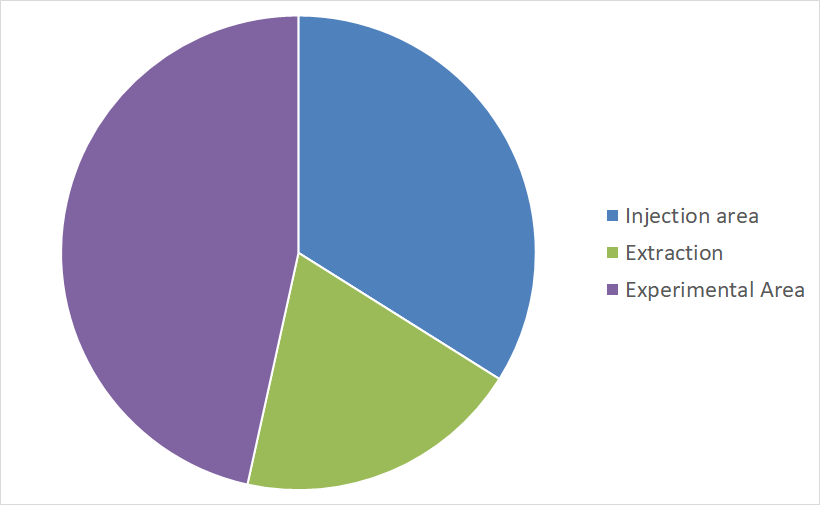}
        \end{center}
    \end{subfigure}
   \begin{subfigure}{0.44\linewidth}
\begin{tabular}{lr}
\hline\hline
\textbf{Item} &  \textbf{cost }    \\ 
\textbf{} &  \textbf{[MCHF]}    \\ 
\hline
Injection &	4.76 \\
Extraction &	2.74 \\
Experimental Area & 	6.53 \\
\hline
\textbf{Sum} & 14.04 \\
\hline\hline 
TT61 Capital maintenance &	0.63 \\
\hline
\end{tabular}
    \end{subfigure}
    \caption{Summary of the civil engineering cost estimate.}
\label{fig:cost_SPSCE} 
\end{center}
\end{figure}

Additional 'below the line' costs for civil engineering for the maintenance of TT61 in order to put the tunnel back into service are estimated as shown in Fig.~\ref{fig:cost_SPSCE} but excluded from the civil engineering and overall project cost estimates.

\subsubsection{Ancillary systems}

Cost estimates are required for each discipline included within the infrastructure and civil engineering parts of the study namely: 
\begin{itemize}
    \item Integration;
    \item Heating cooling and ventilation (HVAC);
    \item Electrical;
    \item Radiation protection;
    \item Transport and handling;
    \item Safety engineering;
    \item Access control;
    \item Survey and alignment;
\end{itemize}

Civil engineering costs are separated out as they represent a large proportion of costs. For each of the other areas, subject experts have carried out individual estimates. 

Electrical cost estimates have been necessarily derived from comparable  projects and considering the concept design at this stage of the project.

It should be noted the following items do not form part of the electrical infrastructure estimate and shall be charged to equipment owners budget estimates:

\begin{itemize}

    \item DC cables for the (inter)connection of power converters and magnets;
    \item Control cables and optical fibres for accelerators components and experimental areas;
    \item Control cables and optical fibres for the access system and radiation monitoring;
\end{itemize}

\begin{figure}[!hbt]
\begin{center}
    \begin{subfigure}{0.55\linewidth}
        \begin{center}
           \includegraphics[width=\textwidth,trim={0.15cm 0.15cm 0.15cm 0.15cm},clip]{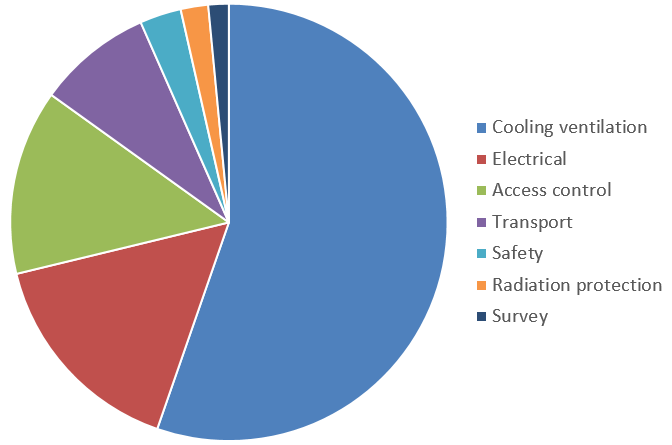}
        \end{center}
    \end{subfigure}
   \begin{subfigure}{0.44\linewidth}
\begin{tabular}{lr}
\hline\hline
\textbf{Item} &  \textbf{cost }    \\ 
\textbf{} &  \textbf{[MCHF]}    \\ 
\hline

Cooling and ventilation &	13.18 \\
Electrical &	3.79 \\
Access control	& 3.26 \\
Transport	& 2.02 \\
Safety	& 0.73 \\
Radiation protection	& 0.49 \\
Survey	& 0.36 \\
\hline
\textbf{Sum} & 23.83 \\
\hline\hline 
\end{tabular}
    \end{subfigure}
    \caption{Summary of the ancillary systems cost estimate.}
\label{fig:cost_ancillary} 
\end{center}
\end{figure}



Additional 'below the line' costs for ancillary systems which would be required to implement eSPS have been identified. These costs are for other associated existing systems requiring modification or upgrade and are estimated as shown in Table~\ref{tab:cost_SPSASNP} but excluded from the above ancillary systems and overall cost estimates.

\begin{table}[!hbt]
\begin{center}
\caption{Table of estimated non-project ancillary systems costs.}
\label{tab:cost_SPSASNP}
\begin{tabular}{lr}
\hline\hline
\textbf{Item} &  \textbf{cost }    \\ 
\textbf{} &  \textbf{[MCHF]}    \\ 
\hline
TT61, TCC6 and T60 safety systems &	0.82 \\
BA7 and related infrastructure safety systems & 0.59 \\
\hline
\textbf{Sum} & 1.41 \\
\hline\hline 
\end{tabular}
\end{center}
\end{table}



\subsection{Compatibilities and dependencies}
The low energy electron beam from the linac or the injector can be used independently from the rest of the accelerator complex. Any experimental program carried in the linac area or in the TT61 tunnel will run separately from the other CERN accelerators. Integration and radiation protection studies demonstrated in particular the possibility of concurrent operation of the linac and access to the n\_ToF target (see Sections~\ref{sec:linac:integration} and \ref{sec:ICE:injection:radprotection}).

For higher energy operation the beam will be accelerated in the SPS and the operation constraints depend on the RF system selected (see Table~\ref{TAB:RFPARAM}). The superconducting cavity bypass system proposed in this study, imposes a switch time between nominal proton operation and electron operation of at least \SI{10}{min} (see Section~\ref{sec:SPS:SRF}). The delivery of the high energy electron beam to the experimental area will need to be done alternately with SPS proton operation but could realistically proceed behind LHC physics stores. The proton physics program making use of the PS or PSB beams will not be impacted by the high energy electron beam production.


\subsection{Towards a technical design report}
\label{sec:towards_tdr}

\paragraph{SPS electron beam handling}
Injection of the electron beam at \SI{3.5}{GeV} in the SPS places strong requirements on the stability of many of many power supplies at very low current. In the TT60 beamline magnets and injection septa, stability at low current and effects of remnant magnetisation will need to be considered and possibly measured. In the ring, existing auxiliary power supplies are no longer capable of operating at very low current as the LEP era hardware was removed, In view of the experience from the from the LEP era, a consolidation program of the existing SPS auxiliary power converters is required to once again control beams at very low energy. The scope and resources needed for such program are not discussed in this study but need addressing before going towards a technical design report (TDR).

Careful measurements with and without beams will be needed to validate the behaviour of the SPS main dipoles and quadrupoles power supplies at very low currents. Tests with proton beams are requested for Run3 to investigate beam dynamics stability and reproducibility between \SI{20}{GeV} and as low as possible, even if the stability at \SI{3.5}{GeV} appears impossible with the current hardware. More detailed studies of the slow extraction concept will also be required. In particular to quantify the effect of power supply stability on the required very low spill rate and mitigation methods.

The internal SPS electron dump concept proposed here will need to be carefully simulated. The existing TIDP collimator device has not been designed to absorb a high brightness electron beam circulating counter-clockwise. Improvements to the existing device or even complete redesign may be needed.

\paragraph{Civil engineering work required at the next stage of project development}

For the project to progress, additional studies will be required as follows:
\begin{itemize}

\item Monitoring and investigation of TT61; 
\item Ground penetrating radar (GPR) scans in B183, TT5, TT4 and TTL2 where cores and breakthroughs are needed; 
\item Ground investigation around the new experimental area, new extraction tunnel and at the location of the CV service building above TT4;
\item Study to minimise the impact on adjacent beam operations and experiments during construction work;
\item A GPR survey to confirm the location of services around the new experimental area;
\item A survey and study of existing drainage systems;
\item An RP study of the area immediately around TT7 and the associated shielding blocks.
\end{itemize}

%% file: include/08-Conclusion/Conclusion.tex
\section{Conclusion}
\label{sec:Conclusion}
The CERN Council adopted in late spring 2020, an update of the European Strategy for Particle Physics~\cite{TheCERNCouncil2020}. Some of its priorities are:
\begin{itemize}
\item Accelerator R\&D including \textit{plasma wakefield acceleration and other high-gradient accelerating structures, bright muon beams, energy recovery linacs};
\item \textit{An electron-positron Higgs factory is the highest-priority next collider}; 
\item The option of a circular \textit{electron-positron Higgs and electroweak factory} is possibly a first stage of CERN's next circular collider, i.e.\ an accelerator requiring CERN increasing its in-house competences on circular electron accelerators;
\item The quest for dark matter including dark sector candidates, and that such experiments should be supported in laboratories in Europe; 
\item Europe should support future neutrino experiments in Japan and the United States. 
\end{itemize}

This conceptual design report describes an infrastructure relevant for all these priorities.  The implementation would make excellent use of the investment made in the CLIC programme and is the natural next step in the development of X-band \textit{high-gradient acceleration} technology.  The multi-GeV electron beam from the linac would drive wakefields in the non-linear regime and would with an independent electron witness bunch demonstrate the applicability of \textit{plasma wakefields for high-gradient acceleration}. The facility would become unique in the world to study collider related challenges, as the only facility with multi-GeV drive bunch and truly independent electron witness bunch.  Addition of a positron witness bunch would make it a complete facility for collider studies.  The 800 MHz super-conducting cavities for the eSPS would be the same type as foreseen for a future \textit{electron-positron Higgs and electroweak factory} as the first stage of a next circular collider at CERN. eSPS could be used for the development and studies of a large number of components and phenomena. The operation of SPS with electrons would train a new generation of CERN staff on circular electron accelerators.


The electron beam delivered by this facility would open a \textit{dark sector research} programme and in particular provide sensitivity to \textit{light dark matter} production significantly beyond the targets predicted by a thermal dark matter origin, and for nature of dark matter particles that are not accessible by direct detection experiments. The future long baseline \textit{neutrino physics} studies need to precisely measure neutrino oscillation probabilities as a function of energy. This critically relies on the ability to model neutrino-nucleus interactions, and this in turn requires input data on electro-nuclear reactions; the beam from this facility would be excellent for such measurements.  In addition, it could serve experiments in \textit{nuclear physics}.  The eSPS could be made operational in about five years, and serve a programme as outlined above. This could start already in LS3 and would operate in parallel and without interference with Run 4 at the LHC.